\documentclass[fleqn,usenatbib,twocolumn]{rasti}

\usepackage{newtxtext,newtxmath}

\usepackage[T1]{fontenc}

\DeclareRobustCommand{\VAN}[3]{#2}
\let\VANthebibliography\thebibliography
\def\thebibliography{\DeclareRobustCommand{\VAN}[3]{##3}\VANthebibliography}

\usepackage{graphicx} 
\usepackage[dvipsnames]{xcolor}
\usepackage{appendix}
\usepackage{amsmath}
\usepackage{subcaption}
\usepackage{threeparttable}
\usepackage{booktabs}
\usepackage{enumitem}
\usepackage{multirow}
\usepackage[utf8]{inputenc}
\usepackage{etoolbox}
\usepackage{tablefootnote}

\hypersetup{
    colorlinks=true,
    linkcolor=NavyBlue, 
    citecolor=NavyBlue,
    urlcolor=NavyBlue
}

\title[The future of high-resolution UV spectroscopy]{The future of high-resolution UV spectroscopy: Science with a UV \'Echelle spectrograph on the Habitable Worlds Observatory, or a dedicated mission}

\author[A.~De~Cia et al.]{
Annalisa~De~Cia,$^{1,*}$
David~Cont,$^{2,*}$
Kevin~Heng,$^{2,3,4,*}$
Boris~Gänsicke,$^{5,*}$
Frank~Grupp,$^{2,6,*}$
Ralf~Bender,$^{2,6,*}$ 
\newauthor Helena~Lamprecht,$^{2,*}$
Frances~H.~Cashman,$^{7,8}$
Hsiao-Wen~Chen,$^{9}$
Valentina~D'Odorico,$^{10,11}$
Kevin~France,$^{12,13,14}$
\newauthor Andrew~Fox,$^{15}$
Jens~Hoeijmakers,$^{16}$
Masahiro~Ikoma,$^{17,18,19,20}$
Wynn~Jacobson-Galán,$^{21}$
Sean~D.~Johnson,$^{22}$
\newauthor Mansi~Kasliwal,$^{21}$
Shingo~Kameda,$^{23,24}$
Varsha~Kulkarni,$^{25}$
Carlo~F.~Manara,$^{1}$
 Antonija~Oklopčić,$^{26}$
\newauthor Cyrielle~Opitom,$^{27}$
Lidia~Oskinova,$^{28}$
Anna~Pala,$^{1}$
Ian~U.~Roederer,$^{29}$
P.~Christian~Schneider,$^{30}$
\newauthor Peter~Scicluna,$^{31,32,33,34}$
Daniel~E.~Welty,$^{7}$
Sascha~Zeegers,$^{26,35}$
Abel~de~Burgos~Sierra,$^{1}$
Miriam~García,$^{36}$
\newauthor Avery~Kim,$^{7}$
Christina~Konstantopoulou,$^{37}$
Roberto Maiolino,$^{38,39,40}$
Daniel~Pauli,$^{41}$
Tanita~Ramburuth-Hurt,$^{42}$
\newauthor Philipp Richter,$^{28}$
Andreas~A.C.~Sander,$^{43,44}$
Tomer~Shenar,$^{45}$
Sergio~Simón-Díaz,$^{46,47}$
O.~Grace~Telford,$^{48}$
\newauthor Anna~Velichko$^{49}$
\\\\\\
Affiliations are listed at the end of the paper.
}

\date{Accepted XXX. Received YYY; in original form ZZZ\\\\$^*$ Project office members.}

\pubyear{\the\year{}}

\begin{document}
\label{firstpage}
\pagerange{\pageref{firstpage}--\pageref{lastpage}}
\maketitle

\begin{abstract}
High-resolution UV spectroscopy serves a diversity of science cases, from small bodies to planets, stars, and galaxies, but is currently limited to the Hubble Space Telescope and bright targets. Major advances require increasing sensitivity by at least one order of magnitude. Here we present the UV science cases for PEGASUS (Planets, Earths, Galaxies, And Stars UV Spectrograph), a UV Échelle high-resolution spectrograph concept, with $R = \lambda/\delta\lambda \sim 100\,000$ (full range 10\,000-140\,000) and covering 90--400\,nm, with a foreseen extension to at least 800\,nm. PEGASUS is ideally suited for the Habitable Worlds Observatory (HWO), enabling transformative science across the UV/optical wavelength ranges. PEGASUS will be unique in high sensitivity (effective area) and high spectral resolution -- an uncharted territory -- as well as robustness, thanks to the simplicity of its design.
Its UV science cases include: I) Formation and evolution of planets and their habitability: properties of exoplanets and atmospheres, protoplanetary disks, Solar System bodies; II) Stellar lives and deaths at their extremes: the first stars and the origin of the elements, compact and massive stars, Supernovae; III) Gas and metals in the baryon cycle of galaxies: the interstellar, circumgalactic, and intergalactic medium and their roles in galaxy growth. These are essential for the Astro Decadal 2020 Survey, Voyage 2050, and HWO. While this paper focuses on high-impact science enabled by UV high-resolution spectroscopy, PEGASUS will extend into the optical regime and lower spectral resolution, making it a multi-purpose, widely used, workhorse spectrograph for HWO. 
\end{abstract}

\begin{keywords}
Instrumentation -- UV Astronomy -- Space Missions
\end{keywords}



\clearpage

\section{Introduction}

High-resolution ultraviolet (UV) spectroscopy serves a broad diversity of current and future science cases including exoplanets and their habitability, young stars and their impact on planet formation \citep[e.g.,][]{espaillat22}, the first stars and the origins of the elements \citep[e.g.][]{Roederer26}, compact and massive stars \citep[e.g.][]{Pala+2022} and the growth of galaxies \citep[e.g.][]{Tumlinson20,Borthakur26,Roman-Duval26}. UV has a key role in astrophysics, because most electronic transitions occur in the UV regime, opening a unique window onto the chemistry of a variety of astrophysical process, from the smaller scales of atoms, molecules, cosmic dust grains to pebbles and small bodies, planets and their atmospheres, compact to massive stars, Population~III (Pop~III) stars, and to even larger scales of galaxy growth and evolution. 

Because the UV is blocked by Earth's atmosphere at $\lambda<$300 nm, the UV is uniquely accessible from space. However, space-based UV spectroscopy is in peril. The only mission that currently offers UV sensitive high-resolution spectroscopy is the the Hubble Space Telescope (HST). UV science is pushing the limits of HST, given the challenge of reaching far-UV (FUV) and near-UV (NUV) sensitivity, even with  the 2.4-m HST mirror. UV high-resolution spectroscopy (e.g.\ with a resolving power of \mbox{$R=\lambda/\delta\lambda \sim$\,100\,000}) with HST is still limited to bright targets. For example, most HST/ULLYSES targets are observed at $R<50\,000$ \citep{Roman-Duval25}. Progress in UV science requires a minimum of ``ten times Hubble'', which could be achieved with a 2--3-m class mission, while 100 times Hubble will require a larger future mission. While the HST lifetime has good probability of reaching well into the 2030s (see the \href{https://hst-docs.stsci.edu/hstos}{HST Overview and Status} webpage), there is no upcoming true successor of HST, with its unique UV spectroscopic capabilities (high sensitivity and spectral resolution), at least for the next 10 years. The international astronomical community needs UV spectroscopic capabilities in the coming decades. 

The need for UV high-resolution spectroscopy has been recognized in the past decade and proposed as a core capability of the previous large space mission concept LUVOIR \citep{LUVOIR19}, in particular with instruments like Ultraviolet Multi-Object Spectrograph \citep[LUMOS][]{France17b} and POLLUX \citep{Bouret18,Muslimov18}.

There is a golden opportunity to install UV spectroscopic capabilities on the planned successor to HST (and in a way also to James Webb Space Telescope and Roman): the Habitable Worlds Observatory (HWO). The HWO instruments that are currently foreseen by NASA are a coronagraph, a multi-object spectrograph, an integral-field spectrograph, a high spatial resolution imager and an additional instrument -- the properties and scope of all of these instruments are still open.

While FUV and NUV high-resolution spectroscopy is a planned capability of HWO \citep[see the NASA HWO Instrument Concept Studies Workshop 2026][]{Sitarski26}, a stand alone high-resolution ($R \sim$\,100\,000) UV-and-optical spectrograph with a simple design and high sensitivity would be complementary to this suite of instruments. Section \ref{sec: comparison HWO} discusses and compares the different HWO UV spectrograph concepts. Simplicity is crucial for a high-resolution UV spectrograph to be optimized for sensitivity: because UV radiation is particularly vulnerable to mirror reflections, specially in the FUV, it is important to minimize the number of optical elements and reflections. Pushing complex instruments to deliver high-resolution spectroscopy in the FUV is a hard challenge, and at the high expenses of sensitivity. 

The most effective solution to address the need for UV high-resolution spectroscopy science is a dedicated instrument, with the minimal amounts of reflections and modes, optimized for sensitivity. At the same time, simple sensitive high-resolution spectrographs can be very robust, long-lived (e.g.\ HST/STIS or Very Large Telescope (VLT) UVES or X-SHOOTER), and represent a workhorse instrument that can serve a broad diversity of scientific cases. 

This paper stems from and extends from \href{https://www.eso.org/sci/meetings/2025/UV-Consortium.html}{``The Future of UV Astronomy in the Age of HWO''} 2025 workshop in Munich (Germany). Here we consider both cases of a dedicated HWO instrument (PEGASUS) or a potential 2-m mirror stand-alone space mission (Theon). The extension to optical ranges can have an important scientific impact, but this paper focuses on the UV part of the spectrum. Section~\ref{sec: overview} presents an overview of the diverse science cases discussed in this paper, as well as the basic technical requirements that are needed. The individual science cases are discussed in more detail in Sections.~\ref{sec: Pillar I} to \ref{sec: Pillar III}, with emphasis on scientific questions that can be uniquely addressed with UV high-resolution spectroscopy. Section~\ref{sec: Astro20} connects these science cases to the Astro Decadal 2020 Survey and the science drivers for HWO. Section~\ref{sec: instrument} discusses the top level requirements and the basic design for a dedicated UV high-resolution spectrograph. We draw our conclusions in Sect.~\ref{sec: conclusions}.

\section{Overview of UV science cases and top-level requirements}
\label{sec: overview}

This paper focuses on UV science and highlights the breadth of science cases that are uniquely accessible with UV spectroscopy. The following science cases are discussed:

\begin{enumerate}[label=Pillar \Roman*.,
       leftmargin=*,
       align=left]
  \item Formation and evolution of planets and their habitability
    \begin{enumerate}
      \item Exoplanets atmospheres and atmospheric escape
      \item Exoplanets bulk composition
      \item Protoplanetary disks
      \item Solar System planetary atmosphere and icy moons
      \item Small bodies in the Solar System
    \end{enumerate}
  \item Stellar lives and deaths at their extremes
    \begin{enumerate}
      \item The first stars and the origin of the elements
      \item Massive stars
      \item White dwarfs and compact binaries
       \item Supernovae
    \end{enumerate}
    \item Gas, dust and metals in the baryon cycle of galaxies
    \begin{enumerate}
      \item Metals in the Interstellar Medium
      \item Cosmic dust
      \item The Circumgalactic and Intergalactic Medium
    \end{enumerate}
\end{enumerate}

Sections ~\ref{sec: Pillar I} to \ref{sec: Pillar III} will describe more details of the individual science cases and the requirements on the instrument capabilities.

Here we offer an overview of the top-level requirements that emerge from the science cases, which define the instrument capabilities and will drive some of the technical specifications of the instrument basic design (Sect.~\ref{sec: instrument}). 

\begin{figure*}
        \centering
        \includegraphics[width=\textwidth]{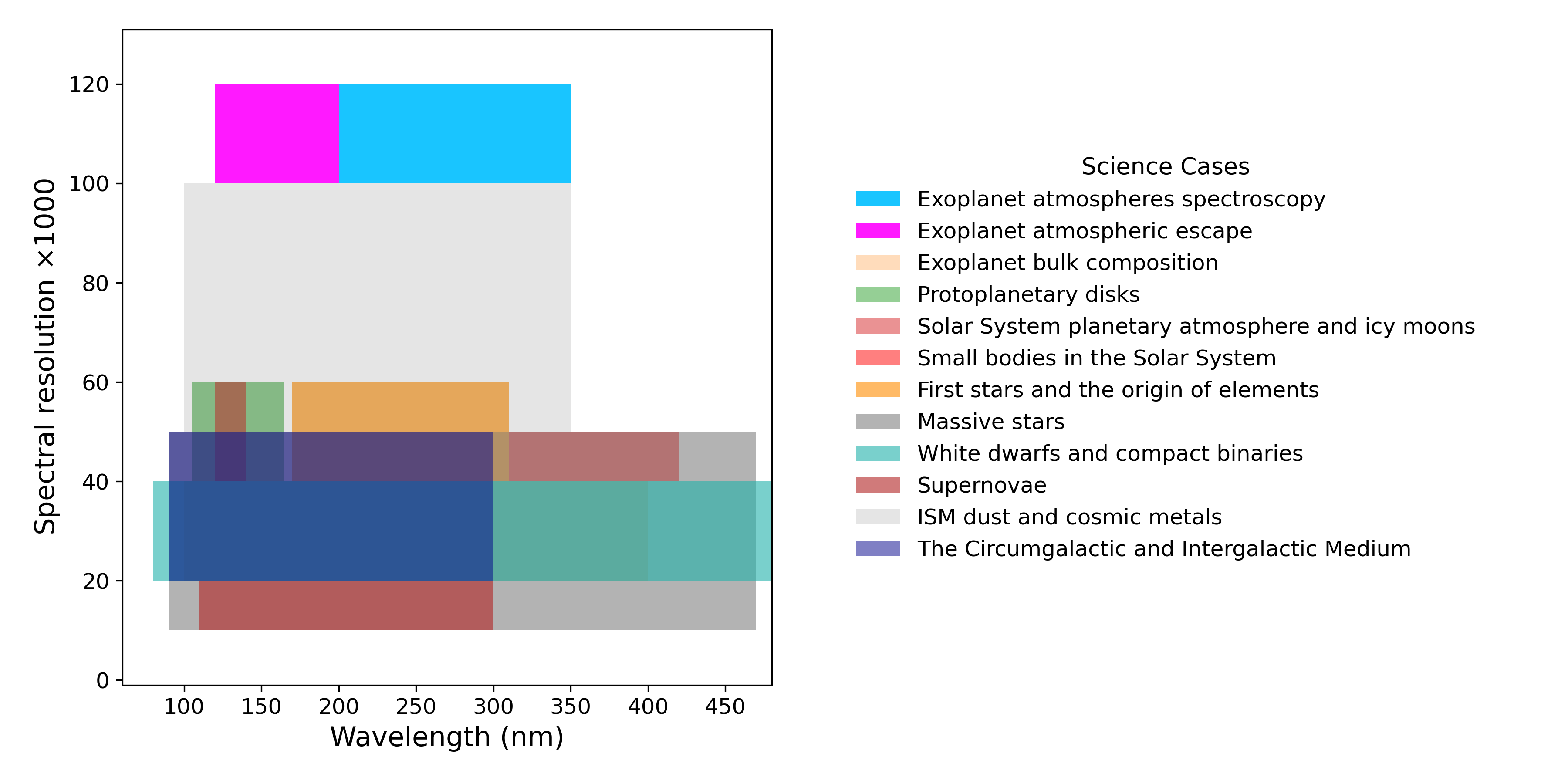}
        \caption{Range of wavelengths and spectral resolutions required by the science cases described in this study. The areas represent the strict requirements, while optimal ranges are listed in Table \ref{table:top-level-requirements}.}
        \label{fig:R_vs_wave}
\end{figure*}

\begin{figure*}
        \centering
        \includegraphics[width=\textwidth]{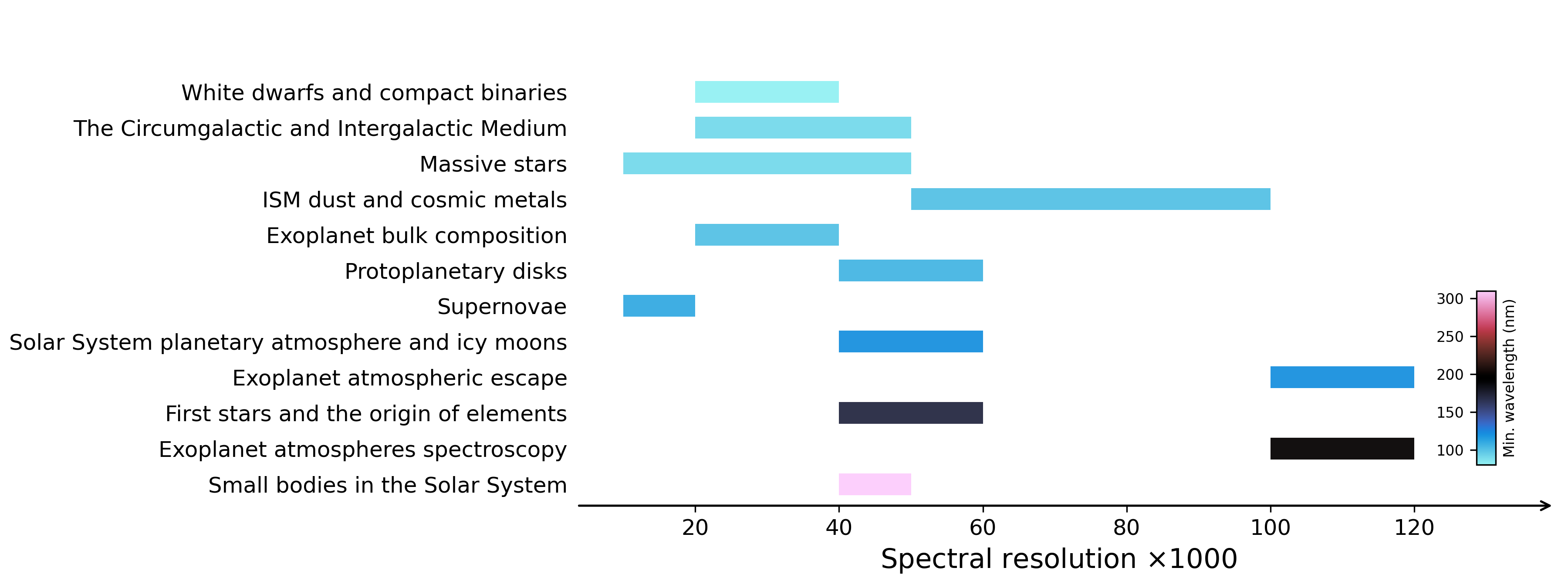}
        \caption{Required ranges of spectral resolutions needed for the individual UV science cases described in the current study. The minimum wavelength required is color coded.}
        \label{fig:R_vs_minwave}
\end{figure*}

The requirements on spectral resolution and wavelength coverage of the individual science cases are listed in Table~\ref{table:top-level-requirements} and shown in Figures~\ref{fig:R_vs_wave} and \ref{fig:R_vs_minwave}. The required spectral resolution and wavelength coverage listed in Table~\ref{table:top-level-requirements} are the ranges that are strictly required to achieve the science case. The optimal (goal) ranges extend the required ranges to lower and higher values. The color map in Fig. \ref{fig:R_vs_minwave} highlights the bluest wavelength required by the science cases. 

Overall, the required spectral resolution ranges from $R=10\,000$ to $R=120\,000$, with an optimal (goal) range of $R=10\,000-200\,000$. The required wavelength coverage is mostly inside the interval \mbox{90--470}~nm, with one of the science cases going down to slightly lower $\lambda=80$~nm and extending to longer wavelengths (up to 700~nm). The overall optimal (goal) wavelength coverage is 80--1000~nm.

\begin{figure*}
        \centering
        \includegraphics[width=\textwidth]{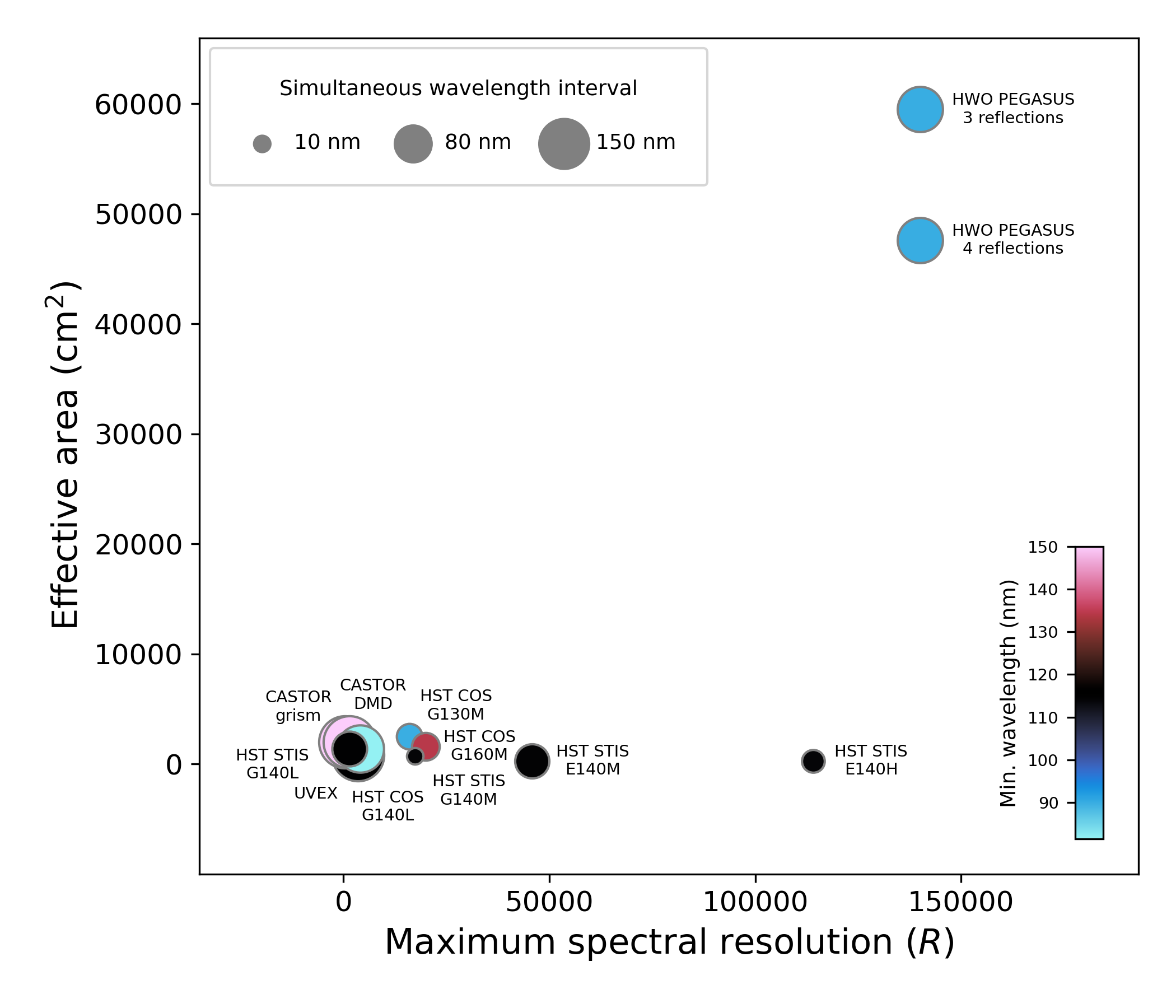}
        \caption{Spectral resolution vs.\ FUV effective area for present and future non-HWO UV space missions in comparison to HWO PEGASUS. The reported values correspond to the maximum spectral resolution and effective area at $\sim$\,150\,nm. The labels encode the UV settings of the different missions/instruments. The size of the scatter points encodes the width of the wavelength interval that can be probed in a single observation in the FUV. For missions/instruments with multiple wavelength settings in the FUV, the circle size corresponds to the wavelength range simultaneously covered by the specific setting.}
        \label{fig:mission_comparison}
\end{figure*}

Figure~\ref{fig:mission_comparison} shows the spectral resolving power $R$ and effective area at 150~nm for present and future UV space missions. The effective area is the geometric area of the mirror corrected by the throughput of the optical system, including the number of internal reflections photons encounter on their way to the detectors. Current and future non-HWO missions are limited either in sensitivity (low effective area) or spectral resolution. For example, HST/STIS has the capability to reach high-spectral resolution, but with a very low effective area, and thus being sensitive only to the brightest targets. Future missions like UVEX or CASTOR will have a limited spectral resolution and lower sensitivity than HST/COS.

The effective area of PEGASUS is calculated assuming an instrument throughput of 22\% (in the FUV; taken from Fig.~\ref{fig:SpecEfficiency}) and considering both cases of three and four reflections in the telescope. A comparison of spectral capabilities among HWO instrument concepts is discussed in a dedicated section (Sect. \ref{sec: comparison HWO}).  

Ground-based high-resolution spectroscopy complements space FUV/NUV, at longer wavelengths, down to the atmospheric cutoff. The Very Large Telescope (VLT) UVES \citep{Dekker00} covers 300-1100~nm with a maximum resolution of $R=80\,000$ in the Blue Arm. The upcoming VLT/CUBES will cover the 300-400~nm wavelength range with $R=20\,000$ and aim at a variety of science cases \citep{Evans23} that are often related to this paper. In the 2030s the Extremely Large Telescope (ELT) high-resolution spectrograph ANDES will cover the wavelength range 400-1800~nm with $R=100\,000$, offering an important scientific complement to the high-resolution spectroscopy capabilities of HWO \citep[e.g.][]{DOdorico24, Roederer24, Palle25}. 

The breadth and depth of the science cases presented in Sections \ref{sec: Pillar I} to \ref{sec: Pillar III} highlight that sensitivity of UV high-resolution spectroscopy is currently the most limiting factor. Transformational advancement in UV science can be enabled by a jump in UV high-resolution spectroscopy sensitivity. Figure~\ref{fig:mission_comparison} put some of these capabilities in context with all other current or future missions that offer UV spectroscopy. HWO PEGASUS aims at uniquely covering the parameter space reaching high spectral resolution and high sensitivity.

\begin{table*}
\caption{Wavelength coverage and spectral resolution (required and optimal intervals) of the individual science cases.}
\label{table:top-level-requirements}
\centering
\renewcommand{\arraystretch}{1.2}
\begin{threeparttable}
\begin{tabular}{l c c c c}
\hline
\noalign{\smallskip}
\multirow{2}{*}{Science case}
& \multicolumn{2}{c}{Wavelength (nm)} & \multicolumn{2}{c}{Spectral resolution ($R$)} \\
& Required & Optimal & Required & Optimal \\
\noalign{\smallskip}
\hline
\noalign{\smallskip}

Exoplanet atmospheres and atmospheric escape    & 120--350 & 100--400 & 100\,000--120\,000 & 20\,000--200\,000 \\
Exoplanet bulk composition                      & 100--400 & 100--900 & 20\,000--40\,000 & 20\,000--80\,000 \\
Protoplanetary disks                            & 105--165 & 100--400 & 40\,000--60\,000 & 40\,000--120\,000 \\
Solar System planetary atmosphere and icy moons & 120--140 & 100--170 & 40\,000--60\,000 & 40\,000--120\,000 \\
Small bodies in the Solar System                & 310--420 & 300--420 & 40\,000--50\,000 & 40\,000--120\,000  \\
First stars and the origin of elements          & 170--310 & 170--360 & 40\,000--60\,000 & 30\,000--100\,000  \\
Massive stars                                   & 90--470  & 90--1000 & 10\,000--50\,000 & 10\,000--100\,000 \\
White dwarfs and compact binaries               & 80--700  & 80--1000 & 20\,000--40\,000 & 10\,000--50\,000 \\
Supernovae                                      & 110--300 & 110--300 & 10\,000--20\,000  & 10\,000--50\,000 \\
Metals in the Interstellar Medium               & 100--350 & 90--450 & 50\,000--100\,000  & 30\,000--200\,000 \\  
Cosmic dust                                     & 100--350 & 90--450 & 50\,000--100\,000  & 30\,000--200\,000 \\  
The Circumgalactic and Intergalactic Medium      & 90--300 & 90--450 & 20\,000--50\,000  & 15\,000--100\,000 \\

\noalign{\smallskip}
\hline
\end{tabular}
{\footnotesize
\textbf{Note.} The upper limits of the optimal spectral resolution intervals correspond to the values at which the information content of the spectra no longer increases.}
\end{threeparttable}
\end{table*}

\section{A dedicated UV high-resolution spectrograph: PEGASUS basic design}
\label{sec: instrument}


We propose PEGASUS (Planets, Earths, Galaxies And Stars UV Spectrograph), a cross-dispersed UV \'Echelle spectrograph, to address the science cases listed in Sect.~\ref{sec: overview} and presented in more detail below in Sects.~\ref{sec: Pillar I} to \ref{sec: Pillar III}. The PEGASUS design is conceived as both a dedicated HWO instrument and a potential standalone space mission with a 2-meter primary mirror (Theon). The optical design is optimized for high photon throughput while maintaining simplicity. No moving parts or mechanisms are employed during regular operations. A focus adjustment mechanism that is only planned to be used during commissioning and in long cadence during operation to account for long term drifts in the overall system is foreseen. No active cooling is required. This minimizes operational complexity and maximizes long-term reliability. 

The wavelength and spectral resolution requirements of the different science cases are summarized in Table~\ref{table:top-level-requirements}. To fulfill these requirements, PEGASUS is designed for continuous wavelength coverage across two arms: a FUV arm spanning 90--200\,nm and a UV-visual (UV-VIS) arm spanning at least 200--400\,nm. The short-wavelength boundary at 90\,nm is set by the Lyman limit, while the long-wavelength boundary coincides with the bluest wavelengths accessible to future ELT/ANDES. Both arms will be optimized to achieve maximum science yield as soon as the final set of science cases is defined and optimization weights are set for cases and wavelength ranges. Thus, PEGASUS and ANDES together will provide continuous high-resolution spectroscopic coverage from the FUV to the near-infrared. An extension of the UV-VIS arm to 800\,nm is planned, depending on detector size and the complexity of the spectrograph camera in the final instrument design. Although wavelengths $>$\,400\,nm are accessible from the ground, space-based high-resolution spectroscopy in the VIS range offers critical advantages, most notably the ability to obtain an absolute flux calibration, which is challenging to achieve from the ground. 
PEGASUS will deliver a maximum spectral resolution of $R=$\,100\,000 in the FUV arm and $R=$\,140\,000 in the UV-VIS arm. Lower resolutions with increased sensitivity can be achieved through on-chip pixel binning, enabled by low read-out noise detectors. Thus, instrument modes with resolutions reduced by integer binning factors relative to the maximum spectral resolution will be available. For binning factors smaller than approximately 1/4 to 1/6, post-processing is predicted to be superior to on-chip binning as the curved shape of the spectral order traces becomes non-negligible during pixel binning. The minimum useful spectral resolution from on-chip pixel binning is thus $R\sim$\,10\,000 in both the FUV and UV-VIS arms.

Incoming photons enter the two instrument arms through separate apertures, one for the FUV and one for the UV-VIS range. This avoids the need for complex UV beam splitters and the associated throughput losses. In both arms, the UV light is fed to the spectrograph via optical fibers, which provide homogeneous illumination of the \'Echelle grating and simplify the overall integration. Hollow waveguides optimized for the spectral regime below 200\,nm are available with high efficiency \citep{Gilliam2021}. Short fibers can be used to achieve constant and well predicted spectrograph illumination. Additionally, the use of fibers allows to decouple the entrance slit from the spectrograph with respect to the common focus. This design reduces integration risks as the FUV and UV-VIS arms of the spectrograph can be tested independently. The use of fibers is potential, and depends on the technology development and performance of UV fibers, which have recently shown promising results \citep{Gilliam2021, Mears2025}.

\begin{figure*}
        \centering
        \includegraphics[width=0.6\linewidth]{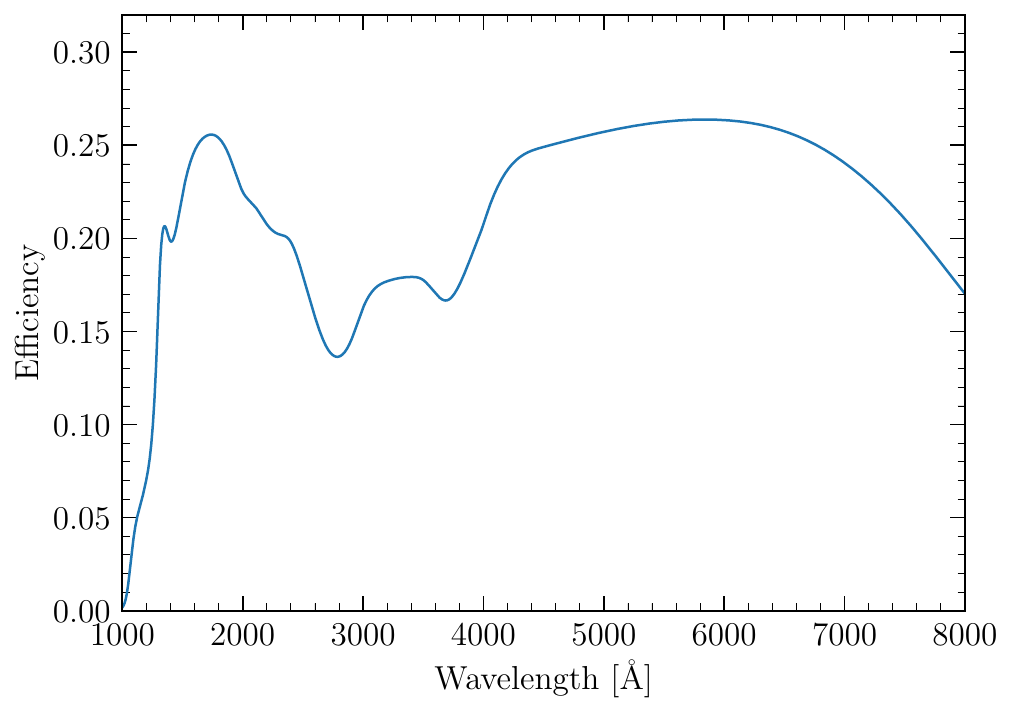}
        \caption{Current best estimate end-to-end instrument efficiency of the PEGASUS spectrograph throughout the planned wavelength coverage, including an extension beyond 4000\,\AA. Margins and contingencies are not included. 
        In the FUV range, at $1500\,$\AA\, $\sim 22\,\%$ is achieved; beyond 4000\,\AA, an efficiency  $>20$\% is reached.}
        \label{fig:SpecEfficiency}
\end{figure*}

The FUV arm employs grating cross-dispersion, while the UV-VIS arm uses prism cross-dispersion. Micro-structured \'Echelle gratings are used in both arms to enhance diffraction efficiency and minimize scattered light. In the FUV arm, a white-pupil design is foreseen to further suppress stray light; alternatively, an additional collimator may be used for the same purpose. Dispersed light is focused onto the detectors by a reflective camera, either of Korsch type (all-mirror design offering full FUV transparency) or incorporating a fused silica/CaF$_2$ field-flattening lens for the UV-VIS arm. Mirror coatings in the FUV arm will make use of advanced Al + AlF$_3$ coatings (see \citealt{tuttle24} for reflectance) to maximize reflectivity at the shortest wavelengths, while Hubble-type Al + MgF$_2$ coatings (reflectance data provided by Manuel Quijada; priv. comm.) will be used in the UV-VIS arm. To maximize detector sensitivity in the UV range, PEGASUS will employ delta-doped CMOS detectors, which are characterized by a high UV quantum efficiency \citep{Hoenk_2022, Skottfeld_2025}. Gradient anti-reflection coatings applied to the detector surfaces will reduce photon losses due to reflections (we assumed a grating efficiency of 0.7 and a cross disperser throughput of 0.8). The described instrument design results in a throughput as shown in Fig.~\ref{fig:SpecEfficiency}.



The HWO PEGASUS effective area, resulting from this throughput, and spectral resolution is put in context of other current and future non-HWO UV missions in Fig.~\ref{fig:mission_comparison}.

\section{Comparison among potential HWO UV spectrographs}
\label{sec: comparison HWO}
Figure \ref{fig:mission_comparison} shows that HWO PEGASUS aims at uniquely cover the parameter space of high spectral resolution and sensitivity, but it does not include other potential HWO instrument concepts as comparison. In this Section we compare HWO PEGASUS with other potential instruments for HWO that have UV spectroscopy. The design of these instruments is under development, and not all instruments or instrument capabilities may in the end be implemented on HWO. 

The HWO instrument concepts that have UV spectroscopy, with different capabilities, and are included in the governmental NASA HWO Instrument Concept Studies so far are an Ultraviolet Multi-Object Spectrograph (UV MOS) and an Ultraviolet Integral Field Unit (UV IFU). An Extra Instrument (XI) concept was not developed by the NASA studies, but left open for community proposals, with capabilities foreseen to align with HWO science goals. We refer to these instruments as described in the NASA HWO Instrument Concept Studies Workshop 2026 \citep{Sitarski26}. In addition, a spectropolarimeter combined with high-resolution spectrograph has been proposed by a European consortium \citep[Pollux,][]{Neiner26}. 

Following the recommendation of the NASA HWO Instrument Concept Studies, we assume the HWO architecture of the Exploratory Analytic Case 5 (EAC5) for the comparison among instruments. The specifications of this telescope design include an inscribed circular pupil diameter of 8.2~m, a focal length of 171~m, an F number of 19, and a maximum delta angle of incidence (the difference in angle of incidence at each mirror) of 11.1 deg. The number of telescope reflections before reaching the different potential instrument bays are 3, for one UV-optimised bay (bay A), or 4 otherwise (bay B, C, D, E). The NASA HWO Instrument Concept Studies have preliminarily assumed UV MOS to be located at bay A and XI to be located at bay E, although this is not binding. In the EAC5 design, bay A has an instrument mass of 500~kg and power of 525~W, and bay E has a mass of 200~kg and power of 200~W. For PEGASUS, we assume a throughput of 22\% at 150~nm (Fig. \ref{fig:SpecEfficiency}).

\begin{table*}
\begin{threeparttable}
    \centering
    \caption{Comparison of UV spectroscopic capabilities for potential HWO instrument/modes concepts. $n$ is the number of telescope mirror reflections of the light beam feeding the instruments. $A_\mathrm{eff}$ is calculated at 150~nm, assuming that 20\% of light is lost at every telescope reflection and a 8.2~m diameter of the primary mirror (EAC5). Values are indicative and their sources described in the text. 
    }
    \begin{tabular}{|l|l|r|l|l|l|}
    \hline 
       Instrument   & $\lambda$ (nm) & Max $R$ & $n$ & $A_\mathrm{eff}$ ($m^2$)& Notes \\
       \hline
       PEGASUS      & 90--1000  & 140\,000 & 3--4 & 4.7--5.9& High-resolution spectrograph, optimized for UV sensitivity.\\
       UV MOS       & 100--1000 &  30\,000 & 3--4 & 2.1--2.7& Multi-object, low $R$.\\
       UV MOS/HRS   & 100--400 &  150\,000 & 3--4 & 1.4--1.7& Nominal mode of MOS, point-source spectroscopy, high $R$.\\
       UV IFU       & 100--400 &    4\,000 & 3--4 & - & Integral field unit, low $R$.\\
       Pollux/Spectroscopy  & 101--1888 & 100\,000 & $4^{a}$ & $2.0^{a}$ & Retractable polarimeter for pure spectroscopy.\\
      \hline 
    \end{tabular} 
    \begin{tablenotes}
    \item[] \textbf{Notes.} $^{a}$ We conservatively adopt 4 telescope reflections for Pollux, assuming no additional reflections are needed to correct for high angles of incidence to enable polarimetry (private communication), given that HWO is an off-axis telescope.
    \end{tablenotes} 
    \label{tab: HWO UV spectrographs}
    \end{threeparttable}
\end{table*}


\subsection{UV MOS}

The UV MOS initial proof-of-concept instrument parameters in the NASA instrument study have a bandpass of 100 - 1000~nm and $R = 500$ - 50~000. The UV MOS concept includes several different gratings with different resolution and wavelength coverage. In the FUV (NUV), the foreseen highest spectral resolution is about $R=30~000$ ($R=20~000$) and simultaneously covers a range $\sim78$~nm ($\sim267$~nm) wide. Many of the characteristics of UV MOS have evolved from the LUMOS concept \citep{France17b}. The UV MOS offers sensitive UV spectroscopy, but with a factor of 3-7 lower spectral resolution than PEGASUS.  

The NASA instrument study of UV MOS also foresees a notional high-resolution spectroscopy (HRS) mode with $R=100~000$, with a bandpass covering 100 - 400~nm. The notional HRS mode was preliminarily included in the governmental NASA HWO Instrument Concept Studies for UV MOS and in the material preliminarily presented by NASA for the call for community HWO Instrument Concept Studies. Current non-governmental HRS studies foresee instrument performances that could be comparable with PEGASUS, with spectral resolution up to $R\sim150~000$ and about 30\% of the throughput of PEGASUS at 150~nm (private communication). 

The UV MOS has been proposed as the instrument which could benefit from the UV-optimised bay, implying 3 reflections before reaching the instrument. In case UV MOS will occupy the HWO UV-optimised bay and PEGASUS will occupy a less optimal bay with an additional telescope reflection, UV MOS/HRS will have about 36\% of the PEGASUS effective area at 150~nm (private communication).  

Overall, the UV MOS/HRS and PEGASUS appear quite comparable, in terms of spectroscopic performance in the UV. The main difference will be the overall complexity of the instrument, with the UV MOS having a full suite of different gratings and the challenge of including multi-object spectroscopy capability. PEGASUS represent a substantially simpler solution to reach high-resolution spectroscopy. The additional advantage of PEGASUS is that it will offer high-resolution spectroscopy covering the optical range (800-1000~nm), while UV MOS/HRS is not planned to extend beyond 400nm.

\subsection{UV IFU}

The Ultraviolet Integral Field Unit (UV IFU), or Ultraviolet Integral Field Spectrograph (UV IFS) is an HWO instrument concept included in the NASA HWO Instrument Concept Studies. Indicately, the UV IFU is foreseen to have a bandpass of 100 - 400~nm and $R=4000$. This spectral resolution is not comparable with the much higher resolution of PEGASUS.

\subsection{Pollux}

The first concept of the Pollux instrument \citep{Muslimov18} had been proposed as one of the potential instruments for LUVOIR \citep{LUVOIR19}. Its science cases prominently feature high-resolution broad band simultaneous spectroscopy as one of the core capabilities of the instrument, together with optional spectropolarimetry \citep{Bouret18,Neiner26}. Pollux is a European instrument concept for a HWO high-resolution spectrograph and spectropolarimeter operating from the far-UV (100 nm) to the NIR \citep[1750~nm,][Le Mignant et al., Proc. SPIE 14146 in prep.]{Neiner26}. 

The coverage of the entire wavelength band is achieved through 5 \'Echelle spectrographs: FUV (100-123~nm), MUV (120-236~nm), NUV (236-438~nm), OPT (438-875~nm), and NIR (875-1750~nm). The FUV arm works singularly and enables point-source spectroscopy or point-source spectropolarimetry at a resolving power of $100\,000$. The other four arms work simultaneously and produce a continuous spectrum from 120 to 1750~nm at a resolving power of $100\,000$ in the MUV and NUV and of $65\,000$ in the OPT and NIR bands. Each of these four arms hosts a dedicated retractable polarimeter. The four arms enable point-source spectroscopy or point-source spectropolarimetry and in the current design the MUV and NUV arms also enable slit spectroscopy.




The off-axis optical design of HWO results in large angles of incidence on the telescope mirrors. At these angles, the mirror reflectivity becomes polarization dependent, complicating the implementation of high-precision polarimetry and potentially requiring additional reflections or polarization optics. The Pollux team is currently developing a design that avoids these additional reflections (private communication). For the purpose of comparing effective areas, we conservatively assume the same four telescope reflections for Pollux as for the other HWO instruments. With the increased spectroscopy throughput provided by the retractible polarimeter, Pollux offers about 40\% of the effective area of PEGASUS, in the FUV ($\sim150$~nm).



In addition to the sensitivity, the main difference between the PEGASUS and Pollux concepts is the complexity of the overall instrument. On one hand, Pollux is a more complex instrument that includes a retractable polarimeter to enable spectroscopy and 5 individual arms overall extending from the FUV to the NIR. On the other hand, PEGASUS is a simpler instrument that focuses on high-resolution spectroscopy and is optimised for sensitivity and robustness (e.g. avoids moving parts).

\section{Science Pillar I. Formation and evolution of planets and their habitability}
\label{sec: Pillar I}

General instruction to the pillar, spanning solar system, exo-planets and protoplanetary discs.


\subsection{Exoplanet atmospheres}
\label{sec: Exoplanet atmospheres}

The discovery of thousands of exoplanets has revealed a striking diversity of planetary sizes, compositions, and orbital environments. Spectroscopic characterization of their atmospheres is needed to shed light on planets’ nature, evolutionary history, and potential to sustain life. A major goal of exoplanet studies is to understand how exoplanetary atmospheres evolve over billions of years since their formation, and to observe evidence of formation and evolutionary processes in the spectrum observable at the present time. In particular, observations aim to constrain chemical abundances or abundance ratios, atmospheric structure (as a function of altitude or in three-dimensions), as well as atmospheric processes (e.g.\ dynamics, chemistry, clouds, interior-atmosphere interactions).

Some of the more pressing open questions in this field are:
\begin{itemize}
    \item What are exoplanet atmospheric compositions and dynamics? 
    \item How do atmospheric escape rates depend on various stellar and planetary properties? 
    \item How does atmospheric escape shape exoplanet populations? How do atmospheres evolve over time? How do we detect hydrogen exospheres and metal escape in dozens of exoplanet systems, down to Earth-sizes? 
    \item How does this influence exoplanet habitability?
\end{itemize}

\subsubsection{High-resolution observations of exoplanet atmospheres}

Spectroscopy of exoplanet atmospheres has until recently been limited to large, close-in gas giants, but the enhanced sensitivity offered by JWST is allowing for smaller planets, including sub-Neptunes and some terrestrial planets to be characterised as well \citep[e.g.][]{Lim2023,Madhu2023,Hu2024}. JWST observes primarily at near to mid-infrared wavelengths, targeting absorption bands of molecules at limited spectral resolutions of $R = \lambda/\Delta \lambda \lesssim 3000$. As most observations of exoplanet atmospheres rely on a time-differential measurement of the flux of the system, equivalent observations are impossible to be attained from the ground due to the variability of the Earth's atmosphere. Instead, the most successful ground-based observations have relied on high-resolution spectroscopy, resolving individual lines of atoms and molecules as well as the planet's time-varying radial velocity due to its orbital motion \citep{Brown2001, Snellen2010}. For chemical species with many observable transitions, and for close-in planets with large orbital velocity, this observing method provides exquisite robustness against false-positive detections and sources of astrophysical or telluric noise. At the same time, these observations are sensitive to km/s motions in the planet's atmospheres, allowing atmospheric dynamics as well as three-dimensional inhomogeneity of the chemical composition of the atmosphere to be directly observed for the hottest gas giants \citep[e.g.][]{Prinoth2022}. By observing lines with different intrinsic strengths, this allows the vertical stratification of the composition and wind-profiles of these planets to be constrained \citep{Seidel2025}.

At optical wavelengths, strong transitions typically require high atmospheric temperatures, liberating refractory metals in the gas phase. As such, high-resolution spectroscopy has been particularly successful for ultra-hot Jupiters \citep{Hoeijmakers2018}. In theory, high-resolution spectroscopy at optical wavelengths is also sensitive to the reflected light \citep[e.g.][]{Martins2013,Hoeijmakers2018b,Vaughan2024}, allowing direct constraints of planetary albedo and scattering properties of non-transiting planets without the need for full phase-curves, yet such observations have until now resulted only in upper limits. At infrared wavelengths, high-resolution spectroscopy has mostly been used to observe molecules (notably CO and H$_2$O), which are also accessible to JWST albeit with loss of velocity-information.

The UV presents unique opportunities for characterizing exoplanet atmospheres at high spectral resolution. Many of the ground-state resonance transitions of common atoms and ions occur at wavelengths below 400 nm, and ground-state transitions are often strong, allowing these lines to probe high altitudes where the atmospheric pressure is low. The UV offers access to transitions of species that are otherwise impossible to access, or only at high temperature (e.g. carbon, magnesium and a variety of metals). Combined with high resolving power, UV spectroscopy will be able to constrain atmospheric dynamics over a wide range of atmospheric altitudes. In addition, several important molecules are mainly accessible in the UV, including SiO, an important cloud precursor species \citep{Lothringer2022} and ozone \citep{Meadows2017}.

As opposed to ground-based observations that are necessarily self-calibrated, space-based high-resolution spectroscopy may retain absolute flux calibration, preserving information of the spectroscopic continuum which is vital for breaking abundance degeneracies derived from absorption lines in transmission spectra \citep[e.g.][]{Heng2017}.

\subsubsection{Atmospheric escape}

How efficiently planets lose their atmospheres and under what conditions they retain them are key open questions. Atmospheric escape is thought to shape the observed population—including features such as the “radius valley” and the scarcity of intermediate-sized planets (the “hot Neptune desert”)—by stripping primordial hydrogen/helium envelopes \citep{Owen2019}. Equally important is the role of escape in determining planetary habitability. The presence and longevity of an atmosphere over geological timescales are necessary to maintain surface conditions suitable for life to develop.
Key scientific questions in this research area include:
\begin{itemize}
    \item \textbf{How do atmospheric escape rates depend on various stellar and planetary properties?} How do stellar UV flux, stellar winds, and planetary mass, radius, composition, orbital history, magnetic field, etc. determine mass-loss rates?

    \item \textbf{How does atmospheric escape shape exoplanet populations?} Can escape explain demographic features such as the radius valley and the hot Neptune desert?

    \item \textbf{How do atmospheres evolve over time?} How does the atmospheric composition of gas-rich planets evolve over time due to preferential escape/retention of certain species? Under what conditions are primordial envelopes lost and secondary atmospheres formed and retained on rocky planets?

\end{itemize}

\begin{figure*}
\begin{center}
\includegraphics[width=1\textwidth]{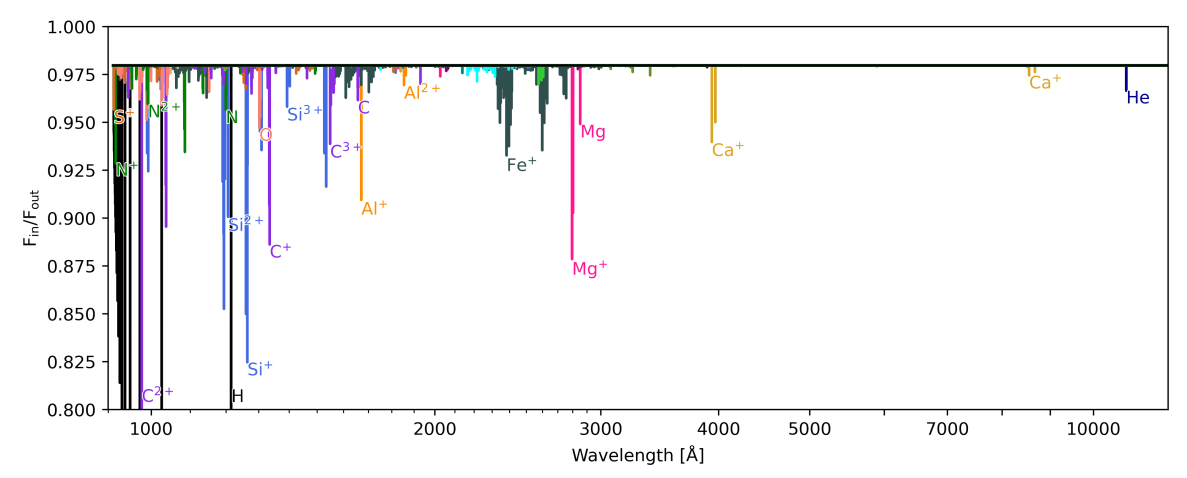}
\end{center}
\caption{Simulated transmission spectrum of an escaping exoplanet atmosphere illustrating a dense array of absorption features across the UV wavelength range. Figure adapted from \citet{Linssen2023}.
\label{fig:exosphere_spectrum}}
\end{figure*}

Transmission spectroscopy of upper atmospheres of exoplanets—where escape occurs—reveals evidence of ongoing atmospheric mass-loss \citep{DosSantos2023}. Ultraviolet observations are uniquely suited to probing these tenuous regions through strong resonance lines of hydrogen (Lyman-$\alpha$), as well as key tracers of heavier elements (e.g., C II, O I, Mg II, Fe II). Compared to longer wavelengths, the UV range offers a richer set of diagnostics of escape (Figure \ref{fig:exosphere_spectrum}), enabling multiple species to be observed simultaneously, which reduces degeneracies in their interpretation and enables more reliable measurement of the atmospheric mass-loss rate.

Until now, the main capability of high-resolution UV spectroscopy has been provided by the STIS instrument aboard the Hubble Space Telescope, yielding detections of ions in the upper atmospheres of hot Jupiters \citep{vidal03,vidal04,sing19}, as well as escaping hydrogen in several Neptune-sized planets \citep[e.g.][]{ehrenreich15}. A claimed detection of carbon in the exosphere of the sub-Neptune $\pi$ Men c suggests that this planet retains a high-mean molecular weight atmosphere \citep{Garcia2021}, which is observationally challenging to measure with other instrumentation.

The sensitivity of current UV facilities, including the Hubble Space Telescope, limits observations to a small number of bright targets. A larger collecting area would facilitate transformative advances. It would enable the detection of atmospheric escape through multiple atomic and ionic species for a statistically significant number of intermediate-sized exoplanets (sub-Neptunes and super-Earths). The evolution of these planets, which straddle the so-called ``radius valley'', is believed to be strongly influenced by atmospheric escape \citep[e.g.]{OwenWu2013}, and observations of atmospheric escape allow for strong implications about their evolutionary history, more so than has been possible for atmospheric escape from gas giants observed until now. A one order-of-magnitude improvement in sensitivity compared to HST would transform atmospheric-escape studies from a handful of case studies into a population-level exploration of planetary evolution.

Furthermore, a larger collecting area would open a new regime in which Earth-sized planets become accessible in the UV, enabling searches for extended hydrogen exospheres, akin to the Earth’s geocorona \citep[e.g.]{Kameda2017, Cherubim2026}, and indirect signatures of water loss \citep{Jura2004}.\\

To achieve the atmospheric-escape science described above, the observations should cover a minimum of 120-200 nm and ideally a 100-400 nm UV bandpass, containing the far-UV escape tracers (especially H I Ly$\alpha$ at 121.6 nm, C II at 133.5 nm, O I near 130.6 nm, and other strong resonance lines) and the near-UV metal lines (e.g., Mg I and Mg II near 280 nm, Fe II lines across the NUV). A velocity resolution of a few km~s$^{-1}$ is needed to resolve line profiles and measure outflow kinematics; this corresponds to a resolving power of roughly $R=50,000-120,000$ (but also higher resolutions are beneficial in order to characterize the line profiles). The minimum useful sensitivity is an S/N of a few per resolution element in the planetary absorption signal; in practice, individual transit spectra will often need stellar S/N values in the hundreds to recover planetary absorption at the percent to sub-percent level. Repeated visits (i.e. observations of multiple transits of the same planet) for co-adding weak signals will likely be necessary. The most relevant targets are relatively nearby stars hosting transiting planets, with photometric B magnitudes in the range $\sim$ 9-13. A potential sample size of up to about 50–100 systems would enable both detailed case studies and population-level constraints on escape rates versus stellar type, irradiation, and planet mass. The observing mode should allow for time-resolved spectroscopy through full transits, along with stellar baseline coverage before and after transit. Individual exposures should be short enough ($\lesssim$100–300~s for typical close-in systems) that the planet’s radial velocity drift during a single integration is smaller than the instrumental velocity resolution, ensuring that the velocity-resolved absorption signal is not smeared by the orbital motion of the planet.

\begin{figure*}
\begin{center}
\includegraphics[width=1\textwidth]{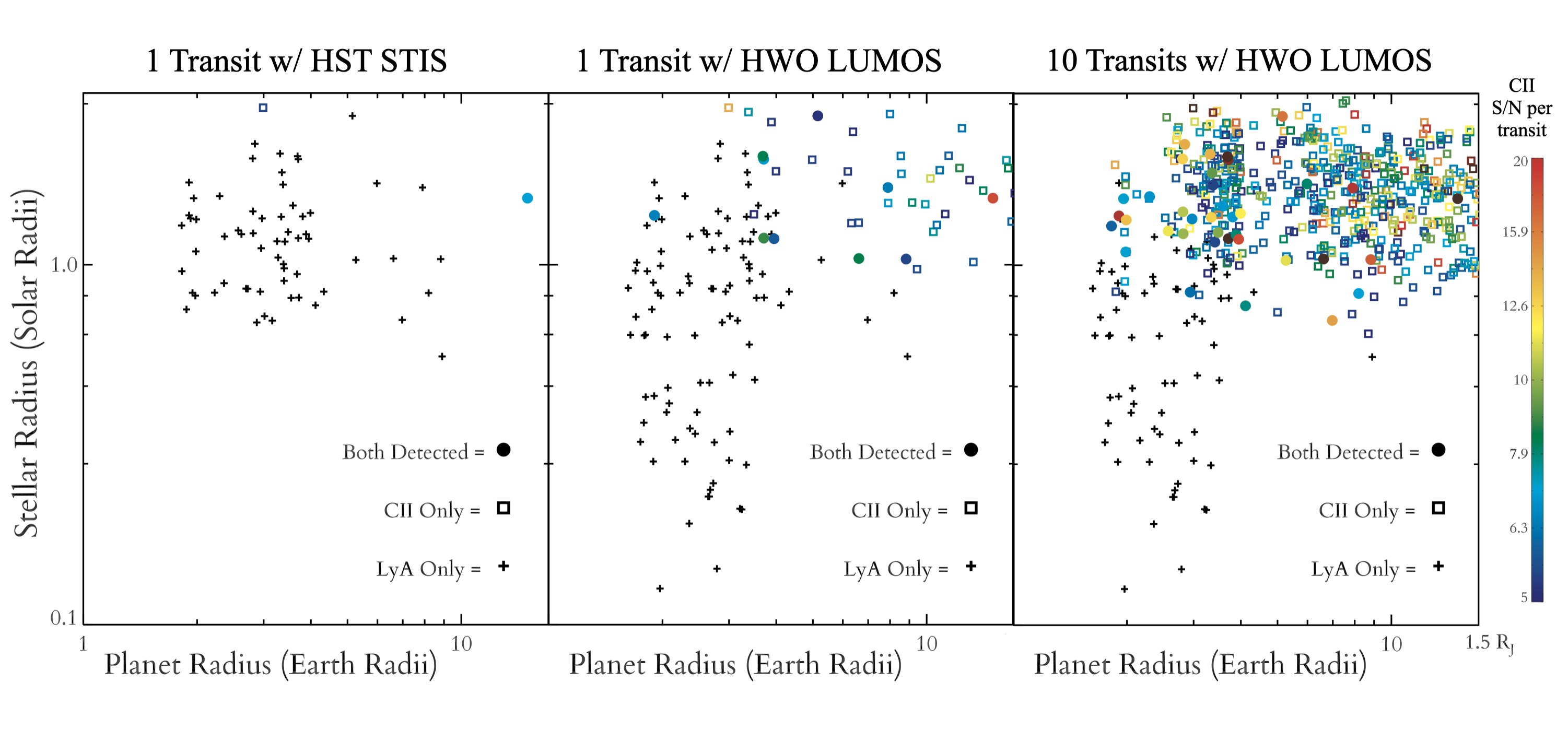}
\end{center}
\caption{Expected detections of escaping hydrogen and/or ionized carbon with HST/STIS and HWO/LUMOS for a simulated population of exoplanets. Figure adapted from \citet{DosSantos2025}.}
\label{fig:atm_escapeHST_HWO}
\end{figure*}

\citet{DosSantos2025} recently simulated the capabilities of an HWO concept with a collecting area about an order-of-magnitude larger than the a hypothetical mission with a 2-meter mirror. Their detection yield in Lyman-$\alpha$ and the C II doublet at 133 nm, for simulated planets of different sizes and host-star types, indicates that the HWO should be able to detect these FUV lines at high S/N ($\sim$5-10) for dozens of exoplanets in a single transit (Figure \ref{fig:atm_escapeHST_HWO}, middle panel). Compared to the current capability of HST/STIS, this is a major leap, enabling detections of hydrogen exospheres in small (down to Earth-sized) planets and detections of metal escape in dozens of systems. Assuming photon-limited observations and all else being equal, PEGASUS should be able to achieve a similar yield by co-adding 10 transits, which represents a substantial but not prohibitive increase in observing time. The detection yield in the NUV, where targets are intrinsically much brighter, should be even higher.

\subsubsection{Requirements}

\textbf{General high-resolution spectroscopy:}
\begin{enumerate}
    \item {\it Wavelength coverage} -- Required: 200--350\,nm; Optimal: 200--400\,nm, to additionally cover the \ion{Ca}{ii} lines and extend the inventory of chemical species.
    \item {\it Spectral resolution} -- Required: 100\,000--120\,000, driven by the desire to resolve atmospheric dynamics via spectral line shapes and to resolve closely packed metal lines;  Optimal: 20\,000--200\,000. The option to go down to spectral resolution of 20\,000 would increase sensitivity, allowing to access particularly faint targets (though with the trade-off of only partially resolving the spectral lines).

    \item {\it Sensitivity} -- Targets to be observed will typically have a V magnitude $\leq 11$.
    \item {\it Stability in flux calibration} -- Allows for continuum calibration and to access planets with small orbital motion. Typically within 10--100\,ppm when averaged over a multi-hour visit (similar to JWST).
\end{enumerate}

\textbf{Atmospheric escape:}
\begin{enumerate}
    \item {\it Wavelength coverage} -- Required: 120--200\,nm; Optimal: 100--400\,nm, driven by specific transitions, including H$\alpha$, carbon, and oxygen. These may only be visible against chromospheric emission lines.
    \item {\it Spectral resolution} -- Required: 100\,000--120\,000, to resolve the profiles of exospheric spectral lines and constrain exospheric kinematics;  Optimal: 50\,000--200\,000.
    \item {\it Sensitivity} -- Targets to be observed have photometric $B$ magnitudes in the range $\sim$\,9--13; the most relevant targets are relatively nearby stars hosting transiting planets.
    \item {\it Stability in flux calibration} -- Required level of within 0.1\% when averaged over a multi-hour visit (transit), to be able to confidently measure percent-level deep transits.
\end{enumerate}

\subsection{The bulk abundances of exoplanets}
One of the fundamental question in exoplanet research~--~\textit{``What are those other worlds made out of?''}~--~is difficult to answer from studies of planets orbiting main-sequence hosts: transit spectroscopy only probes the composition of the planets' atmospheres, and radial velocities plus transit light curves only yield measurements of the bulk densities of planets. Further conclusions on internal structure and bulk composition are model-dependent (e.g. \citealt{dornetal17-1}). These limitations are not restricted to exoplanets: even in the Solar system, we can directly probe only the outer layers of the planets. The most detailed insight into the composition of the Solar system planets comes from the study of \textit{meteorites}, the building blocks from which our planets formed \citep{drake+righter02-1}.

\begin{figure*}
    \centering
    \includegraphics[width=\linewidth]{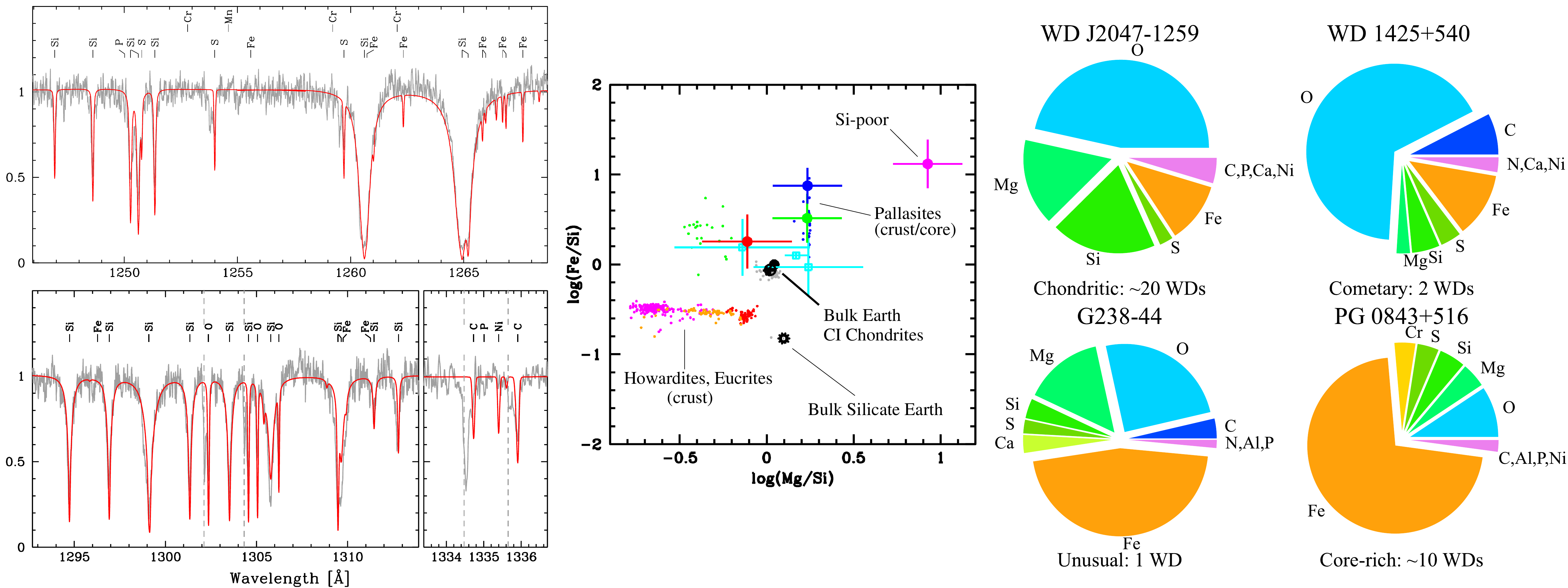}
    \caption{Left: \textit{HST}/COS G130M spectrum ($\lambda/\Delta\lambda\simeq18\,000$) of a debris-enriched white dwarf, illustrating the diagnostic power of ultraviolet spectroscopy for measuring the bulk abundances of exoplanets. Middle: abundance ratios for solar-system meteorites (small dots) and exoplanetesimals (large dots), illustrating the diversity in the abundances among the two types of objects. Right: the $\simeq30$ white dwarfs accreting planetary material have detailed abundance measurements, including water-rich (WDJ2047$-$1257, \citealt{hoskinetal20-1}, icy (WD1425+550, \citealt{xuetal17-1}), rocky but volatile enhanced (G238-44, \citealt{johnsonetal22-1} and (PG0843+516, \citealt{gaensickeetal12-1}) systems. A robust statistical assessment of exoplanet abundances will require to extend this sample to several hundred systems, see Fig.\,\ref{fig:wd_hwo_pegasus}.}
    \label{fig:abundances}
\end{figure*}

\citet{zuckermanetal07-1} demonstrated that the bulk composition of exoplanetary bodies can be accurately measured, analogous to Solar-system meteorite studies, from spectroscopy of white dwarfs enriched with metals by the accretion of planetary debris (Fig.\,\ref{fig:abundances}). The studies carried out so far found a large diversity in the nature of the accreted planetary bodies: whereas many systems resemble the primitive CI chondrites in the Solar system, reflecting the abundances in their protoplanetary discs \citep{swanetal23-1, trierweileretal23-1} some are strongly enhanced in Fe and Ni, indicating differentiation \citep{gaensickeetal12-1, williamsetal25-1}; some in contrast appear to be crust fragments rich in Li, Na, Al, Ca and Ti \citep{hollandsetal21-1}; some exhibit an O-excess implying that they were rich in water or hydrated minerals \citep{farihietal13-2, trierweileretal25-1}; and a few are volatile-rich Kuiper-belt analogues \citep{xuetal17-1, sahuetal25-2}. Perhaps the most impactful discovery is the strongly enhanced abundance of s-process elements (Cu, Nb) in one system, indicative a second-generation exoplanet that formed out of the AGB ejecta (Williams et al. 2026 in press). The existence of second-generation exoplanets has been subject to many theoretical studies \citep{perets10-1, columbaetal23-1}, but their identification has remained elusive until now. 

Key scientific questions that the study of white dwarfs accreting planetary debris addresses include: 

\begin{itemize}
    \item \textbf{Is the solar system representative in terms of its composition?} Initial conditions are a critical input into all planet formation models, and currently the most detailed input for the abundances are measured from solar system meteorites. 
    
    \item \textbf{How common are water and volatile-rich exoplanetesimals?} Water delivery onto rocky planets that formed dry is a key ingredient for their potential habitability \citep{raymondetal04-1, chenetal19-1}. 

    \item \textbf{What light elements are sequestered into planetary cores?} It is well-established that the core of the Earth contains a small amount of elements lighter than those of the iron group \citep{dreibus+palme96-1}, but their nature remains debated \citep{hiroseetal21-1}. 
    
\end{itemize}

Detailed studies have so far been published for $\simeq30$ systems, and while small in number, these are \textit{the most direct measurement of the bulk abundances of exoplanets} and provide critically important inputs into planet formation models \citep{carter-bondetal12-1, bonsoretal23-1}. Whereas identifying white dwarfs has been a challenging task as they are intrinsically faint, and blend in among the much more numerous main-sequence stars and quasars with similar colours. \textit{Gaia}'s accurate astrometry swept away that limitation overnight, identifying a homogenous magnitude-limited sample $\simeq360\,000$ white dwarfs \citep{gentile-fusilloetal21-2}. The selection of debris-enriched white dwarfs is currently carried out as part of the massively multiplexed multi-object low-resolution optical spectroscopic surveys DESI \citep{desi22-1}; SDSS-V \citep{kollmeieretal26-1}; WEAVE \citep{jinetal24-1}; and 4MOST \citep{dejongetal19-1}. Within the next five years, these projects will produce a sample of $\simeq1000$ suitable for detailed bulk abundance studies, paving the pathway to an exoplanetary bulk abundance database comparable in size to our knowledge of solar-system meteorites \citep{nittleretal04-1}.

In conclusion, progress in answering the above questions is now only bottle-necked by the limited access to high-resolution ultraviolet spectroscopy, which contains, by far, the strongest transitions of many important elements, in particular volatiles (C, N, O, P, S). 

\subsubsection{Requirements}

\begin{figure*}
    \centering
    \includegraphics[width=0.7\linewidth]{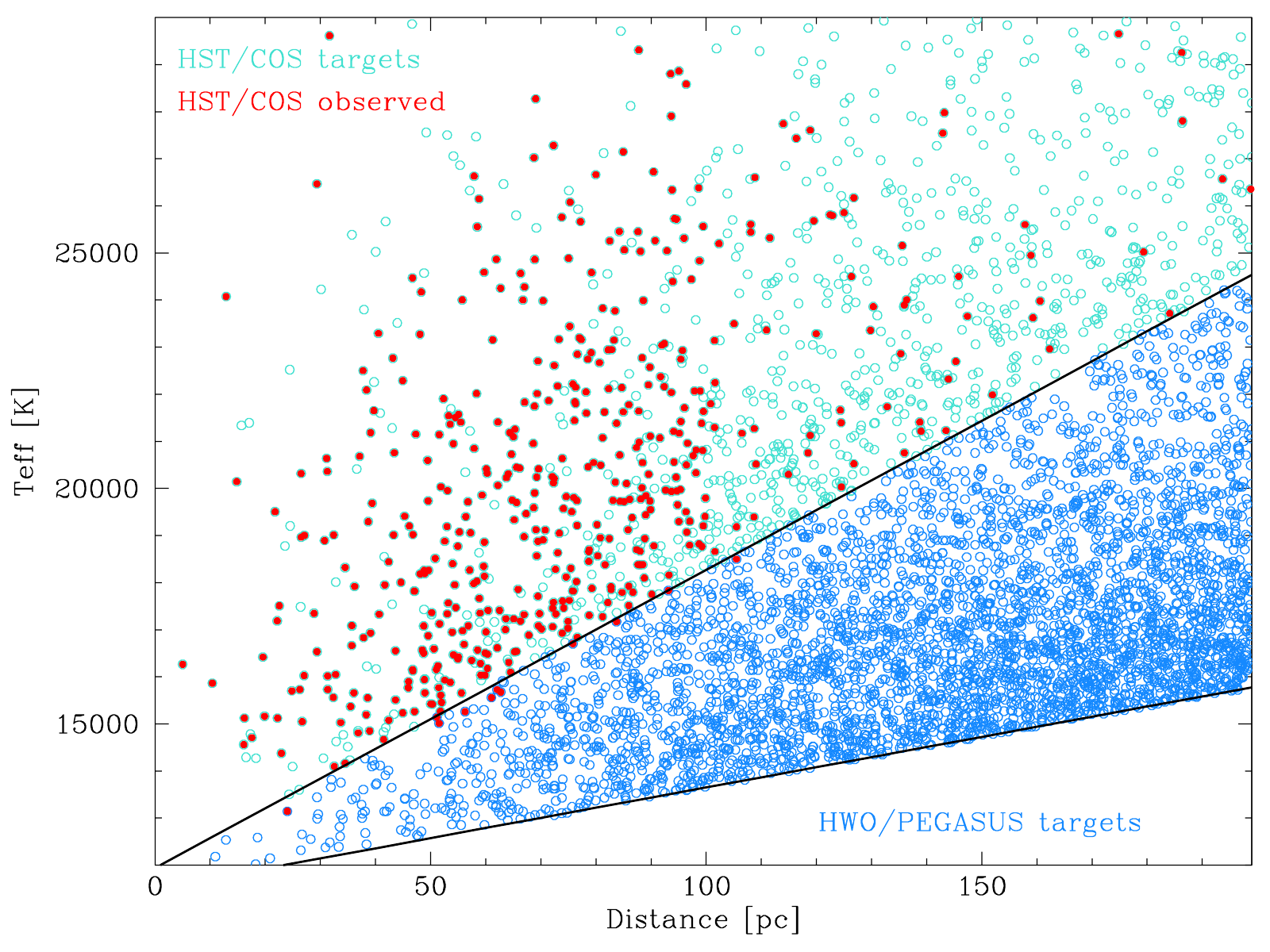}
    \caption{White dwarfs within 200\,pc that are accessible to \textit{HST}/COS G130M spectroscopy (mint circles) and that have already been observed (red dots). The much increased sensitivity of PEGASUS on HWO will enable a detailed statistical study of the bulk abundances of exoplanetary systems by observing systematically several systems spanning a wide range in stellar masses and ages. }
    \label{fig:wd_hwo_pegasus}
\end{figure*}

\begin{enumerate}
    \item {\it Wavelength coverage} -- Required: 100--400\,nm; Optimal: 100--900\,nm. Probing the wavelength range $<400$\,nm is particularly important because it contains the strongest lines of the tracers of planetary cores (Fe, Ni, Cr), mantles (Mg, Si), crusts (Al, Na, Ca, Ti), and volatiles (C, N, O, P, S). These wavelengths also enable the measurement of s-process elements that indicate second-generation planet formation. For relevant line lists see Table~2 of (\citealt{gaensickeetal12-1}; FUV) and Table~2 of (\citealt{kleinetal11-1}; NUV).
    \item {\it Spectral resolution} -- Required: 20\,000--40\,000;  Optimal: 20\,000--80\,000. These values are required to resolve line blends and disentangle the photospheric lines from interstellar transitions \citep[e.g.][]{wilsonetal19-1, kleinetal21-1}.
    \item {\it Sensitivity} -- Targets to be observed have a typical FUV magnitudes in the range 15--18. 
    Measuring the bulk abundances of exoplanetary systems requires spectroscopy of debris-enriched white dwarfs with a S/N ratio of $\simeq30$. Observing a sample of a few hundred systems, necessary both to establish robust statistics on the bulk abundances and to identify outliers, requires reaching that S/N ratio within 2--3\,h exposures with typical continuum flux levels of $5\times10^{-15}\,\mathrm{erg\,cm^{-2}s^{-1}\text{\AA}^{-1}}$. The sensitivity of PEGASUS on a HWO-like mission will enlarge the number of white dwarfs accessible for high-resolution UV spectroscopy by an order of magnitude (Fig.\,\ref{fig:wd_binaries}).
\end{enumerate}

Currently, only \textit{HST} has the capability of partially addressing the above key questions. Because of its sensitivity, COS is the workhorse instrument \citep{gaensickeetal12-1}, however, its resolution of $\simeq18\,000$ is often insufficient to resolve line blends. The E140M grating of STIS overcomes this limitation, however, because if its low throughput only a handful of systems could be observed successfully (e.g. \citealt{wilsonetal19-1, johnsonetal22-1}). Both STIS and COS have very limited capabilities in the near-UV (180$-$320\,nm), either in terms of resolution, wavelength coverage, or throughput, even though this wavelength range is extremely rich in diagnostic transitions (e.g. \citealt{hollandsetal22-1}). 

The NASA mid-ex mission \textit{UVEX} \citep{kulkarnietal21-1} will provide ultraviolet spectroscopy over the wavelength range $\simeq115-265$\,nm with much improved sensitivity compared to \textit{HST}. However, with a resolution of only $\simeq2500$, \textit{UVEX} spectroscopy will not allow the detailed analysis of bulk abundances due to the strong line blending. 
\subsection{Protoplanetary disks}

Protoplanetary disks surrounding the youngest stars (ages $\lesssim$~10 Myr) consist of gas and dust.  While the dust mass is thought to be only 1/100 of the gas mass, its signature has historically been much easier to observe so that the dust content is generally well studied,
particularly in the inner disk, where solar-system like planets form.  Probing the gas content is far more challenging, and requires several different methods: the cold outer-disk gas is now routinely measured by ALMA \citep[e.g.,][]{miotello2023}, while some of the inner disk gas can be seen in CO and H$_{2}$O using high-resolution ground-based spectroscopy at 2~--~5 $\mu$m~\citep{banzatti17}. In addition, mid-IR spectroscopy with $JWST$ recently revealed a wealth of molecules (see, e.g., \citealt{arulanantham24} and references therein), though its spectral and spatial resolution are inferior to those achievable by future UV spectroscopy.

UV spectroscopy provides a unique tool for observing the molecular gas in the inner regions of protoplanetary disks: the strongest electronic band systems of H$_{2}$ and CO reside in the 100 -- 170~nm wavelength range (e.g., \citealt{herczeg02,france11}).  UV-fluorescent H$_{2}$ spectra are sensitive to gas surface densities lower than 10$^{-6}$ g cm$^{-2}$, making them an extremely useful probe of remnant gas at r $<$ 10 AU during the disk dispersal stage after planet cores have formed. In cases where mid-IR CO spectra or traditional accretion diagnostics (e.g. H$\alpha$ equivalent widths) suggest that the inner gas disk has dissipated, far-UV H$_{2}$ observations can offer unambiguous evidence for the presence of a remnant molecular disk and ongoing protostellar mass accretion \citep{ingleby11,france12,arulanantham18,alcala19}.

At the same time, in the innermost regions of disks the very energetic exchanges of material and momentum between the forming star and the disk happen. These accretion and ejection mechanisms have been studied with optical, infrared, and UV spectroscopy for many years \citep[e.g.,][]{Hartmann_2016,pascucci22}. Their properties are studied to determine the underlying mechanism driving the evolution of disks, and thus impacting the formation of planets \citep[e.g.,][]{manara22}. 
UV spectroscopy has been pivotal in studying these mechanisms, as it directly traces the excess emission due to accretion, mainly observable at $\lambda<360$ nm \citep{Calvet1998} and several ejection-tracing lines \citep{xu2021}. 
The recent HST/ULLYSES survey \citep{duval20} has clearly demonstrated the power of UV spectroscopy in this field \citep[e.g.,][]{espaillat22}, and motivates further studies towards less bright and more distant targets.

Some key goals for UV studies of protoplanetary disks and young stars were recently summarized by~\citet{france2026}:
\begin{itemize}
   \item What are the mass outflow rates of protoplanetary disk winds and what mechanisms drive them?
   \item What are the abundances and kinematics of dominant atomic and molecular species in the inner regions of protoplanetary disks? 
   \item How do mass accretion flows change with time? 
 \end{itemize}

\subsubsection{Far-UV Molecular Emission from Protoplanetary Disks}

\begin{figure*}
\begin{center}
\includegraphics[width=0.8\textwidth]{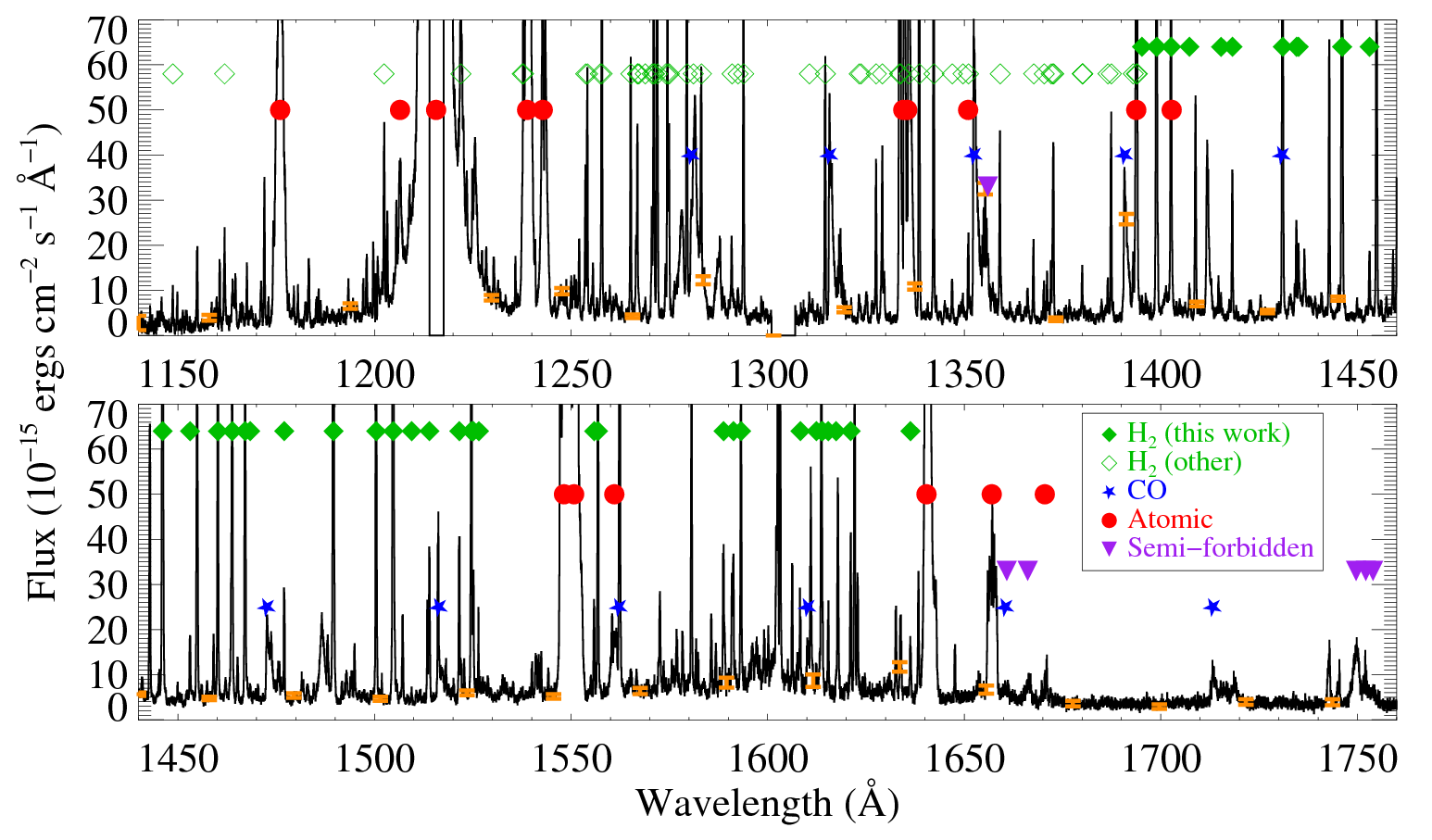}
\end{center}
\caption{A representative far-ultraviolet molecular spectrum of a protoplanetary disk, with H$_2$ and CO fluorescent emission lines shown in green and blue, with strong resonance and semi-forbidden lines noted in red and purple.   
Figure adapted from \citet{france23}.
\label{fig:H2_bump}}
\end{figure*}

The excitation conditions of the gas in protoplanetary disks change with radius. 
Thus, a number of different methods are needed to provide an inventory that constrains theories. Pivotal in this context is the molecular hydrogen content, which comprises the disk's bulk mass. UV H$_2$ lines are powerful diagnostics and ideally complement JWST near- and mid-IR spectral maps (e.g., \citealt{schwarz25} and references therein).
The FUV H$_2$ lines are produced in the warm ($T(H_2)$ $\sim$~1000 K) surface of the inner disk \citep[0.1-10 au,][]{herczeg04,france12,hoadley15} 
and are excited 
by Ly$\alpha$ pumped low-lying excited electronic states (Werner and Lyman bands).   
FUV H$_2$ emission lines are seen in every actively accreting young stellar system (see Figure \ref{fig:H2_bump} top; \citealt{france12,france23}).  
They mostly trace warm gas, but also contain outflow components, too ~\citep{kalscheur25}, quite similar to the somewhat cooler H$_2$ traced by JWST.

UV spectroscopy even provides an (indirect) tracer of water in the inner region since the Ly$\alpha$-driven dissociation of water produces a characteristic, quasi-continuum emission spectrum with a peak near 1600~\AA~that is seen in about half of the Classical T Tauri Stars (CTTSs) and almost 100\% of the systems with cleared inner disks (see Figure \ref{fig:H2_bump} bottom; \citealt{france12,france23}). The exact physical mechanism for this emission feature is still under investigation.
 
Carbon monoxide (CO) is the second most abundant species in protoplanetary disks, and it also shows photo-excited emission features in the far-ultraviolet spectra of several CTTSs~\citep{france11b}. Like the H$_2$ emission lines described above, the CO is also ``pumped'' by accretion-generated Ly$\alpha$ photons, through several rotational transitions of the the $A^1\Pi - X^1 \Sigma^+$ 
(Fourth Positive) electronic transition system.  Unlike the fluorescent H$_2$ emission, the CO traces cooler regions ($T\sim200-500$\,K) of the disk surface~\citep{Schindhelm_2012, arulanantham21}.   The fluorescent CO lines are generally weaker than those of H$_2$, likely due to a combination of lower abundance, less ability to self-shield, and the reduced stretngth of the Ly$\alpha$ at greater distances from the central star.   Detailed kinematic maps of both molecules may be possible using future high-resolution and spectral imaging capabilities aboard the Habitable Worlds Observatory.

\begin{figure}
\begin{center}
\includegraphics[width=\linewidth,trim=15 15 20 5, clip]{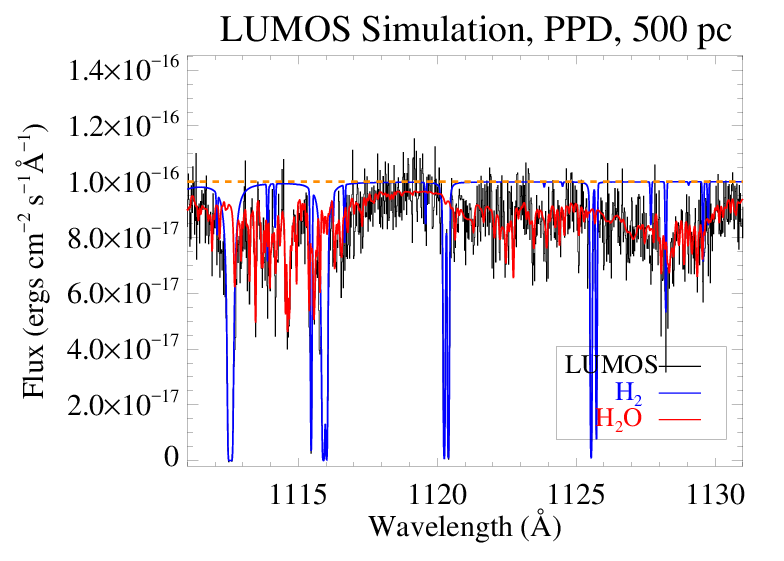}
\end{center}
\caption{Spectral simulation the 111.1 -- 113.2~nm\ spectrum of an edge-on protoplanetary disk from the LUVOIR concept~\citep{2017_LUVOIR}, a spectral region containing strong lines of H$_2$ and H$_2$O \citep{france14,cauley21}.  
\label{fig:CO_emiss}}
\end{figure}   

\subsubsection{UV spectroscopy from accretion and ejection processes in young stars}

Far- and near-ultraviolet (FUV and NUV, respectively) spectroscopy traces continuum and line emission tracing the accretion flows, ejection mechanisms, mass and angular momentum transport, and disk irradiation \citep[e.g.,][]{schneider20}. The continuum emission from the post-shock accretion flow is dominant over the photospheric one at FUV wavelengths, being particularly important to measure both strong and low accretors \citep{ingleby11b,alcala19}. While low-resolution could be sufficient to measure the overall flux in the NUV and FUV, the strong emission lines in the spectrum can impact the observed continuum flux if not properly resolved. These hot FUV lines, for example Si IV, C IV, N V, and He II, are tracers of the hotter gas in the postshock and immediate preshock regions, and are known to have line luminosity correlated with the accretion luminosity \citep{france14,Robinson_2019}.  

P Cygni absorption profiles have been observed in several FUV emission lines, such as N I, Fe II, C II, Si II, Si III, and C IV, when observed at high-resolution \citep{herczeg05,cauley2016}. These profiles trace the winds emitted from the inner regions of disks, and are observed to have similar profiles as other well known wind tracers \citep{espaillat22}.

\subsubsection{Requirements}

A high-sensitivity space observatory with high-resolution capability (R $\geq$ 45,000) enables transformative absorption-line studies of high-inclination ($i$ $>$ 60 degrees) disks. Absorption-line spectroscopy through high-inclination disks, currently limited on HST to a small number of bright stars (e.g., \citealt{roberge00,roberge01,france14,cauley21}), is important because the strongest molecular absorption systems of key disk volatile species such as CO, OH, H$_{2}$O, CO$_{2}$, and CH$_{4}$ have their peak absorption cross-sections in the 100 -- 170~nm range.  Access to wavelengths from 100 -- 115~nm is particularly critical as (1) absorption from cool H$_{2}$ (T $<$ 500 K) is restricted to $\lambda$ $<$ 111~nm, (2) the strongest absorption bands of H$_{2}$O reside between 111 -- 113~nm (Fig. \ref{fig:CO_emiss}), and (3) the CO-dissociation bands reside at $\lambda$ $<$ 108~nm.  
For disk wind studies, a spectral resolving power above 45,000 is needed to resolve the emission lines in the FUV, disentangle them from the continuum, and study their profiles to study winds.

\begin{enumerate}
    \item {\it Wavelength coverage} -- Required: 105--165\,nm; Optimal: 100--400\,nm. 
    \item {\it Spectral resolution} -- Required: 40\,000--60\,000;  Optimal: 40\,000--120\,000. 
    \item {\it Sensitivity} -- Targets to be observed have a typical V magnitudes in the range 10--16.
    \item {\it Instrumental Effective Area} -- Required: 5000\,cm$^{2}$; Optimal: 10\,000\,cm$^{2}$
    \item {\it Instrumental Background Rate} -- Required: 5\,Hz/cm$^{2}$; Optimal: 1\,Hz/cm$^{2}$
    \item {\it Wavelength Stability (1\,h observation)} -- Required: 2\,km\,s$^{-1}$; Optimal: 1\,km\,s$^{-1}$. These values can be achieved through pointing knowledge-based focal plane correction in ground processing.
\end{enumerate}

\begin{figure}
\begin{center}
\includegraphics[width=\linewidth,trim=17 15 20 3, clip]{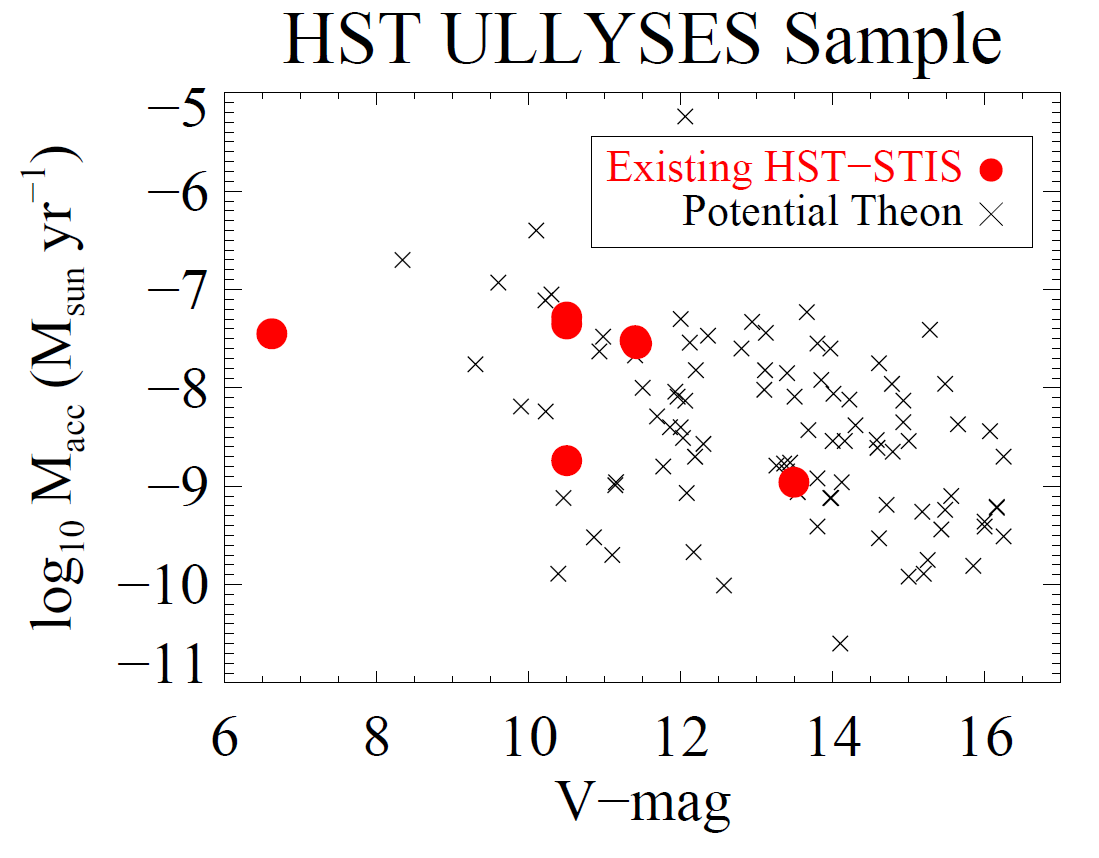}
\end{center}
\caption{Targets in the HST ULLYSES archive \citep{Roman-Duval25}, sorted by V-magnitude and mass-accretion rate.  These observations were obtained with the medium resolution ($R$~$\sim$~14,000 - 20,000) modes (black x's), however, only a small number have been observed at high-resolution ($R$~$>$~40,000; red circles).  The combination of high-resolution and high-sensitivity offered by Theon will enable a factor of $\approx$10 increase in the sample size for high-resolution disk and accretion studies.
\label{fig:highres_archive}}
\end{figure}

\subsection{Solar System planetary atmosphere and icy moons}

Numerous solar system exploration missions have been performed, and they have provided highly precise data through in-situ observations. However, to observe the global planetary atmospheric escape phenomena and its long-term temporal variations, remote sensing via ground-based or space telescopes is still extremely effective; specifically, for observations in the UV spectrum, a space telescope is essential.

One of the primary objectives for solar system science in UV spectroscopy is to investigate the atmospheric escape of terrestrial planets like Mars and Venus to know how these planets lost their water and evolved over billions of years. Using UV spectroscopy to observe emission lines—such as those from hydrogen and oxygen, it would be possible to clarify the mechanisms that determine atmospheric stability and long-term evolution, both of which are critical to a planet's habitability. Another objective is to observe water vapor plumes that erupt from Jupiter’s icy moon, Europa. UV spectroscopy by space telescope enables long-term monitoring to capture the temporal variability and frequency of these eruptions. These data are vital for confirming the existence of a subsurface ocean and characterizing its chemical composition

\subsubsection{Martian exosphere}

Previous observations have successfully detected airglow emissions from hydrogen, oxygen, and carbon atoms in the exosphere. The MAVEN/IUVS instrument has successfully observed exospheric emissions from species such as atomic hydrogen \citep{Chaffin2015}. Moving forward, it will likely become possible to measure the deuterium-to-hydrogen (D/H) ratio with higher sensitivity and to observe nitrogen atoms and ions by extending the lower limit of the wavelength range to approximately 100\,nm.

MAVEN/IUVS has successfully resolved the Lyman-$\alpha$ emissions of H and D \citep {Clarke2017}, providing a breakthrough in our understanding of Martian water loss. Further observations are imperative to fully characterize the planet's atmospheric evolution. The seasonal fluctuations in H and D abundances identified highlight the need for continuous, high-sensitivity monitoring to distinguish between regular seasonal cycles and the impact of stochastic events, such as global dust storms, which can significantly accelerate hydrogen escape. Sustained and more precise measurements of the D/H ratio remain essential for quantifying the fractionation processes at the exobase, allowing scientists to more accurately reconstruct the total volume of water Mars has lost over its geological history.
The observed intensity of H emission and D emission is 3--5\,kR and 0.1--0.3\,kR at the limb. The spectral resolution of 40\,000 is required and 120\,000 is ideal because of the large difference of the intensities of the two (D and H) lines.

While MAVEN has provided unprecedented local data, its observations are inherently limited by its orbital geometry and mission lifespan, making long-term (multiple martian years and solar cycles) monitoring from Earth-orbiting space telescopes highly significant. It is crucial for capturing the full temporal evolution of H and D emissions and for understanding how the D/H ratio responds to infrequent but transformative events like global dust storms. By complementing in-situ observation by orbitors and long-term remote sensing by space telescopes, we can more effectively decouple short-term variability from the long-term trends of atmospheric escape.

\subsubsection{Vapor plume activities on icy moons}

Building on the landmark study \citep{Roth2014}, the discovery of water vapor plumes at Europa’s south pole was achieved using the Hubble Space Telescope’s (HST) Space Telescope Imaging Spectrograph (STIS) to detect far-ultraviolet emissions from atomic oxygen and hydrogen.

The emission intensity of H and O is 500\,R and 60\,R, respectively.

Observing Europa’s plumes with a next-generation Earth-orbiting UV telescope provides a long-term temporal baseline that captures the full evolution of these transient events independently of in-situ mission schedules. While localized probes offer high-resolution snapshots, they cannot provide the continuous, synoptic monitoring required to correlate episodic eruptions with orbital tidal phases or identify multi-year trends in outgassing frequency \citep{Roth2014}. Furthermore, a powerful Earth-based facility allows for the persistent detection of other species across the entire satellite disk, enabling to quantify the total mass-loss rate from the subsurface ocean.

\subsubsection{Requirements}

\begin{enumerate}
    \item {\it Wavelength coverage} -- Required: 120--140\,nm, to measure deuterium and hydrogen lines; Optimal: 100--170\,nm, to include minor species like C and N and their ions.
    \item {\it Spectral resolution} -- Required: 40,000--60,000;  Optimal: 40,000--120,000. 
    \item {\it Sensitivity} -- Targets to be observed have a typical emission intensity in the range 60-5,000 Rayleigh
    \item {\it Angular resolution} -- 0.1 arcsec
\end{enumerate}

\subsection{Small bodies in the Solar System}

 The detection of hundreds of exoplanetary systems has not yet provided a complete understanding of how planets form and how life arises. Small bodies are key to understand the formation and evolution of planetary systems. Comets, in particular, are considered pristine remnants of the planetary formation process. They remained stored and mostly unaltered in very cold reservoirs since their formation. They thus retain crucial insights into the evolution of volatile materials across the protoplanetary disk. Furthermore, from an astrobiological perspective, comets are significant as they may have played a role in delivering water and organic compounds to terrestrial planets, thereby influencing the emergence of life. While solar system comets have been observed for decades, we only recently discovered the first interstellar comet \citep{Williams2017} and know of only three of them so far. Interstellar objects are small bodies formed around other stars and then ejected, likely during planetary migration events. Some of them then enter our solar system, where they provide a unique opportunity to probe in detail material formed in another planetary system. 

\subsubsection{Measuring isotopic ratios in solar system and interstellar comets}

The measurement of isotopic ratios in comets is essential for understanding the conditions that prevailed during the formation of our solar system. These ratios are sensitive to the local conditions in which the comets formed, making them a valuable tool for reconstructing the early history of the planetary systems. Additionally, isotopic ratios can help us determine the role of comets in delivering water to the early Earth. However, due to the complexity of the measurement process, obtaining accurate isotopic ratios in comets remains a significant challenge. Despite the challenges involved, recent decades have seen significant advancements in measuring the isotopic ratios of various elements, including hydrogen, nitrogen, carbon, and oxygen, in solar system comets \citep{Biver2024}.  However, current measurements still suffer from large uncertainties or possible systematic effect. This is for example the case of the D/H ratio, which is crucial to determine the origin of Earth's water. Measurements of the D/H ratio in solar system comets are limited and often suffer from systematics due to different techniques, fields of view, or molecules used. To date, the D/H ratio has only been measured in a few comets using the same technique, and even then, the results are not always consistent \citep{Altwegg2015,Biver2016,Lis2019}. For interstellar comets, isotopic ratios of H, C, and N were only recently measured for 3I/ATLAS, providing key clues on the formation environment of the object and informing the planetary formation history in the Milky Way \citep{Salazar2026,Cordiner2026,Opitom2026}. Determination of isotopic ratio in other interstellar comets in the future will be key to determine their formation conditions. But these might be too faint for the measurements to be made with existing instruments. Current instruments thus do not allow us to gather a large and homogeneous sample of isotopic ratio measurements in solar system and interstellar comets with a single instrument. 

The near-UV, particularly the 300-420 nm range, contains some of the brightest emission lines in cometary spectra. This includes the OH bands around 308 and 312 nm and the CN band around 388 nm that have been used successfully in the past to measure these fours isotopic ratios D/H, $^{14}$N/$^{15}$N, $^{12}$C/$^{13}$C, and $^{16}$O/$^{18}$O \citep{Hutsemekers2008,Manfroid2009}. This range also includes emission from the N$_2^+$ ion, from which isotopic ratio of N in N$_2$ can be determined, which is critical to understand nitrogen chemistry in protoplanetary disks and has so far only been done for one comet with the Rosetta space mission. While some of these measurements can be made from the ground, they can only be done for extremely bright comets, particularly for the D/H, which prevents us from building large homogeneous samples necessary to interpret the values measured for isotopic ratios. At these wavelengths, the Earth atmosphere absorbs a significant fraction of the light, particularly since comets are usually at their brightest when near the Sun and have to be observed at high airmass. A high spectral resolution spectrograph in space covering the near UV would thus revolutionize our measurements of isotopic ratios in solar system and interstellar comets.

\subsubsection{Requirements}

\begin{enumerate}
    \item {\it Wavelength coverage} -- Required: 310--420\,nm; Optimal: 300--420\,nm. 
    \item {\it Spectral resolution} -- Required: 40\,000--50\,000;  Optimal: 40\,000--120\,000. 
    \item {\it Sensitivity} -- Targets to be observed have a typical V magnitudes in the range 5--15.
    \item {\it Additional requirements} -- Ability to track moving targets; large entrance slit/fiber, to gather a much of the light as possible for an extended target.
\end{enumerate}

\section{Science Pillar II. Stellar lives and deaths at their extreme}
\label{sec: Pillar II}

\subsection{The first stars and the origin of the elements}

The first stars, also known as Population~III (Pop~III) stars, 
formed from primordial hydrogen, helium, and lithium,
just a few hundred million years after the Big Bang.
They produced the first metals, 
the first ionizing radiation after the cosmic dark ages,
and perhaps the first black holes.
The first stars are predicted to have been massive and short lived, and no surviving metal-free stars have yet been found in the Milky Way today.
Active searches for the first stars are underway in high-redshift systems
(e.g., \citealt{Saccardi23b,maiolino24,wang24,nakajima25,zackrisson26}).

The metals produced by the first stars in the Milky Way 
may be found in second-generation stars located among the
metal-poor stars in the bulge, halo, and dwarf galaxies
(e.g., \citealt{tumlinson10,frebel15araa,elbadry18}).
The chemical abundance pattern
of each second-generation star can be compared with supernova model predictions
(e.g., \citealt{heger10})
to characterize the nature and end state of one of the first stars (see also Section~\ref{sec:low-z}).
Key insights include
the initial mass function of the first stars that produced metals
(e.g., \citealt{hartwig15,ishigaki18,rossi21}),
detailed properties of the first star supernova explosions
(e.g., \citealt{tominaga14,ezzeddine19,koutsouridou23}),
the low-metallicity tail of the metallicity distribution function
(e.g., \citealt{schorck09,youakim20,bonifacio21}),
and
the birthplaces of the first stars 
(e.g., \citealt{sestito19,vanni24,chiti26}).

Some of the more pressing open questions in this field are: 
\begin{itemize}
   \item What were the chemical products of the first generation of stars? 
   \item What was the mass function of the first stars?
   \item Were low-mass, long-lived metal-free stars able to form?
\end{itemize}

\begin{figure*}
\begin{center}
\includegraphics[width=1.0\textwidth]{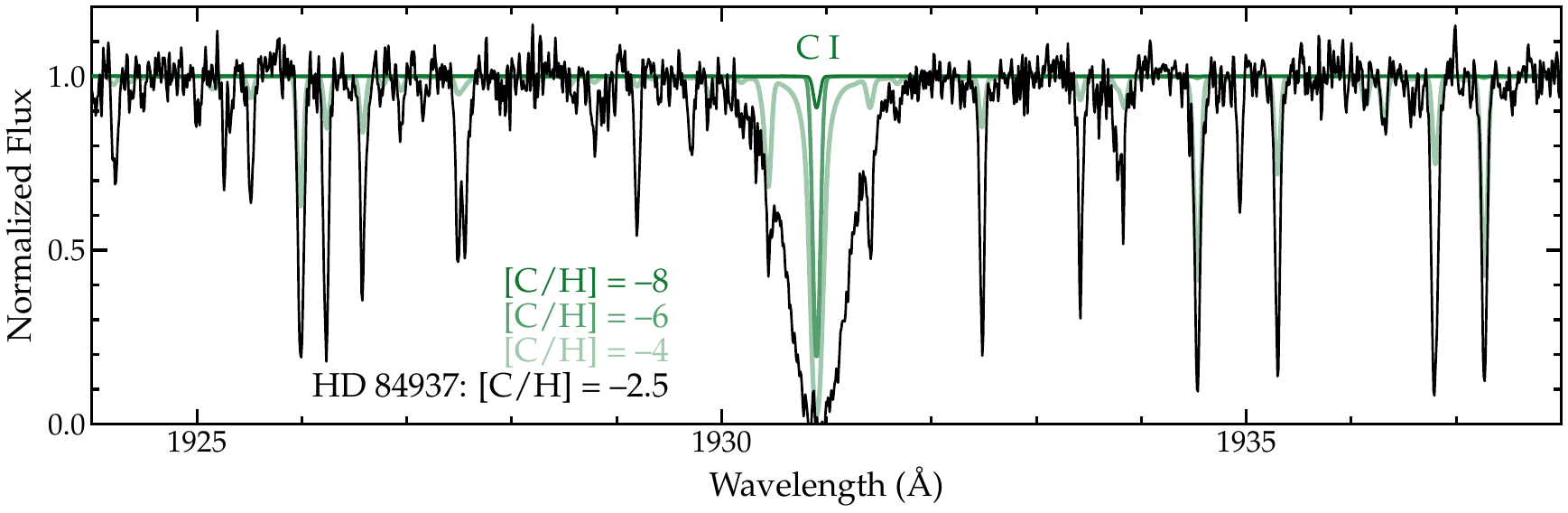}
\end{center}
\caption{Simulated UV spectra covering the C~\textsc{i} line 
at 1930~\AA, analogous to
Figure~2 in \citet{Roederer26}.
The star is a 
typical G-type subgiant star with 
$T_{\rm eff}$ = 5500~K and $\log g$ = 3.5.
These synthetic spectra adopt 
a 1D model atmosphere in local thermodynamic equilibrium (LTE)
and assume that LTE holds 
in the line-forming layers of the atmosphere.
Spectra have been smoothed to a resolving power of 40,000.
The black line is the spectrum of
the bright, F-type metal-poor star 
\mbox{HD~84937}
(HST/STIS/E230H; \citealt{peterson17}).
\label{fig:ump_carbon}}
\end{figure*}

High-resolution UV spectra of low-metallicity stars
present many more lines for analysis than 
optical or near-infrared spectra do.
Typically, only $\approx$~5--10 metals can be detected
in the optical spectra of ultra-metal-poor second-generation stars
with metallicities [Fe/H] $< -4$
(e.g., \citealt{aoki06,caffau12,aguado18}).
These limited chemical inventories 
cannot uniquely identify model characteristics of the Pop~III stars and supernovae, and
Pop~III model constraints improve as more elements are detected
(e.g., \citealt{tominaga14,salvadori19,koutsouridou23,zhang26}).
Some of the elements that offer the
greatest constraining power for model predictions
are only detectable in the UV, 
below the atmospheric cutoff.
These elements include
B~\textsc{i} (with strongest transitions $\lambda\lambda$1825, 2088, 2496),
C~\textsc{i} ($\lambda$1930),
Mg~\textsc{ii} ($\lambda\lambda$2795, 2802),
Si~\textsc{i} ($\lambda\lambda$1850, 2124),
P~\textsc{i} ($\lambda\lambda$1859, 2136),
S~\textsc{i} ($\lambda\lambda$1807, 1820, 1826),
Cr~\textsc{ii} ($\lambda\lambda$2055, 2061, 2065),
Mn~\textsc{ii} ($\lambda\lambda$2576, 2593, 2605),
Fe~\textsc{ii} ($\lambda\lambda$2343, 2382, 2395, 2404, 2585, 2598, 2599, 2607, 2611),
Co~\textsc{ii} ($\lambda\lambda$2286, 2580), 
Ni~\textsc{ii} ($\lambda\lambda$2165, 2216), and
Zn~\textsc{ii} ($\lambda\lambda$2025, 2062).
Figure~\ref{fig:ump_carbon} illustrates a synthesis of the
C~\textsc{i} line at 1930~\AA,
demonstrating that this line would enable the detection of
carbon in a second-generation star if [C/H] $> -8$ or so.
UV and optical spectra combined may enable the detection of 20--25~elements
in second-generation stars.

\begin{figure*}
\begin{center}
\includegraphics[width=0.8\textwidth]{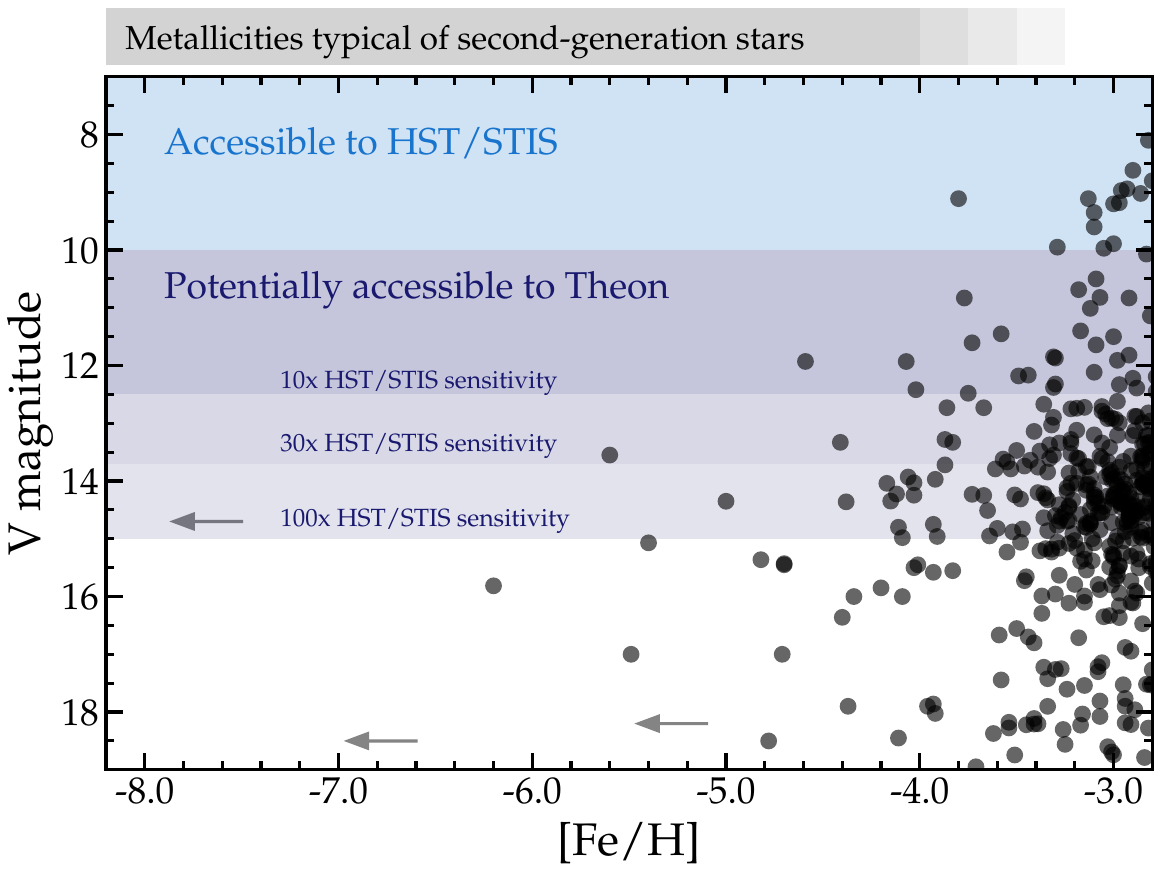}
\end{center}
\caption{Comparison of $V$ magnitudes and
metallicities for known metal-poor stars.
The approximate magnitude ranges accessible
to HST/STIS and Theon are indicated,
along with various estimates of the
sensitivity gain of Theon relative to HST/STIS.
Data collected from JINAbase \citep{abohalima18}
and subsequent literature samples. The left-pointing arrows indicate metallicity upper limits determined from optica spectra.\label{fig:umpstars}}
\end{figure*}   

Figure~\ref{fig:umpstars} illustrates the
potential gain that a UV mission/instrument 10 times (e.g. Theon) or 100 times (e.g. HWO/PEGASUS) more sensitive than HST could offer 
in terms of sample selection.
Each point in Figure~\ref{fig:umpstars} 
marks a known metal-poor star in the Milky Way.
In general,
second-generation stars are more likely to be
found among the lowest metallicities.
Only one potential second-generation star is
accessible to, and has been observed with, STIS
on HST, \mbox{BD~+44$^{\circ}$493}
\citep{placco14b}.
Depending on the sensitivity,
the gain relative to HST could be 
$\approx$~3 additional stars
with [Fe/H] $< -4$ for a 10x gain
to $\approx$~15--20 additional stars
with [Fe/H] $< -4$ for a 100x gain.
Thus, HWO/PEGASUS has the potential to increase the UV high-resolution sample of low-metallicity stars from 1 to $\sim20$. 
Each of these stars may retain the 
metals produced by a single Pop~III supernova,
so the gain in sample size 
translates directly to the number
of Pop~III stars sampled.

 HWO will also enable UV observations of young very-metal-poor stars that could be considered as local analogues of PopIII stars. UV observations of massive stars are discussed in Sect. \ref{sec: massive stars}.

\subsubsection{The r-process and the origin of the heavy elements}

The r-process, or rapid neutron-capture process, is one of the fundamental ways that stars produce the heaviest elements on the periodic table, including silver, gold, platinum, and uranium \citep{Roederer22}. The r-process is responsible for forming many of the heaviest elements by rapidly adding neutrons to lighter atomic nuclei \citep{Burbidge57,Cowan21}. Understanding the r-process helps explain the cosmic origin of elements that are essential to the composition of planets and, by extension, life. The patterns and amounts of r-process elements found in stars can reveal the physical conditions, sites (such as neutron-star mergers or exotic supernovae), and environments where r-process events occurred \citep{Ji18,Yong21}. This provides insight into the processes that shaped the chemical evolution of galaxies \citep{Kobayashi:2020,Molero23}. The abundance patterns of r-process elements in stars can be compared with theoretical models to reveal the nature of the astrophysical sites and the physics governing the r-process \citep{Eichler15,Holmbeck23}. Neutron-star mergers are the one confirmed astrophysical site where r-process nucleosynthesis occurs \citep{Ji16,Drout17,Tanvir17}. Understanding the r-process addresses major themes in astrophysics, such as ``New Messengers and New Physics'' and ``Cosmic Ecosystems,'' as identified by the Astro2020 Decadal Survey.

Currently, the only source of high-resolution UV spectra for r-process studies is HST/STIS \citep{Roederer22}. Although more than 200 r-process-enhanced stars are currently known, only about 10 such stars are bright enough ($V < 10$ mag) to have been observed with STIS over the last 25 years. 
In contrast, heavy elements detectable in optical spectra (e.g., Sr, Y, Zr, Ba, La, Nd, Eu) have been studied in 3,000--6,000 stars, highlighting the limitation of UV observations with HST \citep{Hinkel14}. 
The potential for HWO to unveil the r-process and the origin of the heavy elements is described by \citet{Roederer26b}.

\begin{figure}
    \centering
    \includegraphics[width=1\linewidth]{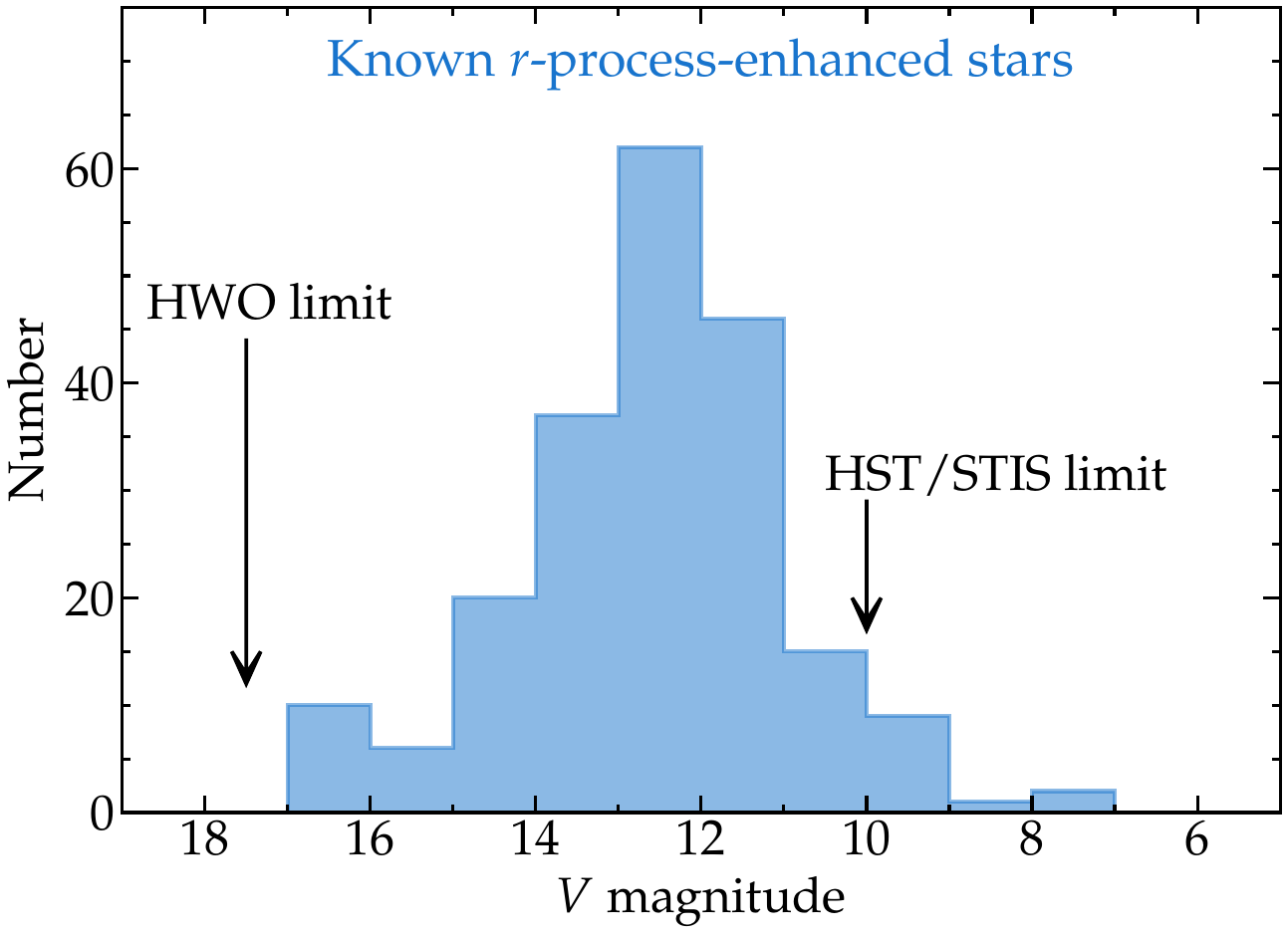}
    \caption{Magnitudes distribution of known stars enhanced by r-process. The limits for HST and HWO are marked. Figure reproduced from \citet{Roederer26b}.}
    \label{fig:maghist}
\end{figure}

Figure \ref{fig:maghist} shows the magnitudes distribution of known stars enhanced by r-process, and the current or future magnitude limits for HST or HWO. HWO can enable UV high-resolution spectroscopy for virtually all known r-process-enhanced stars.

\subsubsection{Requirements}

To first approximation, the requirements for the first stars science case and the r-process science case are the same.
The observing strategy is to collect 
high-resolution and high-S/N UV spectra of
the most metal-poor stars known
to enable the detection of as many lines as possible.
Spectral coverage should span 1700 to 3100~\AA\ 
in as few setups as possible.
(1700~\AA\ is where the photospheric flux drops to zero,
and 3100~\AA\ is accessible from Earth's surface.  Only a limited number of ground-based high-resolution spectrographs cover wavelengths between 3100~\AA\ and 3600~\AA, so this range represents a useful extension.)
A minimum spectral resolving power of approximately
30,000 is needed to detect weak lines, and
100,000 is ideal to fully resolve the line profiles
and distinguish the stellar lines
from any potential interstellar absorption
caused by the same resonance-line transitions in the ISM.~
A minimum S/N level of approximately 50 is necessary 
to detect the weak lines, and 100 is ideal.
The spectra of the target stars are expected to be 
stable across time, so observations could be conducted
at any time and co-added to achieve the desired S/N level.
Typical targets will have $V$ magnitudes in the range of 
approximately 10 to 15, which corresponds to GALEX
$NUV$ magnitudes of approximately 13 to 20 for F- and G-type stars.
Targets would be spread across the sky, so there is no
gain from multi-object spectroscopic capability.
Targets typically would not be found in crowded fields
(neighboring stars within $\approx$~0.1'' with 
$\Delta$mag $<$ 5).
Sample sizes may range from one (individual targets of interest)
to a few tens (a representative sample of r-process-enhanced stars).\\\\

The instrument requirements are summarized as follows: \\

\begin{enumerate}
    \item {\it Wavelength coverage} -- Required: 170--310\,nm; Optimal: 170--360\,nm. 
    \item {\it Spectral resolution} -- Required: 40\,000--60\,000;  Optimal: 30\,000--100\,000. 
    \item {\it Sensitivity} -- Targets to be observed have a typical $V$ magnitude in the range 10--15, corresponding to GALEX NUV magnitudes of approximately 13--20 for F- and G-type stars.
    \item {\it Wavelength stability} -- Maximum wavelength drift allowed over the duration of one observation: $\sim0.01$\,\AA; maximum allowed uncertainty of the wavelength solution: $\sim0.005$\,\AA.
    \item {\it Signal-to-noise} -- Minimum S/N level of approximately 50 to detect weak lines; S/N of 100 is ideal.
    \item {\it Flux} -- Minimum flux value for a given wavelength: $\sim 2 \cdot 10^{-22}$\,erg\,s$^{-1}$\,cm$^{-2}$\,\AA$^{-1}$ (calculated from NUV\,$\sim$\,15 in Vega magnitudes)
\end{enumerate}

\subsection{Massive stars}
\label{sec: massive stars}

\begin{figure*}[!ht]
\begin{center}
\includegraphics[width=0.4\textwidth, angle = 90]{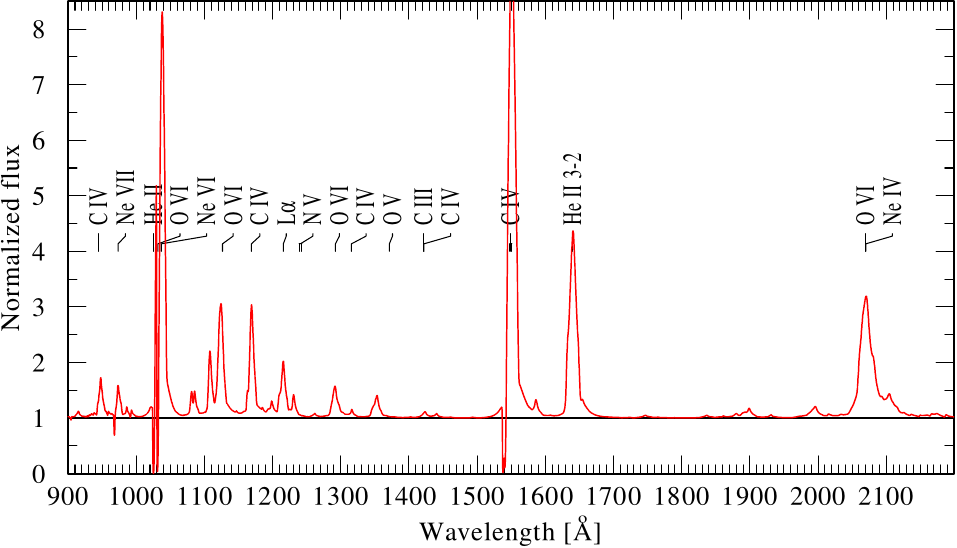}
\end{center}
\caption{Synthetic normalized UV spectrum of a highly evolved  WR  star of carbon sub-type  with parameters $T_\ast=199.5$\,kK, $\log{L/L_\odot}=5.3, \log{\dot{M}}=-5.7\, [M\odot$\,yr$^{-1}]$, $v_{\rm wind}=2000$\,km\,s$^{-1}$, and $Z=0.07\,Z_\odot$ computed with the PoWR  non-LTE stellar atmosphere code (\href{PoWR}{www.astro.physik.uni-potsdam.de/PoWR}).  Such star emits $10^{48.5}$\,s$^{-1}$ of He\,{\sc ii} ionizing photons per second.  }
\label{fig:wruv}
\end{figure*}

Massive stars ($M_{\rm init} > 8\,M_\odot$) are the principal source of stellar feedback in galaxies. They settle onto the main sequence with spectral types earlier than B2 and evolve into blue, yellow, and red supergiants. In the final evolutionary stages, massive stars may shed their envelopes and expose their helium cores. If emission lines dominate their spectra, these hot, helium-rich stars are spectroscopically classified as Wolf-Rayet (WR) type (Fig.\,\ref{fig:wruv}). 
 Massive stars have the highest multiplicity fraction among all stars: {$\approx 70\%$ are expected to interact with companions before undergoing core-collapse \citep{Offner2023,2025NatAs...9.1337S}.}
A massive star's life ends when its core collapses into a neutron star or a black hole. Core collapse may be accompanied by a supernova and a $\gamma$-ray burst. 

During a major part of their lifespan, massive stars are  hot, with the maxima of  spectral energy distributions (SEDs) in the UV. As such, massive stars are the principal cosmic sources of hydrogen-ionizing radiation. Some subtypes of WR stars can also doubly ionize helium \citep{Garnett-1991ApJ...373..458G, Gonzales-Tora2025, Sander2026NatAs..10..290S}. Furthermore, the influence of massive stars is not limited to ionizing radiation. Scattering of UV photons in spectral lines of metal ions accelerates powerful supersonic stellar winds which inject mass, momentum, and energy into the ISM.  
The loss of mass  strongly influences the course of stellar evolution, as well as the type of the final compact object, in both single and binary stars. UV spectroscopy of massive stars is the primary diagnostic tool for measuring stellar ionizing power and determining stellar wind properties.

The rate at which stars lose mass through a radiatively driven wind, $\dot{M}$, depends on the metallicity. This dependency is often assumed in the form $\dot{M}(Z)\propto Z^{-\alpha}$. It is expected that the more metal poor stars have lower $\dot{M}$ and therefore undergo core collapses at a higher mass and faster rotation, producing heavier black holes \citep{Abbott-2016ApJ...818L..22A}. However, the exact relation is neither firmly theoretically predicted nor well constrained observationally, mainly because reliable $\dot{M}$ measurements in low metallicity stars are scarce. 
Realizing this, the  HST director granted discretionary time to build up a UV spectroscopic library of metal poor OB stars. About 300 stars were observed in the nearby LMC ($Z\sim 0.5Z_\odot$) and SMC ($Z\sim 0.2Z_\odot$) galaxies \citep{Vink-2023A&A...675A.154V,2025ApJ...985..109R}. 
At the same time only a handful of UV spectra of stars in lower metallicity galaxies have been obtained so far \citep{Telford-2024ApJ...974...85T,Schoesser25}. The results of these efforts are perplexing -- stellar winds are significantly weaker than predicted by the theory, and stellar properties are often unusual \citep{Rickard-2022A&A...666A.189R,2024A&A...692A..89B,Ramachandran-2024A&A...692A..90R,2025A&A...698A...9F}. However, the sample of low-$Z$ stars spectroscopically studied in the UV is too small to draw firm conclusions and establish a robust empirical scaling. 

In addition to uncertain processes in massive-star evolution such as mixing and mass loss, a fundamental complexity in understanding massive-star evolution is the fact that the majority are expected to interact with companions. This complexity manifests itself in a plethora of insufficiently constrained physical processes such as envelope stripping, mass-transfer, mergers, and common-envelope evolution \citep{2024ARA&A..62...21M}. To achieve new scientific progress, the predictions of evolutionary models must be verified observationally, and this is best accomplished through UV spectroscopy.  Recent examples include attempts to identify and characterize stars stripped of their outer hydrogen-rich envelopes in binary systems. Such stripped stars, if luminous enough, may exhibit WR spectra \citep{Shenar2020}; otherwise, their spectra are dominated by absorption in the optical band \citep{Goetberg2018,2023ApJ...959..125G}.  UV spectroscopy plays a key role in identifying and characterizing such objects and in testing theoretical predictions of mass-loss and mass transfer in binaries at different metallicities
\citep{Shenar2016, Shenar2019,  2022A&A...662A..56K,2022A&A...659A...9P,2023Sci...382.1287D, 2024A&A...683A..37Y, 2026A&A...708A.187M,2023ApJ...959..125G,Ramachandran-2024A&A...692A..90R, Gilkis2023}.

Overall, UV spectroscopy of massive stars is needed to find the answers to the questions of general astrophysical significance. For example, modern simulations bridging the galactic and cosmological scales rely on stellar feedback recipes that require empirical foundations \citep{Schaye-2026MNRAS.548ag375S}. Furthermore, uncovering the origin of the observed black hole mass distribution detected by gravitational-wave observatories requires a significant improvement in our understanding of low-$Z$ massive stars \citep{2025ApJ...979..209V}.
Last but not least, the discovery of UV-luminous and nitrogen-enriched galaxies at high $z$ \citep[e.g.,][]{Cameron23}, preceding the epoch of reionization, raises new questions about the properties of massive stars in the young, low-$Z$ Universe \citep{Naidu-2026OJAp....956033N}.

To address these outstanding challenges, UV spectroscopy of stars in environments that resemble different cosmic epochs is of high priority. Among the most urgent questions are:

\begin{itemize}
    \item {\em How does the wind mass-loss rate scale with metallicity for stars of different masses and evolutionary stages?} Since ionizing photons are effectively absorbed in stellar winds, the number of escaping photons is affected by wind strength. Furthermore, the predictions of stellar evolution models critically depend on mass-loss rate prescriptions. Therefore, the large uncertainties in wind properties, which propagate into population and spectral synthesis models, must be significantly reduced.
    \item {\em What does stellar wind dynamics reveal about stellar physics and the engines driving stellar winds?} Previous IUE and HST UV spectroscopy have revealed that stellar winds of Galactic stars are clumped, inhomogeneous and highly dynamic \citep{1996A&AS..116..257K}. Empiric mass-loss measurements depend on adopted clumping parametrization yet these are currently not known \citep{Oskinova-2007A&A...476.1331O, Brands-2025A&A...697A..54B}. Spectral variability strongly affects empiric mass-loss rate estimates, especially in low-$Z$ stars \citep{Massa-2024ApJ...971..166M, Rickard-2022A&A...666A.189R}, but existing observations are not sufficient to establish systematics. 

    \item {\em How do massive stars evolve, interact, and die at different metallicities?} Spectroscopic analyses by modern 
    non-LTE stellar atmosphere models provide high fidelity stellar parameters \citep{Sander-2024A&A...689A..30S}. Large sample of stars in different galaxies need to be spectroscopically analyzed to construct empiric Hertzsprung-Russel diagram (HRD) of massive star populations in representative galaxies. Critical comparison with evolutionary tracks and population synthesis models is needed to test current stellar evolution models and establish key model parameters.
    
    \item {\em What are the largest stellar masses and how these depend on the star-formation rate (SFR) and metallicity of host galaxies? } 
    The most massive stars known so far reside in dense star clusters within large star forming regions, such as 30\,Dor  \citep{Crowther-2010MNRAS.408..731C}. UV spectroscopy complemented by optical is required to constrain stellar mass by spectroscopic methods and to probe multiplicity status. Spectral modeling is needed to determine feedback parameters of the most massive stars.    
    
    \item {\em Are massive stars responsible for harsh ionizing conditions and abundance patterns in high-$z$ galaxies?} Since individual stars cannot be resolved in galaxies at Mpc distances spectroscopy of compact star clusters in  is the way forward. The HST has already shown that the UV spectra of clusters in nearby galaxies have large diagnostic power \citep{Berg-2022ApJS..261...31B} but farther galaxies with lower metallicities and higher gas content must be probed to bridge between local and high-$z$ Universe. Importantly, nebula spectra must be  analyzed consistently together with the spectra of ionizing sources.    
    
    \item {\em What are the ionizing properties of massive binaries including those hosting accreting compact objects?} UV spectroscopy is the best diagnostic to discover and characterize very hot stars in binaries. Another important sources of  ionizing radiation are high-mass X-ray binaries (HMXB): X-ray luminous black holes and neutron stars accreting matter from their massive star companions \citep{2019A&A...622L..10S}. These systems serve as important test beds for the models of binary evolution and compact object population synthesis \citep{2025A&A...693A..27X}. UV spectroscopy of HMXB  is necessary to determine the properties of donor stars, establish accretion mechanisms, determine accretion disk SEDs, and test the powerful accretion disk outflows \citep{2012ApJ...750..110T,Hainich2020A&A...634A..49H,2024A&A...690A.347R}. 

\end{itemize}

\begin{figure*}[ht!]
\begin{center}
\includegraphics[width=0.7\textwidth]{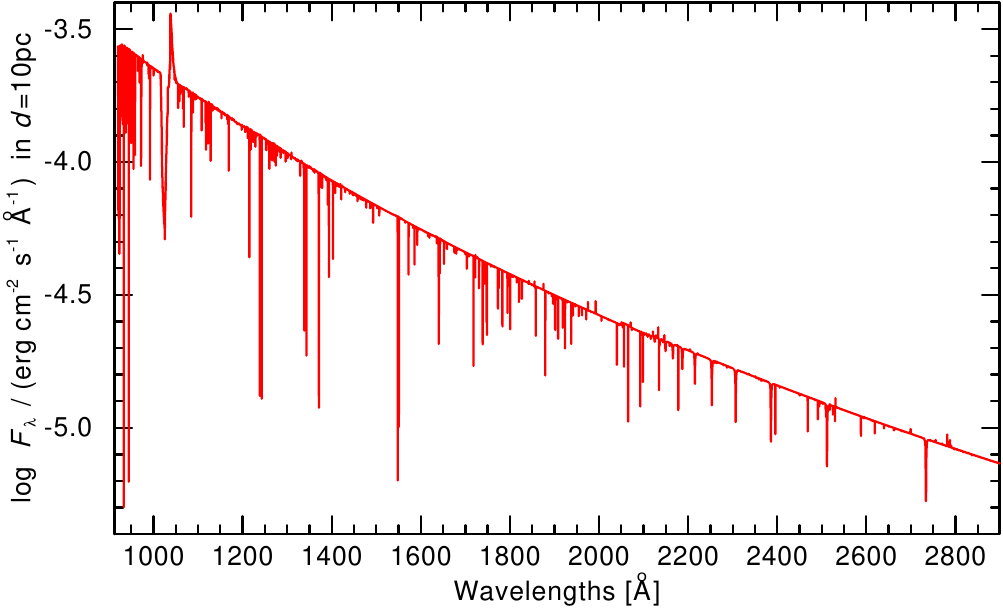}
\end{center}
\caption{Synthetic calibrated UV spectrum of a main-sequence  O-type star 
with $T_\ast=60.2$\,kK, $\log{L/L_\odot}=6.25$ ($M_\ast=120\,M_\odot$), and $Z=0.07\,Z_\odot$ located at a fiducial 10\,pc distance computed with the PoWR  non-LTE stellar atmosphere code (\href{PoWR}{www.astro.physik.uni-potsdam.de/PoWR}).  }
\label{fig:osed}
\end{figure*}   

\subsubsection{Requirements}

Having a broad spectral range of 900--4700\,\AA\ would be a significant advantage. Spectroscopy in this range is necessary 
to determine metallicity and abundances of individual metals. The forest of iron lines is located in the UV; 
further key  diagnostic lines are 
Ne\,{\sc \ion{V}{ii}}\,$\lambda 973$,
O\,{\sc vi}\,$\lambda\lambda 1032, 1038$,
P\,{\sc v}\,$\lambda\lambda 1118, 1128$,
N\,{\sc v}\,$\lambda\lambda 1239, 1243$,
O\,{\sc v}\,$\lambda 1371$,
C\,{\sc iii}\,$\lambda 1175$,
Si\,{\sc iv}\,$\lambda\lambda 1394, 1403$,
C\,{\sc iv}\,$\lambda\lambda 1548, 1550$, He\,{\sc ii}\,$\lambda 1640$,
N\,{\sc iv}\,$\lambda 1718$, 
Al\,{\sc iii} $\lambda\lambda 1854, 1862$, 
Ne\,{\sc VI}\, $\lambda 2070$, and
Mg\,{\sc IV}\, $\lambda 2165$\,\AA\
(Fig.~\ref{fig:wruv}).  An extension of the available UV range bluewards of Ly$\alpha$ would be a big adavntage, e.g.\ the O\,{\sc vi} line is the key  for detecting hot, stripped, non-WR companions in binaries. Including in the required  wavelength range the blue optical (up to $\sim4700$\,\AA) would enable important science by giving access to the important diagnostic lines He\,{\sc ii} $\lambda 4686$\,\AA, complex of  N\,{\sc iii} and C\,{\sc iii} lines at $\approx 4650$\,\AA, C\,{\sc ii} $\lambda 4267$\,\AA,
O\,{\sc vi}\,$\lambda\lambda 3811, 3834$\,\AA, etc.  The H$\alpha$ line is an important tracer of stellar disks and stellar winds, extending the spectral range to 10\,000\,\AA\
would enable simultaneous UV and optical spectroscopy and open a door for new discoveries.

Spectral resolution $>10\,000$ is needed to probe wind structure and dynamics  as well as for detailed modeling of line profiles which contain invaluable information about stellar wind physics, such as wind turbulence and velocity law. 
Spectral resolution beyond 100\,000 will allow unprecedented insights in the wind structure and will help to resolve long standing problems related to stellar wind clumping. The maximum wavelength drift allowed over the duration of one observation as well as uncertainty of the wavelength solution should not exceed 50\% of a spectral resolution element. Finally, in case of binary or multiple systems where both (or more) stars have wind features, it is important to distinguish them by determining the shifts 
and multiplicity features in the P Cyg-type line profiles. 

The high light gathering power is necessary to obtain high S/N spectra with exposures comparable to wind flow-time. 
The S/N\,$\sim 10$ in continuum and $50$ in emission parts of spectral lines is required. To take full advantage of high spectral resolution, the S/N$\sim 100$ in P~Cyg lines is needed. Our science case involves measuring spectra of individual stars at $\propto$\,Mpc distances with V$>20$\,mag  (Fig.\,\ref{fig:lowmet}).  
The fluxes can be easily estimated from model spectral energy distributions (as represented by an example shown in Fig.\,\ref{fig:osed}). Excellent flux calibration on a level of a few per cent is a prerequisite for a quantitative spectroscopy of massive stars. 

Spatial resolution is an important issue for spectroscopy of hot luminous stars residing in star forming regions embedded in nebulae in remote galaxies at Mpc distances.  The required resolution is 0.05--0.1\,arcsec.  To take full advantage of such spatial resolution, narrow slit, or small aperture spectrographs are required. Furthermore, well characterized spatial properties of the detector are necessary  to confidently measure spectra from non-point sources.   

The instrument requirements are summarized as follows: \\

\begin{enumerate}
    \item {\it Wavelength coverage} -- Required: 90--470\,nm; Optimal: 90--1000\,nm. 
    \item {\it Spectral resolution} -- Required: 10\,000--50\,000;  Optimal: 10\,000--100\,000. 
    \item {\it Sensitivity} -- Extragalactic targets are faint, with $V >20$\,mag, while Galactic targets are brights, with $V >1$\,mag. The detector which would allow bright target observations would be highly useful.
    \item {\it Wavelength stability} -- Maximum wavelength drift allowed over the duration of one observation and maximum allowed uncertainty of the wavelength solution: 50\% of one spectral resolution element ($0.5\cdot\lambda\,R^{-1}$).
    \item {\it Signal-to-noise} -- Minimum S/N level of approximately 10 and 50 for continuum and in emission line cores; S/N of 100 in P~Cyg lines is needed.
    \item {\it Flux} -- Excellent flux calibration on a level of a few per cent is needed.
    \item {\it Spatial resolution} -- 0.05--0.1\,arcsec.
\end{enumerate}

\subsection{White dwarfs \& compact binaries}

White dwarf binaries are fundamental astrophysical probes. They provide ideal laboratories to test the models of binary evolution, which also apply to the sources of gravitational waves, whose detection led to the award of the 2017 Nobel Prize in Physics \citep{Abbott+2016a,Abbott+2016b}. Moreover, their final fate is intimately linked to Type Ia Supernovae (SNe\,Ia), i.e. the thermonuclear explosion of a white dwarf following the interaction with a companion star, which have become the fundamental yardsticks on cosmological distance scales  \citep{Branch+1992} and led to the discovery of dark energy \citep{Riess+1998,Perlmutter+1999} and the award of the 2011 Nobel Prize in Physics. Finally, white dwarf binaries play a crucial role in influencing star formation and chemical evolution of the Galaxy by injecting energy into, and enriching, the interstellar medium with material ejected during nova eruptions and SN\,Ia explosions \citep{Matteucci+2009}.

Ultraviolet spectroscopy is crucial for the study of white dwarf in binaries because they are relatively hot ($\gtrsim 9\,000$\,K) and their emission peaks in the UV. In the optical and near-infrared, the emission from the companion stars (in detached system) and/or the accretion flow (in interacting binaries) severely dilutes and often outshines the white dwarf itself. Measurements of their physical parameters strongly rely on UV spectroscopy from the Ly-alpha region down to 3,000\,\AA\ and space-based UV observations is the only method to access and fully characterise the physical properties of these white dwarfs \citep{Parsons+2016,Pala+2017,Pala+2022}. Moreover, several resonance lines of C, N, O and Si are only observable in the UV, making the access to the UV regime fundamental for the study of key phenomena in binary evolution, from compact binary formation processes \citep{Yu+2026}, to stellar mergers \citep{Sahu+2025} and magnetism \citep{Gentile_Fusillo+2018,Wilson+2021}.

\begin{figure*}
    \centering
    \includegraphics[width=\linewidth]{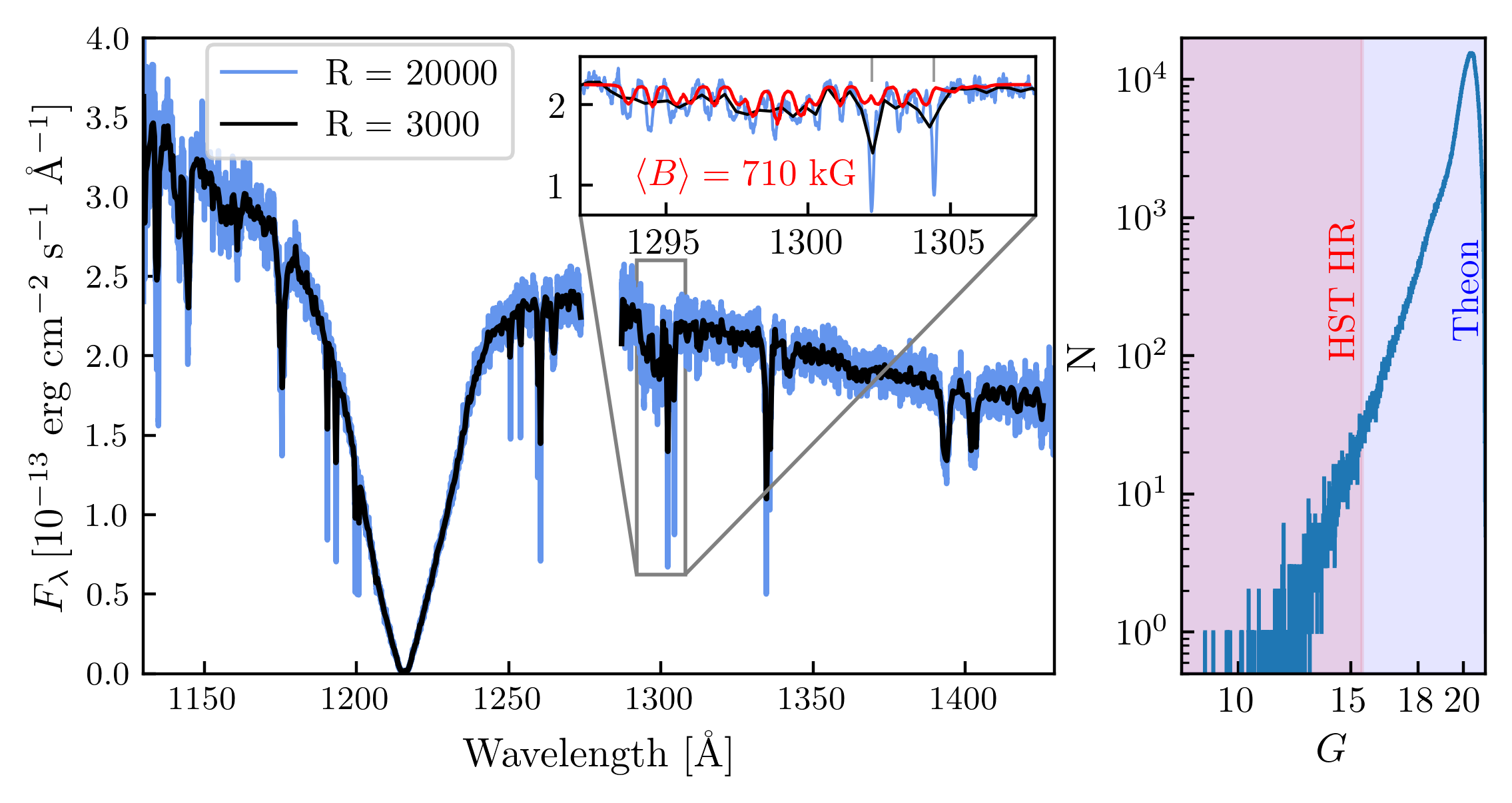}
    \caption{\emph{Left:} \textit{HST}/COS spectrum G130M spectrum of CC\,Cet (light blue, \citealt{Wilson+2021}). The high spectral resolution ($R \simeq 20,000$) allows to resolve Zeeman triplets produced in several spectral lines by a magnetic field of $\simeq 700\,$kG, thus revealing that CC\,Cet has a magnetic field. In the inset, the red line shows the best-fit magnetic model to the \ion{Si}{iii} lines while the grey ticks mark interstellar \ion{O}{i} lines. For a comparison, the same spectrum re-sampled to the resolution of the \textit{HST}/COS G140L ($R \simeq 3000$) is shown in black, highlighting the importance of high-resolution UV observations in order to resolve metal lines, and accurately derive chemical abundances, resolve line splitting due to magnetism, and disentangle the contamination from the ISM. \emph{Right:} magnitude distribution of white dwarfs from \citet{Gentile_Fusillo+2021} highlights the comparative high-resolution sensitivity limits of \textit{HST} and \textit{Theon}. By providing access to fainter objects, which represent the bulk of the population, \textit{Theon} will provide enable the construction of statistically significant, volume-limited samples of white dwarf binaries.}
    \label{fig:wd_binaries}
\end{figure*}

Only UV spectroscopy enables the study of white dwarf binaries in detail, and it is needed to answer several key key scientific questions: 
\begin{enumerate}
    \item \textbf{What mechanisms drive binary evolution?}
    \item \textbf{Which pathways produce successful SN\,Ia explosions?}
    \item \textbf{How do different classes of gravitational-wave sources form?}   
    \item \textbf{How much processed material is returned to the ISM by nova eruptions and SNe\,Ia?}
    \item \textbf{What is the origin of magnetism?}     
\end{enumerate}

\subsubsection{What mechanisms drive binary evolution?}
Orbital angular momentum losses and the common envelope phase are among the key ingredients in the models describing the evolution of all compact binaries. However, the development of such models is currently limited by the lack of (i) accurate observational constraints on the efficiency of the common envelope phase \citep{DiStefano+2023} and of (ii) a complete theoretical framework describing magnetic wind braking \citep{Barraza-Jorquera+2025} (one of the main mechanisms driving angular momentum loss). 

Thanks to their large numbers, proximity, and brightness, the most compact white dwarf binaries (orbital periods ranging from few minutes up to few days) are one of the most powerful tools to constrain these mechanisms. From UV spectroscopic data it is possible to derive masses, effective temperatures, abundances and rotation rates for individual systems. In interacting binaries, UV data also provide a direct measurement of the mass accretion rate onto the white dwarf, which  directly reflects the angular momentum loss rate in the system \citep{Townsley+2009}. By obtaining such parameters for statistically significant volume-limited samples of the different sub-populations of white dwarf binaries, it is possible to derive stringent constraints on the mechanisms of angular momentum loss \citep{Pala+2022} and the efficiency of the common envelope phase \citep{Parsons+2015}, via direct comparisons between observations and binary population synthesis studies.

\subsubsection{Which pathways produce successful SN\,Ia explosions?}
These thermonuclear explosions take place in binaries following the interaction of a white dwarf with its companion. Detached double white dwarfs are particularly relevant because they can give rise to a SN\,Ia by spiralling in under the effect of gravitational radiation until they finally merge, exceeding the Chandrasekhar mass limit \citep{Shen+2024}. White dwarfs in semi-detached systems instead, could surpassed the Chandrasekhar mass limit following accretion from a non-degenerate donor \citep[e.g.][]{Iben+1984}. 
Another possibility is the double-detonation scenario, in which the detonation of a helium layer triggers the detonation of a carbon core, leading to the explosion even at sub-Chandrasekhar masses, possibly leaving behind a supernova survivor \citep{Shen+2018}.

Obtaining accurate white dwarf masses is critically important to probe the response of the white dwarf to both stable and unstable accretion processes. Rotation rates can be used to determine whether a white dwarf is spun up by accretion, possibly close to break–up, allowing the white dwarf to exceed the Chandrasekhar limit without triggering the SN explosion \citep{King+1991}. Finally, abundances allow to reveal and characterise the remnant of such explosion, including merger products \citep{Sahu+2025} and supernova survivors \citep{Werner+2024}.

Only high-resolution ($R \gtrsim 20,000$) UV observations allow to measure these parameters and derive robust observational constraints on the response of the white dwarf to the accretion of mass, angular momentum and energy, in order to finally reveal the pathway to SN\,Ia explosions.

\subsubsection{How do different classes of gravitational-wave sources form?}   
With an estimated population of around 100 million in the Milky Way \citep{Marsh2011}, detached double white dwarfs are the dominant sources of low-frequency gravitational waves in our Galaxy \citep{Korol+2022,Kupfer+2024}, particularly in the millihertz range, where future space-based detectors, such as the \emph{Laser Interferometer Space Antenna} (\emph{LISA}, \citealt{Amaro-Seoane+2023}), the \emph{Tian-Quin} mission \citep{Luo+2016}, and the \emph{Lunar Gravitational-wave Antenna} (\emph{LGWA}, \citealt{Ajith+2025}) will be sensitive. While thousands of these binaries will be individually detected by these missions, many millions will be unresolved and will produce a ``Galactic foreground'', which needs to be precisely modelled since it can overlap with (and potentially mask) the signal from distant cosmological sources.

So far, the identification of detached double white dwarfs has been challenging since they are too compact (orbital separation $\simeq 0.05-2$\,mas) to be directly resolved, and their spectral energy distribution is often dominated by one of the two stellar components and therefore their spectra resemble those of single white dwarfs. Their binary nature can only be revealed thanks to the detections of radial velocity variations, which requires high-resolution ($R \gtrsim 20,000$) spectroscopic observations in order to resolve the narrow Non-Local Thermodynamic Equilibrium (NLTE) core of the H$\alpha$ line \citep{Napiwotzki+2020}. As a consequence of these observational challenges, the current sample of double white dwarf binaries comprises $\simeq\,300$ systems \citep{Munday+2025} and, among these, only a few have a full orbital solution (i.e. masses and orbital periods).

High-resolution phase-resolved spectroscopic observations, from the UV to the H$\alpha$ region, hold the promise to provide major step forwards in the characterisation of double white dwarfs. The radial velocity provides the orbital period of the binary, while the full wavelength coverage is necessary in order to disentangle both stellar components and derive the white dwarf masses \citep{Bours+2015,Gentile_Fusillo+2018}. 
Such observations enable the construction of statistically significant, volume-limited samples of double white dwarfs that allow to measure orbital periods and mass distributions and space densities. These parameters are crucial to accurately estimate the expected number of source detections in order to model the gravitational wave foreground that will limit the sensitivity of future gravitational wave missions \citep{Nelemans2009}. Additionally, these parameters also allow to assess the final fate of the systems — including the possibility of a merger between the two white dwarfs and the potential identification of SN\,Ia progenitors \citep{Marsh+1995}.

\subsubsection{How much processed material is returned to the ISM by nova eruptions and SNe\,Ia?} 
White dwarfs-main sequence (WDMS) binaries with orbital periods of years play a crucial role in Galactic archaeology and are excellent tools for testing current models of star formation and stellar evolution. In such systems, the two stellar components are far apart and they never interacted. Therefore the white dwarf's cooling rate allows for a precise age determination, which can be used to investigate correlations between age-related stellar properties - such as activity, rotation, and metallicity - by studying their companion stars \citep[e.g.][]{Rebassa-Mansergas+2021,Rebassa-Mansergas+2023,Raddi+2022,Chiti+2024}. 

Instead, in mass-transferring white dwarf binaries, interaction often leads to the release of both kinetic energy and (potentially chemically enriched) material in the surrounding interstellar medium (e.g. during nova eruptions or SN\,Ia explosions).

The spectra of these binaries are composite, with the hot white dwarfs dominating in the UV, and the main sequence stars emitting mostly in the optical and near-infrared. Therefore, UV observations are needed to (i) characterise the white dwarf, and (ii) reveal its presence in those systems in which the main sequence star completely outshines the white dwarf in the optical \citep[e.g.][]{Parsons+2016}.
In this way, it is possible to build large ($\simeq 1.5$\,kpc) volume-limited samples, that also provide the scale heights and the possibility to investigate the formation history of different populations of white dwarf binaries \citep{Patterson1984} and their connection to different metallicity environments (halo vs. thin and thick disc), which directly affects the yields of processed material returned to the ISM by nova eruptions and SNe\,Ia \citep{Matteucci+2009}. 

\subsubsection{What is the origin of magnetism?}
Magnetic fields play a fundamental role in the evolution of both single and binary stars, yet their origin remains poorly understood. A puzzling difference in the incidence of magnetic fields has been observed across different white dwarf binary populations, being as high as $\simeq 30\%$ in mass-exchanging white dwarf binaries \citep{Pala+2020}, although this high frequency is not reflected in their parent population of white dwarf-main sequence binaries. 
Several mechanisms have been proposed to explain the presence of these magnetic fields, including: Ap and Bp progenitors that preserve a fossil magnetic field while becoming white dwarfs \citep{Angel+1981}; interaction during the common-envelope phase \citep{Tout+2008}; a rotation-driven dynamo processes occurring as the white dwarf core crystallises \citep{Schreiber+2021}; generation within the convective core of the main-sequence progenitor, with the fields later diffusing to the surface as the resulting white dwarf cools \citep{Camisassa+2024}. Moreover, magnetic fields serve as a ``smoking gun'' for stellar mergers \citep{Burleigh+1999,Hollands+2020,Sahu+2025}.
Understanding these fields is therefore essential for reconstructing the historical evolution of a system, as they provide critical constraints on binary evolution pathways and the specific channels leading to SNe\,Ia.

UV observations are indispensable for such study, because optical data are often insufficient to constrain field strengths. This is particularly true for the most highly magnetized systems, where spectral lines can be washed out in the optical regime, leaving the UV as the only window to detect their signatures \citep{Gaensicke+2001,Caiazzo+2021}. Furthermore, high-resolution spectroscopy is required to resolve Zeeman splitting in weaker fields ($< 100$\,kG) and to break the degeneracy between magnetic line splitting and rotational broadening \citep[e.g.][]{Wilson+2021}.

\subsubsection{Requirements}
The upcoming \emph{Gaia} Data Releases 4 and 5 (December 2026 and late 2030, respectively) and the advent of wide-area multi-object spectroscopic facilities such as 4MOST (4-metre Multi-Object Spectroscopic Telescope, \citealt{deJong+2019}), WEAVE (William Herschel Telescope Enhanced Area Velocity Explorer, \citealt{Jin+2024}), SDSS-V (Sloan Digital Sky Survey - V, \citealt{Kollmeier+2025}) and DESI (Dark Energy Spectroscopic Instrument, \citealt{Cooper+2023}), along with deep, time-domain surveys like the Large Synoptic Survey Telescope (LSST, \citealt{Ivezic+2019}), will identify hundreds of thousands of white dwarf binaries. These will allow to study fainter and more distance targets, thereby probing substantially larger volumes and a comprehensive picture of white dwarf binaries at different evolutionary stages \citep{Pala+2025}.

In order to be able to study in details these discovery and build statistically significant samples, an UV facility with high sensitivity is needed. This should have the capability to achieve signal-to-noise ratios $\mathit{SNR} \gtrsim 10$ in short exposures ($\simeq 1-5\,$min, necessary to avoid orbital smearing for the most compact binaries), at 1300\,\AA\ for $F_\lambda \simeq 10^{-16}\,\mathrm{erg}\,\mathrm{cm}^{-2}\,\mathrm{s}^{-1}\,\text{\AA}^{-1}$. High sensitivity is especially crucial at the shortest wavelengths, down to $\simeq 800\,$\AA, to obtain the blue wing of the Ly-$\alpha$ necessary to constrain the chemical abundances and the white dwarf mass. In order to obtain coverage up to the H-$\alpha$ or Ca triplet lines, a wavelength coverage of up to 7000--10\,000\,\AA\ is required. Finally, high spectral resolution in the range $R \sim $\,20\,000--40\,000 is needed to measure accurate radial velocities and rotation rates ($v \sin(i) \gtrsim 100$\,km/s), resolve line splitting due to magnetism, and disentangle the contamination from the ISM (Figure~\ref{fig:wd_binaries}).

The instrument requirements are summarized as follows: \\

\begin{enumerate}
    \item {\it Wavelength coverage} -- Required: 80--700\,nm; Optimal: 80--1000\,nm. 
    \item {\it Spectral resolution} -- Required: 20\,000--40\,000;  Optimal: 10\,000--50\,000. 
    \item {\it Sensitivity} -- Targets to be observed have typical FUV magnitudes in the range 15--21. 
    \item {\it Signal-to-noise} -- S/N\,$>10$ should be achievable in short exposures (1--5\,min duration).
\end{enumerate}

\subsection{Supernovae}

\begin{figure*}
\centering
\includegraphics[width=\textwidth]{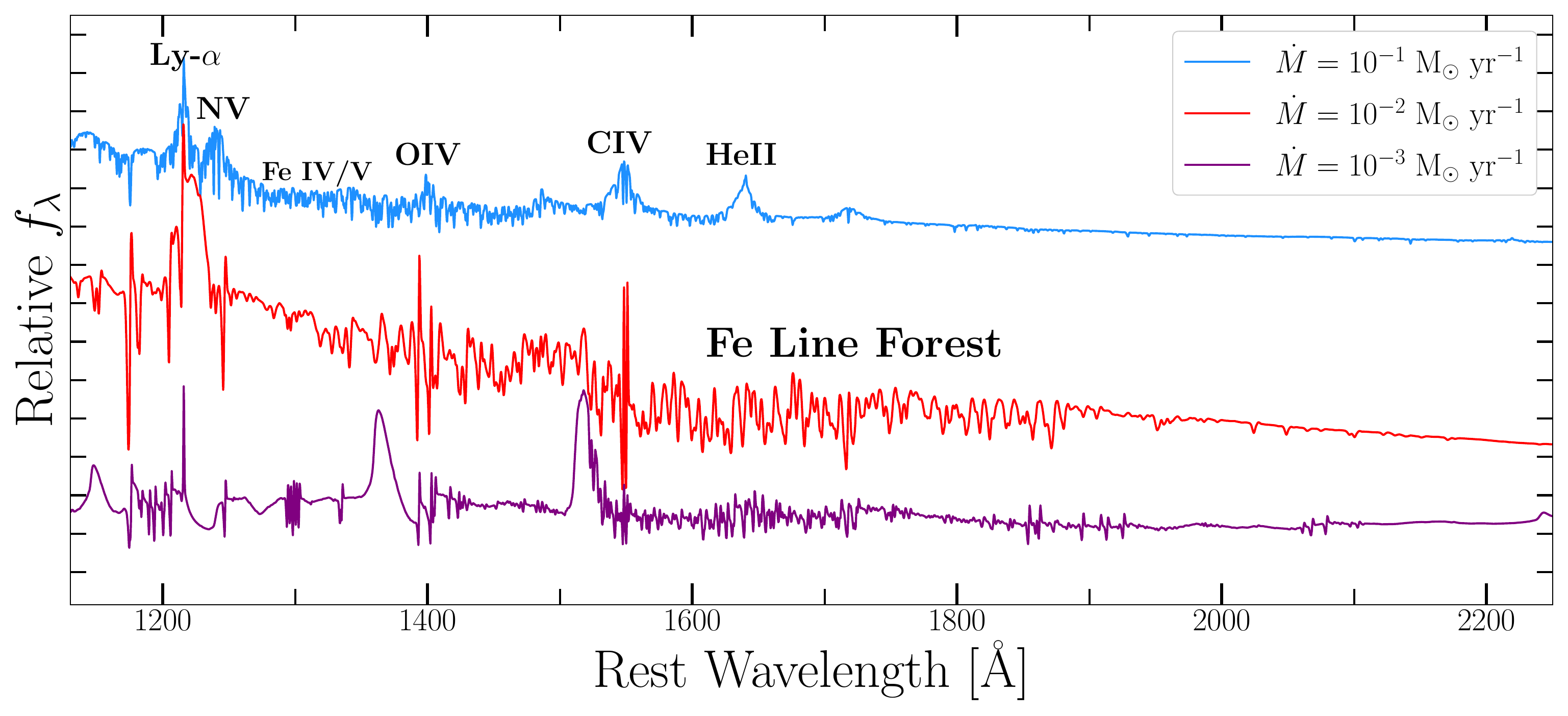}
\caption{\textit{Left:} Model spectra for a type II supernova exploding from a progenitor star with varying mass-loss rates at $\sim 5$~days post-explosion. High resolution UV spectroscopy will detect and resolve the $\sim$1000s of narrow, low and high-ionization emission lines from shock interaction that can directly constrain the composition and temperature of the confined CSM. To date, no such high-resolution UV spectroscopic observations exist of such a young, CSM-interacting SNe. \label{fig:IIn_spec} }
\end{figure*}

\subsubsection{The Final Moments of Massive Star Evolution}

A paramount issue in astrophysics is constraining how the lives of massive stars stars end. This avenue of study has a direct impact on the observed diversity of core-collapse supernovae (SNe), compact object formation, and element creation in the universe. Progressing our understanding of late-stage massive star evolution can be accomplished by probing the composition and structure of the circumstellar material (CSM) surrounding these stars in the final years before explosion. This CSM is comprised of material synthesized during different stages of nuclear burning, and is enriched as the progenitor star loses mass via winds and/or violent outbursts \citep{smith14}. Understanding the structure of this CSM provides needed constraints on the final stages of stellar instability before core collapse and the proposed mechanisms for both dynamic \citep[e.g., gravity waves, super-Eddington winds, etc.][]{owocki04, owocki17, quataert16, fuller17,Wu21} as well as secular \citep[e.g., steady-state winds][]{Beasor20} mass-loss. 

Mass-loss mechanisms during the final stages of massive star evolution remains almost entirely in stellar evolutionary models. However, in the era of all sky transient surveys, ultra-rapid spectroscopic observations of core-collapse SNe have become a powerful tool in understanding the circumstellar environment of pre-SN systems in the final years before explosion \citep{Gal-Yam14, Groh14, Yaron17, bruch21}. We can now identify prominent emission lines (i.e., ``flash'' or ``IIn-like'' features) in very early-time SN spectra from the recombination of photo-ionized CSM after shock breakout. These emission lines are direct evidence of CSM surrounding the progenitor star, comprised of elements ejected during enhanced mass-loss episodes right before explosion. The strength of these features is a robust tracer of the progenitor's chemical composition, identity and recent mass-loss at small distances $r < 10^{15}$~cm from the explosion \citep{Yaron17, dessart17}. Depending of the CSM density and extent, these features can be short-lived, evolving into broad lines forming in the high-velocity ($v_{\rm ej} \sim 10^{4}$~km~s$^{-1}$) outer ejecta layers. Following shock breakout from the stellar surface, reprocessing of the SN radiation by the unshocked, slow CSM produces both narrow emission lines in the early-time SN spectra as well as a boost in luminosity that is most evident in the UV bands \citep{dessart17,Morozova17, Haynie21}. Combining early-time UV/optical light curves and spectroscopy allows for precise constraints to be made on the composition, density and extent of the CSM surrounding the progenitor star at the time of explosion.

\subsubsection{UV Observations as a Unique Tracer of Supergiant Star Explosions}

Presently, the landscape of SN II UV observations remains sparse and unconstrained at early phases ($\sim$days to weeks after explosion). However, given that the UV holds the majority of the SN flux and numerous spectral lines, $\lambda < 3000~$\AA{} are arguably the most essential wavelengths to study both the CSM interaction and the SN ejecta structure. {\it Swift}-UVOT observations show that most SNe~II are bright in the UV, making them ideal candidates for early-time UV spectroscopy \citep{Terreran2022,WJG24a,WJG24b}. Furthermore, objects that interact with dense CSM at early-times can be $\gtrsim$2~mag more luminous in UV bands than SNe~II from red supergiants with weak mass-loss rates, as well as stay UV-bright for longer periods of time. While UV photometry has revealed diversity amongst young SNe~II \citep{WJG24a,WJG24b}, UV spectroscopy is necessary to confirm the presence of narrow emission lines from shock interaction with CSM. Numerical simulations with the radiative transfer code {\tt CMFGEN} show that spectroscopic signatures of such interaction are the most prominent in the UV (Fig. \ref{fig:IIn_spec}). Detection and modeling of narrow UV emission lines, derived from the progenitor star wind, can constrain the surface abundance of the supergiant progenitor star and its mass-loss rate in the final years before explosion, as well as probe the physics of shock interaction within the CSM. 

\subsubsection{The Future of Supernova UV Spectroscopy}

The primary limitations of current UV spectroscopic efforts for SNe are resolution, depth and cadence. To date, high resolution spectroscopy has not been performed on CSM-interacting SNe despite the forest of narrow ($v \approx 20-50$~km~s$^{-1}$) emission and absorption lines that arise from the pre-shock gas derived from the progenitor wind (e.g., see Fig. \ref{fig:IIn_spec}). Resolving these narrow features will enable a detailed re-construction of the CSM density and ionization stratification as well as mass loss history and identity of the progenitor star. Higher resolution UV spectroscopy would also enable a detailed analysis of the host galaxy environments of these CSM-interacting SNe in order to derive metallicity and local ISM densities/abundances. For cadence, the current UV instruments on {\it HST} are rarely re-pointed at new, young SNe ($\sim$1 ultra-rapid ToO opportunity per cycle) and, consequently, no CSM-interacting SN has been observed in far-UV wavelengths at early-times (i.e., $<1$~week post-explosion). Lastly, the sensitivity of {\it HST} limits follow-up UV spectroscopy of SNe that are quite nearby ($D < 50$~Mpc) and UV bright ($m < 20$~mag) \citep{wjg25, Bostroem26}. UV instruments with increased sensitivity will be able to monitor CSM-interacting SNe for longer timescales and detect more distant events to better constrain UV diversity in large samples.

High-resolution, UV spectroscopy of SNe with instruments that have higher sensitivity than {\it HST} will enable fundamental questions to be answered about the origins and properties of CSM-interacting SNe: 
\begin{enumerate}
    \item \textbf{What mechanisms drive mass-loss in massive stars?}
    \item \textbf{What is the elemental abundance and ionization of CSM in SNe?}
    \item \textbf{What is the UV spectral line diversity of SNe?}   
    \item \textbf{How does UV emission contribute to the shock power budget of SNe interacting with CSM?}
    \item \textbf{What are the ISM properties of SN host environments?}     
\end{enumerate}

\subsubsection{Requirements}

The instrument requirements are summarized as follows: \\

\begin{enumerate}
    \item {\it Wavelength coverage} -- Required: 110--300\,nm; Optimal: 110--300\,nm. 
    \item {\it Spectral resolution} -- Required: 10,000--20,000;  Optimal: 10,000--50,000. 
    \item {\it Sensitivity} -- Targets are supernovae with $V$-band magnitudes in the range 15--24 mag.
\end{enumerate}

\section{Science Pillar III. Gas, dust and metals in the baryon cycle of galaxies}
\label{sec: Pillar III}


\newcommand{\feii}{\mbox{Fe\,{\sc ii}}}
\newcommand{\znii}{\mbox{Zn\,{\sc ii}}}
\newcommand{\zni}{\mbox{Zn\,{\sc i}}}
\newcommand{\crii}{\mbox{Cr\,{\sc ii}}}
\newcommand{\tiii}{\mbox{Ti\,{\sc ii}}}
\newcommand{\siii}{\mbox{Si\,{\sc ii}}}
\newcommand{\sii}{\mbox{Si\,{\sc i}}}
\newcommand{\sui}{\mbox{S\,{\sc i}}}
\newcommand{\suii}{\mbox{S\,{\sc ii}}}
\newcommand{\mgi}{\mbox{Mg\,{\sc i}}}
\newcommand{\hi}{\mbox{H\,{\sc i}}}
\newcommand{\hii}{\mbox{H\,{\sc ii}}}
\newcommand{\hh}{\mbox{H}$_2$}
\newcommand{\niii}{\mbox{Ni\,{\sc ii}}}
\newcommand{\alii}{\mbox{Al\,{\sc ii}}}
\newcommand{\aliii}{\mbox{Al\,{\sc iii}}}
\newcommand{\cii}{\mbox{C\,{\sc ii}}}
\newcommand{\civ}{\mbox{C\,{\sc iv}}}
\newcommand{\ci}{\mbox{C\,{\sc i}}}
\newcommand{\nii}{\mbox{N\,{\sc ii}}}
\newcommand{\oi}{\mbox{O\,{\sc i}}}
\newcommand{\coii}{\mbox{Co\,{\sc ii}}}
\newcommand{\caii}{\mbox{Ca\,{\sc ii}}}
\newcommand{\pii}{\mbox{P\,{\sc ii}}}
\newcommand{\mgii}{\mbox{Mg\,{\sc ii}}}
\newcommand{\mnii}{\mbox{Mn\,{\sc ii}}}
\newcommand{\cuii}{\mbox{Cu\,{\sc ii}}}
\newcommand{\kri}{\mbox{Kr\,{\sc i}}}
\newcommand{\geii}{\mbox{Ge\,{\sc ii}}}
\newcommand{\cli}{\mbox{Cl\,{\sc i}}}
\newcommand{\lya}{Ly-$\alpha$}

\begin{figure*}
    \centering
    \includegraphics[width=1.0\textwidth]{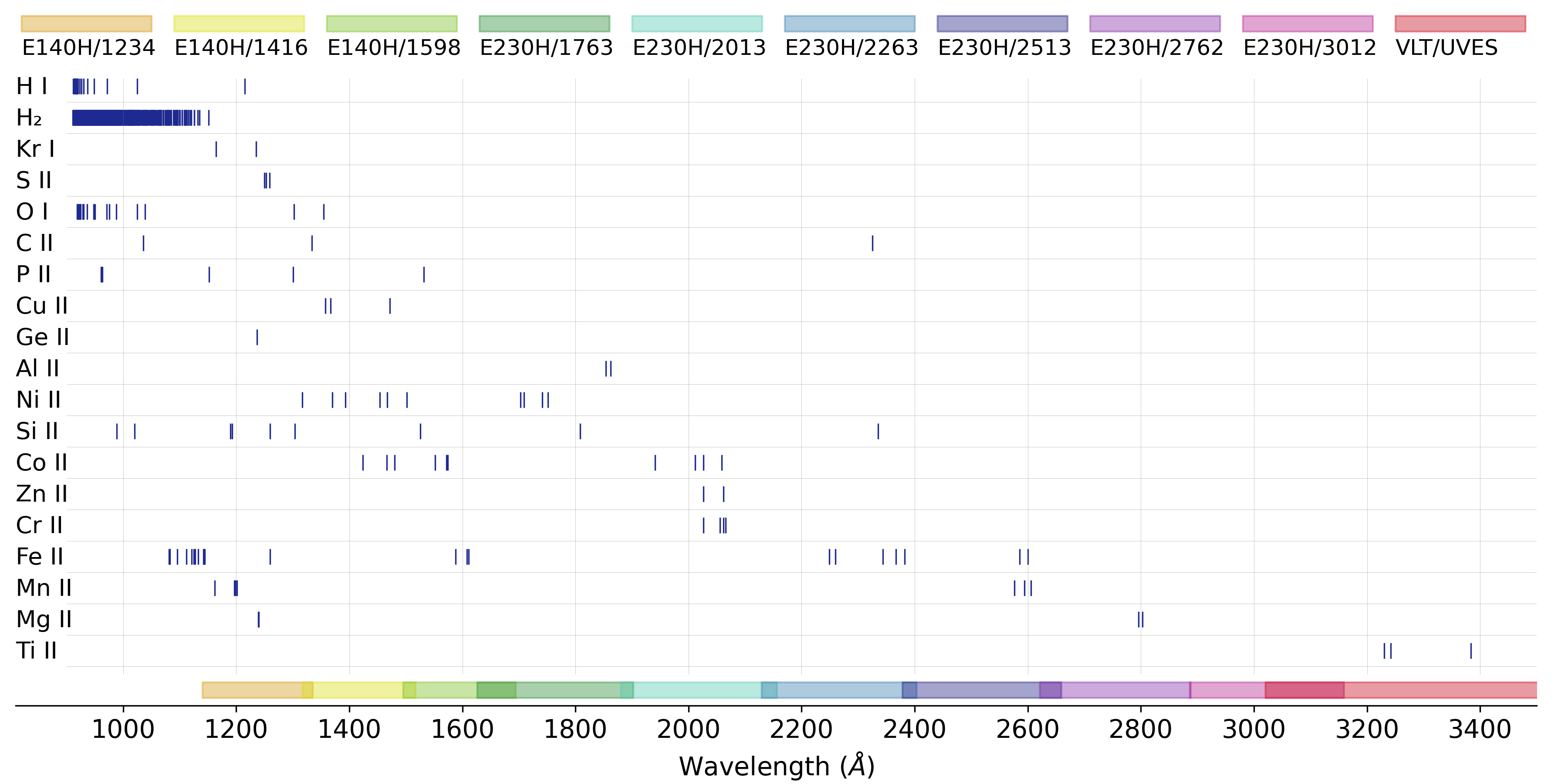}
    \caption{Rest-frame wavelengths of key atomic and molecular transitions used to probe the ISM in the UV. The vertical lines mark transitions of \hi{}, H$_2$ and metal species commonly used for abundance and depletion measurements. Coloured bars indicate the wavelength coverage of the HST/STIS E140H and E230H configurations and VLT/UVES, illustrating the broad spectral range required to access all the elements. Although ions like \cii{} or \siii{} have multiple transitions, many of them are saturated and only the weakest are suitable for studies of chemical abundances.}
    \label{fig: UV lines}
\end{figure*}

\subsection{Metals in the Interstellar Medium}
\label{sec:ISM}

The Interstellar Medium (ISM) is a fundamental component of galaxies that links processes at different scales, from the formation of molecules and cosmic dust grains to stars and galaxies themselves \citep[e.g.][]{Draine11,Maiolino19}. Cosmic, chemically pristine gas accretes on galaxies, providing fuel for the formation of new stars. In turn, stars and their evolutionary phenomena (e.g. supernovae) produce metals, and some of them are recycled back in the ISM and chemically enrich the new population of stars \citep[e.g.][]{Tinsley79,Kobayashi20,Matteucci21}. Grains of cosmic dust, which contain about half of all metals of the ISM, are key coolants for the formation of stars, catalyse the production of molecules, and constitute the building blocks of planets. Thus, ISM metals and dust are fundamental for the chemical evolution of galaxies, stars, planets, and molecules. 

UV spectroscopy has a fundamental role for the study of the ISM metals and dust, because most electronic transitions occur in the UV regime, making UV spectroscopy a unique probe of the ISM chemistry. Figure \ref{fig: UV lines} highlights the richness in the UV of electronic transitions from the ISM warm neutral medium \citep[as well as other ISM phases, see e.g.,][]{Tumlinson17}. Indeed, UV absorption-line spectroscopy of hot stars in the Milky Way has been one of the key science cases for Hubble, allowing detailed study of ISM abundances for many elements \citep[C, N, O, S, P, Si, Cl, Cu, Ca, Na, K, Ti, Mn, Fe, Co, Ni, Zn, Kr, e.g.][]{Savage96,Jenkins09}. Large amounts of the refractory metals are missing from the observable gas-phase ISM, because they are instead incorporated in dust grains, a phenomenon called dust depletion \citep[e.g.][]{Field74,Savage96,Jenkins09,DeCia16}. Yet, the chemical properties of the neutral ISM in its full complexity are still vastly unknown, and UV observations of hot stars are still limited to the brightest ones with low extinction in the Milky Way (and mostly within the Solar neighborhood for high-resolution UV spectroscopy), the Magellanic Clouds \citep{Tchernyshyov15,Jenkins17,Roman-Duval19,Roman-Duval20,DeCia24}, and more rarely in some metal-poor nearby system \citep{Hamanowicz24}.

Some of the most pressing open questions are:
\begin{itemize}
\item \textbf{Metal mixing and the ISM complexity.} How do metals mix and distribute in the ISM? On what spacial and time scales? What are the chemical properties of ISM clouds? Are there cloud-to-cloud differences? How do stars enrich the surrounding ISM? What does the ISM complexity tell us about the chemical evolution of galaxies?
\item \textbf{The low-metallicity Universe.} What are the chemical properties of the neutral ISM in low-metalllicity galaxies? What are the physical conditions of the cold ISM, specially at low metallicity? How do detailed studies of the local low-metallicity Universe enable a more meaningful understanding of the distant Universe?
 \end{itemize}

\subsubsection{Metal mixing and the ISM complexity.}

The ISM is a complex, multi-phase, dynamic medium shaped by many processes that affect its physical conditions, spacial distribution, dust content and properties, metallicity, relative abundances \citep{Draine11}. Until recently, UV absorption-line studies of the neutral ISM in the Milky Way and local galaxies have assumed that the ISM chemical properties are uniform, because of the challenge of characterizing dust depletion from the observed abundances [$X$/H] \citep[e.g.,][]{Savage96,Jenkins09,Ritchey23} and the fact the the Milky Way rotates.

\begin{figure*}
\centering
    \includegraphics[width=\textwidth, trim=0 0 0 0, clip]{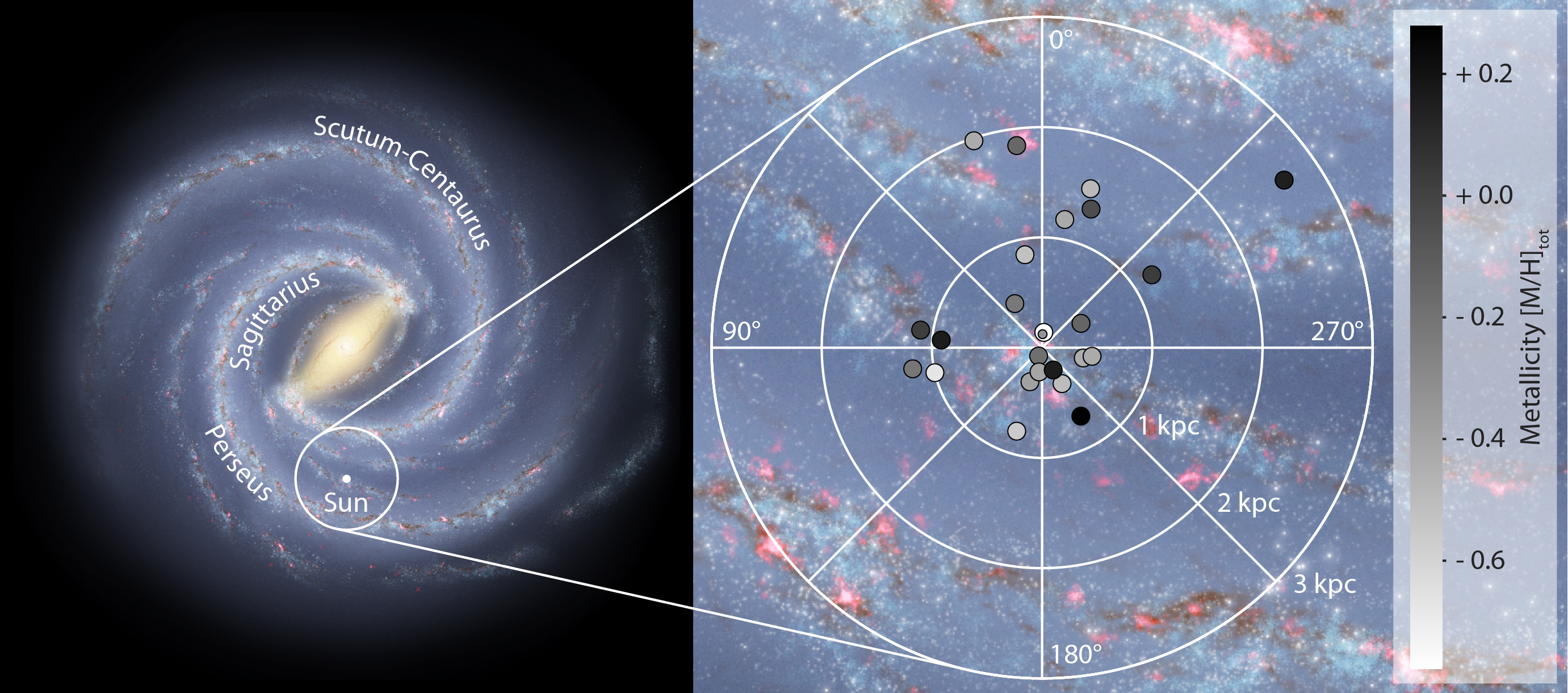}
\caption{Metallicity variations in the neutral ISM. Pockets of low metallicity, somewhere along the line of sight to the labeled targets, were estimated from HST/STIS echelle spectra with mid-resolution \citep[E140M, E230M,][]{DeCia21}.}
\label{fig:metvar}
\end{figure*}

\begin{figure}
    \centering
    \includegraphics[width=1\linewidth, trim=0 0 0 0, clip]{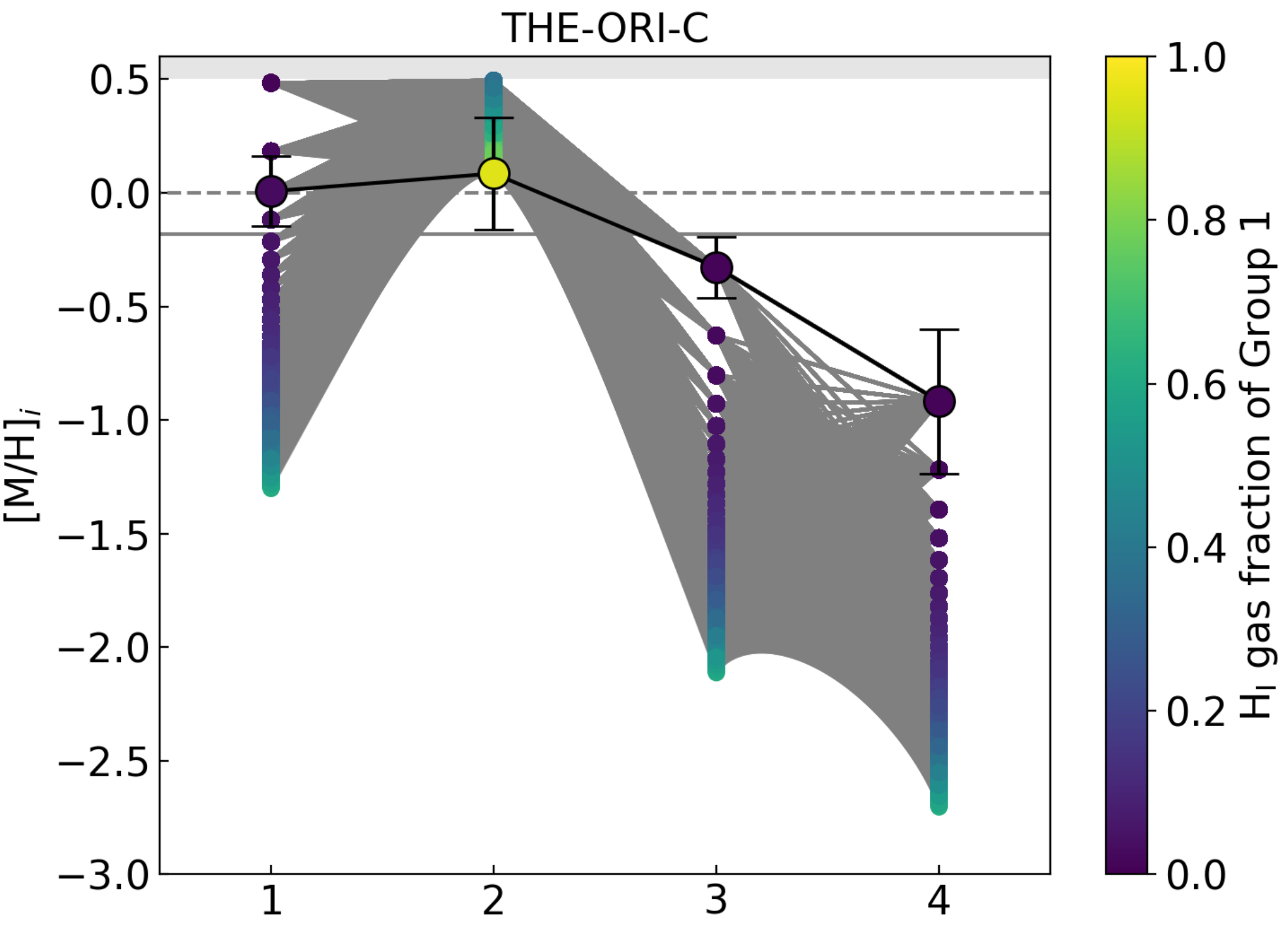}
\caption{Cloud-to-cloud metallicty variations are needed to reproduce the possible combinations of metallicity in individual ISM components (1 to 4) of the absorption line profiles towards $\theta$ Ori$^1$ C, based on high-resolution HST/STIS spectra \citep[E140H, E230H,][]{Ramburuth-Hurt25,Ramburuth-Hurt26b}. Component 2 is associated with the Orion Nebula itself \citep{Price01,Ramburuth-Hurt26}.}
\label{fig:metvar components}
\end{figure}

Variations in metallicities in the Solar neighborhood, however, have been suggested from HST/STIS mid-resolution ($R=30~000$) spectra \citep{DeCia21}, and more convincingly with HST/STIS high-resolution ($R=114~000$) spectra and an agnostic approach \citep{Ramburuth-Hurt25,Ramburuth-Hurt26b}. This is illustrated in Figs. \ref{fig:metvar} and \ref{fig:metvar components}. The recent HST Large Programme "STIS-ISM" (GO 17703) is pushing the HST limits and expanding STIS high-resolution studies to a slightly larger sample. While HST UV high-resolution spectroscopy can be extremely powerful, it's limitation is sensitivity, allowing the study of only stars brighter than roughly 10-11 mag in V band, i.e. mostly in the Solar neighborhood, within a few kpc.

\begin{figure*}
    \centering
    \includegraphics[width=1\linewidth,trim=0 0 0 0,clip]{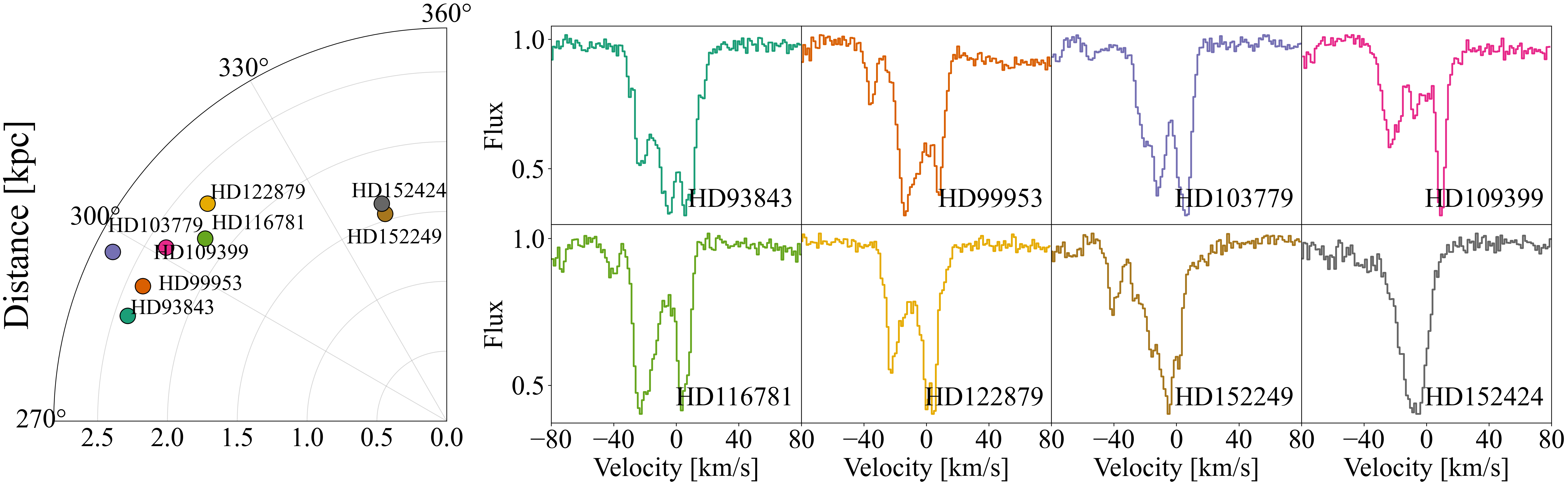}
    \caption{\feii{} $\lambda$2260 absorption-lines (right) towards 8 STIS-ISM OB stars located within $\sim2.8$ kpc from us (left). Individual components of the complex line profiles are associated with different ISM clouds (or groups of clouds) along the line of sight.}
    \label{fig:complexISM}
\end{figure*}

The complexity of the ISM is highlighted in Fig. \ref{fig:complexISM}. In absorption-line spectroscopy, multiple clouds along the line of sight cause absorption at different velocities, resulting in complex, multi-component line profiles (Fig. \ref{fig:complexISM}, right panel). These individual components/clouds may have different chemical properties and they can be studied by measuring the relative abundances of metals in these individual components. Different clouds may have different amounts of dust \citep[e.g][]{Welty20}, different metallicities \citep{Ramburuth-Hurt25,Ramburuth-Hurt26b}, different chemical patterns. In particular, some of these clouds could be enriched by the products of specific stellar populations or phenomena, like (very) massive stars or supernovae that enrich the surrounding medium. 

\begin{figure}
    \centering
    \includegraphics[width=1\linewidth, trim=0 0 0 200, clip]{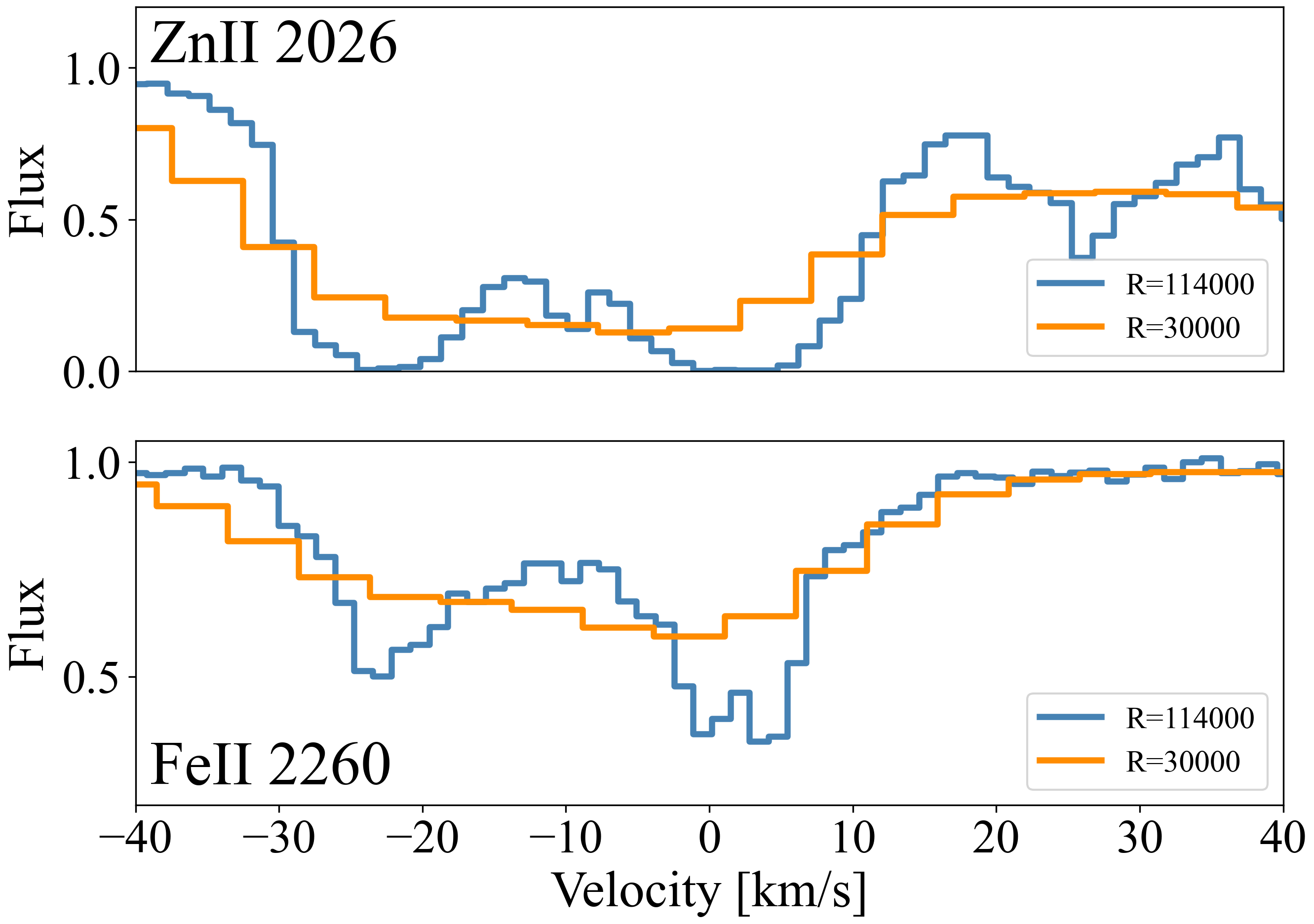}
    \caption{Observed HST/STIS \feii{} $\lambda$2260 absorption line at high-resolution ($R=114~000$, blue). The orange curve shows how this line would appear if observed with a lower resolution. Characterising narrow individual components of the ISM absorption line profiles requires high-resolution ($R\gtrsim100~000$).}
    \label{fig:ISM_highres}
\end{figure}

Figure \ref{fig:complexISM} (left panel) shows the locations of a sub-sample of STIS-ISM OB stars observed with HST/STIS at its highest resolution ($R=114~000$). In principle, if two targets are at a sufficiently close angular separation, it is possible to do differential studies of the line profiles (tomography). ISM absorption studies in tomography are a potential future avenue for isolating and characterizing the chemical properties of specific clouds \citep[e.g.,][]{Price01}, as well as their variability in time \citep{Price01b}. While such studies have been hampered so far by the limited sensitivity of UV high-resolution on HST, they have the important potential to open the study of the interplay between stars and the ISM, both in terms of stellar feedback and metal yields. Figure \ref{fig:ISM_highres} shows that resolving the narrow individual components of the ISM absorption line profiles requires high-resolution, at least $R\gtrsim100~000$.

How metals mix in galaxies is vastly unknown, yet it dramatically affects their chemical evolution of galaxies \citep[e.g.,][]{Matteucci21}. The survival of pockets of low metallicity (or high-metallicity) gas has also strong implications on several other fields of astrophysics, for example the observability of Pair Instability Supernovae \citep[e.g.,][]{Cikota17} or the search for chemical signatures of PopIII stars \citep[e.g.,][]{DOdorico24}. A deeper understanding of ISM metal mixing and cloud-to-cloud variations in chemical properties of the ISM is needed to truly advance our understanding of the interplay between stars and ISM and the chemical evolution of galaxies. 

\subsubsection{Physical conditions in the cold ISM}

Absorption lines from neutral carbon (\ci{}), carbon monoxide (CO), molecular hydrogen (\hh{}), and other molecular species, accessible primarily in the UV, provide powerful and complementary diagnostics of the physical conditions in the cold neutral medium (CNM), tracing the critical transition from diffuse atomic to molecular gas. 
The populations of the \ci{} fine-structure levels are sensitive to the gas density and thermal pressure through collisional and radiative excitation \citep{Jenkins79,Jenkins01,Jenkins11}, while \hh{} and CO constrain the kinetic temperature, ultraviolet radiation field, molecular fraction, and degree of shielding \citep[e.g.,][]{Snow06, Sonnentrucker07, Sheffer08}. 
Together, these tracers provide a comprehensive view of the thermal, chemical, and dynamical state of the ISM and its dependence on metallicity, dust shielding, and the ambient radiation field. 
Their combined analysis can thus yield fundamental insights into the structure and evolution of diffuse and translucent clouds in both the Milky Way \citep[e.g.,][]{Jenkins01,Jenkins11} and lower-metallicity systems such as the Magellanic Clouds \citep[e.g.][]{Welty16, Kosenko24, Balashev25, Tchernyshyov25}.

Theoretical models predict that lower metallicity environments are characterized by systematically different thermal pressures, cloud structures, and carbon chemistry -- owing to reduced dust shielding and enhanced ultraviolet radiation fields.
Those differences produce significant changes in the \cii{} - \ci{} - CO transition and the relative distributions of \ci{}, CO, and \hh{} \citep[e.g.][]{Wolfire03, Krumholz09, Glover12, Sternberg14, Bialy16}. 
Initial high-resolution UV spectra of a small number of sight lines in the Magellanic Clouds have indeed revealed evidence for higher thermal pressures, stronger radiation fields, and systematically weaker CO absorption than in the Milky Way, consistent with the theoretical expectations, while also suggesting a more complex interplay between metallicity, cloud structure, and carbon chemistry \citep{Welty16, Balashev25, Tchernyshyov25}. 
Detailed tests of the model predictions for low-metallicity systems will require high signal-to-noise, high-resolution ($R\gtrsim100\,000$) UV spectroscopy capable of detecting the intrinsically weak CO absorption, resolving its rotational structure, and disentangling the often complex velocity structure of the \ci{} lines, for a significantly larger sample of sight lines. 
Such observations are essential for establishing how the physical conditions, carbon chemistry, dust depletion, and ultimately the conditions leading to star formation vary across different galactic environments.

\subsubsection{The low-metallicity Universe}
\label{sec:low-z}

Unveiling the drivers of galaxy growth is one of the main scientific priorities of the Astro Decadal Survey 2020. While it is challenging to directly study the ISM of the early Universe, nearby analogues of high-$z$ systems can help better understand galaxy growth and the build up of metals and dust in the early Universe. For example, understanding whether the elements that dust is composed of at low metallicity are different from those in our Galaxy's dust may help explain the apparent rapid build-up of dust in the early Universe \citep[e.g.,][]{Rowlands2014}. Another example is the study of massive and very massive stars \citep[e.g.,][]{Crowther16}, which may be common in regions of very recent star formation at low metallicities, and may play a fundamental role in the enrichment of the early Universe. For example, this may be related to the recent JWST discoveries of N-enhancements at high-$z$ \citep[e.g.,][]{Cameron23,Marques-Chaves24}. To this end, the study of low-metallicity, star-forming, gas-rich, well resolved nearby dwarf galaxies has a special role to bridge the study of the early Universe.   

Current HST spectra of OB stars are currently limited to Milky Way (up to the highest spectral resolution of STIS), the Magellanic Clouds (mostly with mid-resolution), and out to few sources in more distant galaxies. \citet{Hamanowicz24} pushed the limits of HST/COS to low-metallicity dwarfs by measuring the ISM metal abundances towards 18 OB stars in IC1613 and Sextans A.   

At a distance of 18 Mpc, I Zw 18 is undergoing a very recent and viguros burst of star formation and is one of the most metal-poor star-forming galaxies known locally \citep{Izotov04,Aloisi07}. This makes it a unique nearby laboratory for studying stellar evolution and feedback, as well as the physics and properties of the ISM, including cosmic dust, under primitive conditions. While ISM absorption-line studies of I Zw 18 OB stars are not in reach of HST, higher sensitivity missions (Theon, HWO) should aim at obtaining at least moderate S/N ($\sim20$) and moderate spectral resolution (down to $R\sim30~000$) studies of some OB stars in I Zw 18 ($22 < V < 30$ mag). This would represent a leap forward in the study of the ISM, including dust, at low metallicities, and its implications for the early Universe. At even further distances, HST can study the host galaxies of Superluminous Supernovae in UV absorption \citep[e.g.][]{Yan18}, which have low metallicity and high star formation. The mid-resolution UV spectroscopy (currently offered by STIS and COS) is crucial to go beyond the limits of HST and maximise HWO sensitivity.

\subsubsection{Requirements}

The study of small-scale variations of chemical properties of the neutral ISM requires UV high-resolution spectroscopy ($R=100\,000$) covering 100-350nm. An extension to $R=50\,000$ is required for the study of fainter targets in low-metallicity, nearby galaxies. Covering the bluest wavelengths is necessary for the study of \hh{}. The target wavelength coverage and resolution are 91-450~nm (from the Ly limit to the blue end of ELT/ANDES wavelength coverage) and $R=30\,000$ - $200\,000$, where the lower resolution mode is needed for the extension to fainter targets and the higher resolution mode aims at fully resolving the sub-km/s individual components and matching the VLT/ESPRESSO Ultra-high-resolution mode.

The requirement on the sensitivity for the UV high-resolution ($R=100000$) spectrograph for the HWO is to obtain $S/N\sim50$ spectra of 50-100 OB stars in the Magellanic Clouds ($11<V<18$ mag), and at least 10-20 of the brightest OB stars in Sextans A ($22<V<30$ mag). A lower-resolution $R=30~000$ - $50~000$, high-sensitivity mode will be necessary for ISM absorption-line studies of I Zw 18.

The instrument requirements are summarized as follows: \\

\begin{enumerate}
    \item {\it Wavelength coverage} -- Required: 100--350\,nm; Optimal: 90--450\,nm. 
    \item {\it Spectral resolution} -- Required: 50\,000--100\,000;  Optimal: 30\,000--200\,000. 
    \item {\it Sensitivity} -- Targets are OB stars with $V$-band magnitudes in the range 5--28 mag.
\end{enumerate}

\subsubsection{Comparison of future capabilities}

\begin{figure*}
    \centering
    \includegraphics[width=0.7\linewidth]{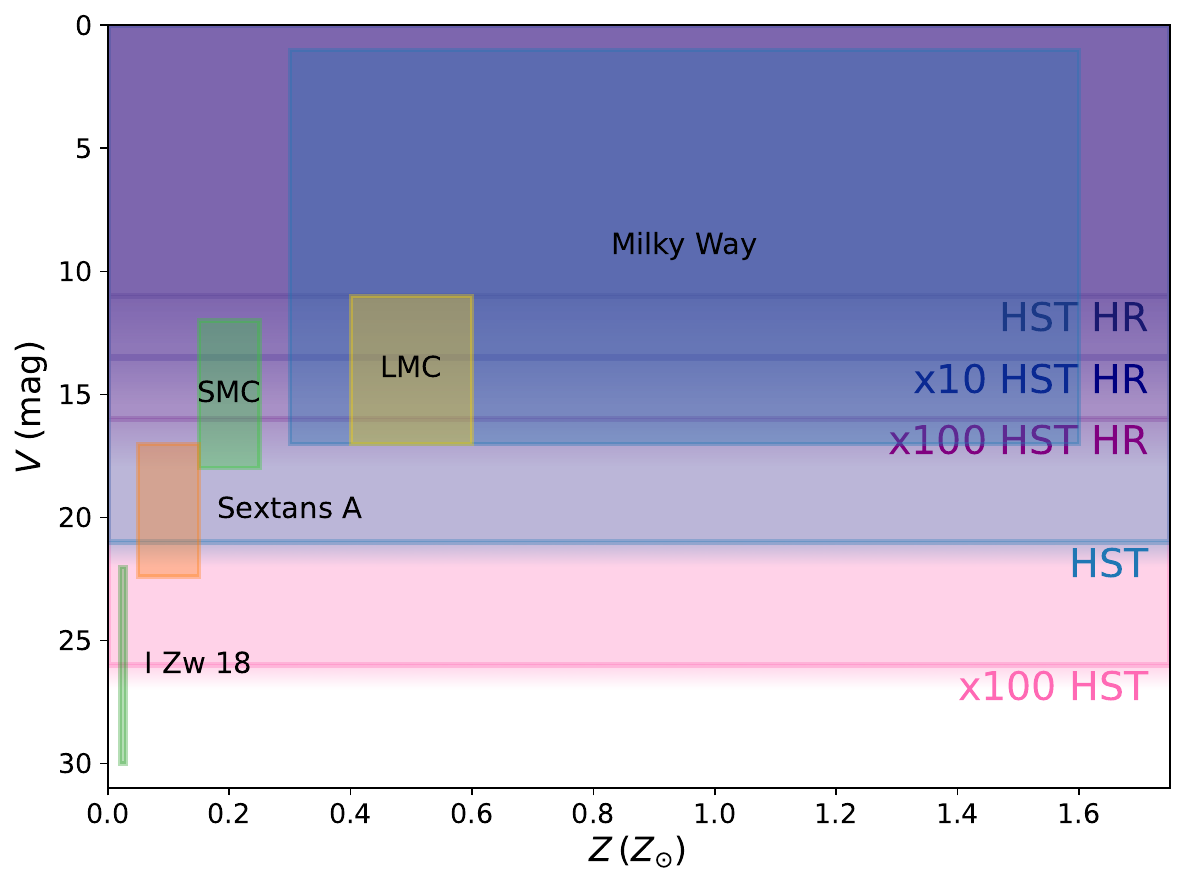}
    \caption{Pushing UV absorption-line spectroscopy of Local Group galaxies to lower metallicity. The ranges of metallicities and V magnitudes of OB stars in the Milky Way, SMC, LMC, Sextans A and I Zw18 are highlighted. The sensitivities to UV absorption-line spectroscopy of OB stars of different instruments are roughly indicated by the shaded areas: "HST HR" represents the HST/STIS high-resolution echelle (E140H and E230H), "HST" represents the lower resolution HST UV spectroscopy modes (both STIS and COS). The magnitude limits of 10x (e.g. Theon) and 100x (HWO/PEGASUS) more sensitive missions are shown, as comparison. The sensitivity of HWO/PEGASUS high-resolution $R=100\,000$ spectroscopy is indicated as "x100 HST HR".}
    \label{fig:lowmet}
\end{figure*}

Figure \ref{fig:lowmet} highlights the ranges of metallicities and V magnitudes of OB stars in the Milky Way \citep[based on data from][]{deBurgos23}, SMC, LMC \citep[e.g.][]{Roman-Duval20}, Sextans A \citep{Lorenzo22,Hamanowicz24} and I Zw18 \citep{Izotov04}. The sensitivity to UV absorption-line spectroscopy of OB stars for HST/STIS E140H and E230H (HST HR) and the lower resolution HST UV spectroscopy modes (both STIS and COS) are indicatively marked, as well as a gain of 10 (e.g. Theon) and 100 times (e.g. HWO) in sensitivity.

Carbon and Silicon are among the most difficult abundances (and depletions) to measure. Even with 100$\times$ greater FUV sensitivity than HST/STIS at R$\sim$100\,000, \citet{Roman-Duval26} calculated that the lowest metallicity galaxy where detections would be feasible is the SMC, where 10--100 sources might be measured in a reasonable time. This nevertheless represents a quantum leap in our understanding of their abundances, which as noted is currently limited to $\sim$10 lines of sight in the Galaxy.

For other elements with more-easily detected lines (e.g. Si, Mg, Fe) the picture is even more positive. Many of these lines are detectable at SNR $>$ 20 in the spectra of OB stars at distances up to 10\,Mpc, assuming sensitivity 100$\times$ better than HST/STIS at R$\sim$100\ 000.
In such a scenario, it would be feasible to measure $\gtrsim100$ lines-of-sight per galaxy within that volume. 
Even for sensitivity only 10$\times$ better, it would still be feasible to measure significant samples of 10--30 lines-of-sight per galaxy, which again represents an enormous improvement over the current state-of-the-art - the complete lack of such observations.

\subsection{Cosmic dust}

Cosmic dust is a fundamental component of the ISM and plays a central role in the evolution of galaxies. About half of all the metals of the ISM are locked into dust grains, making dust an important reservoir of metals in the Universe. The incorporation of metals from the gas-phase of the ISM into dust grains, known as dust depletion \citep{Field74,Savage96,Jenkins09,DeCia16,Konstantopoulou24b}, alters the observed gas-phase metal abundances in the ISM and provides a powerful probe of the chemical composition of dust \citep{Draine11,Jenkins14,Mattsson19,Roman-Duval22,Konstantopoulou24a}. Dust plays a crucial role in many physical processes that drive galaxy formation: it causes efficient gas cooling which is essential for the  formation of new stars, catalyzes the production of molecules, and constitutes the building blocks of planets \citep[e.g.][]{draine03ar,Draine11b}. Thus, dust grains are fundamental for the chemical evolution of galaxies, stars, planets, and molecules. 

Cosmic dust is composed of silicates and carbonaceous dust grains, as well as iron-rich grains \citep[e.g.][]{Jenkins14,Roman-Duval22,Konstantopoulou24a}. However, the detailed composition of dust remains one of the major questions of ISM studies. For instance, the amounts of key elements such as C, Si, and O in dust, as well as which grain species and in which amounts are still poorly constrained. Furthermore, the dominant mechanisms responsible for dust formation, grain growth and destruction remain still uncertain \citep[e.g.,][]{draine03ar,Dwek16,Galliano18,Decleir26}. Understanding how dust composition evolves from the Milky Way to lower metallicity systems is  essential for constraining the origin and evolution of cosmic dust, as well as galaxy evolution. 

UV spectroscopy has a fundamental role for the study of cosmic dust, offering a detailed overview of its chemical composition.
Most ionic transitions that help us tracing the key elements in dust, including C, O, Si, Mg, Fe, Ni, Ti, Zn, Kr, and others, lie in the UV (Fig. \ref{fig: UV lines}). Therefore, with UV absorption-line spectroscopy we can uniquely determine the chemical abundances of ISM gas-phase metals and dust, and thus investigate how dust properties vary across different environments. By measuring the chemical abundances and quantifying the fraction of each element that is removed from the gas-phase and incorporated into dust through dust depletion we can constrain a number of cosmic dust properties, such as the dust composition, dust-to-gas ratio and dust-to-metal ratio \citep[e.g.][]{Jenkins14,DeCia16,Roman-Duval21,Konstantopoulou24a}. These properties provide us with strong constrains on the mechanisms by which cosmic dust is formed and destroyed and how it evolves across galactic environments and through cosmic time. 

Some of the most pressing open questions are:
\begin{itemize}
\item \textbf{Carbon and the composition of dust.} What is the composition of dust? What is the overall amount of C, Si, O, and other elements in dust? 
\item \textbf{Chemically-enhanced 3D/4D dust maps.} Can dust depletion enhance 3D/4D maps of dust extinction with the information of the chemical properties in individual clouds? 
\item \textbf{Diffuse Interstellar Bands and C-rich grains molecules.} What are the sources of Diffuse Interstellar Bands? What are the carriers of C-rich dust grains? 
 \end{itemize}

\subsubsection{Carbon and the composition of dust}
Carbon is a fundamental component of life, and one of the main elements that build up cosmic dust, with carbon-rich dust making up probably large fractions of dust in the Milky Way. Yet, the overall amount of C in dust in the diffuse ISM is largely unknown. How much of all dust is made of C-rich dust exactly, and what is the overall amount of C in dust in the diffuse ISM? The study of dust depletion from UV absorption-line spectroscopy can uniquely answer these questions, through the detection of absorption lines of \cii{} and other elements. However, \cii{} is hard to constrain, because out of the two available transitions, one is too strong and the other is very weak. So far, HST/STIS has only 5 detections of \cii{}] $\lambda$ 2325 in the diffuse ISM, from high-resolution \'Echelle gratings \citep{Sofia04}. A few previous HST/GHRS \cii{}] $\lambda$ 2325 detections exist, but with large uncertainties \citep{Sofia97,Sofia98}. High-resolution ($R\gtrsim100~000$) UV spectroscopy and high S/N ($\gtrsim200$) is needed to be able to detect \cii{}] $\lambda$ 2325.
\begin{figure}
    \centering
    \includegraphics[width=1\linewidth]{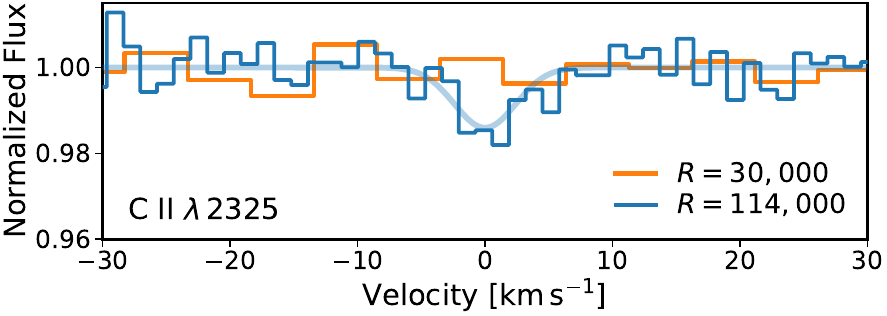}
    \caption{Synthetic \cii{} absorption from gas with log N(\hi) $= 21$ and solar metallicity, assuming $S/N=200$, spectral resolution $R=50~000$ (blue) and $R=100~000$ (orange), and spectral sampling of HST/STIS E230M and E230H, respectively. The high-resolution is necessary for \cii{} detection.}
    \label{fig:C}
\end{figure}
Figure \ref{fig:C} shows that a lower resolution of $R=50~000$ smears the lines and is insufficient for the weak \cii{} detection. Even if the depletion of carbon is quite low, C cosmic abundance is so high that even small variations in the C depletion lead to large variations in the measurements of the overall amount of C in dust. Going beyond the current limitations to solidly measure C in the ISM metals and dust in at least 100 systems in the Milky Way ($20\times$ more than currently available), and at least 10 systems in the Magellanic Clouds requires a significantly more sensitive UV spectrograph than HST \citep{Roman-Duval26}. This is necessary to provide the first statistically robust determination on the amount of C in cosmic dust and determine the overall dust composition.

Similarly, there is a paucity of measurements of \siii{} from the $\lambda$ 2335 weak line \citep{Jenkins09}. Robust O measurements remain also somewhat scarce because they rely on the weak \oi{} $\lambda$ 1355 transition. In the Magellanic Clouds, there are very few constraints of \oi{} and \kri{}. These volatile elements are fundamental for anchoring the ISM measurements of metallicity and dust depletion.  With an improvement in sensitivity of a factor of 10 compared to HST, tens of lines-of-sight in the Galaxy could be measured effectively (with well-constrained column densities of Si, O and Kr from their weak lines), while with 100$\times$ better sensitivity a sample of hundreds of galactic OB stars and tens of extragalactic OB stars could be used to measure gas- and dust-phase abundances. 

Dust composition can be constrained with the study of the depletion of several elements \citep[e.g. C, N, O, S, P, Si, Cl, Cu, Ca, Na, K, Ti, Mn, Fe, Co, Ni, Zn, Kr,][]{Jenkins14,Roman-Duval22,Konstantopoulou24a} in the warm neutral medium, not only C and O. In particular, Ti is highly refractory, and thus is one of the strongest depleted elements, but it is also an $\alpha$-element. These properties make Ti a crucial element for reliably determining dust depletion, as well as for disentangling any potential enhancements of $\alpha$ elements \citep[e.g.][]{DeCia24,Velichko24}. Aluminium is also an interesting element to probe in the ISM, because it is heavily depleted into dust, and its enhancement may related to the metal yields of (vary) massive stars \citep[e.g.,][]{Saccardi23}. Most of the transitions of interest are in the far and near UV, making the broad UV coverage essential for a comprehensive understanding of dust composition.

\subsubsection{Chemically-enhanced 3D/4D dust maps}

One of the next frontiers for the study of cosmic dust is combining two previously separate probes: dust depletion with dust extinction. On the one hand, the recent 3D dust maps \citep{Dharmawardena24,Edenhofer24} allow to pinpoint the physical distances to dusty clouds along lines of sight. On the other hand, the study of dust depletion can add the unique information of the chemical composition of these dusty clouds, which corresponds to the individual velocity components that can be identified in the absorption-line profiles, if the spectral resolution is sufficiently high. An example of the implementation of this technique is provided by \citet{Ramburuth-Hurt26}, who isolated the properties of the warm neutral medium of the Orion Nebula itself (finding supersolar metallicity) and derived new insights into the chemical properties of the surface of the Local Bubble \citep[a supernova-driven, low-density cavity around the Sun][]{Fuchs06,Zucker22,ONeill24}. By comparing the morphology of 3D dust maps at fixed distances to the morphology of HI emission maps at fixed velocities, it is further possible to construct 4D maps that assign distances to individual gas cloud velocity components \cite{Soler+2019, Soler+2025}. Together, these complementary tracers of dust and gas in the ISM will enable coherent 4D maps, including crucial information on the chemical composition of individual, parsec-scale gas clouds in the Milky Way.

\subsubsection{Diffuse Interstellar Bands and C-rich grains molecules.}

The composition of carbonaceous dust carriers are currently unknown. High-resolution spectra reveal narrow absorption features within Diffuse Interstellar Bands (DIBs), which are linked to small carbonaceous particles \citep[e.g.,][]{Foing1994, Campbell2015}. Most DIBs have been observed at wavelength larger than 3200\AA~\citep{Hobbs2009,Cox17}. The detection of DIBs in the UV may help to determine the composition of the obscuring material. So far, only a few shallow features have been detected in the UV \citep{Destree2009}. This is surprising, since the UV wavelength region contains important carbon transitions, e.g. the $\pi$ -> $\pi$* transitions found in aromatic and graphitic carbon materials, which cause absorption features between 2000 and 2600 \AA{} \citep[e.g.,][]{Menella96, Gadallah11}. Shallow and narrow UV DIBs (i.e., $< 1$ \AA{}) may have been overlooked by current instruments, such as COS and STIS onboard HST, since they fall below their detection limits. Therefore, we need a higher spectral resolution in the UV in order to observe the weak and increasingly narrow ISM features and to link these features to models based on laboratory studies and ab initio calculations.

\begin{figure}
    \centering
    \includegraphics[width=1\linewidth]{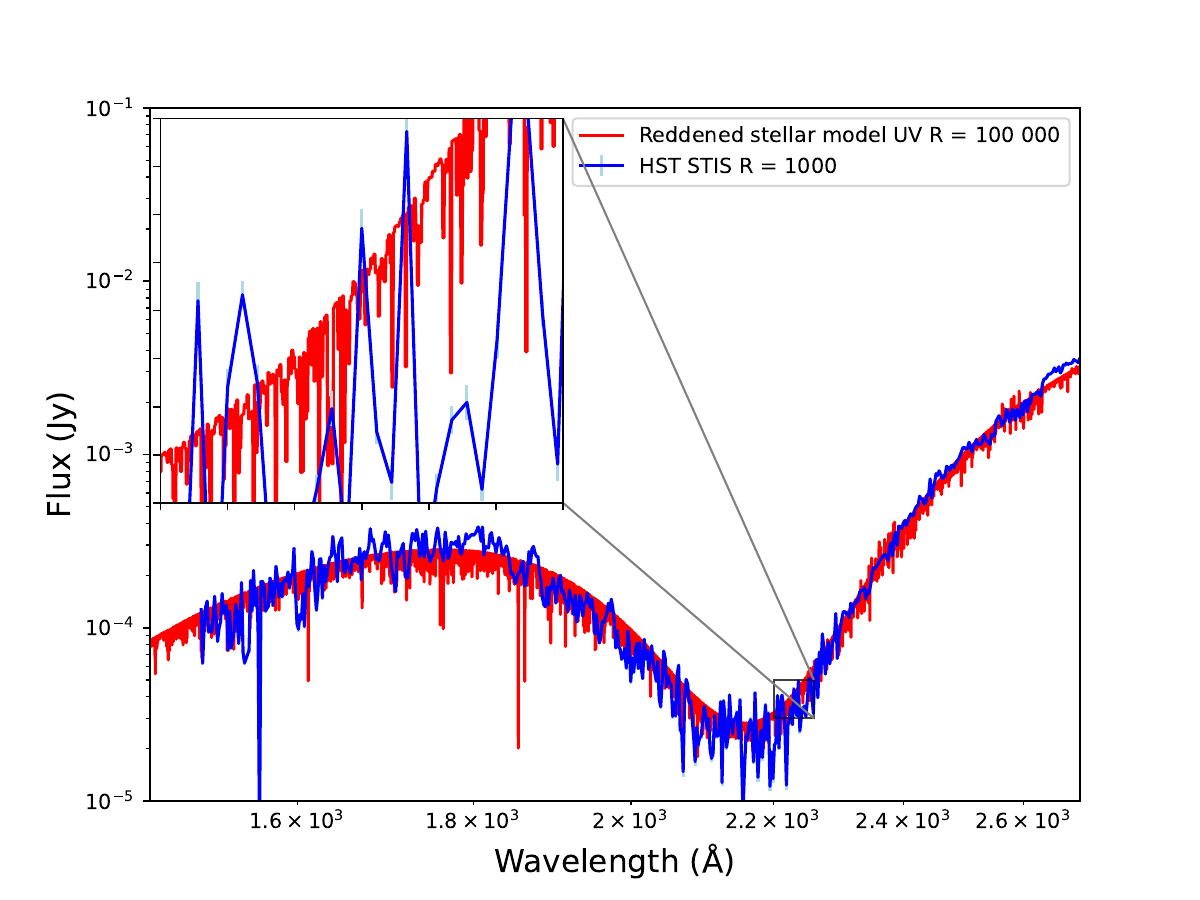}
    \caption{The 2175\AA\ feature of an O8V star in the Galaxy at 3 kpc taken  from~\citet{Zeegers2025}, observed with HST STIS and shown with R=1000 in blue. We cannot detect narrow and shallow dust features. A detailed model is shown in red with R = 100 000 required to observe narrow extinction features and gas lines. }
    \label{fig:uvspectrum}
\end{figure}

In the UV, dust grains are responsible for an overall absorption of radiation, as well as the broad extinction feature at 2175\,\AA, as shown in Figure~\ref{fig:uvspectrum}. The extinction continues to rise toward shorter wavelengths. This FUV rise most likely shows a combined effect of silicates and carbonaceous particles~\citep{Mishra2017}. The extinction at 2175\,\AA~is attributed to small carbonaceous particles~\citep[e.g.,][]{Stecher1965}. A detailed understanding of the UV dust extinction is also important to study the evolution of galaxies. The shape of the extinction can be parametrized by the total-to-selective extinction ratio: $R(V)= A(V)/E(B-V)$ and is linked to the grain size, where low values of $R(V)$ are linked to small dust size and vice versa. 

Within the Galaxy, we can observe variations in the extinction curve ($R(V)~\sim 2-5$) and the strength of the 2175\,\AA~feature, but the central wavelengths of the absorption does not shift~\citep{Valencic04}. In galaxies of the Local Group a wide variety in extinction shape can be observed \citep[e.g.,][]{Gordon03}. For example, the 2175\,\AA~feature is absent in most of the Small Magellanic cloud sightlines and a recent study of M33 by \citet{Gordon2026} shows large variations in the shape of the UV extinction. We can currently explore the UV ISM up to 5.0 magnitudes of extinction, as detailed in Fig. \ref{fig:uvlimitsavrv}. Larger extinction results in too much absorption in the 2175 \AA{} region, which prevents tracing the shape and depth of the broad extinction feature. In Figure~\ref{fig:uvspectrum} we show a UV spectrum with the HST STIS instrument of a O8V star and a simulated observation of the narrow features possible with a resolving power of $R = 100\,000$. Connecting the variation in the overall shape of the extinction curves with narrow extinction features will enable us to observe processing of interstellar dust in detail in the Galaxy and galaxies in the Local Group. This will greatly enhance our understanding of the composition and evolution of dust in the ISM. 

\subsubsection{Requirements}

Determining the properties of cosmic dust, such as dust composition and dust depletion, is observationally demanding because it requires measurements of numerous weak and narrow absorption lines that cover a wide range of wavelengths in the UV.
Measuring the dust depletions of elements, such as C, Si require access to intrinsically weak transitions, including \cii{}] $\lambda$ 2325, \siii{} $\lambda$ 2335. Obtaining reliable measurements of these lines is only possible with $R \gtrsim\,100~000$ UV spectroscopy and high-S/N ($\sim200$) spectra of OB stars. Such observations are currently limited to only a handful of sightlines in the Milky Way, making it difficult to robustly constrain the dust composition and the C contribution to the overall dust budget. An extension to lower spectral resolution ($R\sim50~000$) is required for the observations of fainter sources (see Sect. \ref{sec:ISM}). An extension to higher spectral resolution ($R\sim200~000$) would be desired, to study dust depletion in cold neutral clouds (possibly related to star-forming regions), which are often associated with very narrow (with sub-km/s widths) components in the velocity profiles.   

Obtaining extragalactic measurements of C, Si, O, and Kr depletions remain a major observational challenge, because they are only feasible with an instrument 100$\times$ more sensitive than HST/STIS in the far UV. A quantitative statistical understanding of the dependence of depletions and dust composition on metallicity requires similarly high sensitivity, while lower sensitivities would only enable the construction of a limited qualitative picture. A comprehensive characterization of the dust depletion requires a broad wavelength coverage that spans the full range from 100 to 350~nm. This is necessary for the detection of transitions of the different metals (C, O, Mg, Si, Fe, Ni, Ti, Zn, Kr). At least up to 350~nm coverage is essential to measure the strong \tiii{} lines, which are important probes of dust depletion and are critical for disentangling dust depletion and nucleosynthesis effects, such as $\alpha$-element enhancements. The aim for sensitivity is that high-resolution observations of OB stars with V-band magnitudes down to $\sim18$ mag will be in reach.  

The instrument would also enable studies of DIBs in the UV. Possible UV DIB features may have an FWHM of$\sim24\,\mathrm{m}$\AA~\citep{Destree2009} near 1300\AA. Therefore, in order to observe these DIBs as well as narrow gas lines, we need a resolution of $R \sim 30~000 -100~000$. DIB features are usually expected to have an absorption depth of 1-3\% of the flux. Therefore, these shallow features require signal-to-noise ratios between 50 - 300. Broad features, such as the 2175\,\AA~feature, do not require a high resolution, but are challenging to observe along sight lines with large dust extinction. The combined observations of dust depletion and extinction and UV DIBs would provide crucial constrains on the nature of C-grains and the carriers of the 2175\,\AA~feature, which are currently unknown.

The instrument requirements are summarized as follows: \\

\begin{enumerate}
    \item {\it Wavelength coverage} -- Required: 100--350\,nm; Optimal: 90--450\,nm. 
    \item {\it Spectral resolution} -- Required: 50\,000--100\,000;  Optimal: 30\,000--200\,000. 
    \item {\it Sensitivity} -- Targets to be observed are OB stars with a typical V magnitude in the range 5--28 mag.
\end{enumerate}

\subsubsection{Comparison of future capabilities}

\begin{figure}
    \centering
    \includegraphics[width=1\linewidth]{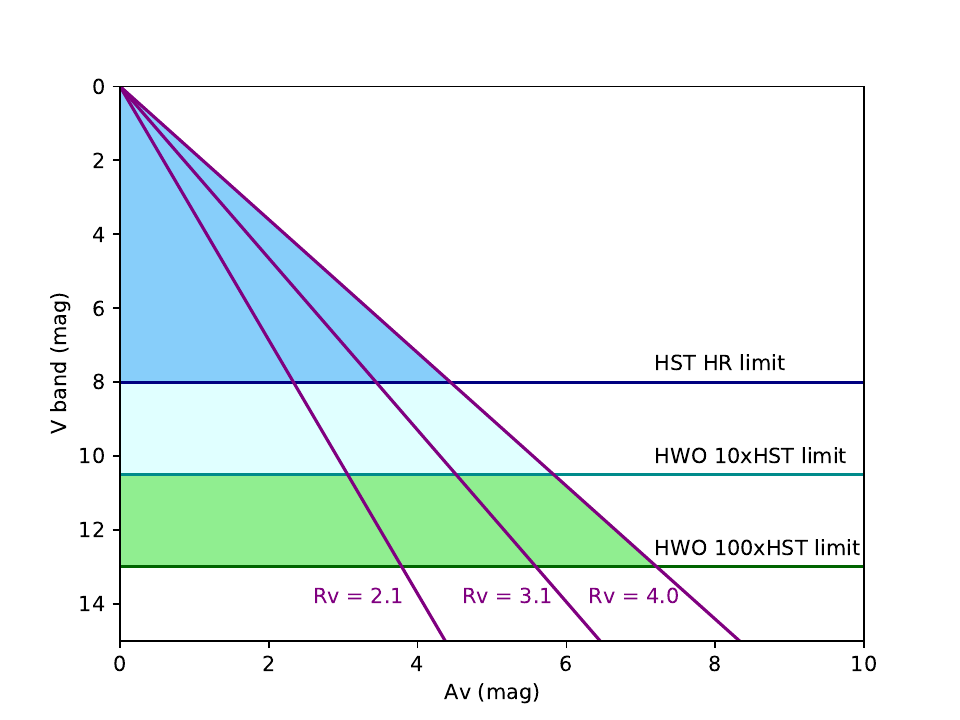}
    \caption{Observing narrow Galactic ISM features in the FUV, such as the unidentified features near 1300~\AA~\citep{Destree2009}, for a Galactic O star. The limits given by $R_V$ parameter, i.e. the ratio of total to selective extinction, are given by the purple lines. HST HR limit represents the high resolution limit for HST COS~\citep{Green12, Fronig2011}. With 10 to 100x the sensitivity of HST, we can probe 5 magnitudes deeper and explore narrow dust and gas features in denser environments of the ISM, which give us crucial insights in the chemistry of these environments.}
    \label{fig:uvlimitsavrv}
\end{figure}

Increasing the UV sensitivity by a factor of 10 (e.g. Theon) with respect to HST will allow reaching OB stars with 2.5 magnitudes fainter, and increase the number of observed targets, probably by tens. A more transformational advance will require an increase of sensitivity of more than 100$\times$ with respect to HST, enable the observations of targets 5 magnitudes fainter. This will make UV high-resolution spectroscopy of OB stars in nearby galaxies (mostly in the Magellanic Clouds, and out to Sextans A) in reach. Figure \ref{fig:lowmet} shows rough sensitivity limits to high-resolution (HR) and low-resolution HST UV spectroscopy of OB stars with HST and HWO (100$\times$ more sensitive).

 In Figure~\ref{fig:uvlimitsavrv} we show the UV sensitivity needed to observe DIBs, given the limits for extinction ($A_V$), $V$-band magnitudes, and the $R_V$ parameter, i.e. the ratio of total to selective extinction, for an O star. The currently existing UV instruments, namely COS and STIS of the HST are limited to observations of stars with $A_V$ up to 4.0 magnitudes for average ISM extinction with $R(V)\sim3.1$ and $A_V$ up to 2.0 magnitudes for $R(V)\sim2.1$~\citep{Snow2011,Destree2009, Fronig2011, Green12,Welty2020}. A next generation UV spectrograph with substantial sensitivity improvement would push these observational limits. For instance, a dedicated mission with a sensitivity $10\times$ HST (e.g. Theon) or $100\times$ better than HST (e.g. HWO) would make possible the observations of objects with up to 2.5 - 5 magnitudes deeper, as well as the study of fainter/more distant objects. This will enable us to study dust in the diffuse ISM in galaxies of the Local Group, as well as dense environments in the Galaxy.

\subsection{The Circumgalactic and Intergalactic Medium}

\newcommand{\ovi}{O~{\sc VI}}
\newcommand{\hmol}{H$_2$}
\newcommand{\kms}{\,km\,s$^{-1}$}
\newcommand{\heii}{He~{\sc II}}
\newcommand{\nitrogeni}{\mbox{N\,{\sc i}}}

Like planets and stars, galaxies have atmospheres, which play essential roles in their evolution. Galactic atmospheres are extended, diffuse, multi-phase, and massive, containing a significant fraction of all cosmic baryons and metals \citep{Putman2012, Werk2014, Peeples2014, Tumlinson2017, peroux2020, Chen:2026}. Known as the circumgalactic medium (CGM), these atmospheres represent a source of fuel for future star formation and
a sink for the metal-rich by-products of past star formation. They form the interface between the more diffuse and filamentary intergalactic medium (IGM), which is the reservoir of pristine gas, and the ISM. Inside the CGM, inflows and outflows cross paths and baryons are circulated between galaxies and their environments, with physical state and enrichment levels that are highly sensitive to the physical processes that govern galaxy evolution \citep[][]{Faucher-Giguere:2023}. 

Developing a full understanding of gas flows, their origins, and their fate is highlighted in the American 2020 Astronomy Decadal Survey (Astro2020) as a priority area 
{\it “Unveiling the Hidden Drivers of Galaxy Growth”} 
and forms a key scientific recommendation for a large-aperture ultraviolet (UV)/optical space telescope. 
IGM-CGM science is therefore recognized by the international community as an important area of modern astrophysics \citep[see][]{Borthakur2025}.

\begin{figure*}
\includegraphics[width=1.0\textwidth]{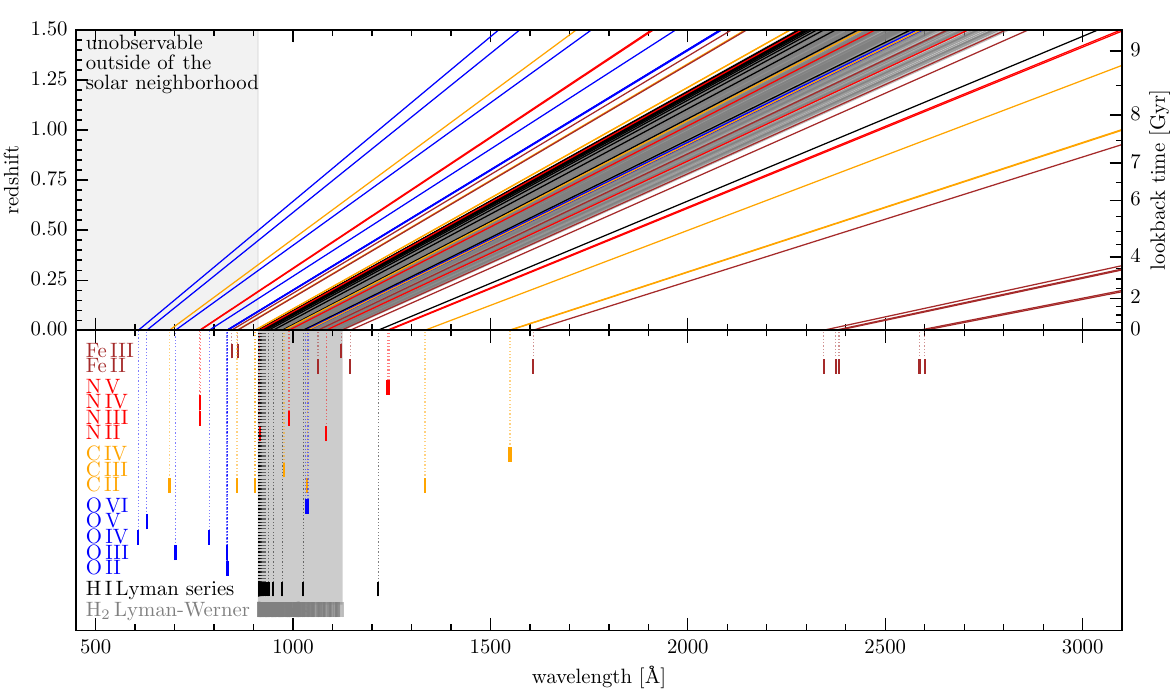}
\caption{UV absorption lines commonly used in IGM-CGM science. The bottom panel shows the rest-frame wavelengths of key transitions of the Lyman and Werner bands of H$_2$ (grey), the Lyman series of H\,I (black), and heavy element ions chosen to highlight different nucleosynthetic processes, with oxygen from core collapse supernova ejecta (blue), carbon from core collapse supernova ejecta as well as AGB star winds (orange), nitrogen from AGB star winds (red), and iron from both core collapse supernova and Type Ia supernova ejecta (brown). The rest-wavelength of each transition is marked with a vertical dash and a dotted line (or grey band for H$_2$) connecting to the top panel which shows the corresponding observed-frame wavelength as a function of redshift (left axis) and cosmological lookback time (right axis). 
Together, these lines make the UV the crucial bandpass for observing and characterizing 
the IGM-CGM across multiple gas phases; no other bandpass has the same density of resonant spectral lines from ground-state transitions per Angstrom.}
\label{fig:cgm_wavelength}
\end{figure*}

Ultraviolet observations have always been the backbone of IGM-CGM science at low redshift. The UV contains a wealth of spectral lines (see Figure~\ref{fig:cgm_wavelength}), including
the \hi\ Lyman series from Lyman-$\alpha$ at 1215\AA\ down to the Lyman limit at 912\AA, suites of resonance lines of cosmically abundant metal ions of diverse nucleosynthetic origins, and the Werner and Lyman bands of molecular hydrogen (\hmol).
The UV therefore covers a wide range of tracers of molecular, neutral, low-ion, to high-ion gas
\citep[see Table 1 in][]{Borthakur2025}
and is therefore {\it the} key wavelength range necessary for studying the diffuse Universe.
Several generations of UV telescopes and instruments, from IUE, Copernicus, FUSE, to HST, have anchored our understanding of the CGM by providing a significant observational dataset of these absorption lines. But even Hubble UV observations of the IGM-CGM have their limitations, and new facilities are needed to drive the field forward, as we now discuss.

\subsubsection{Open questions in CGM/IGM science}

Some of the more pressing open questions in this field are:
\begin{itemize}
    \item How do galaxies grow?
    \item How do metals distribute around galaxies?
    \item How do molecules distribute around galaxies?
    \item Are \hh{} outflows common at the disk-halo interface and in the CGM?
    \item How far from the disk can they survive in the CGM, and what factors promote or inhibit their survival?
    \item Do \hh{} outflows have any connection to the observed neutral \hi{} and low-ionization metal outflows?
    \item What is the location of Galactic and Magellanic high-velocity clouds?
    \item What is the thermodynamic state of the CGM?
    \item Can we chemically trace the origins of the gas and feedback? 
    \item Can we characterise the chemical and kinematical properties of infalling gas and outflows?
    \item How did the \heii\ reionization proceed and when was it completed?
\end{itemize}

Here we highlight several CGM science areas where further UV observations 
are needed:\\

{\bf 1) \ovi\ absorption.} Oxygen is the most abundant cosmic metal. Of all the ionization states of oxygen, \ovi\ is the most accessible in the UV, with a doublet at 1031, 1037\AA. 
\ovi\ has a demonstrated and well-studied presence in the CGM around diverse galaxies from dwarfs to massive systems
\citep{Sembach2003, Tripp08, Thom08, Chen09, Tumlinson2011, Fox2013, Danforth16, Pointon2017, Zahedy19, Tchernyshyov2022, Qu2024, Mishra:2024} and provides a unique tracer of collisionally-ionized coronal gas at ``transition temperatures'' of $T\gtrsim10^5$\,K. This is the highest-temperature gas that can be probed in absorption in the UV (higher temperature tracers like O~{\sc VII} and O~{\sc VIII} lie in the X-ray), meaning that \ovi\ observations are needed for a complete census of baryons in any given CGM.
Warm-hot gas in the CGM can be created by a variety of processes, including shocks, turbulent mixing, and non-equilibrium cooling \citep{Gnat2007, Oppenheimer2013, Oppenheimer2016, Ji2019}. Moreover, significant gains in UV spectral coverage and sensitivity can enable observations of still hotter phases ($T\approx 10^6$ K) at the virial temperature expected for massive galaxies through extreme UV doublets, such as Ne\,VIII at 770, 780 \AA\ \citep[][]{Burchett:2019, Haislmaier:2021} and even Mg\,X at 609, 624 \AA\ \citep[][]{Qu:2016}.
Future observations of the warm-hot phase through \ovi, and more, across a range of redshift, galaxy type, and impact parameter are needed to probe the spatial extent, small-scale structure, and baryonic content of warm-hot CGM gas and constrain modern CGM simulations, which have the spatial resolution to make detailed predictions on \ovi\ \citep{Lochhaas2026, Lucchini2026}.\looseness=-1

{\bf 2) \hmol\ absorption.} Molecular hydrogen is readily observable in the UV via the Werner and Lyman ro-vibrational bands between 912\AA\ and 1180\AA. \hmol\ is a key tracer of the Disk-Halo (DH) Interface, which represents the boundary between the ISM and the CGM, and where gas densities tend to be higher than in the diffuse CGM, and ionized gas must condense to become neutral and ultimately molecular. This is a critical phase change in the baryon cycle, which must occur if CGM gas is ultimately able to power star formation in the disk. 
\hmol\ has been detected in several high-velocity clouds (HVCs) around the Milky Way \citep{Richter2001, Cashman2021, Tchernyshyov2022b} and in DLA systems at $z$=0.4-0.6 that traces the CGM environment of an early-type galaxy \citep{Muzahid16, Zahedy20, Boettcher2021}. 
More observations of \hmol\ in the DH interface are needed to characterize this process. 
Observations of \hmol\ in the CGM will address many questions: (1) Are \hmol\ outflows common at the disk-halo interface and in the CGM? (2) How far from the disk can they survive in the CGM, and what factors promote or inhibit their survival?  (3) Do \hmol\ outflows have any connection to the observed neutral H~{\sc I}  and low-ionization metal outflows \citep{McClure2013,Lockman2016,DiTeodoro2020,Lockman2020}? 

{\bf 3) Locating Galactic and Magellanic HVCs.} HVCs are an observational tracer of the inner CGM of the Milky Way and the gaseous structures surrounding the Magellanic Clouds. 
While we have a good understanding of the kinematics, chemical abundances, and ionization state of these HVCs \citep{Wakker1997, Lehner2012, Fox2014, Richter2017}, their distances have long remained unknown, and are generally only constrained in a statistical sense \citep{Lehner2022}, rather than cloud-by-cloud. 
Lack of distance information leads to substantial uncertainties in the total HVC mass and mass flow rate \citep{Fox2019}.
Progress is being made in bracketing HVC location by identifying them in absorption in the spectra of halo stars at known distances, typically blue horizontal branch (BHB) stars \citep{Bish2019, Mishra2025}. 
However, HST/COS is only able to observe BHBs at reasonable S/N ratios out to $\approx$10--20 kpc \citep{Bish2021, Werk2019}, not far enough away to probe any gas clouds in the extended Milky Way CGM or the Magellanic Clouds. 
More distant BHBs are too faint for HST spectroscopy, requiring prohibitively high exposure times. 
A key science goal for future UV capabilities will be reaching BHBs down to FUV magnitudes of $\approx$20--21 and therefore distances out to $\approx$50--100 kpc, needed to probe gas in the outer halo and map the gaseous Milky Way halo out to the Magellanic Clouds and beyond.

{\bf 4) Thermodynamic state of the CGM.} The temperature, density, and turbulent nature of the CGM are ideal probes of the feedback processes that regulate galaxy evolution, whose physical mechanisms remain poorly understood. For example, while the hot-wind mediated supernova feedback implemented in most simulations of star-forming galaxies predict CGM dominated by a warm-hot phase ($T\sim10^5-10^6$ K), cosmic ray feedback results in a dramatically different structure, with most of the CGM in a significantly cooler ($T\sim10^4$ K), but still highly ionized and diffuse, volume-filling medium \citep[][]{Ji:2020, Bieri:2026}. Moreover, the build-up of stable, rotating disks in galaxies may be driven by a CGM phase transition as cool, turbulence-dominated CGM at $z>1$ transformed into warm-hot galactic atmospheres supported primarily by gas pressure at lower redshift \citep[][]{Stern:2021}. High-resolution spectroscopy of multi-phase, multi-element suites covering a range of atomic masses of CGM features observable in the UV provides unique constraints on gas temperature and turbulent line broadening. When paired with photoionization modeling from different ions of the same element, these data enable gas density inferences and constrain gas pressure for insights into the relative importance of thermal versus non-thermal support. The feasibility of these methods is demonstrated at Cosmic Noon with echelle spectrographs on large, ground-based telescopes \citep{Rauch96,Rudie:2019}, but UV capability is critical to extending insights to the two-thirds of cosmic history over which galaxies transformed into their current, more quiescent states. Moreover, the relatively thin Ly$\alpha$ forest at $z<1$ provides access to a much richer suite of UV features than is available at higher redshifts. Observations with HST are providing first insights \citep{Chen00,Keeney13,Werk2014, Zahedy21, Qu2024, Sameer:2024, Kumar:2024}, but significantly advancing our understanding of the thermodynamics of the CGM across galaxy mass and redshift requires new UV capability.

\begin{figure*}
    \centering
    \includegraphics[width=0.7\linewidth]{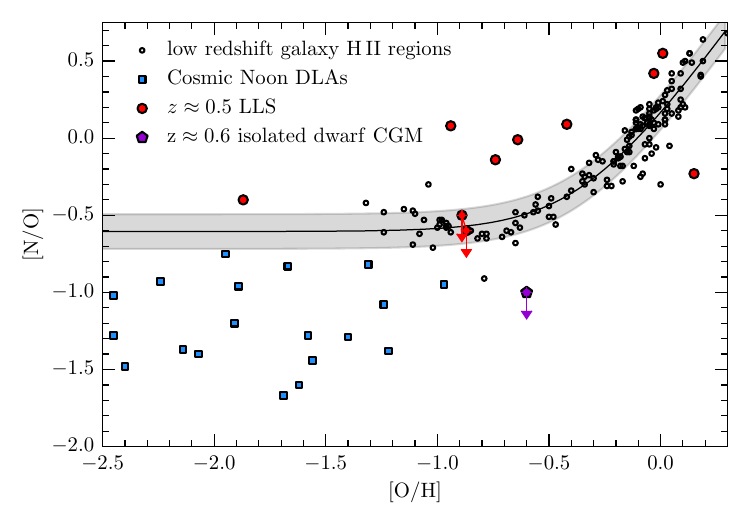}\vspace{-1em}
    \caption{[N/O] versus [O/H] (updated from \citealt{Zahedy21} and using \citealt{Asplund21} solar abundances) in different gaseous environments  including DLAs at Cosmic Noon
    \citep{Petitjean08,Pettini08}, Lyman-Limit Systems at $z\approx 0.5$ 
    \citep{Zahedy21}, and the ionized CGM of a dwarf galaxy at $z\approx 0.6$ 
    \citep[][]{Johnson:2026}, all compared to \hii\ regions in low-$z$ galaxies \citep[][]
    {Berg:2012, Pilyugin:2014, Berg:2016} with a black curve and gray 
    band visualizing the \hii\ region mean trend and scatter. The LLS [N/O] abundances largely fall above the \hii\ regions, consistent with chemically mature ISM that escaped from galaxies and then mixed with more pristine gas. The low-metallicity DLAs and isolated dwarf CGM [N/O] fall below the \hii\ regions, consistent core-collapse supernova ejecta expectations \citep[][]{Nunez:2022, Johnson:2023}. For a few LLS without O ion measurements, we use Si, Mg, or S and assume solar $\alpha$-element abundance patterns.}
    \label{fig: N/O CGM}
\end{figure*}

{\bf 5) Chemical tracers of gas origins and feedback.} 
The relative abundances of chemical elements with different nucleosynthetic origins act as cosmic clocks \citep[][]{Tinsley:1979}, enabling insights into gas enrichment history and timing of outflows relative to star formation. The far-UV is particularly rich in nucleosynthetic tracers, for example: (1) oxygen, ejected by core-collapse supernovae almost immediately after star-formation, (2) nitrogen and carbon, which arise from AGB star winds hundreds of millions of years later, and (3) elements with contributions from Type Ia supernovae that explode billions of years later, such as sulfur and especially iron \citep[][]{Kobayashi:2020}. When observations of ions of diagnostic elements are paired with H\,I measurements from the Lyman series, the combination of chemical maturity from relative abundances such as [N/O or [Fe/O] and metallicity such as [O/H] provides direct insights into both gas origins and subsequent gas mixing. For example, systems with high [C/O] or [N/O] but low metallicity indicate enrichment from outflows from galaxies with the high metallicity AGB stars needed for significant secondary production of these elements  \citep[][]{Lehner:2016, Zahedy21, Kumar:2024} and then subsequent dilution by the more pristine intergalactic medium \citep[][]{Wotta:2016, Chen:2019, Fox:2026}. Super-solar [Fe/O] signifies late-time feedback driving Type Ia supernovae products into the CGM \citep[e.g.,][]{DeCia12,Zahedy17,Zahedy19}. On the other hand, very low [N/O] at high metallicity indicates core-collapse supernovae ejecta that escaped to the CGM without mixing with ISM, indicative of an outflow with low mass-loading \citep[][]{Johnson:2026}. Figure \ref{fig: N/O CGM} shows the [N/O] as a function of metallicity in the diffuse CGM and star-forming regions. While diagnostic relative abundances provide unique insights into the gas flows that govern galaxy evolution, the number of available measurements from HST in the ionized phases that dominate CGM detections are few and far between. Gaining insights in statistical samples without significant selection effects requires new UV capability with both higher sensitivity and higher spectral resolution.

\subsubsection{Low-$z$ galaxies and Damped Lyman-$\alpha$ absorbers}

Another powerful application of UV spectroscopy is probing the ISM and CGM of galaxies that lie along the lines of sight to distant quasars. This technique offers a relatively unbiased view of the chemical and dynamical evolution of galaxies compared to emission-based galaxy surveys, since the absorption line strengths are independent of the galaxy luminosities or star formation rates. Especially important for this purpose are the damped Lyman-$\alpha$ (DLA) absorbers \citep{Wolfe05}. DLAs are the most gas-rich quasar absorption systems, with H I column densities log $N_{\rm H I}$ $\ge$ 20.3 cm$^{-2}$. DLAs mark the primary neutral gas reservoirs available for star formation \citep{Wolfe05}, carry the majority of metals over the first half of cosmic history \citep{Peroux20} and enable determinations of the cosmic \hi{} density, $\Omega_{H I}$ \citep{Rao06,Noterdaeme12}. 
DLAs and sub-damped Lyman-$\alpha$ (sub-DLAs, 19.0 $\le$ log $N_{\rm H I}$ $<$ 20.3 cm$^{-2}$) trace gas neutral in the ISM and CGM of galaxies.

Besides \hi{}, DLAs show absorption lines of various heavy elements, and provide the best constraints on the metallicity, metal (relative) abundances and dust content of the ISM and inner CGM of distant galaxies \citep[e.g.][]{Rafelski12,DeCia18b,Poudel17,Konstantopoulou24a}. Measurements of volatile elements such as O, P, S, Zn in DLAs is especially important to constrain the intrinsic metallicity. These volatile elements, together with the more refractory elements (e.g., Si, Fe, Mn, Cr) also allow  studies of dust depletion in DLAs \citep[e.g.][]{Ledoux02,Vladilo02,DeCia16}. 
While studies of DLA element abundances exist over $0<z<5.5$, they have been limited by small sample sizes and the challenges of obtaining high-resolution spectroscopy for faint distant quasars, coupled with the increasingly denser Lyman forest with increasing redshift. 

\begin{figure*}
    \centering
    \includegraphics[width=\linewidth]{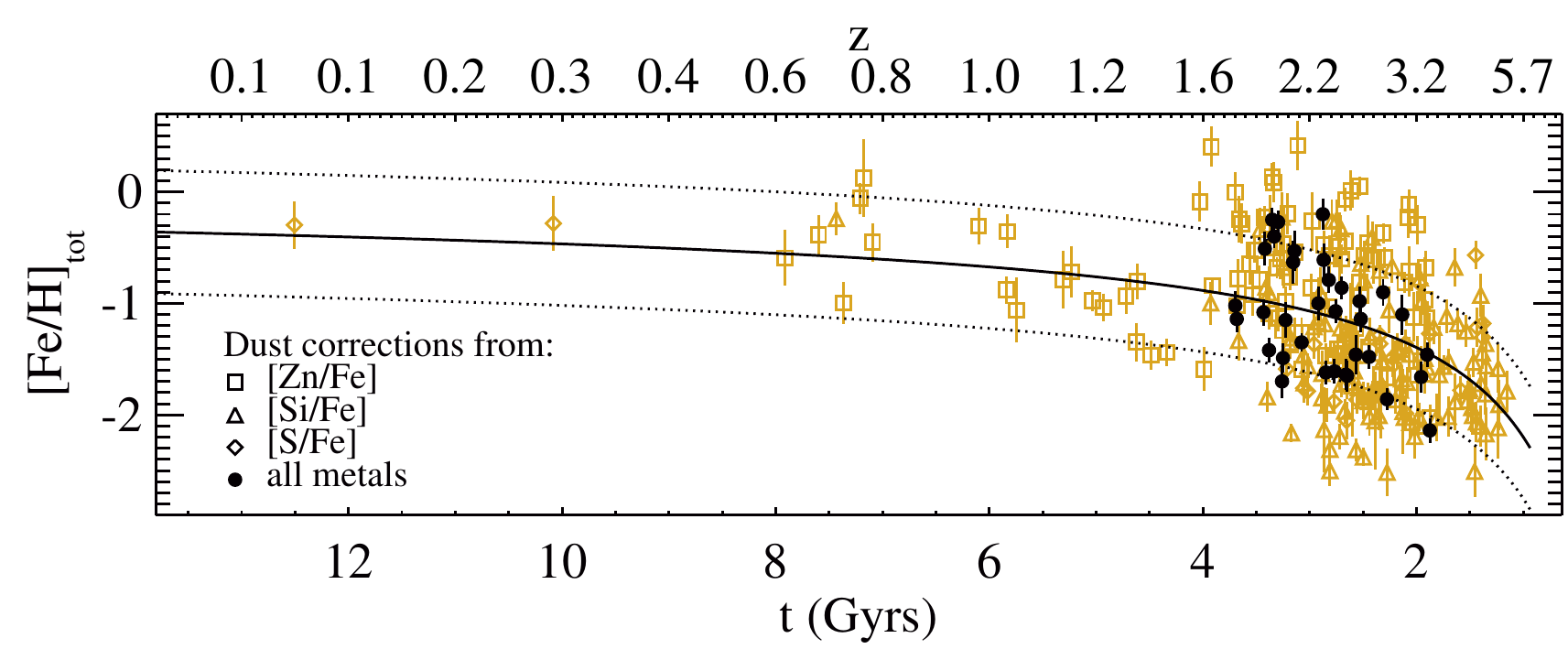}
    \caption{DLA metallicities, after correcting for dust depletion \citep[adapted from][]{DeCia18b}. More sensitive UV high-resolution spectroscopy is needed to fill the low-$z$ desert of DLAs with good metallicity measurements and constrain the chemical properties of the ISM/CGM of nearby galaxies.}
    \label{fig:met_DLAs}
\end{figure*}

UV spectroscopy of large samples of low-redshift DLAs is essential to advance this field for a number of reasons: (1) The absorption lines of the dominant ions of the key elements at redshifts $z<1$ lie primarily in the UV. (2) The contamination from the Lyman forest is far less severe at $z<1$. (3) So far there are only about 10 DLAs with well-constrained metallicity at $z<1$ \citep{DeCia18b}, which spans the last $\sim8$~Gyr of cosmic history. This is illustrated in Figure \ref{fig:met_DLAs}. (4) The detailed information on the kinematics and chemical properties of the neutral ISM/CGM that can be determined in absorption is complementary to study of galaxies in emission. Galaxies associated with DLAs at $z<1$ are easier to detect in emission, making it easier to connect the absorption and emission view  of galaxy evolution. (5) While the evolution of $\Omega_{\rm H I}$ is believed to be relatively flat across $0 < z < 5$, it is currently not well-constrained at $z <1$, due the small sizes of DLA samples allowed by HST UV studies \citep[e.g.][]{Rao06}. 

While some UV spectroscopic studies of DLAs at $z <1$ have been conducted with HST (and reveal significant differences relative to the predictions of cosmic chemical evolution models), they have been limited to small samples because of both the relatively small aperture of HST and the relatively modest resolution of HST STIS and COS, often in the range $R\sim10~000$ - 15~000 for the modes realistic for quasar observations \citep[e.g.][]{Kulkarni05,Battisti12, Som15}.  

High-resolution UV spectroscopy with $R$$\sim$50~000-100~000 with a large space telescope will revolutionize the study of the build up of galaxies through low-$z$ DLAs in a number of ways:

(1) More reliable metal column density determinations due to the better ability to resolve for blends with \hi{} Lyman-forest lines and metal lines of other intervening galaxies along the sight line.

(2) Improved ability to correct for line saturation due to both better sampling of the absorption line profiles, and the better sensitivity to detect weaker transitions (crucial for species such as \cii{} and \oi{} where only strong, saturated lines are measurable with the current HST modes). 

(3) Capability of characterizing individual (groups of) ISM/CGM clouds as individual narrow components in the absorption-line profiles. Dissecting these galaxies in their individual components along the line of sight will be key to study the complexity of the multi-phase ISM/CGM, for example by measuring cloud-to-cloud variations of metal relative abundances, dust depletion, and kinematics in the gas within and around these galaxies. This could be key to characterise infalling or outflowing gas \citep[e.g.,][]{Bouche13}. 

(3) More robust determination of \hi{} column densities by more accurate measurements of Lyman-$\alpha$ line profiles, as well as the coverage of higher-order Lyman series lines to help distinguish true DLAs from blends of multiple lower-$N_{\rm H I}$ absorbers. (Current HST COS G130M grating modes in the lack sufficient sensitivity to cover even Lyman-$\beta$ for $z<0.2$.) This would enable better determination of $\Omega_{H I}$ and help to pin down the low-$z$ end of the evolution of $\Omega_{H I}$. 

(4) Sensitivity to detect weak fine-structure metal transitions (e.g., for \cii{}*, \oi{}*, \siii{}*) that can offer constraints on the cooling rate, electron density etc.

(5) Sensitivity to detect molecular transitions (e.g., \hh{} Lyman and Werner bands, CH, CO) that can offer constraints on the physical conditions in the absorbing gas (e.g., density, UV radiation field intensity). 

(6) More robust determinations of the gas kinematics to measure the velocity dispersion, Doppler $b$ parameters, constraints on bulk vs. thermal motions (by comparing $b$ parameters for a range of elements), and assess the presence of gas within the galaxy disk, disk/halo interface, outflows etc. 

Overall, connecting the detailed chemical and physical properties of the ISM and CGM of galaxies at $z<1$ determined in absorption with the more traditional imaging studies (in emission) of the DLA galaxies (also improved due to the increased sensitivity for detecting DLA host galaxies and any tidal structures around them enabled by a larger UV/optical space telescope) would provide a far better understanding of the interactions between the stellar and gaseous components of galaxies, and between galaxies and their CGM.  

\subsubsection{The IGM and the \heii\ reionization process}

The epoch of helium reionization marks the final baryonic phase transition that influenced the thermal and ionization state of the IGM. While hydrogen reionization ends by $z \simeq 5.3$ \citep[e.g.][]{Bosman22}, the completion of \heii\  reionization is likely delayed to $z\sim3$, when quasars become numerous enough to supply the required hard ($E = h_{\rm P} \nu > 54.4$ eV) UV photons \citep[e.g.][]{Compostella13,Compostella14}.
There remains significant uncertainty in the precise timing and morphology of \heii\ reionization, as the detailed conditions of the IGM during and after \heii\ reionization depend on several poorly constrained parameters of the high-redshift quasar population (e.g., their duty cycle, spectral energy distribution, opening angle and escape fraction of ionizing photons), and the frequency and structure of self-shielding absorbers (Lyman limit and larger column density absorption systems). 
Typically, several generations of quasars are required to fully reionize a given region, resulting in a rich thermal and ionization structure of the gas \citep[e.g.][]{Compostella13,Compostella14,Garaldi19}.

The details of the \heii\ reionization process have become even more relevant in recent times due to the large number of high-$z$ AGNs identified by JWST \citep[e.g.][]{Maiolino24a} and the discussion on their contribution to the \hi\ reionization process \citep[e.g.][]{Madau24}. An important contribution from quasars/AGNs could imply an \heii\ reionization happening too early in cosmic history, at variance with the observational results. 

The \heii\ reionization epoch can be studied directly via spectroscopy of intergalactic \heii\ Lyman-$\alpha$ absorption ($\lambda_{\rm rest} = 303.78$ \AA) toward far-UV (FUV)-bright quasars at $z > 2$. Unfortunately, this approach is challenging due to the need of space-based observations and also to the paucity of bright quasars for which the far-UV is not extinguished by intervening  optically thick \hi\ absorption systems. 
The latest study on this topic \citep{Worseck19} is based on a sample of 25 quasar spectra observed with HST/COS, with very different resolving powers and signal-to-noise ratios (S/N). The lines of sight cover the redshift range $2.3 \le z \le 3.85$  and show a large scatter in optical depth at all redshifts above $z\sim 2.6$, testifying the extreme patchiness of the reionization process. 

While other quasar candidates were selected to have far-UV flux \citep[e.g.][]{Worseck11,Syphers12} and would be needed to increase the statistics and beat cosmic variance, they are generally too faint to be followed up with HST/COS. A spectrograph with the characteristics of wavelength coverage and sensitivity as foreseen for PEGASUS would be ideal to carry out this science case and make a significant step forward.  The minimum resolving power required to resolve the \heii\ Lyman-$\alpha$ forest is $R\sim20\,000$ but $R\sim 40\,000-50\,000$ could be desirable. The wavelength range $900-2000$ \AA\ would allow to study the  \heii\ Lyman-$\alpha$ forest in the redshift range $z\sim2-4$, critical to investigate the final phases of percolation of the He~III bubbles. An extension to the optical wavelength range would allow to measure the \hi\ Lyman-$\alpha$ forest corresponding to the \heii\ one: the ratio of the two optical depths and its evolution with redshift are a probe of the contribution of quasars to the \hi\ reionization \citep{Garaldi19}.

\subsubsection{Requirements}

Probing the multi-phase CGM in local galaxies requires coverage of the entire far-UV bandpass (912--1700\,\AA) to access the full range of molecular, neutral, low-ion, intermediate-ion, and high-ion spectral lines. In particular, the \ovi\ science case requires coverage down to 1000\,\AA\ to cover the \ovi\ doublet at 1031, 1037\,\AA. The \hmol\ science case requires coverage down to the Lyman Limit at 912\,\AA, as the Lyman and Werner bands (as well as the \hi\ Lyman series) extend across the entire far-UV. The CGM thermodynamics and chemistry science cases require access to extreme UV transitions of oxygen (\footnote{Neutral species such as \oi{}, \nitrogeni{}, \ci{} etc. are also accessible in the far UV, but mostly trace the ISM (Section \ref{sec:ISM}).}II, III, IV, V, VI), carbon (II, III, IV), nitrogen (II, III, IV, V), sulfur (II, III, IV, V, VI), and iron (II, III), many of which are at rest-frame wavelengths less than 1100\,\AA. Far-UV spectral coverage is therefore critical for expanding the cosmic volume available for study. An additional extension of the wavelength coverage to 3000\,\AA\ would be beneficial, but not strictly required. Such an extension would allow covering the Mg II doublet at 2796, 2803\,\AA, as well as several other cool-gas transitions, including Zn II at 2026, 2062\,\AA\ and Cr II at 2056, 2062, 2066\,\AA. 

Additionally, extending towards redder wavelengths is required to observe systems at intermediate redshift, as the targeted spectral lines are shifted to significantly longer wavelengths (Fig. \ref{fig:cgm_wavelength}). The required redshift coverage for this is 90-300~nm, with an optimal extension to 450~nm to reach the bluest effective coverage of ELT/ANDES and avoid redshift gaps. 

The same requirements in wavelength coverage would apply also for the \heii\ reionization science case. The extension to optical ranges would allow to make a direct comparison between the \heii\ and \hi\ Lyman-$\alpha$ forests along the same lines of sight.  

A minimum spectral resolution $R\approx20\,000$ (FWHM$\,\approx\,$15\,km\,s$^{-1}$) is needed to resolve warm-hot CGM absorption components and separate them from ISM gas in the CGM of local galaxies. However, higher resolutions of $R\,\approx\,40\,000-50\,000$ are required for measuring thermal line broadening in cooler metal ions, resolving complex line profiles, kinematic modeling, and probing the small-scale structure in the CGM, particularly for the denser phases traced by intermediate/low ions, neutral atoms, and H$_2$. Further, measurements of relative abundances often require detection of absorption features across a wide range of column densities, which necessitates high spectral resolution for the strong lines that are outside of the linear regime, but not fully saturated. The same spectral resolution constraints apply also to the \heii\ reionization science case. A resolution of $R\sim50\,000$ is required for the characterisation of the chemical and kinematic properties of individual absorption components in DLAs, with the goal of separating CGM infalling gas or outflows, or ISM in these galaxies. The resolution of the colder and narrower components will require, optimally, a resolution of $R\sim80\,000-100\,000$.

\begin{figure}
    \centering
    \includegraphics[width=\linewidth,trim=0 0 1200 0, clip]{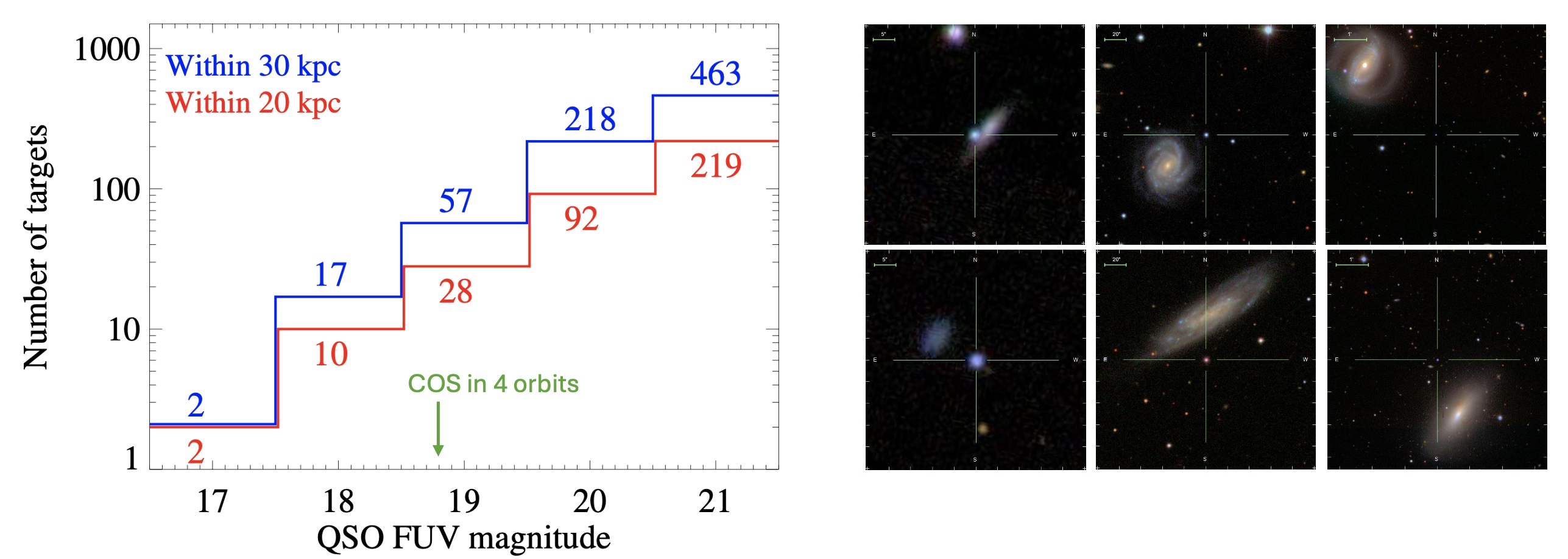}
    \caption{Number of galaxy-QSO pairs as a function of UV magnitude of the background QSO. Figure taken from \citet{Borthakur2025}.}
    \label{fig:QSOs}
\end{figure}

Future UV facilities should exceed HST/COS effective area and sensitivity. HST/COS offers ``blue modes'' with the G130M grating, which cover FUV wavelengths down to the Lyman limit. However, the effective area of these modes is very low below 1100\,\AA, making it very expensive to cover \ovi, \hmol, the higher-order H~{\sc I} Lyman series, and multi-phase, multi-element tracers.
COS G130M observations can reach a GALEX FUV magnitude of 18.8 at S/N=10 (per resolution element) within 4 orbits ($\approx$3 hrs) at 1250\,\AA\ (where COS sensitivity peaks).
Moreover, observing weaker features from less abundant elements such as nitrogen, sulfur, and iron requires $S/N$ levels that are achievable with COS only for the brightest handful of sightlines. The sensitivity requirements for future UV capabilities will be reaching down to FUV magnitudes of 20–21 at $S/N=10$ (per resolution element) in a reasonable time.
The QSO luminosity function rises sharply at faint UV magnitudes, so that going one magnitude deeper would dramatically increase the number of QSO observable in the UV by a factor of $\approx$4 \citep[Fig. \ref{fig:QSOs}, taken from][]{Borthakur2025} and expand the available redshift range, opening significant discovery space.  

The instrument requirements are summarized as follows: \\

\begin{enumerate}
    \item {\it Wavelength coverage} -- Required: 90–300\,nm; Optimal: 90–450\,nm. 
    \item {\it Spectral resolution} -- Required: 20\,000--50\,000;  Optimal: 15\,000--100\,000. 
    \item {\it Sensitivity} -- Targets at FUV magnitudes $<20-21$ should be observed with S/N of 10 or better.
\end{enumerate}

\looseness=-1

\section{Scientific questions}
\label{sec: sci questions}

Some of the prime scientific products of this paper are the pressing open questions for the advancement of the various science cases, as described above (Sect. \ref{sec: Pillar I} to \ref{sec: Pillar III}). Below we summarize these questions.

\begin{enumerate}[label=Pillar \Roman*.,
       leftmargin=*,
       align=left]
  \item Formation and evolution of planets and their habitability
    \begin{enumerate}
      \item \textbf{Exoplanets atmospheres and atmospheric escape.} What are exoplanet atmospheric compositions and dynamics? How do atmospheric escape rates depend on various stellar and planetary properties? How does atmospheric escape shape exoplanet populations? How do atmospheres evolve over time? How do we detect hydrogen exospheres and metal escape in dozens of exoplanet systems, down to Earth-sizes? How do atmospheric escape rates depend on various stellar and planetary properties? How does atmospheric escape shape exoplanet populations? How do atmospheres evolve over time? How does all this influence exoplanet habitability?

      \item \textbf{Exoplanets bulk composition}. Is the solar system representative in terms of its composition? How common are water and volatile-rich exo-planetesimals? What light elements are sequestered into planetary cores?\\
      
      \item \textbf{Protoplanetary disks.}  What are the mass outflow rates of protoplanetary disk winds and what mechanisms drive them? What are the abundances and kinematics of dominant atomic and molecular species in the inner regions of protoplanetary disks? How do mass accretion flows change with time?  \\

    \end{enumerate}
    
  \item Stellar lives and deaths at their extremes
    \begin{enumerate}
    
      \item \textbf{The first stars and the origin of the elements.} What were the chemical products of the first generation of stars? What was the mass function of the first stars? Were low-mass, long-lived metal-free stars able to form?
      
      \item \textbf{Massive stars.} How does the wind mass-loss rate scale with metallicity for stars of different masses and evolutionary stages? What does stellar wind dynamics reveal about stellar physics and the engines driving stellar winds? How do massive stars evolve, interact, and die at different metallicities? What are the largest stellar masses? How do these depend on the star-formation rate (SFR) and metallicity of host galaxies? Are massive stars responsible for harsh ionizing conditions and abundance patterns in high-$z$ galaxies? What are the ionizing properties of massive binaries including those hosting accreting compact objects?

      \item \textbf{White dwarfs and compact binaries.} What mechanisms drive binary evolution? Which pathways produce successful SN Ia explosions? How do different classes of gravitational-wave sources form? How much processed material is returned to the ISM by nova eruptions and SNe Ia? What is the origin of magnetism?

       \item \textbf{Supernovae} What mechanisms drive mass-loss in massive stars? What is the elemental abundance and ionization of the circumstellar medium in SNe? What is the UV spectral line diversity of SNe? How does UV emission contribute to the shock power budget of SNe interacting with the circumstellar medium? What are the ISM properties of SN host environments?     

    \end{enumerate}

    \item Gas, dust and metals in the baryon cycle of galaxies
    \begin{enumerate}
  
      \item \textbf{Metals in the Interstellar Medium.} How do metals mix and distribute in the ISM? On what spacial and time scales? What are the chemical properties of ISM clouds? Are there cloud-to-cloud differences? How do stars enrich the surrounding ISM? What does the ISM complexity tell us about the chemical evolution of galaxies? What are the chemical properties of the neutral ISM in low-metallicity galaxies? What are the physical conditions of the cold ISM, specially at low metallicity? How do detailed studies of the local low-metallicity Universe enable a more meaningful understanding of the distant Universe?\\

      \item \textbf{Cosmic dust.} What is the composition of dust? What is the overall amount of C, Si, O, and other elements in dust? Can dust depletion enhance 3D/4D maps of dust extinction with the information of the chemical properties in individual clouds? What are the sources of Diffuse Interstellar Bands? What are the carriers of C-rich dust grains?\\

      \item \textbf{The Circumgalactic and Intergalactic Medium.} How do galaxies grow? How do metals distribute around galaxies? How do molecules distribute around galaxies and how far do they survive? Do \hh{} outflows have any connection to the observed neutral \hi{} and low-ionization metal outflows? What is the location of Galactic and Magellanic high-velocity clouds? What is the thermodynamic state of the circumgalactic medium? Can we chemically trace the origins of the gas and feedback? Can we characterise the chemical and kinematical properties of infalling gas and outflows? How did the \heii\ reionization proceed and when was it completed?

    \end{enumerate}
\end{enumerate}

\section{Connection to Voyage 2050, the Astro Decadal 2020 survey and HWO science.}
\label{sec: Astro20}

The science that a future high-resolution UV spectrograph will address spans a breadth of topics, from solar system and small bodies, exoplanets formation, their atmospheres and potential habitability, compact stellar object to massive star evolution, Supernovae, the origin of elements from PopIII stars to second-generation stars and gravitational-wave sources, the distribution and properties of gas, metals and dust in the interstellar, circumgalactic, and intergalactic medium. All of these topics are closely related to the main scientific themes of ESA Voyage 2050, the USA Astro 2020 Decadal Survey and HWO science.

\begin{itemize}
    \item The exoplanets bulk composition, atmospheres and their habitability, solar system and small bodies science cases are part of the Voyage 2050 Themes \textit{"Moons of the Giant Planets"} and \textit{"From Temperate Exoplanets to the Milky Way"}, the Astro 2020 Theme \textit{"Worlds and Suns in Context"} and the HWO Themes \textit{"Discovering Living Worlds"} and \textit{"Understanding the Solar System in its Galactic Context"}.

    \item  The origin of the elements in PopIII and second-generation stars science cases, as well as the supernovae and compact and massive stars science cases are part of the Astro 2020 Theme \textit{"Cosmic Ecosystems"} and the HWO Theme \textit{"Following the Evolution of the Elements Over Cosmic Time"}. 

    \item The white dwarfs and compact binaries and origin of (r-process) elements science cases, with their gravitational wave aspect, are part of the Astro 2020 Theme \textit{"New Messengers and New Physics"} and the Voyage 2025 Theme \textit{"New Physical Probes of the Early Universe"}. 

    \item The ISM (metals and dust), CGM and IGM science cases are part of the Astro 2020 Theme \textit{"Cosmic Ecosystems"} (and its \textit{"Unveiling the Hidden Drivers of Galaxy Growth"}), the HWO Theme: "Uncovering the Drivers of Galaxy Growth" and the Voyage 2050 Themes \textit{"The Role of the Multiphase ISM in Star Formation and Galaxy Evolution"} and \textit{"Probing the Large Scale IGM in the Local Universe through Absorption Lines in the UV and X-rays"}. 

\end{itemize}

The science cases described in this paper are grouped into three main science pillars: "Formation and evolution of planets and their habitability", "Stellar lives and deaths at their extremes", "Gas and metals in the baryon cycle of galaxies". These pillars are closely connected to the three Astro 2020 Themes: "Worlds and Suns in Context", "Cosmic Ecosystems","Unveiling the Hidden Drivers of Galaxy Growth".

The HWO science cases have been formulated through 10 top-level scientific questions \citep{Arney26b}, 7 of which are closely related to the science cases enabled by high-resolution UV spectroscopy (Sect. \ref{sec: Pillar I} to \ref{sec: Pillar III}). These questions are: How is matter recycled by galaxies? How does our solar system fit among other planetary systems? How do stars and planets form? When were the heavy elements forged? Are icy worlds in the outer solar system habitable? 
Is there life on exoplanets? How do the most massive stars live and die? High-resolution UV spectroscopy brings a unique way to address these questions.

Overall, sensitive high-resolution UV spectroscopy is fully in scope with Voyage 2050, Astro 2020 and HWO scientific goals, and extends these goals to a vastly uncharted territory, taking advantage of the unique high resolution and high effective area of HWO PEGASUS (Fig. \ref{fig:mission_comparison}).

\section{Summary and conclusions}
\label{sec: conclusions}

The science enabled by a future high-resolution UV spectrograph spans a broad range of astrophysical questions that naturally group into three scientific pillars.I) Formation and evolution of planets and their habitability: Exoplanets atmospheres and atmospheric escape; Exoplanets bulk composition; Protoplanetary disks; Solar System planetary atmosphere and icy moons; Small bodies in the Solar System. II) Stellar lives and deaths at their extremes: The first stars and the origin of the elements; Massive stars; White dwarfs and compact binaries; Supernovae. III) Gas and metals in the baryon cycle of galaxies: Metals in the Interstellar Medium; Cosmic dust; The Circumgalactic and Intergalactic Medium.

\subsection{Key scientific questions}

Among many scientific open questions arising in these fields (Sect.~\ref{sec: sci questions}), the most pressing fundamental questions that lead to high-impact science and breakthroughs are: What are exoplanet atmospheric compositions and dynamics? How does this influence exoplanet habitability? Is the Solar System representative in terms of its composition? What are the chemical products of the first generation of stars? What are the largest stellar masses? Are massive stars responsible for harsh ionizing conditions and abundance patterns in high-z galaxies? What mechanisms drive binary evolution? How do different classes of gravitational-wave sources form? What drives mass loss in massive stars? How do Supernovae interact with the circumstellar medium? How do metals mix and distribute inside and around galaxies? What is the composition of cosmic dust? How do galaxies grow? What is the composition of cosmic dust? How do detailed studies of the local Universe enable a more meaningful understanding of the distant Universe? 

The scientific themes addressed by HWO/PEGASUS are fully aligned with the scientific priorities of Voyage 2050, the Astro2020 Decadal Survey and the Habitable Worlds Observatory (HWO). Each science pillar and individual science case directly supports one or more of the key science objectives identified by these strategic programs.

\subsection{Requirements for UV spectroscopy}

The requirements of the individual science cases and their prioritization drive the overall top-level requirements of a future UV instrument: the spectral resolution should cover at least $10~000 \lesssim R \lesssim 120~000$, the wavelength coverage should span at least $90\lesssim \lambda \lesssim 470$ nm in the UV. The extension to redder wavelengths (at least up to 700-800~nm) is required by one of the UV science cases and is foreseen to enable a broader range of science, which is not the focus of this paper, but will make a future instrument widely used. The possibility of rapid Target of Opportunity observations is required by the Supernova science case. 

The breadth and depth of the science cases presented here highlight that sensitivity of UV high-resolution spectroscopy is currently the most limiting factor for progress. Transformational advancement in UV science can be enabled by a jump in sensitivity of UV high-resolution spectroscopy. A future UV spectroscopic mission should target the parameter space of high spectral resolution and high effective area in the FUV (Fig. \ref{fig:mission_comparison}), so far an uncharted territory. 

\subsection{An instrument concept: PEGASUS}

To achieve the ambitious scientific goals, the most favorable solution is a simple instrument dedicated to UV high-resolution ($R\sim100~000$) spectroscopy, optimized for sensitivity and robustness. Extension to lower spectral resolutions and longer wavelengths is also foreseen, as it will further broaden the scientific scope and make this a widely used instrument.   

PEGASUS will be an \'Echelle cross-dispersed high-resolution spectrograph. To optimize the throughput and reliability, no moving parts or mechanisms are included. No cooling is foreseen. Two arms are currently foreseen: a UV arm spanning indicatively from 90-200~nm and a UV+VIS arm spanning indicatively 200-400~nm, with a foreseen extension to at least 800~nm. The maximum resolving power foreseen are $R=140~000$ in the red arm, and $R=100~000$ in the blue arm. The use of low read-out noise detectors will enable efficient binning reaching lower resolution ($R\sim50~000$ and down to $10\,000$) and higher sensitivity. The two arms will be fed through two separate apertures (i.e. one for the FUV and one for the UV-VIS) to avoid complex UV beam splitters and loss of throughput. The spectrograph is foreseen to be fiber-fed, providing homogeneous illumination and easing the integration topology, and taking advantage of recent FUV fiber technology. The need for a white pupil to correct from FUV stray light or collimators is under assessment.

Other HWO instrument concepts offering high-resolution UV spectroscopy include UVMOS and Pollux. PEGASUS provides the highest UV sensitivity while maintaining a dedicated focus on high-resolution UV spectroscopy, avoiding design trade-offs that could compromise this capability or its performance.

HWO/PEGASUS aims at uniquely reaching high spectral resolution and high sensitivity in the FUV/NUV, a currently unexplored territory, to enable the breakthrough UV science cases. Furthermore, PEGASUS aims at also providing mid-resolution spectroscopy well into the optical ranges, making it a versatile and robust instrument that can serve a wide community, a workhorse \'Echelle spectrograph.
\\\\

\begin{flushleft}
\textbf{Contributions:\\}
Annalisa~De~Cia wrote the \textit{Introduction}, \textit{Overview of UV science and required instrument capability}, \textit{Comparison among potential HWO UV spectrographs}, \textit{Connection to the Astro Decadal 2020 survey and HWO science}, and \textit{Summary and conclusions} sections. She also co-designed the overall structure of the science pillars and the manuscript. 
David~Cont coordinated the collection of contributions, assisted in the writing process, and generated the figures on the requirements of the PEGASUS instrument and its comparison with other UV missions. 
Kevin~Heng designed the overall structure of the manuscript and assisted in the writing process. 
Boris~Gänsicke co-designed the overall structure of the manuscript and assisted in the writing process.
David~Cont, Helena~Lamprecht, and Frank~Grupp wrote the \textit{A dedicated UV high-resolution spectrograph} section. 
Ralf~Bender provided advice on strategic aspects. 
The individual science case sections were mostly written by the following authors: 
{\it Exoplanet atmospheres} by Jens~Hoeijmakers and Antonija~Oklopčić; 
{\it The bulk abundances of exoplanets} by Boris~Gänsicke; 
{\it Protoplanetary disks} by Kevin~France and Carlo~Manara; 
{\it Solar System planetary atmosphere and icy moons} by Shingo~Kameda and Masahiro~Ikoma; 
{\it Small bodies in the Solar System} by Cyrielle~Opitom; 
{\it The first stars and the origin of the elements} by Ian~Roederer; 
{\it Massive stars} by Lidia Oskinova; 
{\it White dwarfs \& compact binaries} by Anna~Pala; 
{\it Cosmic dust} by Annalisa~De~Cia, Sascha~Zeegers and Peter~Scicluna; 
{\it Metals in the Interstellar Medium} by Annalisa~De~Cia and Dan~Welty; 
{\it The Circumgalactic and Intergalactic Medium} by Andrew~Fox, Sean~Johnson, Frances~Cashman, Valentina D'Odorico, Varsha Kulkarni, Annalisa De Cia, and Hsiao-Wen Chen. 
Further contribution to the science cases were given by: 
Abel~de~Burgos, Miriam~García, Avery~Kim, Christina~Konstantopoulou, Roberto~Maiolino, Daniel~Pauli, Tanita~Ramburuth-Hurt, Philipp Richter, Andreas~A.C.~Sander, Tomer~Shenar, Sergio~Simón-Díaz, O.~Grace~Telford, Anna~Velichko.
\end{flushleft}

\section*{Acknowledgements}



\begin{flushleft}
\textbf{Endorsed by:\\}
Marjorie Decleir, European Space Agency (ESA), ESA Office, Space Telescope Science Institute (STScI), 3700 San Martin Drive, Baltimore, MD 21218, USA;\\
Jean-Michel~Désert, Leibniz Institute for Astrophysics Potsdam (AIP), An der Sternwarte 16, 14482 Potsdam, Germany;\\
Avishay Gal-Yam, Weizmann Institute of Science, Rehovot 76100, Israel;\\
Rolf-Peter~Kudritzki, Faculty of Physics, Ludwig Maximilian University, Scheinerstrasse 1, D-81679, Munich, Bavaria, Germany;\\
C\'edric Ledoux, European Southern Observatory, Chile, Alonso de C\'ordova 3107, Casilla 19001, Vitacura, Santiago 19, Chile;\\
Pasquier Noterdaeme, Institut d'Astrophysique de Paris, 98bis bd Arago, 75014 Paris, France;\\
C\'eline Peroux, European Southern Observatory, Karl-Schwarzschild-Str.\ 2, 85748 Garching bei M\"unchen, Germany,\\ Laboratoire d’Astrophysique de Marseille, OAMP, Université Aix-Marseille \& CNRS, 38 rue Frédéric Joliot Curie, 13388 Marseille cedex 13, France;\\
Patrick~Petitjean, Institut d'Astrophysique de Paris, 98bis bd Arago, 75014 Paris, France;\\
Katja Poppenhaeger, Leibniz Institute for Astrophysics Potsdam (AIP), An der Sternwarte 16, 14482 Potsdam, Germany;\\
Francesca Primas, European Southern Observatory, Karl-Schwarzschild-Str.\ 2, 85748 Garching bei M\"unchen, Germany.\\
Maria Bergemann, Max Planck Institute for Astronomy, Koenigstuhl 17, 69117, Heidelberg, Germany. \\

\end{flushleft}

\section{Data Availability}

The data underlying this article are publicly available from the ESO Science Archive Facility (including Pr. IDs 115.284Z, 106.20Y7, 102.C-0699) and the HST Mikulski Archive for Space Telescopes (including GO 17703).



\bibliographystyle{rasti}
\bibliography{science_cases/exoplanets_composition, science_cases/exoplanets_atmospheres, science_cases/protoplanetary_disks, science_cases/sol_sys_and_small_bodies, science_cases/massive_stars, science_cases/origin_elements,
science_cases/WD_and_compact_binaries,
science_cases/solar_system,
main,
science_cases/ISM_dust,
science_cases/CGM,
science_cases/supernovae,
science_cases/top_level_requirements}

@misc{Arney26b,
  author       = {Arney, Giada},
  title        = {Habitable Worlds Observatory Science},
  month        = may,
  year         = 2026,
  publisher    = {Zenodo},
  doi          = {10.5281/zenodo.20042719},
  url          = {https://doi.org/10.5281/zenodo.20042719},
}

@ARTICLE{Roman-Duval25,
       author = {{Roman-Duval}, Julia and {Fischer}, William J. and {Fullerton}, Alexander W. and {Taylor}, Jo and {Plesha}, Rachel and {Proffitt}, Charles and {Monroe}, TalaWanda and {Fischer}, Travis C. and {Aloisi}, Alessandra and {Bouret}, Jean-Claude and et al.},
        title = "{The UV Legacy Library of Young Stars as Essential Standards (ULLYSES) Large Director's Discretionary Program with Hubble. I. Goals, Design, and Initial Results}",
      journal = {\apj},
         year = 2025,
        month = may,
       volume = {985},
       number = {1},
          eid = {109},
        pages = {109},
          doi = {10.3847/1538-4357/adc45b},
archivePrefix = {arXiv},
       eprint = {2504.05446},
 primaryClass = {astro-ph.SR},
       adsurl = {https://ui.adsabs.harvard.edu/abs/2025ApJ...985..109R}
}

@ARTICLE{Tripp08,
       author = {{Tripp}, Todd M. and {Sembach}, Kenneth R. and {Bowen}, David V. and {Savage}, Blair D. and {Jenkins}, Edward B. and {Lehner}, Nicolas and {Richter}, Philipp},
        title = "{A High-Resolution Survey of Low-Redshift QSO Absorption Lines: Statistics and Physical Conditions of O VI Absorbers}",
      journal = {\apjs},
         year = 2008,
        month = jul,
       volume = {177},
       number = {1},
        pages = {39-102},
          doi = {10.1086/587486},
archivePrefix = {arXiv},
       eprint = {0706.1214},
 primaryClass = {astro-ph},
       adsurl = {https://ui.adsabs.harvard.edu/abs/2008ApJS..177...39T}
}

@misc{Sitarski26,
  author       = {Sitarski, Breann and
                  Noecker, Charley and
                  Scowen, Paul},
  title        = {HWO Instrument Studies to Date},
  month        = may,
  year         = 2026,
  publisher    = {Zenodo},
  doi          = {10.5281/zenodo.20149408},
  url          = {https://doi.org/10.5281/zenodo.20149408},
}

@INPROCEEDINGS{France17b,
       author = {{France}, Kevin and {Fleming}, Brian and {West}, Garrett and {McCandliss}, Stephan R. and {Bolcar}, Matthew R. and {Harris}, Walter and {Moustakas}, Leonidas and {O'Meara}, John M. and {Pascucci}, Ilaria and {Rigby}, Jane and {Schiminovich}, David and {Tumlinson}, Jason},
        title = "{The LUVOIR Ultraviolet Multi-Object Spectrograph (LUMOS): instrument definition and design}",
    booktitle = {Society of Photo-Optical Instrumentation Engineers (SPIE) Conference Series},
         year = 2017,
       editor = {{Siegmund}, Oswald H.},
       series = {Society of Photo-Optical Instrumentation Engineers (SPIE) Conference Series},
       volume = {10397},
        month = aug,
          eid = {1039713},
        pages = {1039713},
          doi = {10.1117/12.2272025},
archivePrefix = {arXiv},
       eprint = {1709.06141},
 primaryClass = {astro-ph.IM},
       adsurl = {https://ui.adsabs.harvard.edu/abs/2017SPIE10397E..13F}
}

@ARTICLE{Neiner26,
       author = {{Neiner}, Coralie and {Bouret}, Jean-Claude and {Fossati}, Luca and {le Mignant}, David and {Muslimov}, Eduard and {Gomez de Castro}, Ana Ines and {Marin}, Fr{\'e}d{\'e}ric},
        title = "{The Pollux European instrument concept for HWO: a high-resolution spectrograph and spectropolarimeter from the far-UV to the near-IR}",
      journal = {arXiv e-prints},
         year = 2026,
        month = feb,
          eid = {arXiv:2602.09828},
        pages = {arXiv:2602.09828},
          doi = {10.48550/arXiv.2602.09828},
archivePrefix = {arXiv},
       eprint = {2602.09828},
 primaryClass = {astro-ph.IM},
       adsurl = {https://ui.adsabs.harvard.edu/abs/2026arXiv260209828N}
}

@INPROCEEDINGS{Bouret18,
       author = {{Bouret}, Jean-Claude and {Neiner}, Coralie and {G{\'o}mez de Castro}, Ana I. and {Evans}, Chris and {Gaensicke}, Boris and {Shore}, Steve and {Fossati}, Luca and {Gry}, C{\'e}cile and {Charlot}, St{\'e}phane and {Marin}, Fr{\'e}d{\'e}ric and et al.},
        title = "{The science case for POLLUX: a high-resolution UV spectropolarimeter onboard LUVOIR}",
    booktitle = {Space Telescopes and Instrumentation 2018: Ultraviolet to Gamma Ray},
         year = 2018,
       editor = {{den Herder}, Jan-Willem A. and {Nikzad}, Shouleh and {Nakazawa}, Kazuhiro},
       series = {Society of Photo-Optical Instrumentation Engineers (SPIE) Conference Series},
       volume = {10699},
        month = jul,
          eid = {106993B},
        pages = {106993B},
          doi = {10.1117/12.2312621},
archivePrefix = {arXiv},
       eprint = {1805.10021},
 primaryClass = {astro-ph.IM},
       adsurl = {https://ui.adsabs.harvard.edu/abs/2018SPIE10699E..3BB}
}

@ARTICLE{LUVOIR19,
       author = {{The LUVOIR Team}},
        title = "{The LUVOIR Mission Concept Study Final Report}",
      journal = {arXiv e-prints},
         year = 2019,
        month = dec,
          eid = {arXiv:1912.06219},
        pages = {arXiv:1912.06219},
          doi = {10.48550/arXiv.1912.06219},
archivePrefix = {arXiv},
       eprint = {1912.06219},
 primaryClass = {astro-ph.IM},
       adsurl = {https://ui.adsabs.harvard.edu/abs/2019arXiv191206219T}
}

@INPROCEEDINGS{Borthakur26,
       author = {{Borthakur}, Sanchayeeta and {Burchett}, Joseph N. and {Cashma}, Frances and {Fox}, Andrew J. and {Zheng}, Yong and {French}, David M. and {Bordoloi}, Rongmon and {Koplitz}, Brad},
        title = "{Characterizing Gas Flows Through Observations of the Disk-Circumgalactic Medium Interface with the Habitable Worlds Observatory}",
    booktitle = {Astronomical Society of the Pacific Conference Series},
         year = 2026,
       editor = {{Lee}, Janice C. and {Noviello}, Jessica and {LaMassa}, Stephanie and {Postman}, Marc},
       series = {Astronomical Society of the Pacific Conference Series},
       volume = {542},
        month = feb,
        pages = {233},
          doi = {10.26624/ZTBQ1389},
       adsurl = {https://ui.adsabs.harvard.edu/abs/2026ASPC..542..233B}
}

@INPROCEEDINGS{Tumlinson20,
       author = {{Tumlinson}, J.},
        title = "{LUVOIR Resolves the Baryon Cycle: A New Discovery Space for UV Spectroscopy}",
    booktitle = {American Astronomical Society Meeting Abstracts \#235},
         year = 2020,
       series = {American Astronomical Society Meeting Abstracts},
       volume = {235},
        month = jan,
          eid = {171.24},
        pages = {171.24},
       adsurl = {https://ui.adsabs.harvard.edu/abs/2020AAS...23517124T}
}

@INPROCEEDINGS{Roederer26,
       author = {{Roederer}, Ian U. and {Ezzeddine}, Rana and {Sobeck}, Jennifer S.},
        title = "{Habitable Worlds Observatory: The Nature of the First Stars}",
    booktitle = {Astronomical Society of the Pacific Conference Series},
         year = 2026,
       editor = {{Lee}, Janice C. and {Noviello}, Jessica and {LaMassa}, Stephanie and {Postman}, Marc},
       series = {Astronomical Society of the Pacific Conference Series},
       volume = {542},
        month = feb,
        pages = {73},
          doi = {10.26624/UKXV6779},
       adsurl = {https://ui.adsabs.harvard.edu/abs/2026ASPC..542...73R}
}

@INPROCEEDINGS{Muslimov18,
       author = {{Muslimov}, Eduard and {Bouret}, Jean-Claude and {Neiner}, Coralie and {L{\'o}pez Ariste}, Arturo and {Ferrari}, Marc and {Viv{\`e}s}, S{\'e}bastien and {Hugot}, Emmanuel and {Grange}, Robert and {Lombardo}, Simona and {Lopes}, Louise and {Costeraste}, Josiane and {Brachet}, Frank},
        title = "{POLLUX: a UV spectropolarimeter for the LUVOIR space telescope project}",
    booktitle = {Space Telescopes and Instrumentation 2018: Ultraviolet to Gamma Ray},
         year = 2018,
       editor = {{den Herder}, Jan-Willem A. and {Nikzad}, Shouleh and {Nakazawa}, Kazuhiro},
       series = {Society of Photo-Optical Instrumentation Engineers (SPIE) Conference Series},
       volume = {10699},
        month = jul,
          eid = {1069906},
        pages = {1069906},
          doi = {10.1117/12.2310133},
archivePrefix = {arXiv},
       eprint = {1805.09067},
 primaryClass = {astro-ph.IM},
       adsurl = {https://ui.adsabs.harvard.edu/abs/2018SPIE10699E..06M}
}

@INPROCEEDINGS{Gilliam2021,
       author = {{Gilliam}, Wesley and {Fleming}, Brian T. and {Vorobiev}, Dmitry and {Winter}, Bartlomiej and {Wadsworth}, William and {Birks}, Tim},
        title = "{Far-ultraviolet optical fibers for instrumentation in the sub-200 nm regime}",
    booktitle = {UV/Optical/IR Space Telescopes and Instruments: Innovative Technologies and Concepts X},
         year = 2021,
       editor = {{Barto}, Allison A. and {Breckinridge}, James B. and {Stahl}, H. Philip},
       series = {Society of Photo-Optical Instrumentation Engineers (SPIE) Conference Series},
       volume = {11819},
        month = aug,
          eid = {118190L},
        pages = {118190L},
          doi = {10.1117/12.2594196},
       adsurl = {https://ui.adsabs.harvard.edu/abs/2021SPIE11819E..0LG}
}

@ARTICLE{Palle25,
       author = {{Palle}, Enric and {Biazzo}, Katia and {Bolmont}, Emeline and {Molli{\`e}re}, Paul and {Poppenhaeger}, Katja and {Birkby}, Jayne and {Brogi}, Matteo and {Chauvin}, Gael and {Chiavassa}, Andrea and {Hoeijmakers}, Jens and et al.},
        title = "{Ground-breaking exoplanet science with the ANDES spectrograph at the ELT}",
      journal = {Experimental Astronomy},
         year = 2025,
        month = jun,
       volume = {59},
       number = {3},
          eid = {29},
        pages = {29},
          doi = {10.1007/s10686-025-10000-4},
archivePrefix = {arXiv},
       eprint = {2311.17075},
 primaryClass = {astro-ph.IM},
       adsurl = {https://ui.adsabs.harvard.edu/abs/2025ExA....59...29P}
}

@ARTICLE{Roederer24,
       author = {{Roederer}, Ian U. and {Alvarado-G{\'o}mez}, Juli{\'a}n D. and {Allende Prieto}, Carlos and {Adibekyan}, Vardan and {Aguado}, David S. and {Amado}, Pedro J. and {Amazo-G{\'o}mez}, Eliana M. and {Baratella}, Martina and {Barnes}, Sydney A. and {Bensby}, Thomas and et al.},
        title = "{The discovery space of ELT-ANDES. Stars and stellar populations}",
      journal = {Experimental Astronomy},
         year = 2024,
        month = apr,
       volume = {57},
       number = {2},
          eid = {17},
        pages = {17},
          doi = {10.1007/s10686-024-09938-8},
archivePrefix = {arXiv},
       eprint = {2311.16320},
 primaryClass = {astro-ph.IM},
       adsurl = {https://ui.adsabs.harvard.edu/abs/2024ExA....57...17R}
}

@ARTICLE{Evans23,
       author = {{Evans}, Chris and {Cristiani}, Stefano and {Opitom}, Cyrielle and {Cescutti}, Gabriele and {D'Odorico}, Valentina and {Alcal{\'a}}, Juan Manuel and {Alencar}, Silvia H.~P. and {Balashev}, Sergei and {Barbuy}, Beatriz and {Bastian}, Nate and {Battino}, Umberto and {Cambianica}, Pamela and {Carini}, Roberta and {Carter}, Brad and {Cassisi}, Santi and {Vaz Castilho}, Bruno and {Christlieb}, Norbert and {Cooke}, Ryan and {Covino}, Stefano and {Cremonese}, Gabriele and {Cunha}, Katia and {da Silva}, Andr{\'e} R. and {D'Elia}, Valerio and {De Cia}, Annalisa and {De Silva}, Gayandhi and {Diaz}, Marcos and {Di Marcantonio}, Paolo and {Ernandes}, Heitor and {Fitzsimmons}, Alan and {Franchini}, Mariagrazia and {G{\"a}nsicke}, Boris T. and {Genoni}, Matteo and {Giribaldi}, Riano E. and {Grazian}, Andrea and {Hansen}, Camilla Juul and {La Forgia}, Fiorangela and {Lazzarin}, Monica and {Marcolino}, Wagner and {Marconi}, Marcella and {Migliorini}, Alessandra and {Noterdaeme}, Pasquier and {Pereira}, Claudio and {Pilecki}, Bogumil and {Quirrenbach}, Andreas and {Randich}, Sofia and {Rossi}, Silvia and {Smiljanic}, Rodolfo and {Snodgrass}, Colin and {St{\"u}rmer}, Julian and {Trost}, Andrea and {Vanzella}, Eros and {Ventura}, Paolo and {Wright}, Duncan and {Zafar}, Tayyaba},
        title = "{The CUBES science case}",
      journal = {Experimental Astronomy},
         year = 2023,
        month = feb,
       volume = {55},
       number = {1},
        pages = {1-57},
          doi = {10.1007/s10686-022-09864-7},
archivePrefix = {arXiv},
       eprint = {2208.01677},
 primaryClass = {astro-ph.IM},
       adsurl = {https://ui.adsabs.harvard.edu/abs/2023ExA....55....1E}
}

@INPROCEEDINGS{Dekker00,
       author = {{Dekker}, Hans and {D'Odorico}, Sandro and {Kaufer}, Andreas and {Delabre}, Bernard and {Kotzlowski}, Heinz},
        title = "{Design, construction, and performance of UVES, the echelle spectrograph for the UT2 Kueyen Telescope at the ESO Paranal Observatory}",
    booktitle = {Optical and IR Telescope Instrumentation and Detectors},
         year = 2000,
       editor = {{Iye}, Masanori and {Moorwood}, Alan F.},
       series = {Society of Photo-Optical Instrumentation Engineers (SPIE) Conference Series},
       volume = {4008},
        month = aug,
        pages = {534-545},
          doi = {10.1117/12.395512},
       adsurl = {https://ui.adsabs.harvard.edu/abs/2000SPIE.4008..534D}
      }

@inproceedings{Mears2025,
author = {Robbie Mears and Dmitry Vorobiev and Kerrianne Harrington and Briana Indahl and James M Stone and Brian Fleming and Tim A Birks and William J Wadsworth},
booktitle = {Conference on Lasers and Electro-Optics/Europe (CLEO/Europe 2025) and European Quantum Electronics Conference (EQEC 2025)},
journal = {Conference on Lasers and Electro-Optics/Europe (CLEO/Europe 2025) and European Quantum Electronics Conference (EQEC 2025)},
pages = {ce\_2\_3},
publisher = {Optica Publishing Group},
title = {Anti-resonant hollow core optical fibres for the vacuum-ultraviolet},
year = {2025},
url = {https://opg.optica.org/abstract.cfm?URI=CLEO_Europe-2025-ce_2_3},
doi = {10.1364/CLEO_EUROPE.2025.ce_2_3},
}

@inproceedings{Skottfeld_2025,
author = {Jesper Skottfelt and Chiaki Crews and Ben Dryer and Angaraj Duara and David J. Hall and Michael W. J. Hubbard and Martin J. Prest and Joan Requena and Konstantion D. Stefanov and Douglas Jordan and April D. Jewell and Charles Shapiro and Nathan Bush and Michael E. Hoenk},
title = {{Detector developments for UV space missions}},
volume = {13625},
booktitle = {UV, X-Ray, and Gamma-Ray Space Instrumentation for Astronomy XXIV},
editor = {Oswald H. Siegmund and Keri Hoadley},
organization = {International Society for Optics and Photonics},
publisher = {SPIE},
pages = {136250P},
year = {2025},
doi = {10.1117/12.3063649},
URL = {https://doi.org/10.1117/12.3063649}
}

@inproceedings{Hoenk_2022,
author = {Michael E. Hoenk and April D. Jewell and Gillian Kyne and John Hennessy and Todd Jones and Samuel Cheng and Shouleh Nikzad and David Morris and Katherine Lawrie and Jesper Skottfelt},
title = {{2D-doped silicon detectors for UV/optical/NIR and x-ray astronomy}},
volume = {12191},
booktitle = {X-Ray, Optical, and Infrared Detectors for Astronomy X},
editor = {Andrew D. Holland and James Beletic},
organization = {International Society for Optics and Photonics},
publisher = {SPIE},
pages = {1219113},
year = {2022},
doi = {10.1117/12.2631542},
URL = {https://doi.org/10.1117/12.2631542}
}

@article{Johnson:2023,
	author = {{Johnson}, James W. and {Weinberg}, David H. and {Vincenzo}, Fiorenzo and {Bird}, Jonathan C. and {Griffith}, Emily J.},
	journal = {\mnras},
	month = mar,
	number = {1},
	pages = {782-803},
	title = {{Empirical constraints on the nucleosynthesis of nitrogen}},
	volume = {520},
	year = 2023}

@article{Nunez:2022,
	author = {{Nu{\~n}ez}, Evan H. and {Kirby}, Evan N. and {Steidel}, Charles C.},
	journal = {\apj},
	month = mar,
	number = {1},
	pages = {64},
	title = {{Empirical Constraints on Core-collapse Supernova Yields Using Very Metal-poor Damped Ly{\ensuremath{\alpha}} Absorbers}},
	volume = {927},
	year = 2022}

@article{Berg:2012,
	author = {{Berg}, Danielle A. and {Skillman}, Evan D. and {Marble}, Andrew R. and {van Zee}, Liese and {Engelbracht}, Charles W. and {Lee}, Janice C. and {Kennicutt}, Robert C., Jr. and {Calzetti}, Daniela and {Dale}, Daniel A. and {Johnson}, Benjamin D.},
	journal = {\apj},
	month = aug,
	number = {2},
	pages = {98},
	title = {{Direct Oxygen Abundances for Low-luminosity LVL Galaxies}},
	volume = {754},
	year = 2012}

@article{Pilyugin:2014,
	author = {{Pilyugin}, L.~S. and {Grebel}, E.~K. and {Kniazev}, A.~Y.},
	journal = {\aj},
	month = jun,
	number = {6},
	pages = {131},
	title = {{The Abundance Properties of Nearby Late-type Galaxies. I. The Data}},
	volume = {147},
	year = 2014}

@article{Berg:2016,
	author = {{Berg}, Danielle A. and {Skillman}, Evan D. and {Henry}, Richard B.~C. and {Erb}, Dawn K. and {Carigi}, Leticia},
	journal = {\apj},
	month = aug,
	number = {2},
	pages = {126},
	title = {{Carbon and Oxygen Abundances in Low Metallicity Dwarf Galaxies}},
	volume = {827},
	year = 2016}

@ARTICLE{Thom08,
       author = {{Thom}, C. and {Chen}, Hsiao-Wen},
        title = "{A STIS Survey for O VI Absorption Systems at 0.12 < z lesssim 0.5. I. The Statistical Properties of Ionized Gas}",
      journal = {\apj},
         year = 2008,
        month = aug,
       volume = {683},
       number = {1},
        pages = {22-32},
          doi = {10.1086/587976},
archivePrefix = {arXiv},
       eprint = {0801.2380},
 primaryClass = {astro-ph},
       adsurl = {https://ui.adsabs.harvard.edu/abs/2008ApJ...683...22T}
}

@ARTICLE{Rauch96,
       author = {{Rauch}, M. and {Sargent}, W.~L.~W. and {Womble}, D.~S. and {Barlow}, T.~A.},
        title = "{Temperature and Kinematics of C IV Absorption Systems}",
      journal = {\apjl},
         year = 1996,
        month = aug,
       volume = {467},
        pages = {L5},
          doi = {10.1086/310187},
archivePrefix = {arXiv},
       eprint = {astro-ph/9606041},
 primaryClass = {astro-ph},
       adsurl = {https://ui.adsabs.harvard.edu/abs/1996ApJ...467L...5R}
}

@ARTICLE{Muzahid16,
       author = {{Muzahid}, Sowgat and {Kacprzak}, Glenn G. and {Charlton}, Jane C. and {Churchill}, Christopher W.},
        title = "{Molecular Hydrogen Absorption from the Halo of a z {\ensuremath{\sim}} 0.4 Galaxy}",
      journal = {\apj},
         year = 2016,
        month = may,
       volume = {823},
       number = {1},
          eid = {66},
        pages = {66},
          doi = {10.3847/0004-637X/823/1/66},
archivePrefix = {arXiv},
       eprint = {1601.06782},
 primaryClass = {astro-ph.GA},
       adsurl = {https://ui.adsabs.harvard.edu/abs/2016ApJ...823...66M}
}

@ARTICLE{Pettini08,
       author = {{Pettini}, Max and {Zych}, Berkeley J. and {Steidel}, Charles C. and {Chaffee}, Fred H.},
        title = "{C, N, O abundances in the most metal-poor damped Lyman alpha systems}",
      journal = {\mnras},
         year = 2008,
        month = apr,
       volume = {385},
       number = {4},
        pages = {2011-2024},
          doi = {10.1111/j.1365-2966.2008.12951.x},
archivePrefix = {arXiv},
       eprint = {0712.1829},
 primaryClass = {astro-ph},
       adsurl = {https://ui.adsabs.harvard.edu/abs/2008MNRAS.385.2011P}
}

@ARTICLE{Bosman22,
       author = {{Bosman}, Sarah E.~I. and {Davies}, Frederick B. and {Becker}, George D. and {Keating}, Laura C. and {Davies}, Rebecca L. and {Zhu}, Yongda and {Eilers}, Anna-Christina and {D'Odorico}, Valentina and {Bian}, Fuyan and {Bischetti}, Manuela and et al.},
        title = "{Hydrogen reionization ends by z = 5.3: Lyman-{\ensuremath{\alpha}} optical depth measured by the XQR-30 sample}",
      journal = {\mnras},
         year = 2022,
        month = jul,
       volume = {514},
       number = {1},
        pages = {55-76},
          doi = {10.1093/mnras/stac1046},
archivePrefix = {arXiv},
       eprint = {2108.03699},
 primaryClass = {astro-ph.CO},
       adsurl = {https://ui.adsabs.harvard.edu/abs/2022MNRAS.514...55B}
}

@ARTICLE{Bouche13,
       author = {{Bouch{\'e}}, N. and {Murphy}, M.~T. and {Kacprzak}, G.~G. and {P{\'e}roux}, C. and {Contini}, T. and {Martin}, C.~L. and {Dessauges-Zavadsky}, M.},
        title = "{Signatures of Cool Gas Fueling a Star-Forming Galaxy at Redshift 2.3}",
      journal = {Science},
         year = 2013,
        month = jul,
       volume = {341},
       number = {6141},
        pages = {50-53},
          doi = {10.1126/science.1234209},
archivePrefix = {arXiv},
       eprint = {1306.0134},
 primaryClass = {astro-ph.CO},
       adsurl = {https://ui.adsabs.harvard.edu/abs/2013Sci...341...50B}
}

@ARTICLE{Chen00,
       author = {{Chen}, Hsiao-Wen and {Prochaska}, Jason X.},
        title = "{The Origin of a Chemically Enriched Ly{\ensuremath{\alpha}} Absorption System at Z = 0.167}",
      journal = {\apjl},
         year = 2000,
        month = nov,
       volume = {543},
       number = {1},
        pages = {L9-L13},
          doi = {10.1086/318179},
archivePrefix = {arXiv},
       eprint = {astro-ph/0009001},
 primaryClass = {astro-ph},
       adsurl = {https://ui.adsabs.harvard.edu/abs/2000ApJ...543L...9C}
}

@ARTICLE{Chen09,
       author = {{Chen}, Hsiao-Wen and {Mulchaey}, John S.},
        title = "{Probing The Intergalactic Medium-Galaxy Connection At z < 0.5. I. A Galaxy Survey In Qso Fields And A Galaxy-Absorber Cross-Correlation Study}",
      journal = {\apj},
         year = 2009,
        month = aug,
       volume = {701},
       number = {2},
        pages = {1219-1242},
          doi = {10.1088/0004-637X/701/2/1219},
archivePrefix = {arXiv},
       eprint = {0906.3293},
 primaryClass = {astro-ph.CO},
       adsurl = {https://ui.adsabs.harvard.edu/abs/2009ApJ...701.1219C}
}

@ARTICLE{Compostella13,
       author = {{Compostella}, Michele and {Cantalupo}, Sebastiano and {Porciani}, Cristiano},
        title = "{The imprint of inhomogeneous He II reionization on the H I and He II Ly{\ensuremath{\alpha}} forest}",
      journal = {\mnras},
         year = 2013,
        month = nov,
       volume = {435},
       number = {4},
        pages = {3169-3190},
          doi = {10.1093/mnras/stt1510},
archivePrefix = {arXiv},
       eprint = {1306.5745},
 primaryClass = {astro-ph.CO},
       adsurl = {https://ui.adsabs.harvard.edu/abs/2013MNRAS.435.3169C}
}

@ARTICLE{Compostella14,
       author = {{Compostella}, Michele and {Cantalupo}, Sebastiano and {Porciani}, Cristiano},
        title = "{AGN-driven helium reionization and the incidence of extended He III regions at redshift z > 3}",
      journal = {\mnras},
         year = 2014,
        month = dec,
       volume = {445},
       number = {4},
        pages = {4186-4196},
          doi = {10.1093/mnras/stu2035},
archivePrefix = {arXiv},
       eprint = {1407.1316},
 primaryClass = {astro-ph.CO},
       adsurl = {https://ui.adsabs.harvard.edu/abs/2014MNRAS.445.4186C}
}

@ARTICLE{Danforth16,
       author = {{Danforth}, Charles W. and {Keeney}, Brian A. and {Tilton}, Evan M. and {Shull}, J. Michael and {Stocke}, John T. and {Stevans}, Matthew and {Pieri}, Matthew M. and {Savage}, Blair D. and {France}, Kevin and {Syphers}, David and {Smith}, Britton D. and {Green}, James C. and {Froning}, Cynthia and {Penton}, Steven V. and {Osterman}, Steven N.},
        title = "{An HST/COS Survey of the Low-redshift Intergalactic Medium. I. Survey, Methodology, and Overall Results}",
      journal = {\apj},
         year = 2016,
        month = feb,
       volume = {817},
       number = {2},
          eid = {111},
        pages = {111},
          doi = {10.3847/0004-637X/817/2/111},
archivePrefix = {arXiv},
       eprint = {1402.2655},
 primaryClass = {astro-ph.CO},
       adsurl = {https://ui.adsabs.harvard.edu/abs/2016ApJ...817..111D}
}

@ARTICLE{DeCia12,
       author = {{De Cia}, A. and {Ledoux}, C. and {Fox}, A.~J. and {Vreeswijk}, P.~M. and {Smette}, A. and {Petitjean}, P. and {Bj{\"o}rnsson}, G. and {Fynbo}, J.~P.~U. and {Hjorth}, J. and {Jakobsson}, P.},
        title = "{Rapid-response mode VLT/UVES spectroscopy of super iron-rich gas exposed to GRB 080310. Evidence of ionization in action and episodic star formation in the host}",
      journal = {\aap},
         year = 2012,
        month = sep,
       volume = {545},
          eid = {A64},
        pages = {A64},
          doi = {10.1051/0004-6361/201218884},
archivePrefix = {arXiv},
       eprint = {1207.6102},
 primaryClass = {astro-ph.CO},
       adsurl = {https://ui.adsabs.harvard.edu/abs/2012A&A...545A..64D}
}

@ARTICLE{Garaldi19,
       author = {{Garaldi}, Enrico and {Compostella}, Michele and {Porciani}, Cristiano},
        title = "{The Goldilocks problem of the quasar contribution to reionization}",
      journal = {\mnras},
         year = 2019,
        month = mar,
       volume = {483},
       number = {4},
        pages = {5301-5314},
          doi = {10.1093/mnras/sty3414},
archivePrefix = {arXiv},
       eprint = {1809.10144},
 primaryClass = {astro-ph.CO},
       adsurl = {https://ui.adsabs.harvard.edu/abs/2019MNRAS.483.5301G}
}

@ARTICLE{Keeney13,
       author = {{Keeney}, Brian A. and {Stocke}, John T. and {Rosenberg}, Jessica L. and {Danforth}, Charles W. and {Ryan-Weber}, Emma V. and {Shull}, J. Michael and {Savage}, Blair D. and {Green}, James C.},
        title = "{HST/COS Spectra of Three QSOs That Probe the Circumgalactic Medium of a Single Spiral Galaxy: Evidence for Gas Recycling and Outflow}",
      journal = {\apj},
         year = 2013,
        month = mar,
       volume = {765},
       number = {1},
          eid = {27},
        pages = {27},
          doi = {10.1088/0004-637X/765/1/27},
archivePrefix = {arXiv},
       eprint = {1301.4242},
 primaryClass = {astro-ph.CO},
       adsurl = {https://ui.adsabs.harvard.edu/abs/2013ApJ...765...27K}
}

@ARTICLE{Maiolino24a,
       author = {{Maiolino}, Roberto and {Scholtz}, Jan and {Curtis-Lake}, Emma and {Carniani}, Stefano and {Baker}, William and {de Graaff}, Anna and {Tacchella}, Sandro and {{\"U}bler}, Hannah and {D'Eugenio}, Francesco and {Witstok}, Joris and et al.},
        title = "{JADES: The diverse population of infant black holes at 4 < z < 11: Merging, tiny, poor, but mighty}",
      journal = {\aap},
         year = 2024,
        month = nov,
       volume = {691},
          eid = {A145},
        pages = {A145},
          doi = {10.1051/0004-6361/202347640},
archivePrefix = {arXiv},
       eprint = {2308.01230},
 primaryClass = {astro-ph.GA},
       adsurl = {https://ui.adsabs.harvard.edu/abs/2024A&A...691A.145M}
}

@ARTICLE{Madau24,
       author = {{Madau}, Piero and {Giallongo}, Emanuele and {Grazian}, Andrea and {Haardt}, Francesco},
        title = "{Cosmic Reionization in the JWST Era: Back to AGNs?}",
      journal = {\apj},
         year = 2024,
        month = aug,
       volume = {971},
       number = {1},
          eid = {75},
        pages = {75},
          doi = {10.3847/1538-4357/ad5ce8},
archivePrefix = {arXiv},
       eprint = {2406.18697},
 primaryClass = {astro-ph.CO},
       adsurl = {https://ui.adsabs.harvard.edu/abs/2024ApJ...971...75M}
}

@ARTICLE{Worseck19,
       author = {{Worseck}, G{\'a}bor and {Davies}, Frederick B. and {Hennawi}, Joseph F. and {Prochaska}, J. Xavier},
        title = "{The Evolution of the He II-ionizing Background at Redshifts 2.3 < z < 3.8 Inferred from a Statistical Sample of 24 HST/COS He II Ly{\ensuremath{\alpha}} Absorption Spectra}",
      journal = {\apj},
         year = 2019,
        month = apr,
       volume = {875},
       number = {2},
          eid = {111},
        pages = {111},
          doi = {10.3847/1538-4357/ab0fa1},
archivePrefix = {arXiv},
       eprint = {1808.05247},
 primaryClass = {astro-ph.GA},
       adsurl = {https://ui.adsabs.harvard.edu/abs/2019ApJ...875..111W}
}

@ARTICLE{Worseck11,
       author = {{Worseck}, G{\'a}bor and {Prochaska}, J. Xavier},
        title = "{GALEX Far-ultraviolet Color Selection of UV-bright High-redshift Quasars}",
      journal = {\apj},
         year = 2011,
        month = feb,
       volume = {728},
       number = {1},
          eid = {23},
        pages = {23},
          doi = {10.1088/0004-637X/728/1/23},
archivePrefix = {arXiv},
       eprint = {1004.3347},
 primaryClass = {astro-ph.CO},
       adsurl = {https://ui.adsabs.harvard.edu/abs/2011ApJ...728...23W}
}

@ARTICLE{Syphers12,
       author = {{Syphers}, David and {Anderson}, Scott F. and {Zheng}, Wei and {Meiksin}, Avery and {Schneider}, Donald P. and {York}, Donald G.},
        title = "{HST/COS Observations of Thirteen New He II Quasars}",
      journal = {\aj},
         year = 2012,
        month = apr,
       volume = {143},
       number = {4},
          eid = {100},
        pages = {100},
          doi = {10.1088/0004-6256/143/4/100},
archivePrefix = {arXiv},
       eprint = {1202.0236},
 primaryClass = {astro-ph.CO},
       adsurl = {https://ui.adsabs.harvard.edu/abs/2012AJ....143..100S}
}

@ARTICLE{Noterdaeme12,
       author = {{Noterdaeme}, P. and {Petitjean}, P. and {Carithers}, W.~C. and {P{\^a}ris}, I. and {Font-Ribera}, A. and {Bailey}, S. and {Aubourg}, E. and {Bizyaev}, D. and {Ebelke}, G. and {Finley}, H. and {Ge}, J. and {Malanushenko}, E. and {Malanushenko}, V. and {Miralda-Escud{\'e}}, J. and {Myers}, A.~D. and {Oravetz}, D. and {Pan}, K. and {Pieri}, M.~M. and {Ross}, N.~P. and {Schneider}, D.~P. and {Simmons}, A. and {York}, D.~G.},
        title = "{Column density distribution and cosmological mass density of neutral gas: Sloan Digital Sky Survey-III Data Release 9}",
      journal = {\aap},
         year = 2012,
        month = nov,
       volume = {547},
          eid = {L1},
        pages = {L1},
          doi = {10.1051/0004-6361/201220259},
archivePrefix = {arXiv},
       eprint = {1210.1213},
 primaryClass = {astro-ph.CO},
       adsurl = {https://ui.adsabs.harvard.edu/abs/2012A&A...547L...1N}
}

@ARTICLE{Rao06,
       author = {{Rao}, Sandhya M. and {Turnshek}, David A. and {Nestor}, Daniel B.},
        title = "{Damped Ly{\ensuremath{\alpha}} Systems at z<1.65: The Expanded Sloan Digital Sky Survey Hubble Space Telescope Sample}",
      journal = {\apj},
         year = 2006,
        month = jan,
       volume = {636},
       number = {2},
        pages = {610-630},
          doi = {10.1086/498132},
archivePrefix = {arXiv},
       eprint = {astro-ph/0509469},
 primaryClass = {astro-ph},
       adsurl = {https://ui.adsabs.harvard.edu/abs/2006ApJ...636..610R}
}

@article{Haislmaier:2021,
	author = {{Haislmaier}, Karl J. and {Tripp}, Todd M. and {Katz}, Neal and {Prochaska}, J. Xavier and {Burchett}, Joseph N. and {O'Meara}, John M. and {Werk}, Jessica K.},
	journal = {\mnras},
	month = apr,
	number = {4},
	pages = {4993-5037},
	title = {{The COS Absorption Survey of Baryon Harbors: unveiling the physical conditions of circumgalactic gas through multiphase Bayesian ionization modelling}},
	volume = {502},
	year = 2021}

@article{Qu:2016,
	author = {{Qu}, Z. and {Bregman}, J.~N.},
	journal = {\apj},
	month = dec,
	pages = {189},
	title = {{A Hot Gaseous Galaxy Halo Candidate with Mg X Absorption}},
	volume = 832,
	year = 2016}

@article{Burchett:2019,
	author = {{Burchett}, J.~N. and {Tripp}, T.~M. and {Prochaska}, J.~X. and {Werk}, J.~K. and {Tumlinson}, J. and {Howk}, J.~C. and {Willmer}, C.~N.~A. and {Lehner}, N. and {Meiring}, J.~D. and {Bowen}, D.~V. and {Bordoloi}, R. and {Peeples}, M.~S. and {Jenkins}, E.~B. and {O'Meara}, J.~M. and {Tejos}, N. and {Katz}, N.},
	journal = {\apjl},
	month = jun,
	pages = {L20},
	title = {{The COS Absorption Survey of Baryon Harbors (CASBaH): Warm{\ndash}Hot Circumgalactic Gas Reservoirs Traced by Ne VIII Absorption}},
	volume = 877,
	year = 2019}

@article{Chen:2019,
	author = {{Chen}, H.-W. and {Johnson}, S.~D. and {Straka}, L.~A. and {Zahedy}, F.~S. and {Schaye}, J. and {Muzahid}, S. and {Bouch{\'e}}, N. and {Cantalupo}, S. and {Marino}, R.~A. and {Wendt}, M.},
	journal = {\mnras},
	month = mar,
	pages = {431-441},
	title = {{Characterizing circumgalactic gas around massive ellipticals at z {\ap} 0.4 - III. The galactic environment of a chemically pristine Lyman limit absorber}},
	volume = 484,
	year = 2019}

@article{Fox:2026,
	author = {{Fox}, Andrew J. and {Mishra}, Sapna and {Cashman}, Frances H. and {French}, David M. and {Richter}, Philipp and {Bordoloi}, Rongmon and {Lehner}, Nicolas and {Tumlinson}, Jason and {Borthakur}, Sanchayeeta},
	journal = {arXiv e-prints},
	month = jan,
	pages = {arXiv:2601.00437},
	title = {{Low Metallicity Gas on the Outskirts of the Local Group: the Circumgalactic Medium of Sextans B}},
	year = 2026}

@article{Wotta:2016,
	author = {{Wotta}, C.~B. and {Lehner}, N. and {Howk}, J.~C. and {O'Meara}, J.~M. and {Prochaska}, J.~X.},
	journal = {\apj},
	month = nov,
	pages = {95},
	title = {{Low-metallicity Absorbers Account for Half of the Dense Circumgalactic Gas at z {\lsim} 1}},
	volume = 831,
	year = 2016}

@article{Johnson:2026,
	author = {{Johnson}, Sean D. and {Mishra}, Nishant and {Muzahid}, Sowgat and {Rudie}, Gwen C. and {Zahedy}, Fakhri S. and {Qu}, Zhijie and {Faucher-Gigu{\`e}re}, Claude-Andr{\'e} and {Stern}, Jonathan and {Li}, Jennifer I.-Hsiu and {Fuller}, Elise and {Cantalupo}, Sebastiano and {Chen}, Hsiao-Wen and {Kadri}, Ahmad and {Kumar}, Suyash and {Liu}, Zhuoqi (Will) and {Walth}, Gregory},
	journal = {\apjl},
	month = jan,
	number = {2},
	pages = {L30},
	title = {{MUSEQuBES: Physical Conditions, Origins, and Multielement Abundances of the Circumgalactic Medium of an Isolated, Star-forming Dwarf Galaxy at z = 0.57}},
	volume = {996},
	year = 2026}

@article{Kumar:2024,
	author = {{Kumar}, Suyash and {Chen}, Hsiao-Wen and {Qu}, Zhijie and {Chen}, Mandy C. and {Zahedy}, Fakhri S. and {Johnson}, Sean D. and {Muzahid}, Sowgat and {Cantalupo}, Sebastiano},
	journal = {The Open Journal of Astrophysics},
	month = oct,
	pages = {94},
	title = {{On the Nature of the C IV-bearing Circumgalactic Medium at z 1}},
	volume = {7},
	year = 2024}

@article{Lehner:2016,
	author = {{Lehner}, Nicolas and {O'Meara}, John M. and {Howk}, J. Christopher and {Prochaska}, J. Xavier and {Fumagalli}, Michele},
	journal = {\apj},
	month = dec,
	number = {2},
	pages = {283},
	title = {{The Cosmic Evolution of the Metallicity Distribution of Ionized Gas Traced by Lyman Limit Systems}},
	volume = {833},
	year = 2016}

@article{Kobayashi:2020,
	author = {{Kobayashi}, Chiaki and {Karakas}, Amanda I. and {Lugaro}, Maria},
	journal = {\apj},
	month = sep,
	number = {2},
	pages = {179},
	title = {{The Origin of Elements from Carbon to Uranium}},
	volume = {900},
	year = 2020}

@article{Tinsley:1979,
	author = {{Tinsley}, B.~M.},
	journal = {\apj},
	month = may,
	pages = {1046-1056},
	title = {{Stellar lifetimes and abundance ratios in chemical evolution.}},
	volume = {229},
	year = 1979}

@article{Rudie:2019,
	author = {{Rudie}, Gwen C. and {Steidel}, Charles C. and {Pettini}, Max and {Trainor}, Ryan F. and {Strom}, Allison L. and {Hummels}, Cameron B. and {Reddy}, Naveen A. and {Shapley}, Alice E.},
	journal = {\apj},
	month = nov,
	number = {1},
	pages = {61},
	title = {{Column Density, Kinematics, and Thermal State of Metal-bearing Gas within the Virial Radius of z {\ensuremath{\sim}} 2 Star-forming Galaxies in the Keck Baryonic Structure Survey}},
	volume = {885},
	year = 2019}

@article{Sameer:2024,
	author = {{Sameer} and {Charlton}, Jane C. and {Wakker}, Bart P. and {Kacprzak}, Glenn G. and {Nielsen}, Nikole M. and {Churchill}, Christopher W. and {Richter}, Philipp and {Muzahid}, Sowgat and {Ho}, Stephanie H. and {Nateghi}, Hasti and {Rosenwasser}, Benjamin and {Narayanan}, Anand and {Ganguly}, Rajib},
	journal = {\mnras},
	month = jun,
	number = {4},
	pages = {3827-3854},
	title = {{Cloud-by-cloud multiphase investigation of the circumgalactic medium of low-redshift galaxies}},
	volume = {530},
	year = 2024}

@article{Stern:2021,
	author = {{Stern}, Jonathan and {Faucher-Gigu{\`e}re}, Claude-Andr{\'e} and {Fielding}, Drummond and {Quataert}, Eliot and {Hafen}, Zachary and {Gurvich}, Alexander B. and {Ma}, Xiangcheng and {Byrne}, Lindsey and {El-Badry}, Kareem and {Angl{\'e}s-Alc{\'a}zar}, Daniel and {Chan}, T.~K. and {Feldmann}, Robert and {Kere{\v{s}}}, Du{\v{s}}an and {Wetzel}, Andrew and {Murray}, Norman and {Hopkins}, Philip F.},
	journal = {\apj},
	month = apr,
	number = {2},
	pages = {88},
	title = {{Virialization of the Inner CGM in the FIRE Simulations and Implications for Galaxy Disks, Star Formation, and Feedback}},
	volume = {911},
	year = 2021}

@article{Bieri:2026,
	author = {{Bieri}, Rebekka and {Pakmor}, R{\"u}diger and {van de Voort}, Freeke and {Talbot}, Rosie Y. and {Werhahn}, Maria and {Pfrommer}, Christoph and {Springel}, Volker},
	journal = {\mnras},
	month = apr,
	number = {2},
	pages = {stag216},
	title = {{Unveiling the impact of cosmic rays on the disc sizes and outflows from dwarf scales to galaxy groups}},
	volume = {547},
	year = 2026}

@article{Ji:2020,
	author = {{Ji}, Suoqing and {Chan}, T.~K. and {Hummels}, Cameron B. and {Hopkins}, Philip F. and {Stern}, Jonathan and {Kere{\v{s}}}, Du{\v{s}}an and {Quataert}, Eliot and {Faucher-Gigu{\`e}re}, Claude-Andr{\'e} and {Murray}, Norman},
	journal = {\mnras},
	month = aug,
	number = {4},
	pages = {4221-4238},
	title = {{Properties of the circumgalactic medium in cosmic ray-dominated galaxy haloes}},
	volume = {496},
	year = 2020}

@article{Mishra:2024,
	author = {{Mishra}, Nishant and {Johnson}, Sean D. and {Rudie}, Gwen C. and {Chen}, Hsiao-Wen and {Schaye}, Joop and {Qu}, Zhijie and {Zahedy}, Fakhri S. and {Boettcher}, Erin T. and {Cantalupo}, Sebastiano and {Chen}, Mandy C. and {Faucher-Gigu{\'e}re}, Claude-Andr{\'e} and {Greene}, Jenny E. and {Li}, Jennifer I.-Hsiu and {Liu}, Zhuoqi (Will) and {Lopez}, Sebastian and {Petitjean}, Patrick},
	journal = {\apj},
	month = nov,
	number = {1},
	pages = {149},
	title = {{The Cosmic Ultraviolet Baryon Survey (CUBS). IX. The Enriched Circumgalactic and Intergalactic Medium Around Star-forming Field Dwarf Galaxies Traced by O VI Absorption}},
	volume = {976},
	year = 2024}

@article{Faucher-Giguere:2023,
	author = {{Faucher-Gigu{\`e}re}, Claude-Andr{\'e} and {Oh}, S. Peng},
	journal = {\araa},
	month = aug,
	pages = {131-195},
	title = {{Key Physical Processes in the Circumgalactic Medium}},
	volume = {61},
	year = 2023}

@inproceedings{Chen:2026,
	author = {{Chen}, Hsiao-Wen and {Zahedy}, Fakhri S.},
	booktitle = {Encyclopedia of Astrophysics, Volume 4},
	month = jan,
	pages = {370-400},
	title = {{The circumgalactic medium}},
	volume = {4},
	year = 2026}

@ARTICLE{Gnat2007,
       author = {{Gnat}, Orly and {Sternberg}, Amiel},
        title = "{Time-dependent Ionization in Radiatively Cooling Gas}",
      journal = {\apjs},
         year = 2007,
        month = feb,
       volume = {168},
       number = {2},
        pages = {213-230},
          doi = {10.1086/509786},
archivePrefix = {arXiv},
       eprint = {astro-ph/0608181},
 primaryClass = {astro-ph},
       adsurl = {https://ui.adsabs.harvard.edu/abs/2007ApJS..168..213G}
}

@ARTICLE{Oppenheimer2013,
       author = {{Oppenheimer}, Benjamin D. and {Schaye}, Joop},
        title = "{Non-equilibrium ionization and cooling of metal-enriched gas in the presence of a photoionization background}",
      journal = {\mnras},
         year = 2013,
        month = sep,
       volume = {434},
       number = {2},
        pages = {1043-1062},
          doi = {10.1093/mnras/stt1043},
archivePrefix = {arXiv},
       eprint = {1302.5710},
 primaryClass = {astro-ph.CO},
       adsurl = {https://ui.adsabs.harvard.edu/abs/2013MNRAS.434.1043O}
}

@ARTICLE{Lucchini2026,
    author = {{Lucchini}, Scott and {Abramson}, Cecilia and {Hummels}, Cameron and {Conroy}, Charlie and {Hernquist}, Lars and {Smith}, Aaron},
        title = "{ENhanced Galactic Atmospheres With Arepo: Resolving the CGM at 200 pc with the ENGAWA Simulations}",
      journal = {arXiv e-prints},
         year = 2026,
        month = mar,
          eid = {arXiv:2603.05584},
        pages = {arXiv:2603.05584},
          doi = {10.48550/arXiv.2603.05584},
archivePrefix = {arXiv},
       eprint = {2603.05584},
 primaryClass = {astro-ph.GA},
       adsurl = {https://ui.adsabs.harvard.edu/abs/2026arXiv260305584L}
}

@ARTICLE{Wakker1997,
       author = {{Wakker}, B.~P. and {van Woerden}, H.},
        title = "{High-Velocity Clouds}",
      journal = {\araa},
         year = 1997,
        month = jan,
       volume = {35},
        pages = {217-266},
          doi = {10.1146/annurev.astro.35.1.217},
       adsurl = {https://ui.adsabs.harvard.edu/abs/1997ARA&A..35..217W}
}

@ARTICLE{Ji2019,
       author = {{Ji}, Suoqing and {Oh}, S. Peng and {Masterson}, Phillip},
        title = "{Simulations of radiative turbulent mixing layers}",
      journal = {\mnras},
         year = 2019,
        month = jul,
       volume = {487},
       number = {1},
        pages = {737-754},
          doi = {10.1093/mnras/stz1248},
archivePrefix = {arXiv},
       eprint = {1809.09101},
 primaryClass = {astro-ph.GA},
       adsurl = {https://ui.adsabs.harvard.edu/abs/2019MNRAS.487..737J}
}

@ARTICLE{Bish2019,
       author = {{Bish}, Hannah V. and {Werk}, Jessica K. and {Prochaska}, J. Xavier and {Rubin}, Kate H.~R. and {Zheng}, Yong and {O'Meara}, John M. and {Deason}, Alis J.},
        title = "{Galactic Gas Flows from Halo to Disk: Tomography and Kinematics at the Milky Way{\textquoteright}s Disk-Halo Interface}",
      journal = {\apj},
         year = 2019,
        month = sep,
       volume = {882},
       number = {2},
          eid = {76},
        pages = {76},
          doi = {10.3847/1538-4357/ab3414},
archivePrefix = {arXiv},
       eprint = {1907.09459},
 primaryClass = {astro-ph.GA},
       adsurl = {https://ui.adsabs.harvard.edu/abs/2019ApJ...882...76B}
}

@ARTICLE{Fox2013,
       author = {{Fox}, Andrew J. and {Lehner}, Nicolas and {Tumlinson}, Jason and {Howk}, J. Christopher and {Tripp}, Todd M. and {Prochaska}, J. Xavier and {O'Meara}, John M. and {Werk}, Jessica K. and {Bordoloi}, Rongmon and {Katz}, Neal and {Oppenheimer}, Benjamin D. and {Dav{\'e}}, Romeel},
        title = "{The High-ion Content and Kinematics of Low-redshift Lyman Limit Systems}",
      journal = {\apj},
         year = 2013,
        month = dec,
       volume = {778},
       number = {2},
          eid = {187},
        pages = {187},
          doi = {10.1088/0004-637X/778/2/187},
archivePrefix = {arXiv},
       eprint = {1310.6267},
 primaryClass = {astro-ph.CO},
       adsurl = {https://ui.adsabs.harvard.edu/abs/2013ApJ...778..187F}
}

@ARTICLE{Lochhaas2026,
       author = {{Lochhaas}, Cassandra and {Peeples}, Molly S. and {O'Shea}, Brian W. and {Tumlinson}, Jason and {Corlies}, Lauren and {Saeedzadeh}, Vida and {Lehner}, Nicolas and {Wright}, Anna C. and {Werk}, Jessica K. and {Trapp}, Cameron W. and {Augustin}, Ramona and {Acharyya}, Ayan and {Smith}, Britton D. and {Vargas}, Carlos J.},
        title = "{Figuring Out Gas \& Galaxies in Enzo (FOGGIE). XI. Circumgalactic O VI Emission Traces Clumpy Inflowing Recycled Gas}",
      journal = {\apj},
         year = 2026,
        month = mar,
       volume = {1000},
       number = {1},
          eid = {104},
        pages = {104},
          doi = {10.3847/1538-4357/ae458f},
archivePrefix = {arXiv},
       eprint = {2510.25844},
 primaryClass = {astro-ph.GA},
       adsurl = {https://ui.adsabs.harvard.edu/abs/2026ApJ..1000..104L}
}

@ARTICLE{Sembach2003,
       author = {{Sembach}, K.~R. and {Wakker}, B.~P. and {Savage}, B.~D. and {Richter}, P. and {Meade}, M. and {Shull}, J.~M. and {Jenkins}, E.~B. and {Sonneborn}, G. and {Moos}, H.~W.},
        title = "{Highly Ionized High-Velocity Gas in the Vicinity of the Galaxy}",
      journal = {\apjs},
         year = 2003,
        month = may,
       volume = {146},
       number = {1},
        pages = {165-208},
          doi = {10.1086/346231},
archivePrefix = {arXiv},
       eprint = {astro-ph/0207562},
 primaryClass = {astro-ph},
       adsurl = {https://ui.adsabs.harvard.edu/abs/2003ApJS..146..165S}
}

@ARTICLE{Oppenheimer2016,
       author = {{Oppenheimer}, Benjamin D. and {Crain}, Robert A. and {Schaye}, Joop and {Rahmati}, Alireza and {Richings}, Alexander J. and {Trayford}, James W. and {Tumlinson}, Jason and {Bower}, Richard G. and {Schaller}, Matthieu and {Theuns}, Tom},
        title = "{Bimodality of low-redshift circumgalactic O VI in non-equilibrium EAGLE zoom simulations}",
      journal = {\mnras},
         year = 2016,
        month = aug,
       volume = {460},
       number = {2},
        pages = {2157-2179},
          doi = {10.1093/mnras/stw1066},
archivePrefix = {arXiv},
       eprint = {1603.05984},
 primaryClass = {astro-ph.GA},
       adsurl = {https://ui.adsabs.harvard.edu/abs/2016MNRAS.460.2157O}
}

@ARTICLE{Bish2021,
       author = {{Bish}, Hannah V. and {Werk}, Jessica K. and {Peek}, Joshua and {Zheng}, Yong and {Putman}, Mary},
        title = "{The QuaStar Survey: Detecting Hidden Low-velocity Gas in the Milky Way's Circumgalactic Medium}",
      journal = {\apj},
         year = 2021,
        month = may,
       volume = {912},
       number = {1},
          eid = {8},
        pages = {8},
          doi = {10.3847/1538-4357/abeb6b},
archivePrefix = {arXiv},
       eprint = {2010.03610},
 primaryClass = {astro-ph.GA},
       adsurl = {https://ui.adsabs.harvard.edu/abs/2021ApJ...912....8B}
}

@ARTICLE{Boettcher2021,
       author = {{Boettcher}, Erin and {Chen}, Hsiao-Wen and {Zahedy}, Fakhri S. and {Cooper}, Thomas J. and {Johnson}, Sean D. and {Rudie}, Gwen C. and {Chen}, Mandy C. and {Petitjean}, Patrick and {Cantalupo}, Sebastiano and {Cooksey}, Kathy L. and {Faucher-Gigu{\`e}re}, Claude-Andr{\'e} and {Greene}, Jenny E. and {Lopez}, Sebastian and {Mulchaey}, John S. and {Penton}, Steven V. and {Putman}, Mary E. and {Rafelski}, Marc and {Rauch}, Michael and {Schaye}, Joop and {Simcoe}, Robert A. and {Walth}, Gregory L.},
        title = "{The Cosmic Ultraviolet Baryon Survey (CUBS). II. Discovery of an H$_{2}$-bearing DLA in the Vicinity of an Early-type Galaxy at z = 0.576}",
      journal = {\apj},
         year = 2021,
        month = may,
       volume = {913},
       number = {1},
          eid = {18},
        pages = {18},
          doi = {10.3847/1538-4357/abf0a0},
archivePrefix = {arXiv},
       eprint = {2010.11958},
 primaryClass = {astro-ph.GA},
       adsurl = {https://ui.adsabs.harvard.edu/abs/2021ApJ...913...18B}
}

@ARTICLE{Borthakur2025,
       author = {{Borthakur}, Sanchayeeta and {Burchett}, Joseph N. and {Cashman}, Frances and {Fox}, Andrew J. and {Zheng}, Yong and {French}, David M. and {Bordoloi}, Rongmon and {Koplitz}, Brad},
        title = "{Characterizing gas flows through observations of the disk-circumgalactic medium interface with the Habitable Worlds Observatory}",
      journal = {Journal of Astronomical Telescopes, Instruments, and Systems},
         year = 2025,
        month = oct,
       volume = {11},
          eid = {042207},
        pages = {042207},
          doi = {10.1117/1.JATIS.11.4.042207},
archivePrefix = {arXiv},
       eprint = {2506.10517},
 primaryClass = {astro-ph.GA},
       adsurl = {https://ui.adsabs.harvard.edu/abs/2025JATIS..11d2207B}
}

@ARTICLE{Cashman2021,
       author = {{Cashman}, Frances H. and {Fox}, Andrew J. and {Savage}, Blair D. and {Wakker}, Bart P. and {Krishnarao}, Dhanesh and {Benjamin}, Robert A. and {Richter}, Philipp and {Ashley}, Trisha and {Jenkins}, Edward B. and {Lockman}, Felix J. and {Bordoloi}, Rongmon and {Kim}, Tae-Sun},
        title = "{Molecular Gas within the Milky Way's Nuclear Wind}",
      journal = {\apjl},
         year = 2021,
        month = dec,
       volume = {923},
       number = {1},
          eid = {L11},
        pages = {L11},
          doi = {10.3847/2041-8213/ac3cbc},
archivePrefix = {arXiv},
       eprint = {2112.03335},
 primaryClass = {astro-ph.GA},
       adsurl = {https://ui.adsabs.harvard.edu/abs/2021ApJ...923L..11C}
}

@ARTICLE{DiTeodoro2020,
       author = {{Di Teodoro}, Enrico M. and {McClure-Griffiths}, N.~M. and {Lockman}, Felix J. and {Armillotta}, Lucia},
        title = "{Cold gas in the Milky Way's nuclear wind}",
      journal = {\nat},
         year = 2020,
        month = aug,
       volume = {584},
       number = {7821},
        pages = {364-367},
          doi = {10.1038/s41586-020-2595-z},
archivePrefix = {arXiv},
       eprint = {2008.09121},
 primaryClass = {astro-ph.GA},
       adsurl = {https://ui.adsabs.harvard.edu/abs/2020Natur.584..364D}
}

@ARTICLE{Fox2014,
       author = {{Fox}, Andrew J. and {Wakker}, Bart P. and {Barger}, Kathleen A. and {Hernandez}, Audra K. and {Richter}, Philipp and {Lehner}, Nicolas and {Bland-Hawthorn}, Joss and {Charlton}, Jane C. and {Westmeier}, Tobias and {Thom}, Christopher and {Tumlinson}, Jason and {Misawa}, Toru and {Howk}, J. Christopher and {Haffner}, L. Matthew and {Ely}, Justin and {Rodriguez-Hidalgo}, Paola and {Kumari}, Nimisha},
        title = "{The COS/UVES Absorption Survey of the Magellanic Stream. III. Ionization, Total Mass, and Inflow Rate onto the Milky Way}",
      journal = {\apj},
         year = 2014,
        month = jun,
       volume = {787},
       number = {2},
          eid = {147},
        pages = {147},
          doi = {10.1088/0004-637X/787/2/147},
archivePrefix = {arXiv},
       eprint = {1404.5514},
 primaryClass = {astro-ph.GA},
       adsurl = {https://ui.adsabs.harvard.edu/abs/2014ApJ...787..147F}
}

@ARTICLE{Fox2019,
       author = {{Fox}, Andrew J. and {Richter}, Philipp and {Ashley}, Trisha and {Heckman}, Timothy M. and {Lehner}, Nicolas and {Werk}, Jessica K. and {Bordoloi}, Rongmon and {Peeples}, Molly S.},
        title = "{The Mass Inflow and Outflow Rates of the Milky Way}",
      journal = {\apj},
         year = 2019,
        month = oct,
       volume = {884},
       number = {1},
          eid = {53},
        pages = {53},
          doi = {10.3847/1538-4357/ab40ad},
archivePrefix = {arXiv},
       eprint = {1909.05561},
 primaryClass = {astro-ph.GA},
       adsurl = {https://ui.adsabs.harvard.edu/abs/2019ApJ...884...53F}
}

@ARTICLE{Lehner2012,
       author = {{Lehner}, N. and {Howk}, J.~C. and {Thom}, C. and {Fox}, A.~J. and {Tumlinson}, J. and {Tripp}, T.~M. and {Meiring}, J.~D.},
        title = "{High-velocity clouds as streams of ionized and neutral gas in the halo of the Milky Way}",
      journal = {\mnras},
         year = 2012,
        month = aug,
       volume = {424},
       number = {4},
        pages = {2896-2913},
          doi = {10.1111/j.1365-2966.2012.21428.x},
archivePrefix = {arXiv},
       eprint = {1203.2626},
 primaryClass = {astro-ph.GA},
       adsurl = {https://ui.adsabs.harvard.edu/abs/2012MNRAS.424.2896L}
}

@ARTICLE{Lehner2022,
       author = {{Lehner}, Nicolas and {Howk}, J. Christopher and {Marasco}, Antonino and {Fraternali}, Filippo},
        title = "{Intermediate- and high-velocity clouds in the Milky Way - I. Covering factors and vertical heights}",
      journal = {\mnras},
         year = 2022,
        month = jul,
       volume = {513},
       number = {3},
        pages = {3228-3240},
          doi = {10.1093/mnras/stac987},
archivePrefix = {arXiv},
       eprint = {2202.05848},
 primaryClass = {astro-ph.GA},
       adsurl = {https://ui.adsabs.harvard.edu/abs/2022MNRAS.513.3228L}
}

@ARTICLE{Lockman2016,
       author = {{Lockman}, Felix J. and {McClure-Griffiths}, N.~M.},
        title = "{Tracing the Milky Way Nuclear Wind with 21cm Atomic Hydrogen Emission}",
      journal = {\apj},
         year = 2016,
        month = aug,
       volume = {826},
       number = {2},
          eid = {215},
        pages = {215},
          doi = {10.3847/0004-637X/826/2/215},
archivePrefix = {arXiv},
       eprint = {1605.01140},
 primaryClass = {astro-ph.GA},
       adsurl = {https://ui.adsabs.harvard.edu/abs/2016ApJ...826..215L}
}

@ARTICLE{Lockman2020,
       author = {{Lockman}, Felix J. and {Di Teodoro}, Enrico M. and {McClure-Griffiths}, N.~M.},
        title = "{Observation of Acceleration of H I Clouds within the Fermi Bubbles}",
      journal = {\apj},
         year = 2020,
        month = jan,
       volume = {888},
       number = {1},
          eid = {51},
        pages = {51},
          doi = {10.3847/1538-4357/ab55d8},
archivePrefix = {arXiv},
       eprint = {1911.06864},
 primaryClass = {astro-ph.GA},
       adsurl = {https://ui.adsabs.harvard.edu/abs/2020ApJ...888...51L}
}

@ARTICLE{McClure2013,
       author = {{McClure-Griffiths}, N.~M. and {Green}, J.~A. and {Hill}, A.~S. and {Lockman}, F.~J. and {Dickey}, J.~M. and {Gaensler}, B.~M. and {Green}, A.~J.},
        title = "{Atomic Hydrogen in a Galactic Center Outflow}",
      journal = {\apjl},
         year = 2013,
        month = jun,
       volume = {770},
       number = {1},
          eid = {L4},
        pages = {L4},
          doi = {10.1088/2041-8205/770/1/L4},
archivePrefix = {arXiv},
       eprint = {1304.7538},
 primaryClass = {astro-ph.GA},
       adsurl = {https://ui.adsabs.harvard.edu/abs/2013ApJ...770L...4M}
}

@ARTICLE{Mishra2025,
       author = {{Mishra}, Sapna and {Fox}, Andrew J. and {Smoker}, J.~V. and {Lucchini}, Scott and {D'Onghia}, Elena},
        title = "{The Distance to the Magellanic Stream: Constraints from Optical Absorption along Stellar Sight Lines}",
      journal = {\apj},
         year = 2025,
        month = may,
       volume = {984},
       number = {2},
          eid = {104},
        pages = {104},
          doi = {10.3847/1538-4357/adc68a},
archivePrefix = {arXiv},
       eprint = {2503.14368},
 primaryClass = {astro-ph.GA},
       adsurl = {https://ui.adsabs.harvard.edu/abs/2025ApJ...984..104M}
}

@ARTICLE{Peeples2014,
       author = {{Peeples}, Molly S. and {Werk}, Jessica K. and {Tumlinson}, Jason and {Oppenheimer}, Benjamin D. and {Prochaska}, J. Xavier and {Katz}, Neal and {Weinberg}, David H.},
        title = "{A Budget and Accounting of Metals at z \raisebox{-0.5ex}\textasciitilde 0: Results from the COS-Halos Survey}",
      journal = {\apj},
         year = 2014,
        month = may,
       volume = {786},
       number = {1},
          eid = {54},
        pages = {54},
          doi = {10.1088/0004-637X/786/1/54},
archivePrefix = {arXiv},
       eprint = {1310.2253},
 primaryClass = {astro-ph.CO},
       adsurl = {https://ui.adsabs.harvard.edu/abs/2014ApJ...786...54P}
}

@ARTICLE{peroux2020,
       author = {{P{\'e}roux}, C{\'e}line and {Howk}, J. Christopher},
        title = "{The Cosmic Baryon and Metal Cycles}",
      journal = {\araa},
         year = 2020,
        month = aug,
       volume = {58},
        pages = {363-406},
          doi = {10.1146/annurev-astro-021820-120014},
archivePrefix = {arXiv},
       eprint = {2011.01935},
 primaryClass = {astro-ph.GA},
       adsurl = {https://ui.adsabs.harvard.edu/abs/2020ARA&A..58..363P}
}

@ARTICLE{Pointon2017,
       author = {{Pointon}, Stephanie K. and {Nielsen}, Nikole M. and {Kacprzak}, Glenn G. and {Muzahid}, Sowgat and {Churchill}, Christopher W. and {Charlton}, Jane C.},
        title = "{The Impact of the Group Environment on the O VI Circumgalactic Medium}",
      journal = {\apj},
         year = 2017,
        month = jul,
       volume = {844},
       number = {1},
          eid = {23},
        pages = {23},
          doi = {10.3847/1538-4357/aa7743},
archivePrefix = {arXiv},
       eprint = {1706.03895},
 primaryClass = {astro-ph.GA},
       adsurl = {https://ui.adsabs.harvard.edu/abs/2017ApJ...844...23P}
}

@ARTICLE{Putman2012,
       author = {{Putman}, M.~E. and {Peek}, J.~E.~G. and {Joung}, M.~R.},
        title = "{Gaseous Galaxy Halos}",
      journal = {\araa},
         year = 2012,
        month = sep,
       volume = {50},
        pages = {491-529},
          doi = {10.1146/annurev-astro-081811-125612},
archivePrefix = {arXiv},
       eprint = {1207.4837},
 primaryClass = {astro-ph.GA},
       adsurl = {https://ui.adsabs.harvard.edu/abs/2012ARA&A..50..491P}
}

@ARTICLE{Qu2024,
       author = {{Qu}, Zhijie and {Chen}, Hsiao-Wen and {Johnson}, Sean D. and {Rudie}, Gwen C. and {Zahedy}, Fakhri S. and {DePalma}, David and {Schaye}, Joop and {Boettcher}, Erin T. and {Cantalupo}, Sebastiano and {Chen}, Mandy C. and {Faucher-Gigu{\`e}re}, Claude-Andr{\'e} and {Li}, Jennifer I.-Hsiu and {Mulchaey}, John S. and {Petitjean}, Patrick and {Rafelski}, Marc},
        title = "{The Cosmic Ultraviolet Baryon Survey (CUBS). VII. On the Warm-hot Circumgalactic Medium Probed by O VI and Ne VIII at 0.4 {\ensuremath{\lesssim}} z {\ensuremath{\lesssim}} 0.7}",
      journal = {\apj},
         year = 2024,
        month = jun,
       volume = {968},
       number = {1},
          eid = {8},
        pages = {8},
          doi = {10.3847/1538-4357/ad410b},
archivePrefix = {arXiv},
       eprint = {2402.08016},
 primaryClass = {astro-ph.GA},
       adsurl = {https://ui.adsabs.harvard.edu/abs/2024ApJ...968....8Q}
}

@ARTICLE{Richter2001,
       author = {{Richter}, Philipp and {Sembach}, Kenneth R. and {Wakker}, Bart P. and {Savage}, Blair D.},
        title = "{Molecular Hydrogen in High-Velocity Clouds}",
      journal = {\apjl},
         year = 2001,
        month = dec,
       volume = {562},
       number = {2},
        pages = {L181-L184},
          doi = {10.1086/338050},
archivePrefix = {arXiv},
       eprint = {astro-ph/0110279},
 primaryClass = {astro-ph},
       adsurl = {https://ui.adsabs.harvard.edu/abs/2001ApJ...562L.181R}
}

@ARTICLE{Richter2017,
       author = {{Richter}, P. and {Nuza}, S.~E. and {Fox}, A.~J. and {Wakker}, B.~P. and {Lehner}, N. and {Ben Bekhti}, N. and {Fechner}, C. and {Wendt}, M. and {Howk}, J.~C. and {Muzahid}, S. and {Ganguly}, R. and {Charlton}, J.~C.},
        title = "{An HST/COS legacy survey of high-velocity ultraviolet absorption in the Milky Way's circumgalactic medium and the Local Group}",
      journal = {\aap},
         year = 2017,
        month = nov,
       volume = {607},
          eid = {A48},
        pages = {A48},
          doi = {10.1051/0004-6361/201630081},
archivePrefix = {arXiv},
       eprint = {1611.07024},
 primaryClass = {astro-ph.GA},
       adsurl = {https://ui.adsabs.harvard.edu/abs/2017A&A...607A..48R}
}

@ARTICLE{Tchernyshyov2022,
       author = {{Tchernyshyov}, Kirill and {Werk}, Jessica K. and {Wilde}, Matthew C. and {Prochaska}, J. Xavier and {Tripp}, Todd M. and {Burchett}, Joseph N. and {Bordoloi}, Rongmon and {Howk}, J. Christopher and {Lehner}, Nicolas and {O'Meara}, John M. and {Tejos}, Nicolas and {Tumlinson}, Jason},
        title = "{The CGM$^{2}$ Survey: Circumgalactic O VI from Dwarf to Massive Star-forming Galaxies}",
      journal = {\apj},
         year = 2022,
        month = mar,
       volume = {927},
       number = {2},
          eid = {147},
        pages = {147},
          doi = {10.3847/1538-4357/ac450c},
archivePrefix = {arXiv},
       eprint = {2110.13167},
 primaryClass = {astro-ph.GA},
       adsurl = {https://ui.adsabs.harvard.edu/abs/2022ApJ...927..147T}
}

@ARTICLE{Tchernyshyov2022b,
       author = {{Tchernyshyov}, Kirill},
        title = "{A Detection of H$_{2}$ in a High-velocity Cloud toward the Large Magellanic Cloud}",
      journal = {\apj},
         year = 2022,
        month = jun,
       volume = {931},
       number = {2},
          eid = {78},
        pages = {78},
          doi = {10.3847/1538-4357/ac68e0},
archivePrefix = {arXiv},
       eprint = {2204.09066},
 primaryClass = {astro-ph.GA},
       adsurl = {https://ui.adsabs.harvard.edu/abs/2022ApJ...931...78T}
}

@ARTICLE{Tumlinson2011,
       author = {{Tumlinson}, J. and {Thom}, C. and {Werk}, J.~K. and {Prochaska}, J.~X. and {Tripp}, T.~M. and {Weinberg}, D.~H. and {Peeples}, M.~S. and {O'Meara}, J.~M. and {Oppenheimer}, B.~D. and {Meiring}, J.~D. and {Katz}, N.~S. and {Dav{\'e}}, R. and {Ford}, A.~B. and {Sembach}, K.~R.},
        title = "{The Large, Oxygen-Rich Halos of Star-Forming Galaxies Are a Major Reservoir of Galactic Metals}",
      journal = {Science},
         year = 2011,
        month = nov,
       volume = {334},
       number = {6058},
        pages = {948},
          doi = {10.1126/science.1209840},
archivePrefix = {arXiv},
       eprint = {1111.3980},
 primaryClass = {astro-ph.CO},
       adsurl = {https://ui.adsabs.harvard.edu/abs/2011Sci...334..948T}
}

@ARTICLE{Tumlinson2017,
       author = {{Tumlinson}, Jason and {Peeples}, Molly S. and {Werk}, Jessica K.},
        title = "{The Circumgalactic Medium}",
      journal = {\araa},
         year = 2017,
        month = aug,
       volume = {55},
       number = {1},
        pages = {389-432},
          doi = {10.1146/annurev-astro-091916-055240},
archivePrefix = {arXiv},
       eprint = {1709.09180},
 primaryClass = {astro-ph.GA},
       adsurl = {https://ui.adsabs.harvard.edu/abs/2017ARA&A..55..389T}
}

@ARTICLE{Werk2014,
       author = {{Werk}, Jessica K. and {Prochaska}, J. Xavier and {Tumlinson}, Jason and {Peeples}, Molly S. and {Tripp}, Todd M. and {Fox}, Andrew J. and {Lehner}, Nicolas and {Thom}, Christopher and {O'Meara}, John M. and {Ford}, Amanda Brady and {Bordoloi}, Rongmon and {Katz}, Neal and {Tejos}, Nicolas and {Oppenheimer}, Benjamin D. and {Dav{\'e}}, Romeel and {Weinberg}, David H.},
        title = "{The COS-Halos Survey: Physical Conditions and Baryonic Mass in the Low-redshift Circumgalactic Medium}",
      journal = {\apj},
         year = 2014,
        month = sep,
       volume = {792},
       number = {1},
          eid = {8},
        pages = {8},
          doi = {10.1088/0004-637X/792/1/8},
archivePrefix = {arXiv},
       eprint = {1403.0947},
 primaryClass = {astro-ph.CO},
       adsurl = {https://ui.adsabs.harvard.edu/abs/2014ApJ...792....8W}
}

@ARTICLE{Werk2019,
       author = {{Werk}, Jessica K. and {Rubin}, K.~H.~R. and {Bish}, H.~V. and {Prochaska}, J.~X. and {Zheng}, Y. and {O'Meara}, J.~M. and {Lenz}, D. and {Hummels}, C. and {Deason}, A.~J.},
        title = "{The Nature of Ionized Gas in the Milky Way Galactic Fountain}",
      journal = {\apj},
         year = 2019,
        month = dec,
       volume = {887},
       number = {1},
          eid = {89},
        pages = {89},
          doi = {10.3847/1538-4357/ab54cf},
archivePrefix = {arXiv},
       eprint = {1904.11014},
 primaryClass = {astro-ph.GA},
       adsurl = {https://ui.adsabs.harvard.edu/abs/2019ApJ...887...89W}
}

@ARTICLE{Kulkarni05,
       author = {{Kulkarni}, Varsha P. and {Fall}, S. Michael and {Lauroesch}, James T. and {York}, Donald G. and {Welty}, Daniel E. and {Khare}, Pushpa and {Truran}, James W.},
        title = "{Hubble Space Telescope Observations of Element Abundances in Low-Redshift Damped Ly{\ensuremath{\alpha}} Galaxies and Implications for the Global Metallicity-Redshift Relation}",
      journal = {\apj},
         year = 2005,
        month = jan,
       volume = {618},
       number = {1},
        pages = {68-90},
          doi = {10.1086/425956},
archivePrefix = {arXiv},
       eprint = {astro-ph/0409234},
 primaryClass = {astro-ph},
       adsurl = {https://ui.adsabs.harvard.edu/abs/2005ApJ...618...68K}
}

@ARTICLE{Battisti12,
       author = {{Battisti}, A.~J. and {Meiring}, J.~D. and {Tripp}, T.~M. and {Prochaska}, J.~X. and {Werk}, J.~K. and {Jenkins}, E.~B. and {Lehner}, N. and {Tumlinson}, J. and {Thom}, C.},
        title = "{The First Observations of Low-redshift Damped Ly{\ensuremath{\alpha}} Systems with the Cosmic Origins Spectrograph: Chemical Abundances and Affiliated Galaxies}",
      journal = {\apj},
         year = 2012,
        month = jan,
       volume = {744},
       number = {2},
          eid = {93},
        pages = {93},
          doi = {10.1088/0004-637X/744/2/93},
archivePrefix = {arXiv},
       eprint = {1110.4557},
 primaryClass = {astro-ph.CO},
       adsurl = {https://ui.adsabs.harvard.edu/abs/2012ApJ...744...93B}
}

@ARTICLE{Som15,
       author = {{Som}, Debopam and {Kulkarni}, Varsha P. and {Meiring}, Joseph and {York}, Donald G. and {P{\'e}roux}, Celine and {Lauroesch}, James T. and {Aller}, Monique C. and {Khare}, Pushpa},
        title = "{Hubble Space Telescope Observations of Sub-damped Ly{\ensuremath{\alpha}} Absorbers at z < 0.5, and Implications for Galaxy Chemical Evolution}",
      journal = {\apj},
         year = 2015,
        month = jun,
       volume = {806},
       number = {1},
          eid = {25},
        pages = {25},
          doi = {10.1088/0004-637X/806/1/25},
archivePrefix = {arXiv},
       eprint = {1502.01989},
 primaryClass = {astro-ph.CO},
       adsurl = {https://ui.adsabs.harvard.edu/abs/2015ApJ...806...25S}
}

@ARTICLE{Zahedy17,
       author = {{Zahedy}, Fakhri S. and {Chen}, Hsiao-Wen and {Gauthier}, Jean-Ren{\'e} and {Rauch}, Michael},
        title = "{On the radial profile of gas-phase Fe/{\ensuremath{\alpha}} ratio around distant galaxies}",
      journal = {\mnras},
         year = 2017,
        month = apr,
       volume = {466},
       number = {1},
        pages = {1071-1081},
          doi = {10.1093/mnras/stw3124},
archivePrefix = {arXiv},
       eprint = {1611.09874},
 primaryClass = {astro-ph.GA},
       adsurl = {https://ui.adsabs.harvard.edu/abs/2017MNRAS.466.1071Z}
}

@article{Zahedy19,
	author = {{Zahedy}, F.~S. and {Chen}, H.-W. and {Johnson}, S.~D. and {Pierce}, R.~M. and {Rauch}, M. and {Huang}, Y.-H. and {Weiner}, B.~D. and {Gauthier}, J.-R.},
	journal = {MNRAS},
	month = apr,
	number = {2},
	pages = {2257-2280},
	title = {{Characterizing Circumgalactic Gas around Massive Ellipticals at z\~{}0.4 - II. Physical Properties and Elemental Abundances}},
	volume = {484},
	year = 2019}

@ARTICLE{Zahedy20,
       author = {{Zahedy}, Fakhri S. and {Chen}, Hsiao-Wen and {Boettcher}, Erin and {Rauch}, Michael and {French}, K. Decker and {Zabludoff}, Ann I.},
        title = "{Evidence for Late-time Feedback from the Discovery of Multiphase Gas in a Massive Elliptical at z = 0.4}",
      journal = {\apjl},
         year = 2020,
        month = nov,
       volume = {904},
       number = {1},
          eid = {L10},
        pages = {L10},
          doi = {10.3847/2041-8213/abc48d},
archivePrefix = {arXiv},
       eprint = {2009.14232},
 primaryClass = {astro-ph.GA},
       adsurl = {https://ui.adsabs.harvard.edu/abs/2020ApJ...904L..10Z}
}

@ARTICLE{Zahedy21,
       author = {{Zahedy}, Fakhri S. and {Chen}, Hsiao-Wen and {Cooper}, Thomas M. and {Boettcher}, Erin and {Johnson}, Sean D. and {Rudie}, Gwen C. and {Chen}, Mandy C. and {Cantalupo}, Sebastiano and {Cooksey}, Kathy L. and {Faucher-Gigu{\`e}re}, Claude-Andr{\'e} and {Greene}, Jenny E. and {Lopez}, Sebastian and {Mulchaey}, John S. and {Penton}, Steven V. and {Petitjean}, Patrick and {Putman}, Mary E. and {Rafelski}, Marc and {Rauch}, Michael and {Schaye}, Joop and {Simcoe}, Robert A. and {Walth}, Gregory L.},
        title = "{The Cosmic Ultraviolet Baryon Survey (CUBS) - III. Physical properties and elemental abundances of Lyman-limit systems at $z < 1$}",
      journal = {\mnras},
         year = 2021,
        month = sep,
       volume = {506},
       number = {1},
        pages = {877-902},
          doi = {10.1093/mnras/stab1661},
archivePrefix = {arXiv},
       eprint = {2106.04608},
 primaryClass = {astro-ph.GA},
       adsurl = {https://ui.adsabs.harvard.edu/abs/2021MNRAS.506..877Z}
}

@ARTICLE{Kosenko24,
       author = {{Kosenko}, D.~N. and {Balashev}, S.~A. and {Klimenko}, V.~V.},
        title = "{Cold diffuse interstellar medium of Magellanic Clouds: II. Physical conditions from excitation of C I and H$_{2}$}",
      journal = {\mnras},
         year = 2024,
        month = mar,
       volume = {528},
       number = {3},
        pages = {5065-5079},
          doi = {10.1093/mnras/stae354},
archivePrefix = {arXiv},
       eprint = {2309.01599},
 primaryClass = {astro-ph.GA},
       adsurl = {https://ui.adsabs.harvard.edu/abs/2024MNRAS.528.5065K}
}

@ARTICLE{Gadallah11,
       author = {{Gadallah}, K.~A.~K. and {Mutschke}, H. and {J{\"a}ger}, C.},
        title = "{UV irradiated hydrogenated amorphous carbon (HAC) materials as a carrier candidate of the interstellar UV bump at 217.5 nm}",
      journal = {\aap},
         year = 2011,
        month = apr,
       volume = {528},
          eid = {A56},
        pages = {A56},
          doi = {10.1051/0004-6361/201015542},
       adsurl = {https://ui.adsabs.harvard.edu/abs/2011A&A...528A..56G}
}

@ARTICLE{Balashev25,
       author = {{Balashev}, S.~A. and {Kosenko}, D.~N. and {Noterdaeme}, P.},
        title = "{First detections of CO absorption in the Magellanic Clouds and direct measurement of the CO-to-H$_{2}$ ratio}",
      journal = {\aap},
         year = 2025,
        month = apr,
       volume = {696},
          eid = {L16},
        pages = {L16},
          doi = {10.1051/0004-6361/202452913},
archivePrefix = {arXiv},
       eprint = {2503.12516},
 primaryClass = {astro-ph.GA},
       adsurl = {https://ui.adsabs.harvard.edu/abs/2025A&A...696L..16B}
}

@ARTICLE{Jenkins01,
       author = {{Jenkins}, Edward B. and {Tripp}, Todd M.},
        title = "{The Distribution of Thermal Pressures in the Interstellar Medium from a Survey of C I Fine-Structure Excitation}",
      journal = {\apjs},
         year = 2001,
        month = dec,
       volume = {137},
       number = {2},
        pages = {297-340},
          doi = {10.1086/323326},
archivePrefix = {arXiv},
       eprint = {astro-ph/0107177},
 primaryClass = {astro-ph},
       adsurl = {https://ui.adsabs.harvard.edu/abs/2001ApJS..137..297J}
}

@ARTICLE{Jenkins79,
       author = {{Jenkins}, E.~B. and {Shaya}, E.~J.},
        title = "{A survey of interstellar C I: insights on carbon abundances, UV grain albedos, and pressures in the interstellar medium.}",
      journal = {\apj},
         year = 1979,
        month = jul,
       volume = {231},
        pages = {55-72},
          doi = {10.1086/157163},
       adsurl = {https://ui.adsabs.harvard.edu/abs/1979ApJ...231...55J}
}

@ARTICLE{Jenkins11,
       author = {{Jenkins}, Edward B. and {Tripp}, Todd M.},
        title = "{The Distribution of Thermal Pressures in the Diffuse, Cold Neutral Medium of Our Galaxy. II. An Expanded Survey of Interstellar C I Fine-structure Excitations}",
      journal = {\apj},
         year = 2011,
        month = jun,
       volume = {734},
       number = {1},
          eid = {65},
        pages = {65},
          doi = {10.1088/0004-637X/734/1/65},
archivePrefix = {arXiv},
       eprint = {1104.2323},
 primaryClass = {astro-ph.GA},
       adsurl = {https://ui.adsabs.harvard.edu/abs/2011ApJ...734...65J}
}

@ARTICLE{Welty16,
       author = {{Welty}, Daniel E. and {Lauroesch}, James T. and {Wong}, Tony and {York}, Donald G.},
        title = "{Thermal Pressures in the Interstellar Medium of the Magellanic Clouds}",
      journal = {\apj},
         year = 2016,
        month = apr,
       volume = {821},
       number = {2},
          eid = {118},
        pages = {118},
          doi = {10.3847/0004-637X/821/2/118},
archivePrefix = {arXiv},
       eprint = {1603.03801},
 primaryClass = {astro-ph.GA},
       adsurl = {https://ui.adsabs.harvard.edu/abs/2016ApJ...821..118W}
}

@ARTICLE{Tchernyshyov25,
       author = {{Tchernyshyov}, Kirill and {Werk}, Jessica K. and {Roman-Duval}, Julia},
        title = "{An Ultraviolet Study of CO Chemistry in the Magellanic Clouds}",
      journal = {\aj},
         year = 2025,
        month = sep,
       volume = {170},
       number = {3},
          eid = {176},
        pages = {176},
          doi = {10.3847/1538-3881/adf33a},
archivePrefix = {arXiv},
       eprint = {2508.10996},
 primaryClass = {astro-ph.GA},
       adsurl = {https://ui.adsabs.harvard.edu/abs/2025AJ....170..176T}
}

@ARTICLE{Snow06,
       author = {{Snow}, Theodore P. and {McCall}, Benjamin J.},
        title = "{Diffuse Atomic and Molecular Clouds}",
      journal = {\araa},
         year = 2006,
        month = sep,
       volume = {44},
       number = {1},
        pages = {367-414},
          doi = {10.1146/annurev.astro.43.072103.150624},
       adsurl = {https://ui.adsabs.harvard.edu/abs/2006ARA&A..44..367S}
}

@ARTICLE{Sonnentrucker07,
       author = {{Sonnentrucker}, P. and {Welty}, D.~E. and {Thorburn}, J.~A. and {York}, D.~G.},
        title = "{Abundances and Behavior of $^{12}$CO, $^{13}$CO, and C$_{2}$ in Translucent Sight Lines}",
      journal = {\apjs},
         year = 2007,
        month = jan,
       volume = {168},
       number = {1},
        pages = {58-99},
          doi = {10.1086/508687},
archivePrefix = {arXiv},
       eprint = {astro-ph/0608557},
 primaryClass = {astro-ph},
       adsurl = {https://ui.adsabs.harvard.edu/abs/2007ApJS..168...58S}
}

@ARTICLE{Sheffer08,
       author = {{Sheffer}, Y. and {Rogers}, M. and {Federman}, S.~R. and {Abel}, N.~P. and {Gredel}, R. and {Lambert}, D.~L. and {Shaw}, G.},
        title = "{Ultraviolet Survey of CO and H$_{2}$ in Diffuse Molecular Clouds: The Reflection of Two Photochemistry Regimes in Abundance Relationships}",
      journal = {\apj},
         year = 2008,
        month = nov,
       volume = {687},
       number = {2},
        pages = {1075-1106},
          doi = {10.1086/591484},
archivePrefix = {arXiv},
       eprint = {0807.0940},
 primaryClass = {astro-ph},
       adsurl = {https://ui.adsabs.harvard.edu/abs/2008ApJ...687.1075S}
}

@ARTICLE{Wolfire03,
       author = {{Wolfire}, Mark G. and {McKee}, Christopher F. and {Hollenbach}, David and {Tielens}, A.~G.~G.~M.},
        title = "{Neutral Atomic Phases of the Interstellar Medium in the Galaxy}",
      journal = {\apj},
         year = 2003,
        month = apr,
       volume = {587},
       number = {1},
        pages = {278-311},
          doi = {10.1086/368016},
archivePrefix = {arXiv},
       eprint = {astro-ph/0207098},
 primaryClass = {astro-ph},
       adsurl = {https://ui.adsabs.harvard.edu/abs/2003ApJ...587..278W}
}

@ARTICLE{Krumholz09,
       author = {{Krumholz}, Mark R. and {McKee}, Christopher F. and {Tumlinson}, Jason},
        title = "{The Atomic-to-Molecular Transition in Galaxies. II: H I and H$_{2}$ Column Densities}",
      journal = {\apj},
         year = 2009,
        month = mar,
       volume = {693},
       number = {1},
        pages = {216-235},
          doi = {10.1088/0004-637X/693/1/216},
archivePrefix = {arXiv},
       eprint = {0811.0004},
 primaryClass = {astro-ph},
       adsurl = {https://ui.adsabs.harvard.edu/abs/2009ApJ...693..216K}
}

@ARTICLE{Glover12,
       author = {{Glover}, Simon C.~O. and {Clark}, Paul C.},
        title = "{Is molecular gas necessary for star formation?}",
      journal = {\mnras},
         year = 2012,
        month = mar,
       volume = {421},
       number = {1},
        pages = {9-19},
          doi = {10.1111/j.1365-2966.2011.19648.x},
archivePrefix = {arXiv},
       eprint = {1105.3073},
 primaryClass = {astro-ph.GA},
       adsurl = {https://ui.adsabs.harvard.edu/abs/2012MNRAS.421....9G}
}

@ARTICLE{Sternberg14,
       author = {{Sternberg}, Amiel and {Le Petit}, Franck and {Roueff}, Evelyne and {Le Bourlot}, Jacques},
        title = "{H I-to-H$_{2}$ Transitions and H I Column Densities in Galaxy Star-forming Regions}",
      journal = {\apj},
         year = 2014,
        month = jul,
       volume = {790},
       number = {1},
          eid = {10},
        pages = {10},
          doi = {10.1088/0004-637X/790/1/10},
archivePrefix = {arXiv},
       eprint = {1404.5042},
 primaryClass = {astro-ph.GA},
       adsurl = {https://ui.adsabs.harvard.edu/abs/2014ApJ...790...10S}
}

@ARTICLE{Bialy16,
       author = {{Bialy}, Shmuel and {Sternberg}, Amiel},
        title = "{Analytic H I-to-H$_{2}$ Photodissociation Transition Profiles}",
      journal = {\apj},
         year = 2016,
        month = may,
       volume = {822},
       number = {2},
          eid = {83},
        pages = {83},
          doi = {10.3847/0004-637X/822/2/83},
archivePrefix = {arXiv},
       eprint = {1601.02608},
 primaryClass = {astro-ph.GA},
       adsurl = {https://ui.adsabs.harvard.edu/abs/2016ApJ...822...83B}
}

@ARTICLE{Galliano18,
       author = {{Galliano}, Fr{\'e}d{\'e}ric and {Galametz}, Maud and {Jones}, Anthony P.},
        title = "{The Interstellar Dust Properties of Nearby Galaxies}",
      journal = {\araa},
         year = 2018,
        month = sep,
       volume = {56},
        pages = {673-713},
          doi = {10.1146/annurev-astro-081817-051900},
archivePrefix = {arXiv},
       eprint = {1711.07434},
 primaryClass = {astro-ph.GA},
       adsurl = {https://ui.adsabs.harvard.edu/abs/2018ARA&A..56..673G}
}

@ARTICLE{Decleir26,
       author = {{Decleir}, Marjorie and {Gordon}, Karl D. and {De Cia}, Annalisa and {Hensley}, Brandon S. and {Baes}, Maarten and {Chen}, Chian-Chou and {Elyajouri}, Meriem and {Galliano}, Fr{\'e}d{\'e}ric and {Henning}, Thomas and {Jencson}, Jacob and et al.},
        title = "{Summary of Discussion Sessions from ``The Dusty Universe 2025: The Fifth Pandust Conference''}",
      journal = {arXiv e-prints},
         year = 2026,
        month = may,
          eid = {arXiv:2605.18992},
        pages = {arXiv:2605.18992},
          doi = {10.48550/arXiv.2605.18992},
archivePrefix = {arXiv},
       eprint = {2605.18992},
 primaryClass = {astro-ph.GA},
       adsurl = {https://ui.adsabs.harvard.edu/abs/2026arXiv260518992D}
}

@ARTICLE{deBurgos23,
       author = {{de Burgos}, A. and {Sim{\'o}n-D{\'\i}az}, S. and {Urbaneja}, M.~A. and {Negueruela}, I.},
        title = "{The IACOB project. IX. Building a modern empirical database of Galactic O9 - B9 supergiants: Sample selection, description, and completeness}",
      journal = {\aap},
         year = 2023,
        month = jun,
       volume = {674},
          eid = {A212},
        pages = {A212},
          doi = {10.1051/0004-6361/202346179},
archivePrefix = {arXiv},
       eprint = {2305.00305},
 primaryClass = {astro-ph.SR},
       adsurl = {https://ui.adsabs.harvard.edu/abs/2023A&A...674A.212D}
}

@ARTICLE{Mattsson19,
       author = {{Mattsson}, Lars and {De Cia}, Annalisa and {Andersen}, Anja C. and {Petitjean}, Patrick},
        title = "{Dust-depletion sequences in damped Lyman-{\ensuremath{\alpha}} absorbers. II. The composition of cosmic dust, from low-metallicity systems to the Galaxy}",
      journal = {\aap},
         year = 2019,
        month = apr,
       volume = {624},
          eid = {A103},
        pages = {A103},
          doi = {10.1051/0004-6361/201731482},
archivePrefix = {arXiv},
       eprint = {1901.04710},
 primaryClass = {astro-ph.GA},
       adsurl = {https://ui.adsabs.harvard.edu/abs/2019A&A...624A.103M}
}

@ARTICLE{Yan18,
       author = {{Yan}, Lin and {Perley}, D.~A. and {De Cia}, A. and {Quimby}, R. and {Lunnan}, R. and {Rubin}, Kate H.~R. and {Brown}, P.~J.},
        title = "{Far-UV HST  Spectroscopy of an Unusual Hydrogen-poor Superluminous Supernova: SN2017egm}",
      journal = {\apj},
         year = 2018,
        month = may,
       volume = {858},
       number = {2},
          eid = {91},
        pages = {91},
          doi = {10.3847/1538-4357/aabad5},
archivePrefix = {arXiv},
       eprint = {1711.01534},
 primaryClass = {astro-ph.HE},
       adsurl = {https://ui.adsabs.harvard.edu/abs/2018ApJ...858...91Y}
}

@ARTICLE{Lorenzo22,
       author = {{Lorenzo}, M. and {Garcia}, M. and {Najarro}, F. and {Herrero}, A. and {Cervi{\~n}o}, M. and {Castro}, N.},
        title = "{A new reference catalogue for the very metal-poor Universe: +150 OB stars in Sextans A}",
      journal = {\mnras},
         year = 2022,
        month = nov,
       volume = {516},
       number = {3},
        pages = {4164-4179},
          doi = {10.1093/mnras/stac2050},
archivePrefix = {arXiv},
       eprint = {2207.09700},
 primaryClass = {astro-ph.GA},
       adsurl = {https://ui.adsabs.harvard.edu/abs/2022MNRAS.516.4164L}
}

@ARTICLE{Izotov04,
       author = {{Izotov}, Yuri I. and {Thuan}, Trinh X.},
        title = "{Deep Hubble Space Telescope ACS Observations of I Zw 18: a Young Galaxy in Formation}",
      journal = {\apj},
         year = 2004,
        month = dec,
       volume = {616},
       number = {2},
        pages = {768-782},
          doi = {10.1086/424990},
archivePrefix = {arXiv},
       eprint = {astro-ph/0408391},
 primaryClass = {astro-ph},
       adsurl = {https://ui.adsabs.harvard.edu/abs/2004ApJ...616..768I}
}

@ARTICLE{Aloisi07,
       author = {{Aloisi}, A. and {Clementini}, G. and {Tosi}, M. and {Annibali}, F. and {Contreras}, R. and {Fiorentino}, G. and {Mack}, J. and {Marconi}, M. and {Musella}, I. and {Saha}, A. and {Sirianni}, M. and {van der Marel}, R.~P.},
        title = "{I Zw 18 Revisited with HST ACS and Cepheids: New Distance and Age}",
      journal = {\apjl},
         year = 2007,
        month = oct,
       volume = {667},
       number = {2},
        pages = {L151-L154},
          doi = {10.1086/522368},
archivePrefix = {arXiv},
       eprint = {0707.2371},
 primaryClass = {astro-ph},
       adsurl = {https://ui.adsabs.harvard.edu/abs/2007ApJ...667L.151A}
}

@ARTICLE{Cikota17,
       author = {{Cikota}, Aleksandar and {De Cia}, Annalisa and {Schulze}, Steve and {Vreeswijk}, Paul M. and {Leloudas}, Giorgos and {Gal-Yam}, Avishay and {Perley}, Daniel A. and {Cikota}, Stefan and {Kim}, Sam and {Patat}, Ferdinando and et al.},
        title = "{Spatially resolved analysis of superluminous supernovae PTF 11hrq and PTF 12dam host galaxies}",
      journal = {\mnras},
         year = 2017,
        month = aug,
       volume = {469},
       number = {4},
        pages = {4705-4717},
          doi = {10.1093/mnras/stx1110},
archivePrefix = {arXiv},
       eprint = {1705.01948},
 primaryClass = {astro-ph.HE},
       adsurl = {https://ui.adsabs.harvard.edu/abs/2017MNRAS.469.4705C}
}

@ARTICLE{DOdorico24,
       author = {{D'Odorico}, Valentina and {Bolton}, James S. and {Christensen}, Lise and {De Cia}, Annalisa and {Zackrisson}, Erik and {Kordt}, Aron and {Izzo}, Luca and {Li}, Jiangtao and {Maiolino}, Roberto and {Marconi}, Alessandro and et al.},
        title = "{Galaxy formation and symbiotic evolution with the inter-galactic medium in the age of ELT-ANDES}",
      journal = {Experimental Astronomy},
         year = 2024,
        month = dec,
       volume = {58},
       number = {3},
          eid = {21},
        pages = {21},
          doi = {10.1007/s10686-024-09967-3},
archivePrefix = {arXiv},
       eprint = {2311.16803},
 primaryClass = {astro-ph.GA},
       adsurl = {https://ui.adsabs.harvard.edu/abs/2024ExA....58...21D}
}

@ARTICLE{Price01b,
       author = {{Price}, R.~J. and {Crawford}, I.~A. and {Howarth}, I.~D.},
        title = "{{\ensuremath{\delta}} Orionis: further temporal variability and evidence for small-scale structure in the interstellar medium}",
      journal = {\mnras},
         year = 2001,
        month = mar,
       volume = {321},
       number = {3},
        pages = {553-558},
          doi = {10.1046/j.1365-8711.2001.04055.x},
       adsurl = {https://ui.adsabs.harvard.edu/abs/2001MNRAS.321..553P}
}

@ARTICLE{Price01,
       author = {{Price}, R.~J. and {Crawford}, I.~A. and {Barlow}, M.~J. and {Howarth}, I.~D.},
        title = "{An ultra-high-resolution study of the interstellar medium towards Orion}",
      journal = {\mnras},
         year = 2001,
        month = dec,
       volume = {328},
       number = {2},
        pages = {555-582},
          doi = {10.1046/j.1365-8711.2001.04893.x},
       adsurl = {https://ui.adsabs.harvard.edu/abs/2001MNRAS.328..555P}
}

@ARTICLE{Hamanowicz24,
       author = {{Hamanowicz}, Aleksandra and {Tchernyshyov}, Kirill and {Roman-Duval}, Julia and {Jenkins}, Edward B. and {Rafelski}, Marc and {Gordon}, Karl D. and {Zheng}, Yong and {Garcia}, Miriam and {Werk}, Jessica},
        title = "{METAL-Z: Measuring Dust Depletion in Low-metallicity Dwarf Galaxies}",
      journal = {\apj},
         year = 2024,
        month = may,
       volume = {966},
       number = {1},
          eid = {80},
        pages = {80},
          doi = {10.3847/1538-4357/ad307b},
archivePrefix = {arXiv},
       eprint = {2402.18733},
 primaryClass = {astro-ph.GA},
       adsurl = {https://ui.adsabs.harvard.edu/abs/2024ApJ...966...80H}
}

@INPROCEEDINGS{Roman-Duval26,
       author = {{Roman-Duval}, Julia and {Boquien}, Mederic and {Choi}, Yumi},
        title = "{Probing the Variations of Interstellar Dust Abundance and Properties Within and Between Galaxies with HWO UV Spectroscopy in the Local Volume}",
    booktitle = {Astronomical Society of the Pacific Conference Series},
         year = 2026,
       editor = {{Lee}, Janice C. and {Noviello}, Jessica and {LaMassa}, Stephanie and {Postman}, Marc},
       series = {Astronomical Society of the Pacific Conference Series},
       volume = {542},
        month = feb,
        pages = {97},
          doi = {10.26624/DRNL2690},
       adsurl = {https://ui.adsabs.harvard.edu/abs/2026ASPC..542...97R}
}

@ARTICLE{Ritchey23,
       author = {{Ritchey}, Adam M. and {Jenkins}, Edward B. and {Shull}, J. Michael and {Savage}, Blair D. and {Federman}, S.~R. and {Lambert}, David L.},
        title = "{The Distribution of Metallicities in the Local Galactic Interstellar Medium}",
      journal = {\apj},
         year = 2023,
        month = jul,
       volume = {952},
       number = {1},
          eid = {57},
        pages = {57},
          doi = {10.3847/1538-4357/acda25},
archivePrefix = {arXiv},
       eprint = {2301.09743},
 primaryClass = {astro-ph.GA},
       adsurl = {https://ui.adsabs.harvard.edu/abs/2023ApJ...952...57R}
}

@ARTICLE{Konstantopoulou24a,
       author = {{Konstantopoulou}, Christina and {De Cia}, Annalisa and {Ledoux}, C{\'e}dric and {Krogager}, Jens-Kristian and {Mattsson}, Lars and {Watson}, Darach and {Heintz}, Kasper E. and {P{\'e}roux}, C{\'e}line and {Noterdaeme}, Pasquier and {Andersen}, Anja C. and et al.},
        title = "{Dust depletion of metals from local to distant galaxies. II. Cosmic dust-to-metal ratio and dust composition}",
      journal = {\aap},
         year = 2024,
        month = jan,
       volume = {681},
          eid = {A64},
        pages = {A64},
          doi = {10.1051/0004-6361/202347171},
archivePrefix = {arXiv},
       eprint = {2310.07709},
 primaryClass = {astro-ph.GA},
       adsurl = {https://ui.adsabs.harvard.edu/abs/2024A&A...681A..64K}
}

@ARTICLE{Konstantopoulou24b,
       author = {{Konstantopoulou}, Christina and {De Cia}, Annalisa and {Krogager}, Jens-Kristian and {Ledoux}, C{\'e}dric and {Roman-Duval}, Julia and {Jenkins}, Edward B. and {Ramburuth-Hurt}, Tanita and {Velichko}, Anna},
        title = "{DUNE: Dust depletion UNified method across cosmic time and Environments}",
      journal = {\aap},
         year = 2024,
        month = nov,
       volume = {691},
          eid = {A129},
        pages = {A129},
          doi = {10.1051/0004-6361/202451488},
archivePrefix = {arXiv},
       eprint = {2410.06155},
 primaryClass = {astro-ph.GA},
       adsurl = {https://ui.adsabs.harvard.edu/abs/2024A&A...691A.129K}
}

@ARTICLE{Ramburuth-Hurt25,
       author = {{Ramburuth-Hurt}, T. and {De Cia}, A. and {Krogager}, J.-K. and {Ledoux}, C. and {Jenkins}, E. and {Fox}, A.~J. and {Konstantopoulou}, C. and {Velichko}, A. and {Dalla Pola}, L.},
        title = "{Investigating chemical variations between interstellar gas clouds in the solar neighbourhood}",
      journal = {\aap},
         year = 2025,
        month = mar,
       volume = {695},
          eid = {A14},
        pages = {A14},
          doi = {10.1051/0004-6361/202451729},
archivePrefix = {arXiv},
       eprint = {2412.18986},
 primaryClass = {astro-ph.GA},
       adsurl = {https://ui.adsabs.harvard.edu/abs/2025A&A...695A..14R}
}

@ARTICLE{Velichko24,
       author = {{Velichko}, Anna and {De Cia}, Annalisa and {Konstantopoulou}, Christina and {Ledoux}, C{\'e}dric and {Krogager}, Jens-Kristian and {Ramburuth-Hurt}, Tanita},
        title = "{The {\ensuremath{\alpha}}-element enrichment of gas in distant galaxies}",
      journal = {\aap},
         year = 2024,
        month = may,
       volume = {685},
          eid = {A103},
        pages = {A103},
          doi = {10.1051/0004-6361/202348601},
archivePrefix = {arXiv},
       eprint = {2402.15508},
 primaryClass = {astro-ph.GA},
       adsurl = {https://ui.adsabs.harvard.edu/abs/2024A&A...685A.103V}
}

@ARTICLE{DeCia24,
       author = {{De Cia}, Annalisa and {Roman-Duval}, Julia and {Konstantopoulou}, Christina and {Noterdaeme}, Pasquier and {Ramburuth-Hurt}, Tanita and {Velichko}, Anna and {Fox}, Andrew J. and {Ledoux}, C{\'e}dric and {Petitjean}, Patrick and {Jermann}, Iris and et al.},
        title = "{{\ensuremath{\alpha}}-element enhancements in the ISM of the LMC and SMC: Evidence of recent star formation}",
      journal = {\aap},
         year = 2024,
        month = mar,
       volume = {683},
          eid = {A216},
        pages = {A216},
          doi = {10.1051/0004-6361/202346611},
       adsurl = {https://ui.adsabs.harvard.edu/abs/2024A&A...683A.216D}
}

@ARTICLE{Kobayashi20,
       author = {{Kobayashi}, Chiaki and {Karakas}, Amanda I. and {Lugaro}, Maria},
        title = "{The Origin of Elements from Carbon to Uranium}",
      journal = {\apj},
         year = 2020,
        month = sep,
       volume = {900},
       number = {2},
          eid = {179},
        pages = {179},
          doi = {10.3847/1538-4357/abae65},
archivePrefix = {arXiv},
       eprint = {2008.04660},
 primaryClass = {astro-ph.GA},
       adsurl = {https://ui.adsabs.harvard.edu/abs/2020ApJ...900..179K}
}

@INPROCEEDINGS{Roman-Duval20,
       author = {{Roman-Duval}, J.~C. and {De Rosa}, G. and {Proffitt}, C. and {Reid}, I. and {Brown}, T. and {Fullerton}, A. and {Fischer}, W. and {Aloisi}, A. and {Britt}, C. and {Busko}, I. and {Carlberg}, J. and {Fox}, A. and {Frazer}, E. and {James}, B. and {Jedrzejewski}, R. and {Lockwood}, S. and {Monroe}, T. and {Oliveira}, C. and {Plesha}, R. and {Riedel}, A. and {Riley}, A. and {Shaw}, R. and {Smith}, L. and {Sohn}, S. and {Taylor}, J. and {Ubeda}, L. and {Welty}, D.},
        title = "{Overview of the ULLYSES Director's Discretionary HST program}",
    booktitle = {American Astronomical Society Meeting Abstracts \#235},
         year = 2020,
       series = {American Astronomical Society Meeting Abstracts},
       volume = {235},
        month = jan,
          eid = {232.03},
        pages = {232.03},
       adsurl = {https://ui.adsabs.harvard.edu/abs/2020AAS...23523203R}
}

@ARTICLE{Roman-Duval22,
       author = {{Roman-Duval}, Julia and {Jenkins}, Edward B. and {Tchernyshyov}, Kirill and {Clark}, Christopher J.~R. and {De Cia}, Annalisa and {Gordon}, Karl D. and {Hamanowicz}, Aleksandra and {Lebouteiller}, Vianney and {Rafelski}, Marc and {Sandstrom}, Karin and {Werk}, Jessica and {Yanchulova Merica-Jones}, Petia},
        title = "{METAL: The Metal Evolution, Transport, and Abundance in the Large Magellanic Cloud Hubble program. IV. Calibration of Dust Depletions vs Abundance Ratios in the Milky Way and Magellanic Clouds and Application to Damped Lyman-alpha Systems}",
      journal = {arXiv e-prints},
         year = 2022,
        month = jun,
          eid = {arXiv:2206.03639},
        pages = {arXiv:2206.03639},
          doi = {10.48550/arXiv.2206.03639},
archivePrefix = {arXiv},
       eprint = {2206.03639},
 primaryClass = {astro-ph.GA},
       adsurl = {https://ui.adsabs.harvard.edu/abs/2022arXiv220603639R}
}

@ARTICLE{Roman-Duval19,
       author = {{Roman-Duval}, Julia and {Jenkins}, Edward B. and {Williams}, Benjamin and {Tchernyshyov}, Kirill and {Gordon}, Karl and {Meixner}, Margaret and {Hagen}, Lea and {Peek}, Joshua and {Sandstrom}, Karin and {Werk}, Jessica and {Yanchulova Merica-Jones}, Petia},
        title = "{METAL: The Metal Evolution, Transport, and Abundance in the Large Magellanic Cloud Hubble Program. I. Overview and Initial Results}",
      journal = {\apj},
         year = 2019,
        month = feb,
       volume = {871},
       number = {2},
          eid = {151},
        pages = {151},
          doi = {10.3847/1538-4357/aaf8bb},
archivePrefix = {arXiv},
       eprint = {1901.06027},
 primaryClass = {astro-ph.GA},
       adsurl = {https://ui.adsabs.harvard.edu/abs/2019ApJ...871..151R}
}

@ARTICLE{Matteucci21,
       author = {{Matteucci}, Francesca},
        title = "{Modelling the chemical evolution of the Milky Way}",
      journal = {\aapr},
         year = 2021,
        month = dec,
       volume = {29},
       number = {1},
          eid = {5},
        pages = {5},
          doi = {10.1007/s00159-021-00133-8},
archivePrefix = {arXiv},
       eprint = {2106.13145},
 primaryClass = {astro-ph.GA},
       adsurl = {https://ui.adsabs.harvard.edu/abs/2021A&ARv..29....5M}
}

@ARTICLE{Welty20,
       author = {{Welty}, Daniel E. and {Sonnentrucker}, Paule and {Snow}, Theodore P. and {York}, Donald G.},
        title = "{HD 62542: Probing the Bare, Dense Core of a Translucent Interstellar Cloud}",
      journal = {\apj},
         year = 2020,
        month = jul,
       volume = {897},
       number = {1},
          eid = {36},
        pages = {36},
          doi = {10.3847/1538-4357/ab8f8e},
archivePrefix = {arXiv},
       eprint = {2005.10846},
 primaryClass = {astro-ph.GA},
       adsurl = {https://ui.adsabs.harvard.edu/abs/2020ApJ...897...36W}
}

@ARTICLE{Tinsley79,
       author = {{Tinsley}, B.~M.},
        title = "{Stellar lifetimes and abundance ratios in chemical evolution.}",
      journal = {\apj},
         year = 1979,
        month = may,
       volume = {229},
        pages = {1046-1056},
          doi = {10.1086/157039},
       adsurl = {https://ui.adsabs.harvard.edu/abs/1979ApJ...229.1046T}
}

@ARTICLE{Crowther16,
       author = {{Crowther}, Paul A. and {Caballero-Nieves}, S.~M. and {Bostroem}, K.~A. and {Ma{\'\i}z Apell{\'a}niz}, J. and {Schneider}, F.~R.~N. and {Walborn}, N.~R. and {Angus}, C.~R. and {Brott}, I. and {Bonanos}, A. and {de Koter}, A. and {de Mink}, S.~E. and {Evans}, C.~J. and {Gr{\"a}fener}, G. and {Herrero}, A. and {Howarth}, I.~D. and {Langer}, N. and {Lennon}, D.~J. and {Puls}, J. and {Sana}, H. and {Vink}, J.~S.},
        title = "{The R136 star cluster dissected with Hubble Space Telescope/STIS. I. Far-ultraviolet spectroscopic census and the origin of He II {\ensuremath{\lambda}}1640 in young star clusters}",
      journal = {\mnras},
         year = 2016,
        month = may,
       volume = {458},
       number = {1},
        pages = {624-659},
          doi = {10.1093/mnras/stw273},
archivePrefix = {arXiv},
       eprint = {1603.04994},
 primaryClass = {astro-ph.SR},
       adsurl = {https://ui.adsabs.harvard.edu/abs/2016MNRAS.458..624C}
}

@ARTICLE{Asplund21,
       author = {{Asplund}, M. and {Amarsi}, A.~M. and {Grevesse}, N.},
        title = "{The chemical make-up of the Sun: A 2020 vision}",
      journal = {\aap},
         year = 2021,
        month = sep,
       volume = {653},
          eid = {A141},
        pages = {A141},
          doi = {10.1051/0004-6361/202140445},
archivePrefix = {arXiv},
       eprint = {2105.01661},
 primaryClass = {astro-ph.SR},
       adsurl = {https://ui.adsabs.harvard.edu/abs/2021A&A...653A.141A}
}

@ARTICLE{Roman-Duval21,
       author = {{Roman-Duval}, Julia and {Jenkins}, Edward B. and {Tchernyshyov}, Kirill and {Williams}, Benjamin and {Clark}, Christopher J.~R. and {Gordon}, Karl D. and {Meixner}, Margaret and {Hagen}, Lea and {Peek}, Joshua and {Sandstrom}, Karin and {Werk}, Jessica and {Yanchulova Merica-Jones}, Petia},
        title = "{METAL: The Metal Evolution, Transport, and Abundance in the Large Magellanic Cloud Hubble Program. II. Variations of Interstellar Depletions and Dust-to-gas Ratio within the LMC}",
      journal = {\apj},
         year = 2021,
        month = apr,
       volume = {910},
       number = {2},
          eid = {95},
        pages = {95},
          doi = {10.3847/1538-4357/abdeb6},
archivePrefix = {arXiv},
       eprint = {2101.09399},
 primaryClass = {astro-ph.GA},
       adsurl = {https://ui.adsabs.harvard.edu/abs/2021ApJ...910...95R}
}

@ARTICLE{DeCia21,
       author = {{De Cia}, Annalisa and {Jenkins}, Edward B. and {Fox}, Andrew J. and {Ledoux}, C{\'e}dric and {Ramburuth-Hurt}, Tanita and {Konstantopoulou}, Christina and {Petitjean}, Patrick and {Krogager}, Jens-Kristian},
        title = "{Large metallicity variations in the Galactic interstellar medium}",
      journal = {\nat},
         year = 2021,
        month = sep,
       volume = {597},
       number = {7875},
        pages = {206-208},
          doi = {10.1038/s41586-021-03780-0},
archivePrefix = {arXiv},
       eprint = {2109.03249},
 primaryClass = {astro-ph.GA},
       adsurl = {https://ui.adsabs.harvard.edu/abs/2021Natur.597..206D}
}

@ARTICLE{DeCia18b,
       author = {{De Cia}, Annalisa and {Ledoux}, C{\'e}dric and {Petitjean}, Patrick and {Savaglio}, Sandra},
        title = "{The cosmic evolution of dust-corrected metallicity in the neutral gas}",
      journal = {\aap},
         year = 2018,
        month = apr,
       volume = {611},
          eid = {A76},
        pages = {A76},
          doi = {10.1051/0004-6361/201731970},
archivePrefix = {arXiv},
       eprint = {1709.06578},
 primaryClass = {astro-ph.GA},
       adsurl = {https://ui.adsabs.harvard.edu/abs/2018A&A...611A..76D}
}

@ARTICLE{Poudel17,
   author = {{Poudel}, S. and {Kulkarni}, V.~P. and {Morrison}, S. and {P{\'e}roux}, C. and 
	{Som}, D. and {Rahmani}, H. and {Quiret}, S.},
    title = "{Early metal enrichment of gas-rich galaxies at z\~{}5}",
  journal = {ArXiv e-prints},
archivePrefix = "arXiv",
   eprint = {1710.03315},
     year = 2017,
    month = oct,
   adsurl = {http://adsabs.harvard.edu/abs/2017arXiv171003315P}
}

@ARTICLE{Sofia97,
       author = {{Sofia}, Ulysses J. and {Cardelli}, Jason A. and {Guerin}, Kenneth P. and {Meyer}, David M.},
        title = "{Carbon in the Diffuse Interstellar Medium}",
      journal = {\apjl},
         year = 1997,
        month = jun,
       volume = {482},
       number = {1},
        pages = {L105-L108},
          doi = {10.1086/310681},
       adsurl = {https://ui.adsabs.harvard.edu/abs/1997ApJ...482L.105S}
}

@ARTICLE{Sofia98,
       author = {{Sofia}, Ulysses J. and {Fitzpatrick}, Edward L. and {Meyer}, David M.},
        title = "{A Reanalysis of the Carbon Abundance in the Translucent Cloud Toward HD 24534}",
      journal = {\apjl},
         year = 1998,
        month = sep,
       volume = {504},
       number = {1},
        pages = {L47-L49},
          doi = {10.1086/311566},
archivePrefix = {arXiv},
       eprint = {astro-ph/9806401},
 primaryClass = {astro-ph},
       adsurl = {https://ui.adsabs.harvard.edu/abs/1998ApJ...504L..47S}
}

@ARTICLE{Ramburuth-Hurt26,
       author = {{Ramburuth-Hurt}, T. and {De Cia}, A. and {Krogager}, J.-K. and {Ledoux}, C. and {Fox}, A.~J.},
        title = "{Connecting the dusty dots: Dust depletion and extinction of local interstellar clouds}",
      journal = {\aap},
         year = 2026,
        month = jun,
       volume = {710},
          eid = {A64},
        pages = {A64},
          doi = {10.1051/0004-6361/202557895},
archivePrefix = {arXiv},
       eprint = {2604.24752},
 primaryClass = {astro-ph.GA},
       adsurl = {https://ui.adsabs.harvard.edu/abs/2026A&A...710A..64R}
}

@ARTICLE{Ramburuth-Hurt26b,
       author = {{Ramburuth-Hurt}, T. and {De Cia}, A. and {Krogager}, J.-K. and {Ledoux}, C. and {Jenkins}, E. and {Fox}, A.~J. and {Konstantopoulou}, C. and {Velichko}, A. and {Dalla Pola}, L.},
        title = "{Investigating chemical variations between interstellar gas clouds in the solar neighbourhood (Corrigendum)}",
      journal = {\aap},
         year = 2026,
        month = mar,
       volume = {707},
          eid = {C3},
        pages = {C3},
          doi = {10.1051/0004-6361/202659479e},
       adsurl = {https://ui.adsabs.harvard.edu/abs/2026A&A...707C...3R}
}

@ARTICLE{Edenhofer24,
       author = {{Edenhofer}, Gordian and {Zucker}, Catherine and {Frank}, Philipp and {Saydjari}, Andrew K. and {Speagle}, Joshua S. and {Finkbeiner}, Douglas and {En{\ss}lin}, Torsten A.},
        title = "{A parsec-scale Galactic 3D dust map out to 1.25 kpc from the Sun}",
      journal = {\aap},
         year = 2024,
        month = may,
       volume = {685},
          eid = {A82},
        pages = {A82},
          doi = {10.1051/0004-6361/202347628},
archivePrefix = {arXiv},
       eprint = {2308.01295},
 primaryClass = {astro-ph.GA},
       adsurl = {https://ui.adsabs.harvard.edu/abs/2024A&A...685A..82E}
}

@ARTICLE{Gordon03,
       author = {{Gordon}, Karl D. and {Clayton}, Geoffrey C. and {Misselt}, K.~A. and {Landolt}, Arlo U. and {Wolff}, Michael J.},
        title = "{A Quantitative Comparison of the Small Magellanic Cloud, Large Magellanic Cloud, and Milky Way Ultraviolet to Near-Infrared Extinction Curves}",
      journal = {\apj},
         year = 2003,
        month = sep,
       volume = {594},
       number = {1},
        pages = {279-293},
          doi = {10.1086/376774},
archivePrefix = {arXiv},
       eprint = {astro-ph/0305257},
 primaryClass = {astro-ph},
       adsurl = {https://ui.adsabs.harvard.edu/abs/2003ApJ...594..279G}
}

@ARTICLE{Cox17,
       author = {{Cox}, Nick L.~J. and {Cami}, Jan and {Farhang}, Amin and {Smoker}, Jonathan and {Monreal-Ibero}, Ana and {Lallement}, Rosine and {Sarre}, Peter J. and {Marshall}, Charlotte C.~M. and {Smith}, Keith T. and {Evans}, Christopher J. and et al.},
        title = "{The ESO Diffuse Interstellar Bands Large Exploration Survey (EDIBLES) . I. Project description, survey sample, and quality assessment}",
      journal = {\aap},
         year = 2017,
        month = oct,
       volume = {606},
          eid = {A76},
        pages = {A76},
          doi = {10.1051/0004-6361/201730912},
archivePrefix = {arXiv},
       eprint = {1708.01429},
 primaryClass = {astro-ph.GA},
       adsurl = {https://ui.adsabs.harvard.edu/abs/2017A&A...606A..76C}
}

@ARTICLE{Dharmawardena24,
       author = {{Dharmawardena}, T.~E. and {Bailer-Jones}, C.~A.~L. and {Fouesneau}, M. and {Foreman-Mackey}, D. and {Coronica}, P. and {Colnaghi}, T. and {M{\"u}ller}, T. and {Wilson}, A.~G.},
        title = "{All-sky three-dimensional dust density and extinction Maps of the Milky Way out to 2.8 kpc}",
      journal = {\mnras},
         year = 2024,
        month = aug,
       volume = {532},
       number = {3},
        pages = {3480-3498},
          doi = {10.1093/mnras/stae1474},
archivePrefix = {arXiv},
       eprint = {2406.06740},
 primaryClass = {astro-ph.GA},
       adsurl = {https://ui.adsabs.harvard.edu/abs/2024MNRAS.532.3480D}
}

@ARTICLE{Zucker22,
       author = {{Zucker}, Catherine and {Goodman}, Alyssa A. and {Alves}, Jo{\~a}o and {Bialy}, Shmuel and {Foley}, Michael and {Speagle}, Joshua S. and {Gro{\^I}{\texttwosuperior}schedl}, Josefa and {Finkbeiner}, Douglas P. and {Burkert}, Andreas and {Khimey}, Diana and et al.},
        title = "{Star formation near the Sun is driven by expansion of the Local Bubble}",
      journal = {\nat},
         year = 2022,
        month = jan,
       volume = {601},
       number = {7893},
        pages = {334-337},
          doi = {10.1038/s41586-021-04286-5},
archivePrefix = {arXiv},
       eprint = {2201.05124},
 primaryClass = {astro-ph.GA},
       adsurl = {https://ui.adsabs.harvard.edu/abs/2022Natur.601..334Z}
}

@ARTICLE{ONeill24,
       author = {{O'Neill}, Theo J. and {Zucker}, Catherine and {Goodman}, Alyssa A. and {Edenhofer}, Gordian},
        title = "{The Local Bubble Is a Local Chimney: A New Model from 3D Dust Mapping}",
      journal = {\apj},
         year = 2024,
        month = oct,
       volume = {973},
       number = {2},
          eid = {136},
        pages = {136},
          doi = {10.3847/1538-4357/ad61de},
archivePrefix = {arXiv},
       eprint = {2403.04961},
 primaryClass = {astro-ph.GA},
       adsurl = {https://ui.adsabs.harvard.edu/abs/2024ApJ...973..136O}
}

@ARTICLE{Fuchs06,
       author = {{Fuchs}, B. and {Breitschwerdt}, D. and {de Avillez}, M.~A. and {Dettbarn}, C. and {Flynn}, C.},
        title = "{The search for the origin of the Local Bubble redivivus}",
      journal = {\mnras},
         year = 2006,
        month = dec,
       volume = {373},
       number = {3},
        pages = {993-1003},
          doi = {10.1111/j.1365-2966.2006.11044.x},
archivePrefix = {arXiv},
       eprint = {astro-ph/0609227},
 primaryClass = {astro-ph},
       adsurl = {https://ui.adsabs.harvard.edu/abs/2006MNRAS.373..993F}
}

@ARTICLE{Dwek16,
   author = {{Dwek}, E.},
    title = "{Iron: A Key Element for Understanding the Origin and Evolution of Interstellar Dust}",
  journal = {\apj},
archivePrefix = "arXiv",
   eprint = {1605.01957},
     year = 2016,
    month = jul,
   volume = 825,
      eid = {136},
    pages = {136},
      doi = {10.3847/0004-637X/825/2/136},
   adsurl = {http://adsabs.harvard.edu/abs/2016ApJ...825..136D}
}

@ARTICLE{Cameron23,
       author = {{Cameron}, Alex J. and {Katz}, Harley and {Rey}, Martin P. and {Saxena}, Aayush},
        title = "{Nitrogen enhancements 440 Myr after the big bang: supersolar N/O, a tidal disruption event, or a dense stellar cluster in GN-z11?}",
      journal = {\mnras},
         year = 2023,
        month = aug,
       volume = {523},
       number = {3},
        pages = {3516-3525},
          doi = {10.1093/mnras/stad1579},
archivePrefix = {arXiv},
       eprint = {2302.10142},
 primaryClass = {astro-ph.GA},
       adsurl = {https://ui.adsabs.harvard.edu/abs/2023MNRAS.523.3516C}
}

@ARTICLE{Marques-Chaves24,
       author = {{Marques-Chaves}, R. and {Schaerer}, D. and {Kuruvanthodi}, A. and {Korber}, D. and {Prantzos}, N. and {Charbonnel}, C. and {Weibel}, A. and {Izotov}, Y.~I. and {Messa}, M. and {Brammer}, G. and et al.},
        title = "{Extreme N-emitters at high redshift: Possible signatures of supermassive stars and globular cluster or black hole formation in action}",
      journal = {\aap},
         year = 2024,
        month = jan,
       volume = {681},
          eid = {A30},
        pages = {A30},
          doi = {10.1051/0004-6361/202347411},
archivePrefix = {arXiv},
       eprint = {2307.04234},
 primaryClass = {astro-ph.GA},
       adsurl = {https://ui.adsabs.harvard.edu/abs/2024A&A...681A..30M}
}

@ARTICLE{Tumlinson17,
   author = {{Tumlinson}, J. and {Peeples}, M.~S. and {Werk}, J.~K.},
    title = "{The Circumgalactic Medium}",
  journal = {\araa},
     year = 2017,
    month = aug,
   volume = 55,
    pages = {389-432},
      doi = {10.1146/annurev-astro-091916-055240},
   adsurl = {http://adsabs.harvard.edu/abs/2017ARA%26A..55..389T}
}

@ARTICLE{Jenkins17,
   author = {{Jenkins}, E.~B. and {Wallerstein}, G.},
    title = "{Interstellar Gas-phase Element Depletions in the Small Magellanic Cloud: A Guide to Correcting for Dust in QSO Absorption Line Systems}",
  journal = {\apj},
archivePrefix = "arXiv",
   eprint = {1705.02675},
     year = 2017,
    month = apr,
   volume = 838,
      eid = {85},
    pages = {85},
      doi = {10.3847/1538-4357/aa64d4},
   adsurl = {http://adsabs.harvard.edu/abs/2017ApJ...838...85J}
}

@ARTICLE{Maiolino19,
       author = {{Maiolino}, R. and {Mannucci}, F.},
        title = "{De re metallica: the cosmic chemical evolution of galaxies}",
      journal = {\aapr},
         year = 2019,
        month = feb,
       volume = {27},
       number = {1},
          eid = {3},
        pages = {3},
          doi = {10.1007/s00159-018-0112-2},
archivePrefix = {arXiv},
       eprint = {1811.09642},
 primaryClass = {astro-ph.GA},
       adsurl = {https://ui.adsabs.harvard.edu/abs/2019A&ARv..27....3M}
}

@ARTICLE{DeCia16,
   author = {{De Cia}, A. and {Ledoux}, C. and {Mattsson}, L. and {Petitjean}, P. and 
	{Srianand}, R. and {Gavignaud}, I. and {Jenkins}, E.~B.},
    title = "{Dust-depletion sequences in damped Lyman-{$\alpha$} absorbers. A unified picture from low-metallicity systems to the Galaxy}",
  journal = {\aap},
archivePrefix = "arXiv",
   eprint = {1608.08621},
     year = 2016,
    month = dec,
   volume = 596,
      eid = {A97},
    pages = {A97},
      doi = {10.1051/0004-6361/201527895},
   adsurl = {http://adsabs.harvard.edu/abs/2016A%26A...596A..97D}
}

@ARTICLE{Vladilo02,
   author = {{Vladilo}, G.},
    title = "{Chemical abundances of damped Ly alpha systems:. A new method for estimating dust depletion effects}",
  journal = {\aap},
   eprint = {astro-ph/0206048},
     year = 2002,
    month = aug,
   volume = 391,
    pages = {407-415},
      doi = {10.1051/0004-6361:20020822},
   adsurl = {http://adsabs.harvard.edu/abs/2002A%26A...391..407V}
}

@ARTICLE{Peroux20,
       author = {{P{\'e}roux}, C{\'e}line and {Howk}, J. Christopher},
        title = "{The Cosmic Baryon and Metal Cycles}",
      journal = {\araa},
         year = 2020,
        month = aug,
       volume = {58},
        pages = {363-406},
          doi = {10.1146/annurev-astro-021820-120014},
archivePrefix = {arXiv},
       eprint = {2011.01935},
 primaryClass = {astro-ph.GA},
       adsurl = {https://ui.adsabs.harvard.edu/abs/2020ARA&A..58..363P}
}

@ARTICLE{Rafelski12,
   author = {{Rafelski}, M. and {Wolfe}, A.~M. and {Prochaska}, J.~X. and 
	{Neeleman}, M. and {Mendez}, A.~J.},
    title = "{Metallicity Evolution of Damped Ly{$\alpha$} Systems Out to z \~{} 5}",
  journal = {\apj},
archivePrefix = "arXiv",
   eprint = {1205.5047},
 primaryClass = "astro-ph.CO",
     year = 2012,
    month = aug,
   volume = 755,
      eid = {89},
    pages = {89},
      doi = {10.1088/0004-637X/755/2/89},
   adsurl = {http://adsabs.harvard.edu/abs/2012ApJ...755...89R}
}

@ARTICLE{Tchernyshyov15,
   author = {{Tchernyshyov}, K. and {Meixner}, M. and {Seale}, J. and {Fox}, A. and 
	{Friedman}, S.~D. and {Dwek}, E. and {Galliano}, F.},
    title = "{Elemental Depletions in the Magellanic Clouds and the Evolution of Depletions with Metallicity}",
  journal = {\apj},
archivePrefix = "arXiv",
   eprint = {1503.08852},
     year = 2015,
    month = oct,
   volume = 811,
      eid = {78},
    pages = {78},
      doi = {10.1088/0004-637X/811/2/78},
   adsurl = {http://adsabs.harvard.edu/abs/2015ApJ...811...78T}
}

@ARTICLE{Jenkins14,
   author = {{Jenkins}, E.~B.},
    title = "{Depletions of Elements from the Gas Phase: A Guide on Dust Compositions}",
  journal = {ArXiv:1402.4765},
archivePrefix = "arXiv",
   eprint = {1402.4765},
 primaryClass = "astro-ph.GA",
     year = 2014,
    month = feb,
   adsurl = {http://adsabs.harvard.edu/abs/2014arXiv1402.4765J}
}

@BOOK{Draine11,
   author = {{Draine}, B.~T.},
    title = "{Physics of the Interstellar and Intergalactic Medium}",
booktitle = {Physics of the Interstellar and Intergalactic Medium by Bruce T.~Draine.~Princeton University Press, 2011.~ISBN: 978-0-691-12214-4},
     year = 2011,
     publisher= {Princeton University Press},
   adsurl = {http://adsabs.harvard.edu/abs/2011piim.book.....D}
}

@ARTICLE{Field74,
   author = {{Field}, G.~B.},
    title = "{Interstellar abundances: gas and dust.}",
  journal = {\apj},
     year = 1974,
    month = feb,
   volume = 187,
    pages = {453-459},
      doi = {10.1086/152654},
   adsurl = {http://adsabs.harvard.edu/abs/1974ApJ...187..453F}
}

@ARTICLE{Jenkins09,
   author = {{Jenkins}, E.~B.},
    title = "{A Unified Representation of Gas-Phase Element Depletions in the Interstellar Medium}",
  journal = {\apj},
archivePrefix = "arXiv",
   eprint = {0905.3173},
 primaryClass = "astro-ph.GA",
     year = 2009,
    month = aug,
   volume = 700,
    pages = {1299-1348},
      doi = {10.1088/0004-637X/700/2/1299},
   adsurl = {http://adsabs.harvard.edu/abs/2009ApJ...700.1299J}
}

@ARTICLE{Petitjean08,
   author = {{Petitjean}, P. and {Ledoux}, C. and {Srianand}, R.},
    title = "{The nitrogen and oxygen abundances in the neutral gas at high redshift}",
  journal = {\aap},
archivePrefix = "arXiv",
   eprint = {0712.2760},
     year = 2008,
    month = mar,
   volume = 480,
    pages = {349-357},
      doi = {10.1051/0004-6361:20078607},
   adsurl = {http://adsabs.harvard.edu/abs/2008A%26A...480..349P}
}

@ARTICLE{Draine11b,
   author = {{Draine}, B.~T.},
    title = "{On Radiation Pressure in Static, Dusty H II Regions}",
  journal = {\apj},
archivePrefix = "arXiv",
   eprint = {1003.0474},
 primaryClass = "astro-ph.GA",
     year = 2011,
    month = may,
   volume = 732,
    pages = {100-+},
      doi = {10.1088/0004-637X/732/2/100},
   adsurl = {http://adsabs.harvard.edu/abs/2011ApJ...732..100D}
}

@ARTICLE{Wolfe05,
   author = {{Wolfe}, A.~M. and {Gawiser}, E. and {Prochaska}, J.~X.},
    title = "{Damped Ly {$\alpha$} Systems}",
  journal = {\araa},
   eprint = {arXiv:astro-ph/0509481},
     year = 2005,
    month = sep,
   volume = 43,
    pages = {861-918},
      doi = {10.1146/annurev.astro.42.053102.133950},
   adsurl = {http://adsabs.harvard.edu/abs/2005ARA%26A..43..861W}
}

@ARTICLE{Ledoux02,
   author = {{Ledoux}, C. and {Bergeron}, J. and {Petitjean}, P.},
    title = "{Dust depletion and abundance pattern in damped Lyalpha systems: A sample of Mn and Ti abundances at z < 2.2}",
  journal = {\aap},
   eprint = {arXiv:astro-ph/0202134},
     year = 2002,
    month = apr,
   volume = 385,
    pages = {802-815},
      doi = {10.1051/0004-6361:20020198},
   adsurl = {http://adsabs.harvard.edu/abs/2002A%26A...385..802L}
}

@ARTICLE{Saccardi23,
       author = {{Saccardi}, A. and {Vergani}, S.~D. and {De Cia}, A. and {D'Elia}, V. and {Heintz}, K.~E. and {Izzo}, L. and {Palmerio}, J.~T. and {Petitjean}, P. and {Rossi}, A. and {de Ugarte Postigo}, A. and et al.},
        title = "{Dissecting the interstellar medium of a z = 6.3 galaxy. X-shooter spectroscopy and HST imaging of the afterglow and environment of the Swift GRB 210905A}",
      journal = {\aap},
         year = 2023,
        month = mar,
       volume = {671},
          eid = {A84},
        pages = {A84},
          doi = {10.1051/0004-6361/202244205},
archivePrefix = {arXiv},
       eprint = {2211.16524},
 primaryClass = {astro-ph.GA},
       adsurl = {https://ui.adsabs.harvard.edu/abs/2023A&A...671A..84S}
}

@ARTICLE{Savage96,
   author = {{Savage}, B.~D. and {Sembach}, K.~R.},
    title = "{Interstellar Abundances from Absorption-Line Observations with the Hubble Space Telescope}",
  journal = {\araa},
     year = 1996,
   volume = 34,
    pages = {279-330},
      doi = {10.1146/annurev.astro.34.1.279},
   adsurl = {http://adsabs.harvard.edu/abs/1996ARA%26A..34..279S}
}

@ARTICLE{Rowlands2014,
       author = {{Rowlands}, K. and {Gomez}, H.~L. and {Dunne}, L. and {Arag{\'o}n-Salamanca}, A. and {Dye}, S. and {Maddox}, S. and {da Cunha}, E. and {van der Werf}, P.},
        title = "{The dust budget crisis in high-redshift submillimetre galaxies}",
      journal = {\mnras},
         year = 2014,
        month = jun,
       volume = {441},
       number = {2},
        pages = {1040-1058},
          doi = {10.1093/mnras/stu605},
archivePrefix = {arXiv},
       eprint = {1403.2995},
 primaryClass = {astro-ph.GA},
       adsurl = {https://ui.adsabs.harvard.edu/abs/2014MNRAS.441.1040R}
}

@ARTICLE{Sofia04,
       author = {{Sofia}, Ulysses J. and {Lauroesch}, James T. and {Meyer}, David M. and {Cartledge}, Stefan I.~B.},
        title = "{Interstellar Carbon in Translucent Sight Lines}",
      journal = {\apj},
         year = 2004,
        month = apr,
       volume = {605},
       number = {1},
        pages = {272-277},
          doi = {10.1086/382592},
archivePrefix = {arXiv},
       eprint = {astro-ph/0401510},
 primaryClass = {astro-ph},
       adsurl = {https://ui.adsabs.harvard.edu/abs/2004ApJ...605..272S}
}

@ARTICLE{Stecher1965,
       author = {{Stecher}, T.~P. and {Donn}, B.},
        title = "{On Graphite and Interstellar Extinction}",
      journal = {\apj},
         year = 1965,
        month = nov,
       volume = {142},
        pages = {1681},
          doi = {10.1086/148461},
       adsurl = {https://ui.adsabs.harvard.edu/abs/1965ApJ...142.1681S}
}

@ARTICLE{Valencic04,
       author = {{Valencic}, Lynne A. and {Clayton}, Geoffrey C. and {Gordon}, Karl D.},
        title = "{Ultraviolet Extinction Properties in the Milky Way}",
      journal = {\apj},
         year = 2004,
        month = dec,
       volume = {616},
       number = {2},
        pages = {912-924},
          doi = {10.1086/424922},
archivePrefix = {arXiv},
       eprint = {astro-ph/0408409},
 primaryClass = {astro-ph},
       adsurl = {https://ui.adsabs.harvard.edu/abs/2004ApJ...616..912V}
}

@ARTICLE{Gordon2026,
       author = {{Gordon}, Karl D. and {Yanchulova Merica-Jones}, Petia and {Clayton}, Geoffrey C. and {Bohlin}, Ralph and {Decleir}, Marjorie and {Murray}, Claire E. and {Bianchi}, Luciana},
        title = "{Large Variations Seen in First Ultraviolet Spectroscopic M33 Dust Extinction Curves}",
      journal = {\apj},
         year = 2026,
        month = apr,
       volume = {1001},
       number = {2},
          eid = {221},
        pages = {221},
          doi = {10.3847/1538-4357/ae5702},
archivePrefix = {arXiv},
       eprint = {2603.23662},
 primaryClass = {astro-ph.GA},
       adsurl = {https://ui.adsabs.harvard.edu/abs/2026ApJ..1001..221G}
}

@ARTICLE{Zeegers2025,
       author = {{Zeegers}, S.~T. and {Marshall}, Jonathan P. and {Gordon}, Karl D. and {Misselt}, Karl A. and {Otten}, G.~P.~P.~L. and {Bouwman}, Jeroen and {Chiar}, Jean and {Decleir}, Marjorie and {Dharmawardena}, Thavisha and {Kemper}, F. and {Li}, Aigen and {Narang}, Mayank and {Potapov}, Alexey and {Puravankara}, Manoj and {Scicluna}, Peter and {Tyagi}, Himanshu and {Zari}, Eleonora and {Wei}, ChuanYu and {Kaper}, Lex and {Backs}, Frank and {Bromley}, Stefan T. and {Chu}, Laurie and {Costantini}, Elisa and {Geballe}, T.~R. and {Green}, Joel D. and {Gunasekera}, Chamani and {G{\"u}nay}, Burcu and {Henning}, Thomas and {Jones}, Olivia and {Mari{\~n}oso Guiu}, Joan and {McClure}, Melissa and {Pendleton}, Yvonne J. and {Roman-Duval}, Julia C. and {Shenar}, Tomer and {Tielens}, Alexander G.~G.~M. and {Waters}, L.~B.~F.~M.},
        title = "{Investigating Silicate, Carbon, and Water in the Diffuse Interstellar Medium: The First Shots from WISCI}",
      journal = {\apj},
         year = 2025,
        month = jul,
       volume = {987},
       number = {1},
          eid = {25},
        pages = {25},
          doi = {10.3847/1538-4357/add73b},
archivePrefix = {arXiv},
       eprint = {2506.20033},
 primaryClass = {astro-ph.GA},
       adsurl = {https://ui.adsabs.harvard.edu/abs/2025ApJ...987...25Z}
}

@ARTICLE{Mishra2017,
       author = {{Mishra}, Ajay and {Li}, Aigen},
        title = "{Interstellar Silicon Depletion and the Ultraviolet Extinction}",
      journal = {\apj},
         year = 2017,
        month = dec,
       volume = {850},
       number = {2},
          eid = {138},
        pages = {138},
          doi = {10.3847/1538-4357/aa937a},
archivePrefix = {arXiv},
       eprint = {1710.04905},
 primaryClass = {astro-ph.GA},
       adsurl = {https://ui.adsabs.harvard.edu/abs/2017ApJ...850..138M}
}

@ARTICLE{Campbell2015,
       author = {{Campbell}, E.~K. and {Holz}, M. and {Gerlich}, D. and {Maier}, J.~P.},
        title = "{Laboratory confirmation of C$_{60}$$^{+}$ as the carrier of two diffuse interstellar bands}",
      journal = {Natur},
         year = 2015,
        month = jul,
       volume = {523},
       number = {7560},
        pages = {322-323},
          doi = {10.1038/nature14566},
       adsurl = {https://ui.adsabs.harvard.edu/abs/2015Natur.523..322C}
}

@ARTICLE{Foing1994,
       author = {{Foing}, B.~H. and {Ehrenfreund}, P.},
        title = "{Detection of two interstellar absorption bands coincident with spectral features of C$_{60}$$^{+}$}",
      journal = {\nat},
         year = 1994,
        month = may,
       volume = {369},
       number = {6478},
        pages = {296-298},
          doi = {10.1038/369296a0},
       adsurl = {https://ui.adsabs.harvard.edu/abs/1994Natur.369..296F}
}

@ARTICLE{Destree2009,
       author = {{Destree}, Joshua D. and {Snow}, Theodore P.},
        title = "{Unidentified Features in the Ultraviolet Spectrum of X Per}",
      journal = {\apj},
         year = 2009,
        month = may,
       volume = {697},
       number = {1},
        pages = {684-692},
          doi = {10.1088/0004-637X/697/1/684},
archivePrefix = {arXiv},
       eprint = {0903.1593},
 primaryClass = {astro-ph.GA},
       adsurl = {https://ui.adsabs.harvard.edu/abs/2009ApJ...697..684D}
}

@ARTICLE{Hobbs2009,
       author = {{Hobbs}, L.~M. and {York}, D.~G. and {Thorburn}, J.~A. and {Snow}, T.~P. and {Bishof}, M. and {Friedman}, S.~D. and {McCall}, B.~J. and {Oka}, T. and {Rachford}, B. and {Sonnentrucker}, P. and {Welty}, D.~E.},
        title = "{Studies of the Diffuse Interstellar Bands. III. HD 183143}",
      journal = {ApJ},
         year = 2009,
        month = nov,
       volume = {705},
       number = {1},
        pages = {32-45},
          doi = {10.1088/0004-637X/705/1/32},
archivePrefix = {arXiv},
       eprint = {0910.2983},
 primaryClass = {astro-ph.GA},
       adsurl = {https://ui.adsabs.harvard.edu/abs/2009ApJ...705...32H}
}

@INPROCEEDINGS{Snow2011,
       author = {{Snow}, T.~P. and {Destree}, J.~D.},
        title = "{The Diffuse Interstellar Bands in History and in the UV}",
    booktitle = {EAS Publications Series},
         year = 2011,
       editor = {{Joblin}, C. and {Tielens}, A.~G.~G.~M.},
       series = {EAS Publications Series},
       volume = {46},
        month = mar,
    publisher = {EDP},
        pages = {341-347},
          doi = {10.1051/eas/1146035},
       adsurl = {https://ui.adsabs.harvard.edu/abs/2011EAS....46..341S}
}

@ARTICLE{Fronig2011,
       author = {{Froning}, Cynthia S.},
        title = "{Early science results from the Cosmic Origins Spectrograph}",
      journal = {\apss},
         year = 2011,
        month = sep,
       volume = {335},
       number = {1},
        pages = {267-272},
          doi = {10.1007/s10509-011-0763-1},
       adsurl = {https://ui.adsabs.harvard.edu/abs/2011Ap&SS.335..267F}
}

@ARTICLE{Welty2020,
       author = {{Welty}, Daniel E. and {Sonnentrucker}, Paule and {Snow}, Theodore P. and {York}, Donald G.},
        title = "{HD 62542: Probing the Bare, Dense Core of a Translucent Interstellar Cloud}",
      journal = {\apj},
         year = 2020,
        month = jul,
       volume = {897},
       number = {1},
          eid = {36},
        pages = {36},
          doi = {10.3847/1538-4357/ab8f8e},
archivePrefix = {arXiv},
       eprint = {2005.10846},
 primaryClass = {astro-ph.GA},
       adsurl = {https://ui.adsabs.harvard.edu/abs/2020ApJ...897...36W}
}

@ARTICLE{Soler+2019,
       author = {{Soler}, J.~D. and {Beuther}, H. and {Rugel}, M. and {Wang}, Y. and {Clark}, P.~C. and {Glover}, S.~C.~O. and {Goldsmith}, P.~F. and {Heyer}, M. and {Anderson}, L.~D. and {Goodman}, A. and {Henning}, Th. and {Kainulainen}, J. and {Klessen}, R.~S. and {Longmore}, S.~N. and {McClure-Griffiths}, N.~M. and {Menten}, K.~M. and {Mottram}, J.~C. and {Ott}, J. and {Ragan}, S.~E. and {Smith}, R.~J. and {Urquhart}, J.~S. and {Bigiel}, F. and {Hennebelle}, P. and {Roy}, N. and {Schilke}, P.},
        title = "{Histogram of oriented gradients: a technique for the study of molecular cloud formation}",
      journal = {\aap},
         year = 2019,
        month = feb,
       volume = {622},
          eid = {A166},
        pages = {A166},
          doi = {10.1051/0004-6361/201834300},
archivePrefix = {arXiv},
       eprint = {1809.08338},
 primaryClass = {astro-ph.GA},
       adsurl = {https://ui.adsabs.harvard.edu/abs/2019A&A...622A.166S}
}

@ARTICLE{Soler+2025,
       author = {{Soler}, J.~D. and {Molinari}, S. and {Glover}, S.~C.~O. and {Smith}, R.~J. and {Klessen}, R.~S. and {Benjamin}, R.~A. and {Hennebelle}, P. and {Peek}, J.~E.~G. and {Beuther}, H. and {Edenhofer}, G. and {Zari}, E. and {Swiggum}, C. and {Zucker}, C.},
        title = "{Kinetic tomography of the Galactic plane within 1.25 kiloparsecs from the Sun: The interstellar flows revealed by H I and CO line emission and 3D dust}",
      journal = {\aap},
         year = 2025,
        month = mar,
       volume = {695},
          eid = {A222},
        pages = {A222},
          doi = {10.1051/0004-6361/202453022},
archivePrefix = {arXiv},
       eprint = {2411.12257},
 primaryClass = {astro-ph.GA},
       adsurl = {https://ui.adsabs.harvard.edu/abs/2025A&A...695A.222S}
}

@ARTICLE{Branch+1992,
       author = {{Branch}, David and {Tammann}, G.~A.},
        title = "{Type IA supernovae as standard candles.}",
      journal = {\href{https://ui.adsabs.harvard.edu/abs/1992ARA&A..30..359B}{\textcolor{blue}{ARA\&A}}},
         year = 1992,
        month = jan,
       volume = {30},
        pages = {359-389},
          doi = {10.1146/annurev.aa.30.090192.002043},
       adsurl = {https://ui.adsabs.harvard.edu/abs/1992ARA&A..30..359B}
}

@ARTICLE{Riess+1998,
       author = {{Riess}, Adam G. and {Filippenko}, Alexei V. and {Challis}, Peter and
         {Clocchiatti}, Alejandro and {Diercks}, Alan and {Garnavich}, Peter M. and
         {Gilliland}, Ron L. and {Hogan}, Craig J. and {Jha}, Saurabh and
         {Kirshner}, Robert P. and {Leibundgut}, B. and {Phillips}, M.~M. and
         {Reiss}, David and {Schmidt}, Brian P. and {Schommer}, Robert A. and
         {Smith}, R. Chris and {Spyromilio}, J. and {Stubbs}, Christopher and
         {Suntzeff}, Nicholas B. and {Tonry}, John},
        title = "{Observational Evidence from Supernovae for an Accelerating Universe and a Cosmological Constant}",
      journal = {\href{https://ui.adsabs.harvard.edu/abs/1998AJ....116.1009R}{\textcolor{blue}{\aj}}},
         year = 1998,
        month = sep,
       volume = {116},
       number = {3},
        pages = {1009-1038},
          doi = {10.1086/300499},
archivePrefix = {arXiv},
       eprint = {astro-ph/9805201},
 primaryClass = {astro-ph},
       adsurl = {https://ui.adsabs.harvard.edu/abs/1998AJ....116.1009R}
}

@ARTICLE{Perlmutter+1999,
       author = {{Perlmutter}, S. and {Aldering}, G. and {Goldhaber}, G. and
         {Knop}, R.~A. and {Nugent}, P. and {Castro}, P.~G. and {Deustua}, S. and
         {Fabbro}, S. and {Goobar}, A. and {Groom}, D.~E. and {Hook}, I.~M. and
         {Kim}, A.~G. and {Kim}, M.~Y. and {Lee}, J.~C. and {Nunes}, N.~J. and
         {Pain}, R. and {Pennypacker}, C.~R. and {Quimby}, R. and {Lidman}, C. and
         {Ellis}, R.~S. and {Irwin}, M. and {McMahon}, R.~G. and
         {Ruiz-Lapuente}, P. and {Walton}, N. and {Schaefer}, B. and
         {Boyle}, B.~J. and {Filippenko}, A.~V. and {Matheson}, T. and
         {Fruchter}, A.~S. and {Panagia}, N. and {Newberg}, H.~J.~M. and
         {Couch}, W.~J. and {Project}, The Supernova Cosmology},
        title = "{Measurements of {\ensuremath{\Omega}} and {\ensuremath{\Lambda}} from 42 High-Redshift Supernovae}",
      journal = {\href{https://ui.adsabs.harvard.edu/abs/1999ApJ...517..565P}{\textcolor{blue}{\apj}}},
         year = 1999,
        month = jun,
       volume = {517},
       number = {2},
        pages = {565-586},
          doi = {10.1086/307221},
archivePrefix = {arXiv},
       eprint = {astro-ph/9812133},
 primaryClass = {astro-ph},
       adsurl = {https://ui.adsabs.harvard.edu/abs/1999ApJ...517..565P}
}

@ARTICLE{Abbott+2016a,
       author = {{Abbott}, B.~P. and {Abbott}, R. and {Abbott}, T.~D. and
         {Abernathy}, M.~R. and {Acernese}, F. and {Ackley}, K. and {Adams}, C. and
         {Adams}, T. and {Addesso}, P. and {Adhikari}, R.~X. and {Adya}, V.~B. and
         {Affeldt}, C. and {Agathos}, M. and {Agatsuma}, K. and {Aggarwal}, N. and
         {Aguiar}, O.~D. and {Aiello}, L. and {Ain}, A. and {Ajith}, P. and
         {Allen}, B. and {Allocca}, A. and {Altin}, P.~A. and {Anderson}, S.~B. and
         {Anderson}, W.~G. and {Arai}, K. and {Araya}, M.~C. and
         {Arceneaux}, C.~C. and {Areeda}, J.~S. and {Arnaud}, N. and
         {Arun}, K.~G. and {Ascenzi}, S. and {Ashton}, G. and {Ast}, M. and
         {Aston}, S.~M. and {Astone}, P. and {Aufmuth}, P. and {Aulbert}, C. and
         {Babak}, S. and {Bacon}, P. and {Bader}, M.~K.~M. and {Baker}, P.~T. and
         {Baldaccini}, F. and {Ballardin}, G. and {Ballmer}, S.~W. and
         {Barayoga}, J.~C. and {Barclay}, S.~E. and {Barish}, B.~C. and
         {Barker}, D. and {Barone}, F. and {Barr}, B. and {Barsotti}, L. and
         {Barsuglia}, M. and {Barta}, D. and {Bartlett}, J. and {Bartos}, I. and
         {Bassiri}, R. and {Basti}, A. and {Batch}, J.~C. and {Baune}, C. and
         {Bavigadda}, V. and {Bazzan}, M. and {Behnke}, B. and {Bejger}, M. and
         {Belczynski}, C. and {Bell}, A.~S. and {Bell}, C.~J. and
         {Berger}, B.~K. and {Bergman}, J. and {Bergmann}, G. and
         {Berry}, C.~P.~L. and {Bersanetti}, D. and {Bertolini}, A. and
         {Betzwieser}, J. and {Bhagwat}, S. and {Bhandare}, R. and
         {Bilenko}, I.~A. and {Billingsley}, G. and {Birch}, J. and
         {Birney}, R. and {Biscans}, S. and {Bisht}, A. and {Bitossi}, M. and
         {Biwer}, C. and {Bizouard}, M.~A. and {Blackburn}, J.~K. and
         {Blair}, C.~D. and {Blair}, D.~G. and {Blair}, R.~M. and {Bloemen}, S. and
         {Bock}, O. and {Bodiya}, T.~P. and {Boer}, M. and {Bogaert}, G. and
         {Bogan}, C. and {Bohe}, A. and {Bojtos}, P. and {Bond}, C. and
         {Bondu}, F. and {Bonnand}, R. and {Boom}, B.~A. and {Bork}, R. and
         {Boschi}, V. and {Bose}, S. and {Bouffanais}, Y. and {Bozzi}, A. and
         {Bradaschia}, C. and {Brady}, P.~R. and {Braginsky}, V.~B. and
         {Branchesi}, M. and {Brau}, J.~E. and {Briant}, T. and {Brillet}, A. and
         {Brinkmann}, M. and {Brisson}, V. and {Brockill}, P. and
         {Brooks}, A.~F. and {Brown}, D.~A. and {Brown}, D.~D. and
         {Brown}, N.~M. and {Buchanan}, C.~C. and {Buikema}, A. and {Bulik}, T. and
         {Bulten}, H.~J. and {Buonanno}, A. and {Buskulic}, D. and {Buy}, C. and
         {Byer}, R.~L. and {Cadonati}, L. and {Cagnoli}, G. and {Cahillane}, C. and
         {Calder{\'o}n Bustillo}, J. and {Callister}, T. and {Calloni}, E. and
         {Camp}, J.~B. and {Cannon}, K.~C. and {Cao}, J. and {Capano}, C.~D. and
         {Capocasa}, E. and {Carbognani}, F. and {Caride}, S. and
         {Casanueva Diaz}, J. and {Casentini}, C. and {Caudill}, S. and
         {Cavagli{\`a}}, M. and {Cavalier}, F. and {Cavalieri}, R. and
         {Cella}, G. and {Cepeda}, C. and {Cerboni Baiardi}, L. and
         {Cerretani}, G. and {Cesarini}, E. and {Chakraborty}, R. and
         {Chalermsongsak}, T. and {Chamberlin}, S.~J. and {Chan}, M. and
         {Chao}, S. and {Charlton}, P. and {Chassande-Mottin}, E. and
         {Chen}, H.~Y. and {Chen}, Y. and {Cheng}, C. and {Chincarini}, A. and
         {Chiummo}, A. and {Cho}, H.~S. and {Cho}, M. and {Chow}, J.~H. and
         {Christensen}, N. and {Chu}, Q. and {Chua}, S. and {Chung}, S. and
         {Ciani}, G. and {Clara}, F. and {Clark}, J.~A. and {Cleva}, F. and
         {Coccia}, E. and {Cohadon}, P. -F. and {Colla}, A. and
         {Collette}, C.~G. and {Cominsky}, L. and {Constancio}, M., Jr. and
         {Conte}, A. and {Conti}, L. and {Cook}, D. and {Corbitt}, T.~R. and
         {Cornish}, N. and {Corsi}, A. and {Cortese}, S. and {Costa}, C.~A. and
         {Coughlin}, M.~W. and {Coughlin}, S.~B. and {Coulon}, J. -P. and
         {Countryman}, S.~T. and {Couvares}, P. and {Cowan}, E.~E. and
         {Coward}, D.~M. and {Cowart}, M.~J. and {Coyne}, D.~C. and {Coyne}, R. and
         {Craig}, K. and {Creighton}, J.~D.~E. and {Cripe}, J. and
         {Crowder}, S.~G. and {Cumming}, A. and {Cunningham}, L. and
         {Cuoco}, E. and {Dal Canton}, T. and {Danilishin}, S.~L. and
         {D'Antonio}, S. and {Danzmann}, K. and {Darman}, N.~S. and
         {Dattilo}, V. and {Dave}, I. and {Daveloza}, H.~P. and {Davier}, M. and
         {Davies}, G.~S. and {Daw}, E.~J. and {Day}, R. and {DeBra}, D. and
         {Debreczeni}, G. and {Degallaix}, J. and {De Laurentis}, M. and
         {Del{\'e}glise}, S. and {Del Pozzo}, W. and {Denker}, T. and
         {Dent}, T. and {Dereli}, H. and {Dergachev}, V. and {DeRosa}, R. and
         {DeRosa}, R.~T. and {DeSalvo}, R. and {Dhurandhar}, S. and
         {D{\'\i}az}, M.~C. and {Di Fiore}, L. and {Di Giovanni}, M. and
         {Di Lieto}, A. and {Di Pace}, S. and {Di Palma}, I. and
         {Di Virgilio}, A. and {Dojcinoski}, G. and {Dolique}, V. and
         {Donovan}, F. and {Dooley}, K.~L. and {Doravari}, S. and {Douglas}, R. and
         {Downes}, T.~P. and {Drago}, M. and {Drever}, R.~W.~P. and
         {Driggers}, J.~C. and {Du}, Z. and {Ducrot}, M. and {Dwyer}, S.~E. and
         {Edo}, T.~B. and {Edwards}, M.~C. and {Effler}, A. and
         {Eggenstein}, H. -B. and {Ehrens}, P. and {Eichholz}, J. and
         {Eikenberry}, S.~S. and {Engels}, W. and {Essick}, R.~C. and
         {Etzel}, T. and {Evans}, M. and {Evans}, T.~M. and {Everett}, R. and
         {Factourovich}, M. and {Fafone}, V. and {Fair}, H. and {Fairhurst}, S. and
         {Fan}, X. and {Fang}, Q. and {Farinon}, S. and {Farr}, B. and
         {Farr}, W.~M. and {Favata}, M. and {Fays}, M. and {Fehrmann}, H. and
         {Fejer}, M.~M. and {Ferrante}, I. and {Ferreira}, E.~C. and
         {Ferrini}, F. and {Fidecaro}, F. and {Fiori}, I. and {Fiorucci}, D. and
         {Fisher}, R.~P. and {Flaminio}, R. and {Fletcher}, M. and
         {Fournier}, J. -D. and {Franco}, S. and {Frasca}, S. and
         {Frasconi}, F. and {Frei}, Z. and {Freise}, A. and {Frey}, R. and
         {Frey}, V. and {Fricke}, T.~T. and {Fritschel}, P. and {Frolov}, V.~V. and
         {Fulda}, P. and {Fyffe}, M. and {Gabbard}, H.~A.~G. and {Gair}, J.~R. and
         {Gammaitoni}, L. and {Gaonkar}, S.~G. and {Garufi}, F. and {Gatto}, A. and
         {Gaur}, G. and {Gehrels}, N. and {Gemme}, G. and {Gendre}, B. and
         {Genin}, E. and {Gennai}, A. and {George}, J. and {Gergely}, L. and
         {Germain}, V. and {Ghosh}, Archisman and {Ghosh}, S. and
         {Giaime}, J.~A. and {Giardina}, K.~D. and {Giazotto}, A. and
         {Gill}, K. and {Glaefke}, A. and {Goetz}, E. and {Goetz}, R. and
         {Gondan}, L. and {Gonz{\'a}lez}, G. and {Gonzalez Castro}, J.~M. and
         {Gopakumar}, A. and {Gordon}, N.~A. and {Gorodetsky}, M.~L. and
         {Gossan}, S.~E. and {Gosselin}, M. and {Gouaty}, R. and {Graef}, C. and
         {Graff}, P.~B. and {Granata}, M. and {Grant}, A. and {Gras}, S. and
         {Gray}, C. and {Greco}, G. and {Green}, A.~C. and {Groot}, P. and
         {Grote}, H. and {Grunewald}, S. and {Guidi}, G.~M. and {Guo}, X. and
         {Gupta}, A. and {Gupta}, M.~K. and {Gushwa}, K.~E. and
         {Gustafson}, E.~K. and {Gustafson}, R. and {Hacker}, J.~J. and
         {Hall}, B.~R. and {Hall}, E.~D. and {Hammond}, G. and {Haney}, M. and
         {Hanke}, M.~M. and {Hanks}, J. and {Hanna}, C. and {Hannam}, M.~D. and
         {Hanson}, J. and {Hardwick}, T. and {Harms}, J. and {Harry}, G.~M. and
         {Harry}, I.~W. and {Hart}, M.~J. and {Hartman}, M.~T. and
         {Haster}, C. -J. and {Haughian}, K. and {Heidmann}, A. and
         {Heintze}, M.~C. and {Heitmann}, H. and {Hello}, P. and {Hemming}, G. and
         {Hendry}, M. and {Heng}, I.~S. and {Hennig}, J. and
         {Heptonstall}, A.~W. and {Heurs}, M. and {Hild}, S. and {Hoak}, D. and
         {Hodge}, K.~A. and {Hofman}, D. and {Hollitt}, S.~E. and {Holt}, K. and
         {Holz}, D.~E. and {Hopkins}, P. and {Hosken}, D.~J. and {Hough}, J. and
         {Houston}, E.~A. and {Howell}, E.~J. and {Hu}, Y.~M. and {Huang}, S. and
         {Huerta}, E.~A. and {Huet}, D. and {Hughey}, B. and {Husa}, S. and
         {Huttner}, S.~H. and {Huynh-Dinh}, T. and {Idrisy}, A. and {Indik}, N. and
         {Ingram}, D.~R. and {Inta}, R. and {Isa}, H.~N. and {Isac}, J. -M. and
         {Isi}, M. and {Islas}, G. and {Isogai}, T. and {Iyer}, B.~R. and
         {Izumi}, K. and {Jacqmin}, T. and {Jang}, H. and {Jani}, K. and
         {Jaranowski}, P. and {Jawahar}, S. and {Jim{\'e}nez-Forteza}, F. and
         {Johnson}, W.~W. and {Jones}, D.~I. and {Jones}, R. and
         {Jonker}, R.~J.~G. and {Ju}, L. and {K}, Haris and {Kalaghatgi}, C.~V. and
         {Kalogera}, V. and {Kandhasamy}, S. and {Kang}, G. and {Kanner}, J.~B. and
         {Karki}, S. and {Kasprzack}, M. and {Katsavounidis}, E. and
         {Katzman}, W. and {Kaufer}, S. and {Kaur}, T. and {Kawabe}, K. and
         {Kawazoe}, F. and {K{\'e}f{\'e}lian}, F. and {Kehl}, M.~S. and
         {Keitel}, D. and {Kelley}, D.~B. and {Kells}, W. and {Kennedy}, R. and
         {Key}, J.~S. and {Khalaidovski}, A. and {Khalili}, F.~Y. and
         {Khan}, I. and {Khan}, S. and {Khan}, Z. and {Khazanov}, E.~A. and
         {Kijbunchoo}, N. and {Kim}, C. and {Kim}, J. and {Kim}, K. and
         {Kim}, Nam-Gyu and {Kim}, Namjun and {Kim}, Y. -M. and {King}, E.~J. and
         {King}, P.~J. and {Kinzel}, D.~L. and {Kissel}, J.~S. and
         {Kleybolte}, L. and {Klimenko}, S. and {Koehlenbeck}, S.~M. and
         {Kokeyama}, K. and {Koley}, S. and {Kondrashov}, V. and {Kontos}, A. and
         {Korobko}, M. and {Korth}, W.~Z. and {Kowalska}, I. and {Kozak}, D.~B. and
         {Kringel}, V. and {Krishnan}, B. and {Kr{\'o}lak}, A. and
         {Krueger}, C. and {Kuehn}, G. and {Kumar}, P. and {Kuo}, L. and
         {Kutynia}, A. and {Lackey}, B.~D. and {Landry}, M. and {Lange}, J. and
         {Lantz}, B. and {Lasky}, P.~D. and {Lazzarini}, A. and {Lazzaro}, C. and
         {Leaci}, P. and {Leavey}, S. and {Lebigot}, E.~O. and {Lee}, C.~H. and
         {Lee}, H.~K. and {Lee}, H.~M. and {Lee}, K. and {Lenon}, A. and
         {Leonardi}, M. and {Leong}, J.~R. and {Leroy}, N. and {Letendre}, N. and
         {Levin}, Y. and {Levine}, B.~M. and {Li}, T.~G.~F. and {Libson}, A. and
         {Littenberg}, T.~B. and {Lockerbie}, N.~A. and {Logue}, J. and
         {Lombardi}, A.~L. and {Lord}, J.~E. and {Lorenzini}, M. and
         {Loriette}, V. and {Lormand}, M. and {Losurdo}, G. and {Lough}, J.~D. and
         {L{\"u}ck}, H. and {Lundgren}, A.~P. and {Luo}, J. and {Lynch}, R. and
         {Ma}, Y. and {MacDonald}, T. and {Machenschalk}, B. and {MacInnis}, M. and
         {Macleod}, D.~M. and {Maga{\~n}a-Sandoval}, F. and {Magee}, R.~M. and
         {Mageswaran}, M. and {Majorana}, E. and {Maksimovic}, I. and
         {Malvezzi}, V. and {Man}, N. and {Mandel}, I. and {Mandic}, V. and
         {Mangano}, V. and {Mansell}, G.~L. and {Manske}, M. and
         {Mantovani}, M. and {Marchesoni}, F. and {Marion}, F. and
         {M{\'a}rka}, S. and {M{\'a}rka}, Z. and {Markosyan}, A.~S. and
         {Maros}, E. and {Martelli}, F. and {Martellini}, L. and
         {Martin}, I.~W. and {Martin}, R.~M. and {Martynov}, D.~V. and
         {Marx}, J.~N. and {Mason}, K. and {Masserot}, A. and
         {Massinger}, T.~J. and {Masso-Reid}, M. and {Matichard}, F. and
         {Matone}, L. and {Mavalvala}, N. and {Mazumder}, N. and {Mazzolo}, G. and
         {McCarthy}, R. and {McClelland}, D.~E. and {McCormick}, S. and
         {McGuire}, S.~C. and {McIntyre}, G. and {McIver}, J. and
         {McManus}, D.~J. and {McWilliams}, S.~T. and {Meacher}, D. and
         {Meadors}, G.~D. and {Meidam}, J. and {Melatos}, A. and {Mendell}, G. and
         {Mendoza-Gandara}, D. and {Mercer}, R.~A. and {Merilh}, E. and
         {Merzougui}, M. and {Meshkov}, S. and {Messenger}, C. and
         {Messick}, C. and {Meyers}, P.~M. and {Mezzani}, F. and {Miao}, H. and
         {Michel}, C. and {Middleton}, H. and {Mikhailov}, E.~E. and
         {Milano}, L. and {Miller}, J. and {Millhouse}, M. and {Minenkov}, Y. and
         {Ming}, J. and {Mirshekari}, S. and {Mishra}, C. and {Mitra}, S. and
         {Mitrofanov}, V.~P. and {Mitselmakher}, G. and {Mittleman}, R. and
         {Moggi}, A. and {Mohan}, M. and {Mohapatra}, S.~R.~P. and
         {Montani}, M. and {Moore}, B.~C. and {Moore}, C.~J. and {Moraru}, D. and
         {Moreno}, G. and {Morriss}, S.~R. and {Mossavi}, K. and {Mours}, B. and
         {Mow-Lowry}, C.~M. and {Mueller}, C.~L. and {Mueller}, G. and
         {Muir}, A.~W. and {Mukherjee}, Arunava and {Mukherjee}, D. and
         {Mukherjee}, S. and {Mukund}, N. and {Mullavey}, A. and {Munch}, J. and
         {Murphy}, D.~J. and {Murray}, P.~G. and {Mytidis}, A. and
         {Nardecchia}, I. and {Naticchioni}, L. and {Nayak}, R.~K. and
         {Necula}, V. and {Nedkova}, K. and {Nelemans}, G. and {Neri}, M. and
         {Neunzert}, A. and {Newton}, G. and {Nguyen}, T.~T. and
         {Nielsen}, A.~B. and {Nissanke}, S. and {Nitz}, A. and {Nocera}, F. and
         {Nolting}, D. and {Normandin}, M.~E.~N. and {Nuttall}, L.~K. and
         {Oberling}, J. and {Ochsner}, E. and {O'Dell}, J. and {Oelker}, E. and
         {Ogin}, G.~H. and {Oh}, J.~J. and {Oh}, S.~H. and {Ohme}, F. and
         {Oliver}, M. and {Oppermann}, P. and {Oram}, Richard J. and
         {O'Reilly}, B. and {O'Shaughnessy}, R. and {Ottaway}, D.~J. and
         {Ottens}, R.~S. and {Overmier}, H. and {Owen}, B.~J. and {Pai}, A. and
         {Pai}, S.~A. and {Palamos}, J.~R. and {Palashov}, O. and {Palomba}, C. and
         {Pal-Singh}, A. and {Pan}, H. and {Pankow}, C. and {Pannarale}, F. and
         {Pant}, B.~C. and {Paoletti}, F. and {Paoli}, A. and {Papa}, M.~A. and
         {Paris}, H.~R. and {Parker}, W. and {Pascucci}, D. and
         {Pasqualetti}, A. and {Passaquieti}, R. and {Passuello}, D. and
         {Patricelli}, B. and {Patrick}, Z. and {Pearlstone}, B.~L. and
         {Pedraza}, M. and {Pedurand}, R. and {Pekowsky}, L. and {Pele}, A. and
         {Penn}, S. and {Perreca}, A. and {Phelps}, M. and {Piccinni}, O. and
         {Pichot}, M. and {Piergiovanni}, F. and {Pierro}, V. and {Pillant}, G. and
         {Pinard}, L. and {Pinto}, I.~M. and {Pitkin}, M. and {Poggiani}, R. and
         {Popolizio}, P. and {Post}, A. and {Powell}, J. and {Prasad}, J. and
         {Predoi}, V. and {Premachandra}, S.~S. and {Prestegard}, T. and
         {Price}, L.~R. and {Prijatelj}, M. and {Principe}, M. and
         {Privitera}, S. and {Prix}, R. and {Prodi}, G.~A. and {Prokhorov}, L. and
         {Puncken}, O. and {Punturo}, M. and {Puppo}, P. and {P{\"u}rrer}, M. and
         {Qi}, H. and {Qin}, J. and {Quetschke}, V. and {Quintero}, E.~A. and
         {Quitzow-James}, R. and {Raab}, F.~J. and {Rabeling}, D.~S. and
         {Radkins}, H. and {Raffai}, P. and {Raja}, S. and {Rakhmanov}, M. and
         {Rapagnani}, P. and {Raymond}, V. and {Razzano}, M. and {Re}, V. and
         {Read}, J. and {Reed}, C.~M. and {Regimbau}, T. and {Rei}, L. and
         {Reid}, S. and {Reitze}, D.~H. and {Rew}, H. and {Reyes}, S.~D. and
         {Ricci}, F. and {Riles}, K. and {Robertson}, N.~A. and {Robie}, R. and
         {Robinet}, F. and {Rocchi}, A. and {Rolland}, L. and {Rollins}, J.~G. and
         {Roma}, V.~J. and {Romano}, J.~D. and {Romano}, R. and {Romanov}, G. and
         {Romie}, J.~H. and {Rosi{\'n}ska}, D. and {Rowan}, S. and
         {R{\"u}diger}, A. and {Ruggi}, P. and {Ryan}, K. and {Sachdev}, S. and
         {Sadecki}, T. and {Sadeghian}, L. and {Salconi}, L. and {Saleem}, M. and
         {Salemi}, F. and {Samajdar}, A. and {Sammut}, L. and {Sanchez}, E.~J. and
         {Sandberg}, V. and {Sandeen}, B. and {Sanders}, J.~R. and
         {Sassolas}, B. and {Sathyaprakash}, B.~S. and {Saulson}, P.~R. and
         {Sauter}, O. and {Savage}, R.~L. and {Sawadsky}, A. and {Schale}, P. and
         {Schilling}, R. and {Schmidt}, J. and {Schmidt}, P. and {Schnabel}, R. and
         {Schofield}, R.~M.~S. and {Sch{\"o}nbeck}, A. and {Schreiber}, E. and
         {Schuette}, D. and {Schutz}, B.~F. and {Scott}, J. and {Scott}, S.~M. and
         {Sellers}, D. and {Sentenac}, D. and {Sequino}, V. and {Sergeev}, A. and
         {Serna}, G. and {Setyawati}, Y. and {Sevigny}, A. and
         {Shaddock}, D.~A. and {Shah}, S. and {Shahriar}, M.~S. and
         {Shaltev}, M. and {Shao}, Z. and {Shapiro}, B. and {Shawhan}, P. and
         {Sheperd}, A. and {Shoemaker}, D.~H. and {Shoemaker}, D.~M. and
         {Siellez}, K. and {Siemens}, X. and {Sigg}, D. and {Silva}, A.~D. and
         {Simakov}, D. and {Singer}, A. and {Singer}, L.~P. and {Singh}, A. and
         {Singh}, R. and {Singhal}, A. and {Sintes}, A.~M. and
         {Slagmolen}, B.~J.~J. and {Smith}, J.~R. and {Smith}, N.~D. and
         {Smith}, R.~J.~E. and {Son}, E.~J. and {Sorazu}, B. and
         {Sorrentino}, F. and {Souradeep}, T. and {Srivastava}, A.~K. and
         {Staley}, A. and {Steinke}, M. and {Steinlechner}, J. and
         {Steinlechner}, S. and {Steinmeyer}, D. and {Stephens}, B.~C. and
         {Stevenson}, S.~P. and {Stone}, R. and {Strain}, K.~A. and
         {Straniero}, N. and {Stratta}, G. and {Strauss}, N.~A. and
         {Strigin}, S. and {Sturani}, R. and {Stuver}, A.~L. and
         {Summerscales}, T.~Z. and {Sun}, L. and {Sutton}, P.~J. and
         {Swinkels}, B.~L. and {Szczepa{\'n}czyk}, M.~J. and {Tacca}, M. and
         {Talukder}, D. and {Tanner}, D.~B. and {T{\'a}pai}, M. and
         {Tarabrin}, S.~P. and {Taracchini}, A. and {Taylor}, R. and
         {Theeg}, T. and {Thirugnanasambandam}, M.~P. and {Thomas}, E.~G. and
         {Thomas}, M. and {Thomas}, P. and {Thorne}, K.~A. and {Thorne}, K.~S. and
         {Thrane}, E. and {Tiwari}, S. and {Tiwari}, V. and {Tokmakov}, K.~V. and
         {Tomlinson}, C. and {Tonelli}, M. and {Torres}, C.~V. and
         {Torrie}, C.~I. and {T{\"o}yr{\"a}}, D. and {Travasso}, F. and
         {Traylor}, G. and {Trifir{\`o}}, D. and {Tringali}, M.~C. and
         {Trozzo}, L. and {Tse}, M. and {Turconi}, M. and {Tuyenbayev}, D. and
         {Ugolini}, D. and {Unnikrishnan}, C.~S. and {Urban}, A.~L. and
         {Usman}, S.~A. and {Vahlbruch}, H. and {Vajente}, G. and {Valdes}, G. and
         {van Bakel}, N. and {van Beuzekom}, M. and {van den Brand}, J.~F.~J. and
         {van den Broeck}, C. and {Vander-Hyde}, D.~C. and {van der Schaaf}, L. and
         {van Heijningen}, J.~V. and {van Veggel}, A.~A. and {Vardaro}, M. and
         {Vass}, S. and {Vas{\'u}th}, M. and {Vaulin}, R. and {Vecchio}, A. and
         {Vedovato}, G. and {Veitch}, J. and {Veitch}, P.~J. and
         {Venkateswara}, K. and {Verkindt}, D. and {Vetrano}, F. and
         {Vicer{\'e}}, A. and {Vinciguerra}, S. and {Vine}, D.~J. and
         {Vinet}, J. -Y. and {Vitale}, S. and {Vo}, T. and {Vocca}, H. and
         {Vorvick}, C. and {Voss}, D. and {Vousden}, W.~D. and
         {Vyatchanin}, S.~P. and {Wade}, A.~R. and {Wade}, L.~E. and {Wade}, M. and
         {Walker}, M. and {Wallace}, L. and {Walsh}, S. and {Wang}, G. and
         {Wang}, H. and {Wang}, M. and {Wang}, X. and {Wang}, Y. and
         {Ward}, R.~L. and {Warner}, J. and {Was}, M. and {Weaver}, B. and
         {Wei}, L. -W. and {Weinert}, M. and {Weinstein}, A.~J. and {Weiss}, R. and
         {Welborn}, T. and {Wen}, L. and {We{\ss}els}, P. and {Westphal}, T. and
         {Wette}, K. and {Whelan}, J.~T. and {White}, D.~J. and
         {Whiting}, B.~F. and {Williams}, R.~D. and {Williamson}, A.~R. and
         {Willis}, J.~L. and {Willke}, B. and {Wimmer}, M.~H. and {Winkler}, W. and
         {Wipf}, C.~C. and {Wittel}, H. and {Woan}, G. and {Worden}, J. and
         {Wright}, J.~L. and {Wu}, G. and {Yablon}, J. and {Yam}, W. and
         {Yamamoto}, H. and {Yancey}, C.~C. and {Yap}, M.~J. and {Yu}, H. and
         {Yvert}, M. and {Zadro{\.z}ny}, A. and {Zangrando}, L. and
         {Zanolin}, M. and {Zendri}, J. -P. and {Zevin}, M. and {Zhang}, F. and
         {Zhang}, L. and {Zhang}, M. and {Zhang}, Y. and {Zhao}, C. and
         {Zhou}, M. and {Zhou}, Z. and {Zhu}, X.~J. and {Zucker}, M.~E. and
         {Zuraw}, S.~E. and {and} and {Zweizig}, J. and
         {LIGO Scientific Collaboration} and {Virgo Collaboration}},
        title = "{Astrophysical Implications of the Binary Black-hole Merger GW150914}",
      journal = {\href{https://ui.adsabs.harvard.edu/abs/2016ApJ...818L..22A}{\textcolor{blue}{\apjl}}},
         year = 2016,
        month = feb,
       volume = {818},
       number = {2},
          eid = {L22},
        pages = {L22},
          doi = {10.3847/2041-8205/818/2/L22},
archivePrefix = {arXiv},
       eprint = {1602.03846},
 primaryClass = {astro-ph.HE},
       adsurl = {https://ui.adsabs.harvard.edu/abs/2016ApJ...818L..22A}
}

@ARTICLE{Abbott+2016b,
       author = {{Abbott}, B.~P. and {Abbott}, R. and {Abbott}, T.~D. and
         {Abernathy}, M.~R. and {Acernese}, F. and {Ackley}, K. and {Adams}, C. and
         {Adams}, T. and {Addesso}, P. and {Adhikari}, R.~X. and {Adya}, V.~B. and
         {Affeldt}, C. and {Agathos}, M. and {Agatsuma}, K. and {Aggarwal}, N. and
         {Aguiar}, O.~D. and {Aiello}, L. and {Ain}, A. and {Ajith}, P. and
         {Allen}, B. and {Allocca}, A. and {Altin}, P.~A. and {Anderson}, S.~B. and
         {Anderson}, W.~G. and {Arai}, K. and {Araya}, M.~C. and
         {Arceneaux}, C.~C. and {Areeda}, J.~S. and {Arnaud}, N. and
         {Arun}, K.~G. and {Ascenzi}, S. and {Ashton}, G. and {Ast}, M. and
         {Aston}, S.~M. and {Astone}, P. and {Aufmuth}, P. and {Aulbert}, C. and
         {Babak}, S. and {Bacon}, P. and {Bader}, M.~K.~M. and {Baker}, P.~T. and
         {Baldaccini}, F. and {Ballardin}, G. and {Ballmer}, S.~W. and
         {Barayoga}, J.~C. and {Barclay}, S.~E. and {Barish}, B.~C. and
         {Barker}, D. and {Barone}, F. and {Barr}, B. and {Barsotti}, L. and
         {Barsuglia}, M. and {Barta}, D. and {Bartlett}, J. and {Bartos}, I. and
         {Bassiri}, R. and {Basti}, A. and {Batch}, J.~C. and {Baune}, C. and
         {Bavigadda}, V. and {Bazzan}, M. and {Bejger}, M. and {Bell}, A.~S. and
         {Berger}, B.~K. and {Bergmann}, G. and {Berry}, C.~P.~L. and
         {Bersanetti}, D. and {Bertolini}, A. and {Betzwieser}, J. and
         {Bhagwat}, S. and {Bhandare}, R. and {Bilenko}, I.~A. and
         {Billingsley}, G. and {Birch}, J. and {Birney}, R. and {Birnholtz}, O. and
         {Biscans}, S. and {Bisht}, A. and {Bitossi}, M. and {Biwer}, C. and
         {Bizouard}, M.~A. and {Blackburn}, J.~K. and {Blair}, C.~D. and
         {Blair}, D.~G. and {Blair}, R.~M. and {Bloemen}, S. and {Bock}, O. and
         {Boer}, M. and {Bogaert}, G. and {Bogan}, C. and {Bohe}, A. and
         {Bond}, C. and {Bondu}, F. and {Bonnand}, R. and {Boom}, B.~A. and
         {Bork}, R. and {Boschi}, V. and {Bose}, S. and {Bouffanais}, Y. and
         {Bozzi}, A. and {Bradaschia}, C. and {Brady}, P.~R. and
         {Braginsky}, V.~B. and {Branchesi}, M. and {Brau}, J.~E. and
         {Briant}, T. and {Brillet}, A. and {Brinkmann}, M. and {Brisson}, V. and
         {Brockill}, P. and {Broida}, J.~E. and {Brooks}, A.~F. and
         {Brown}, D.~A. and {Brown}, D.~D. and {Brown}, N.~M. and {Brunett}, S. and
         {Buchanan}, C.~C. and {Buikema}, A. and {Bulik}, T. and
         {Bulten}, H.~J. and {Buonanno}, A. and {Buskulic}, D. and {Buy}, C. and
         {Byer}, R.~L. and {Cabero}, M. and {Cadonati}, L. and {Cagnoli}, G. and
         {Cahillane}, C. and {Calder{\'o}n Bustillo}, J. and {Callister}, T. and
         {Calloni}, E. and {Camp}, J.~B. and {Cannon}, K.~C. and {Cao}, J. and
         {Capano}, C.~D. and {Capocasa}, E. and {Carbognani}, F. and
         {Caride}, S. and {Casanueva Diaz}, J. and {Casentini}, C. and
         {Caudill}, S. and {Cavagli{\`a}}, M. and {Cavalier}, F. and
         {Cavalieri}, R. and {Cella}, G. and {Cepeda}, C.~B. and
         {Cerboni Baiardi}, L. and {Cerretani}, G. and {Cesarini}, E. and
         {Chamberlin}, S.~J. and {Chan}, M. and {Chao}, S. and {Charlton}, P. and
         {Chassande-Mottin}, E. and {Cheeseboro}, B.~D. and {Chen}, H.~Y. and
         {Chen}, Y. and {Cheng}, C. and {Chincarini}, A. and {Chiummo}, A. and
         {Cho}, H.~S. and {Cho}, M. and {Chow}, J.~H. and {Christensen}, N. and
         {Chu}, Q. and {Chua}, S. and {Chung}, S. and {Ciani}, G. and
         {Clara}, F. and {Clark}, J.~A. and {Cleva}, F. and {Coccia}, E. and
         {Cohadon}, P. -F. and {Colla}, A. and {Collette}, C.~G. and
         {Cominsky}, L. and {Constancio}, M. and {Conte}, A. and {Conti}, L. and
         {Cook}, D. and {Corbitt}, T.~R. and {Cornish}, N. and {Corsi}, A. and
         {Cortese}, S. and {Costa}, C.~A. and {Coughlin}, M.~W. and
         {Coughlin}, S.~B. and {Coulon}, J. -P. and {Countryman}, S.~T. and
         {Couvares}, P. and {Cowan}, E.~E. and {Coward}, D.~M. and
         {Cowart}, M.~J. and {Coyne}, D.~C. and {Coyne}, R. and {Craig}, K. and
         {Creighton}, J.~D.~E. and {Cripe}, J. and {Crowder}, S.~G. and
         {Cumming}, A. and {Cunningham}, L. and {Cuoco}, E. and
         {Dal Canton}, T. and {Danilishin}, S.~L. and {D'Antonio}, S. and
         {Danzmann}, K. and {Darman}, N.~S. and {Dasgupta}, A. and
         {Da Silva Costa}, C.~F. and {Dattilo}, V. and {Dave}, I. and
         {Davier}, M. and {Davies}, G.~S. and {Daw}, E.~J. and {Day}, R. and
         {De}, S. and {DeBra}, D. and {Debreczeni}, G. and {Degallaix}, J. and
         {De Laurentis}, M. and {Del{\'e}glise}, S. and {Del Pozzo}, W. and
         {Denker}, T. and {Dent}, T. and {Dergachev}, V. and {De Rosa}, R. and
         {DeRosa}, R.~T. and {DeSalvo}, R. and {Devine}, R.~C. and {Dhurand
        har}, S. and {D{\'\i}az}, M.~C. and {Di Fiore}, L. and
         {Di Giovanni}, M. and {Di Girolamo}, T. and {Di Lieto}, A. and
         {Di Pace}, S. and {Di Palma}, I. and {Di Virgilio}, A. and
         {Dolique}, V. and {Donovan}, F. and {Dooley}, K.~L. and {Doravari}, S. and
         {Douglas}, R. and {Downes}, T.~P. and {Drago}, M. and
         {Drever}, R.~W.~P. and {Driggers}, J.~C. and {Ducrot}, M. and
         {Dwyer}, S.~E. and {Edo}, T.~B. and {Edwards}, M.~C. and {Effler}, A. and
         {Eggenstein}, H. -B. and {Ehrens}, P. and {Eichholz}, J. and
         {Eikenberry}, S.~S. and {Engels}, W. and {Essick}, R.~C. and
         {Etzel}, T. and {Evans}, M. and {Evans}, T.~M. and {Everett}, R. and
         {Factourovich}, M. and {Fafone}, V. and {Fair}, H. and {Fairhurst}, S. and
         {Fan}, X. and {Fang}, Q. and {Farinon}, S. and {Farr}, B. and
         {Farr}, W.~M. and {Favata}, M. and {Fays}, M. and {Fehrmann}, H. and
         {Fejer}, M.~M. and {Fenyvesi}, E. and {Ferrante}, I. and
         {Ferreira}, E.~C. and {Ferrini}, F. and {Fidecaro}, F. and {Fiori}, I. and
         {Fiorucci}, D. and {Fisher}, R.~P. and {Flaminio}, R. and
         {Fletcher}, M. and {Fong}, H. and {Fournier}, J. -D. and {Frasca}, S. and
         {Frasconi}, F. and {Frei}, Z. and {Freise}, A. and {Frey}, R. and
         {Frey}, V. and {Fritschel}, P. and {Frolov}, V.~V. and {Fulda}, P. and
         {Fyffe}, M. and {Gabbard}, H.~A.~G. and {Gair}, J.~R. and
         {Gammaitoni}, L. and {Gaonkar}, S.~G. and {Garufi}, F. and {Gaur}, G. and
         {Gehrels}, N. and {Gemme}, G. and {Geng}, P. and {Genin}, E. and
         {Gennai}, A. and {George}, J. and {Gergely}, L. and {Germain}, V. and
         {Ghosh}, Abhirup and {Ghosh}, Archisman and {Ghosh}, S. and
         {Giaime}, J.~A. and {Giardina}, K.~D. and {Giazotto}, A. and
         {Gill}, K. and {Glaefke}, A. and {Goetz}, E. and {Goetz}, R. and
         {Gondan}, L. and {Gonz{\'a}lez}, G. and {Gonzalez Castro}, J.~M. and
         {Gopakumar}, A. and {Gordon}, N.~A. and {Gorodetsky}, M.~L. and
         {Gossan}, S.~E. and {Gosselin}, M. and {Gouaty}, R. and {Grado}, A. and
         {Graef}, C. and {Graff}, P.~B. and {Granata}, M. and {Grant}, A. and
         {Gras}, S. and {Gray}, C. and {Greco}, G. and {Green}, A.~C. and
         {Groot}, P. and {Grote}, H. and {Grunewald}, S. and {Guidi}, G.~M. and
         {Guo}, X. and {Gupta}, A. and {Gupta}, M.~K. and {Gushwa}, K.~E. and
         {Gustafson}, E.~K. and {Gustafson}, R. and {Hacker}, J.~J. and
         {Hall}, B.~R. and {Hall}, E.~D. and {Hamilton}, H. and {Hammond}, G. and
         {Haney}, M. and {Hanke}, M.~M. and {Hanks}, J. and {Hanna}, C. and
         {Hannam}, M.~D. and {Hanson}, J. and {Hardwick}, T. and {Harms}, J. and
         {Harry}, G.~M. and {Harry}, I.~W. and {Hart}, M.~J. and
         {Hartman}, M.~T. and {Haster}, C. -J. and {Haughian}, K. and
         {Healy}, J. and {Heidmann}, A. and {Heintze}, M.~C. and {Heitmann}, H. and
         {Hello}, P. and {Hemming}, G. and {Hendry}, M. and {Heng}, I.~S. and
         {Hennig}, J. and {Henry}, J. and {Heptonstall}, A.~W. and {Heurs}, M. and
         {Hild}, S. and {Hoak}, D. and {Hofman}, D. and {Holt}, K. and
         {Holz}, D.~E. and {Hopkins}, P. and {Hough}, J. and {Houston}, E.~A. and
         {Howell}, E.~J. and {Hu}, Y.~M. and {Huang}, S. and {Huerta}, E.~A. and
         {Huet}, D. and {Hughey}, B. and {Husa}, S. and {Huttner}, S.~H. and
         {Huynh-Dinh}, T. and {Indik}, N. and {Ingram}, D.~R. and {Inta}, R. and
         {Isa}, H.~N. and {Isac}, J. -M. and {Isi}, M. and {Isogai}, T. and
         {Iyer}, B.~R. and {Izumi}, K. and {Jacqmin}, T. and {Jang}, H. and
         {Jani}, K. and {Jaranowski}, P. and {Jawahar}, S. and {Jian}, L. and
         {Jim{\'e}nez-Forteza}, F. and {Johnson}, W.~W. and
         {Johnson-McDaniel}, N.~K. and {Jones}, D.~I. and {Jones}, R. and
         {Jonker}, R.~J.~G. and {Ju}, L. and {K}, Haris and {Kalaghatgi}, C.~V. and
         {Kalogera}, V. and {Kandhasamy}, S. and {Kang}, G. and {Kanner}, J.~B. and
         {Kapadia}, S.~J. and {Karki}, S. and {Karvinen}, K.~S. and
         {Kasprzack}, M. and {Katsavounidis}, E. and {Katzman}, W. and
         {Kaufer}, S. and {Kaur}, T. and {Kawabe}, K. and
         {K{\'e}f{\'e}lian}, F. and {Kehl}, M.~S. and {Keitel}, D. and
         {Kelley}, D.~B. and {Kells}, W. and {Kennedy}, R. and {Key}, J.~S. and
         {Khalili}, F.~Y. and {Khan}, I. and {Khan}, S. and {Khan}, Z. and
         {Khazanov}, E.~A. and {Kijbunchoo}, N. and {Kim}, Chi-Woong and
         {Kim}, Chunglee and {Kim}, J. and {Kim}, K. and {Kim}, N. and
         {Kim}, W. and {Kim}, Y. -M. and {Kimbrell}, S.~J. and {King}, E.~J. and
         {King}, P.~J. and {Kissel}, J.~S. and {Klein}, B. and {Kleybolte}, L. and
         {Klimenko}, S. and {Koehlenbeck}, S.~M. and {Koley}, S. and
         {Kondrashov}, V. and {Kontos}, A. and {Korobko}, M. and {Korth}, W.~Z. and
         {Kowalska}, I. and {Kozak}, D.~B. and {Kringel}, V. and {Krishnan}, B. and
         {Kr{\'o}lak}, A. and {Krueger}, C. and {Kuehn}, G. and {Kumar}, P. and
         {Kumar}, R. and {Kuo}, L. and {Kutynia}, A. and {Lackey}, B.~D. and {Land
        ry}, M. and {Lange}, J. and {Lantz}, B. and {Lasky}, P.~D. and
         {Laxen}, M. and {Lazzarini}, A. and {Lazzaro}, C. and {Leaci}, P. and
         {Leavey}, S. and {Lebigot}, E.~O. and {Lee}, C.~H. and {Lee}, H.~K. and
         {Lee}, H.~M. and {Lee}, K. and {Lenon}, A. and {Leonardi}, M. and
         {Leong}, J.~R. and {Leroy}, N. and {Letendre}, N. and {Levin}, Y. and
         {Lewis}, J.~B. and {Li}, T.~G.~F. and {Libson}, A. and
         {Littenberg}, T.~B. and {Lockerbie}, N.~A. and {Lombardi}, A.~L. and
         {London}, L.~T. and {Lord}, J.~E. and {Lorenzini}, M. and
         {Loriette}, V. and {Lormand}, M. and {Losurdo}, G. and {Lough}, J.~D. and
         {Lousto}, C.~O. and {L{\"u}ck}, H. and {Lundgren}, A.~P. and
         {Lynch}, R. and {Ma}, Y. and {Machenschalk}, B. and {MacInnis}, M. and
         {Macleod}, D.~M. and {Maga{\~n}a-Sandoval}, F. and
         {Maga{\~n}a Zertuche}, L. and {Magee}, R.~M. and {Majorana}, E. and
         {Maksimovic}, I. and {Malvezzi}, V. and {Man}, N. and {Mandel}, I. and
         {Mandic}, V. and {Mangano}, V. and {Mansell}, G.~L. and {Manske}, M. and
         {Mantovani}, M. and {Marchesoni}, F. and {Marion}, F. and
         {M{\'a}rka}, S. and {M{\'a}rka}, Z. and {Markosyan}, A.~S. and
         {Maros}, E. and {Martelli}, F. and {Martellini}, L. and
         {Martin}, I.~W. and {Martynov}, D.~V. and {Marx}, J.~N. and
         {Mason}, K. and {Masserot}, A. and {Massinger}, T.~J. and
         {Masso-Reid}, M. and {Mastrogiovanni}, S. and {Matichard}, F. and
         {Matone}, L. and {Mavalvala}, N. and {Mazumder}, N. and {McCarthy}, R. and
         {McClelland}, D.~E. and {McCormick}, S. and {McGuire}, S.~C. and
         {McIntyre}, G. and {McIver}, J. and {McManus}, D.~J. and {McRae}, T. and
         {McWilliams}, S.~T. and {Meacher}, D. and {Meadors}, G.~D. and
         {Meidam}, J. and {Melatos}, A. and {Mendell}, G. and {Mercer}, R.~A. and
         {Merilh}, E.~L. and {Merzougui}, M. and {Meshkov}, S. and
         {Messenger}, C. and {Messick}, C. and {Metzdorff}, R. and
         {Meyers}, P.~M. and {Mezzani}, F. and {Miao}, H. and {Michel}, C. and
         {Middleton}, H. and {Mikhailov}, E.~E. and {Milano}, L. and
         {Miller}, A.~L. and {Miller}, A. and {Miller}, B.~B. and {Miller}, J. and
         {Millhouse}, M. and {Minenkov}, Y. and {Ming}, J. and {Mirshekari}, S. and
         {Mishra}, C. and {Mitra}, S. and {Mitrofanov}, V.~P. and
         {Mitselmakher}, G. and {Mittleman}, R. and {Moggi}, A. and {Mohan}, M. and
         {Mohapatra}, S.~R.~P. and {Montani}, M. and {Moore}, B.~C. and
         {Moore}, C.~J. and {Moraru}, D. and {Moreno}, G. and {Morriss}, S.~R. and
         {Mossavi}, K. and {Mours}, B. and {Mow-Lowry}, C.~M. and {Mueller}, G. and
         {Muir}, A.~W. and {Mukherjee}, Arunava and {Mukherjee}, D. and
         {Mukherjee}, S. and {Mukund}, N. and {Mullavey}, A. and {Munch}, J. and
         {Murphy}, D.~J. and {Murray}, P.~G. and {Mytidis}, A. and
         {Nardecchia}, I. and {Naticchioni}, L. and {Nayak}, R.~K. and
         {Nedkova}, K. and {Nelemans}, G. and {Nelson}, T.~J.~N. and {Neri}, M. and
         {Neunzert}, A. and {Newton}, G. and {Nguyen}, T.~T. and
         {Nielsen}, A.~B. and {Nissanke}, S. and {Nitz}, A. and {Nocera}, F. and
         {Nolting}, D. and {Normandin}, M.~E.~N. and {Nuttall}, L.~K. and
         {Oberling}, J. and {Ochsner}, E. and {O'Dell}, J. and {Oelker}, E. and
         {Ogin}, G.~H. and {Oh}, J.~J. and {Oh}, S.~H. and {Ohme}, F. and
         {Oliver}, M. and {Oppermann}, P. and {Oram}, Richard J. and
         {O'Reilly}, B. and {O'Shaughnessy}, R. and {Ottaway}, D.~J. and
         {Overmier}, H. and {Owen}, B.~J. and {Pai}, A. and {Pai}, S.~A. and
         {Palamos}, J.~R. and {Palashov}, O. and {Palomba}, C. and
         {Pal-Singh}, A. and {Pan}, H. and {Pankow}, C. and {Pannarale}, F. and
         {Pant}, B.~C. and {Paoletti}, F. and {Paoli}, A. and {Papa}, M.~A. and
         {Paris}, H.~R. and {Parker}, W. and {Pascucci}, D. and
         {Pasqualetti}, A. and {Passaquieti}, R. and {Passuello}, D. and
         {Patricelli}, B. and {Patrick}, Z. and {Pearlstone}, B.~L. and
         {Pedraza}, M. and {Pedurand}, R. and {Pekowsky}, L. and {Pele}, A. and
         {Penn}, S. and {Perreca}, A. and {Perri}, L.~M. and {Pfeiffer}, H.~P. and
         {Phelps}, M. and {Piccinni}, O.~J. and {Pichot}, M. and
         {Piergiovanni}, F. and {Pierro}, V. and {Pillant}, G. and {Pinard}, L. and
         {Pinto}, I.~M. and {Pitkin}, M. and {Poe}, M. and {Poggiani}, R. and
         {Popolizio}, P. and {Post}, A. and {Powell}, J. and {Prasad}, J. and
         {Predoi}, V. and {Prestegard}, T. and {Price}, L.~R. and
         {Prijatelj}, M. and {Principe}, M. and {Privitera}, S. and {Prix}, R. and
         {Prodi}, G.~A. and {Prokhorov}, L. and {Puncken}, O. and {Punturo}, M. and
         {Puppo}, P. and {P{\"u}rrer}, M. and {Qi}, H. and {Qin}, J. and
         {Qiu}, S. and {Quetschke}, V. and {Quintero}, E.~A. and
         {Quitzow-James}, R. and {Raab}, F.~J. and {Rabeling}, D.~S. and
         {Radkins}, H. and {Raffai}, P. and {Raja}, S. and {Rajan}, C. and
         {Rakhmanov}, M. and {Rapagnani}, P. and {Raymond}, V. and
         {Razzano}, M. and {Re}, V. and {Read}, J. and {Reed}, C.~M. and
         {Regimbau}, T. and {Rei}, L. and {Reid}, S. and {Reitze}, D.~H. and
         {Rew}, H. and {Reyes}, S.~D. and {Ricci}, F. and {Riles}, K. and
         {Rizzo}, M. and {Robertson}, N.~A. and {Robie}, R. and {Robinet}, F. and
         {Rocchi}, A. and {Rolland}, L. and {Rollins}, J.~G. and {Roma}, V.~J. and
         {Romano}, J.~D. and {Romano}, R. and {Romanov}, G. and {Romie}, J.~H. and
         {Rosi{\'n}ska}, D. and {Rowan}, S. and {R{\"u}diger}, A. and
         {Ruggi}, P. and {Ryan}, K. and {Sachdev}, S. and {Sadecki}, T. and
         {Sadeghian}, L. and {Sakellariadou}, M. and {Salconi}, L. and
         {Saleem}, M. and {Salemi}, F. and {Samajdar}, A. and {Sammut}, L. and
         {Sanchez}, E.~J. and {Sandberg}, V. and {Sandeen}, B. and {Sand
        ers}, J.~R. and {Sassolas}, B. and {Sathyaprakash}, B.~S. and
         {Saulson}, P.~R. and {Sauter}, O.~E.~S. and {Savage}, R.~L. and
         {Sawadsky}, A. and {Schale}, P. and {Schilling}, R. and {Schmidt}, J. and
         {Schmidt}, P. and {Schnabel}, R. and {Schofield}, R.~M.~S. and
         {Sch{\"o}nbeck}, A. and {Schreiber}, E. and {Schuette}, D. and
         {Schutz}, B.~F. and {Scott}, J. and {Scott}, S.~M. and {Sellers}, D. and
         {Sengupta}, A.~S. and {Sentenac}, D. and {Sequino}, V. and
         {Sergeev}, A. and {Setyawati}, Y. and {Shaddock}, D.~A. and
         {Shaffer}, T. and {Shahriar}, M.~S. and {Shaltev}, M. and
         {Shapiro}, B. and {Shawhan}, P. and {Sheperd}, A. and
         {Shoemaker}, D.~H. and {Shoemaker}, D.~M. and {Siellez}, K. and
         {Siemens}, X. and {Sieniawska}, M. and {Sigg}, D. and {Silva}, A.~D. and
         {Singer}, A. and {Singer}, L.~P. and {Singh}, A. and {Singh}, R. and
         {Singhal}, A. and {Sintes}, A.~M. and {Slagmolen}, B.~J.~J. and
         {Smith}, J.~R. and {Smith}, N.~D. and {Smith}, R.~J.~E. and
         {Son}, E.~J. and {Sorazu}, B. and {Sorrentino}, F. and {Souradeep}, T. and
         {Srivastava}, A.~K. and {Staley}, A. and {Steinke}, M. and
         {Steinlechner}, J. and {Steinlechner}, S. and {Steinmeyer}, D. and
         {Stephens}, B.~C. and {Stevenson}, S.~P. and {Stone}, R. and
         {Strain}, K.~A. and {Straniero}, N. and {Stratta}, G. and
         {Strauss}, N.~A. and {Strigin}, S. and {Sturani}, R. and
         {Stuver}, A.~L. and {Summerscales}, T.~Z. and {Sun}, L. and
         {Sunil}, S. and {Sutton}, P.~J. and {Swinkels}, B.~L. and
         {Szczepa{\'n}czyk}, M.~J. and {Tacca}, M. and {Talukder}, D. and
         {Tanner}, D.~B. and {T{\'a}pai}, M. and {Tarabrin}, S.~P. and
         {Taracchini}, A. and {Taylor}, R. and {Theeg}, T. and {Thirugnanasamband
        am}, M.~P. and {Thomas}, E.~G. and {Thomas}, M. and {Thomas}, P. and
         {Thorne}, K.~A. and {Thrane}, E. and {Tiwari}, S. and {Tiwari}, V. and
         {Tokmakov}, K.~V. and {Toland}, K. and {Tomlinson}, C. and
         {Tonelli}, M. and {Tornasi}, Z. and {Torres}, C.~V. and
         {Torrie}, C.~I. and {T{\"o}yr{\"a}}, D. and {Travasso}, F. and
         {Traylor}, G. and {Trifir{\`o}}, D. and {Tringali}, M.~C. and
         {Trozzo}, L. and {Tse}, M. and {Turconi}, M. and {Tuyenbayev}, D. and
         {Ugolini}, D. and {Unnikrishnan}, C.~S. and {Urban}, A.~L. and
         {Usman}, S.~A. and {Vahlbruch}, H. and {Vajente}, G. and {Valdes}, G. and
         {Vallisneri}, M. and {van Bakel}, N. and {van Beuzekom}, M. and
         {van den Brand}, J.~F.~J. and {Van Den Broeck}, C. and {Vand
        er-Hyde}, D.~C. and {van der Schaaf}, L. and {van Heijningen}, J.~V. and
         {van Veggel}, A.~A. and {Vardaro}, M. and {Vass}, S. and
         {Vas{\'u}th}, M. and {Vaulin}, R. and {Vecchio}, A. and {Vedovato}, G. and
         {Veitch}, J. and {Veitch}, P.~J. and {Venkateswara}, K. and
         {Verkindt}, D. and {Vetrano}, F. and {Vicer{\'e}}, A. and
         {Vinciguerra}, S. and {Vine}, D.~J. and {Vinet}, J. -Y. and
         {Vitale}, S. and {Vo}, T. and {Vocca}, H. and {Vorvick}, C. and
         {Voss}, D.~V. and {Vousden}, W.~D. and {Vyatchanin}, S.~P. and
         {Wade}, A.~R. and {Wade}, L.~E. and {Wade}, M. and {Walker}, M. and
         {Wallace}, L. and {Walsh}, S. and {Wang}, G. and {Wang}, H. and
         {Wang}, M. and {Wang}, X. and {Wang}, Y. and {Ward}, R.~L. and
         {Warner}, J. and {Was}, M. and {Weaver}, B. and {Wei}, L. -W. and
         {Weinert}, M. and {Weinstein}, A.~J. and {Weiss}, R. and {Wen}, L. and
         {We{\ss}els}, P. and {Westphal}, T. and {Wette}, K. and
         {Whelan}, J.~T. and {Whiting}, B.~F. and {Williams}, R.~D. and
         {Williamson}, A.~R. and {Willis}, J.~L. and {Willke}, B. and
         {Wimmer}, M.~H. and {Winkler}, W. and {Wipf}, C.~C. and {Wittel}, H. and
         {Woan}, G. and {Woehler}, J. and {Worden}, J. and {Wright}, J.~L. and
         {Wu}, D.~S. and {Wu}, G. and {Yablon}, J. and {Yam}, W. and
         {Yamamoto}, H. and {Yancey}, C.~C. and {Yu}, H. and {Yvert}, M. and
         {Zadro{\.z}ny}, A. and {Zangrando}, L. and {Zanolin}, M. and
         {Zendri}, J. -P. and {Zevin}, M. and {Zhang}, L. and {Zhang}, M. and
         {Zhang}, Y. and {Zhao}, C. and {Zhou}, M. and {Zhou}, Z. and
         {Zhu}, X.~J. and {Zucker}, M.~E. and {Zuraw}, S.~E. and {Zweizig}, J. and
         {Boyle}, M. and {Hemberger}, D. and {Kidder}, L.~E. and {Lovelace}, G. and
         {Ossokine}, S. and {Scheel}, M. and {Szilagyi}, B. and {Teukolsky}, S. and
         {LIGO Scientific Collaboration} and {VIRGO Collaboration}},
        title = "{GW151226: Observation of Gravitational Waves from a 22-Solar-Mass Binary Black Hole Coalescence}",
      journal = {\href{https://ui.adsabs.harvard.edu/abs/2016PhRvL.116x1103A}{\textcolor{blue}{\prl}}},
         year = 2016,
        month = jun,
       volume = {116},
       number = {24},
          eid = {241103},
        pages = {241103},
          doi = {10.1103/PhysRevLett.116.241103},
archivePrefix = {arXiv},
       eprint = {1606.04855},
 primaryClass = {gr-qc},
       adsurl = {https://ui.adsabs.harvard.edu/abs/2016PhRvL.116x1103A}
}

@ARTICLE{Yu+2026,
       author = {{Yu}, W. and {Pala}, A.~F. and {Kupfer}, T. and {G{\"a}nsicke}, B.~T. and {Koester}, D. and {Belloni}, D. and {Wong}, T.~L.~S. and {Schreiber}, M.~R. and {van Roestel}, J. and {Brown}, A.~J. and {Waagen}, E.~O. and {Gonz{\'a}lez-Carballo}, J.-L. and {Bednarz}, S. and {Bernacki}, K. and {De Martino}, D. and {Fern{\'a}ndez Ma{\~n}anes}, E. and {Gonz{\'a}lez Farf{\'a}n}, R. and {Green}, M.~J. and {Groot}, P.~J. and {Hambsch}, F.-J. and {Knigge}, C. and {Martin-Velasco}, J.-L. and {Morales-Aimar}, M. and {Myers}, G. and {Naves Nogues}, R. and {Poggiani}, R. and {Popowicz}, A. and {Ramsay}, G. and {Reina-Lorenz}, E. and {Rodr{\'\i}guez-Gil}, P. and {Salto-Gonz{\'a}lez}, J.-L. and {Sion}, E.~M. and {Steeghs}, D. and {Szkody}, P. and {Toloza}, O. and {Tovmassian}, G.},
        title = "{The evolutionary history of ultra-compact accreting binaries: I. Chemical abundances and the formation channel of the eclipsing AM CVn system ZTF J225237.05{\ensuremath{-}}051917.4 from HST spectroscopy}",
      journal = {\aap},
         year = 2026,
        month = jan,
       volume = {706},
          eid = {A14},
        pages = {A14},
          doi = {10.1051/0004-6361/202557568},
archivePrefix = {arXiv},
       eprint = {2512.04147},
 primaryClass = {astro-ph.SR},
       adsurl = {https://ui.adsabs.harvard.edu/abs/2026A&A...706A..14Y}
}

@ARTICLE{Sahu+2025,
       author = {{Sahu}, Snehalata and {B{\'e}dard}, Antoine and {G{\"a}nsicke}, Boris T. and {Tremblay}, Pier-Emmanuel and {Koester}, Detlev and {Farihi}, Jay and {Hermes}, J.~J. and {Hollands}, Mark A. and {Cunningham}, Tim and {Redfield}, Seth},
        title = "{A hot white dwarf merger remnant revealed by an ultraviolet detection of carbon}",
      journal = {Nature Astronomy},
         year = 2025,
        month = sep,
       volume = {9},
        pages = {1347-1355},
          doi = {10.1038/s41550-025-02590-y},
archivePrefix = {arXiv},
       eprint = {2508.03811},
 primaryClass = {astro-ph.SR},
       adsurl = {https://ui.adsabs.harvard.edu/abs/2025NatAs...9.1347S}
}

@ARTICLE{Rebassa-Mansergas+2021,
       author = {{Rebassa-Mansergas}, A. and {Maldonado}, J. and {Raddi}, R. and {Knowles}, A.~T. and {Torres}, S. and {Hoskin}, M. and {Cunningham}, T. and {Hollands}, M. and {Ren}, J. and {G{\"a}nsicke}, B.~T. and {Tremblay}, P.-E. and {Castro-Rodr{\'\i}guez}, N. and {Camisassa}, M. and {Koester}, D.},
        title = "{Constraining the solar neighbourhood age-metallicity relation from white dwarf-main sequence binaries}",
      journal = {\mnras},
         year = 2021,
        month = aug,
       volume = {505},
       number = {3},
        pages = {3165-3176},
          doi = {10.1093/mnras/stab1559},
archivePrefix = {arXiv},
       eprint = {2105.13379},
 primaryClass = {astro-ph.SR},
       adsurl = {https://ui.adsabs.harvard.edu/abs/2021MNRAS.505.3165R}
}

@ARTICLE{Shen+2018,
       author = {{Shen}, Ken J. and {Boubert}, Douglas and {G{\"a}nsicke}, Boris T. and {Jha}, Saurabh W. and {Andrews}, Jennifer E. and {Chomiuk}, Laura and {Foley}, Ryan J. and {Fraser}, Morgan and {Gromadzki}, Mariusz and {Guillochon}, James and {Kotze}, Marissa M. and {Maguire}, Kate and {Siebert}, Matthew R. and {Smith}, Nathan and {Strader}, Jay and {Badenes}, Carles and {Kerzendorf}, Wolfgang E. and {Koester}, Detlev and {Kromer}, Markus and {Miles}, Broxton and {Pakmor}, R{\"u}diger and {Schwab}, Josiah and {Toloza}, Odette and {Toonen}, Silvia and {Townsley}, Dean M. and {Williams}, Brian J.},
        title = "{Three Hypervelocity White Dwarfs in Gaia DR2: Evidence for Dynamically Driven Double-degenerate Double-detonation Type Ia Supernovae}",
      journal = {\apj},
         year = 2018,
        month = sep,
       volume = {865},
       number = {1},
          eid = {15},
        pages = {15},
          doi = {10.3847/1538-4357/aad55b},
archivePrefix = {arXiv},
       eprint = {1804.11163},
 primaryClass = {astro-ph.SR},
       adsurl = {https://ui.adsabs.harvard.edu/abs/2018ApJ...865...15S}
}

@ARTICLE{Gentile_Fusillo+2018,
       author = {{Gentile Fusillo}, N.~P. and {Tremblay}, P.-E. and {Jordan}, S. and {G{\"a}nsicke}, B.~T. and {Kalirai}, J.~S. and {Cummings}, J.},
        title = "{Can magnetic fields suppress convection in the atmosphere of cool white dwarfs? A case study on WD2105-820}",
      journal = {\mnras},
         year = 2018,
        month = jan,
       volume = {473},
       number = {3},
        pages = {3693-3699},
          doi = {10.1093/mnras/stx2584},
archivePrefix = {arXiv},
       eprint = {1710.02151},
 primaryClass = {astro-ph.SR},
       adsurl = {https://ui.adsabs.harvard.edu/abs/2018MNRAS.473.3693G}
}

@ARTICLE{Marsh2011,
       author = {{Marsh}, T.~R.},
        title = "{Double white dwarfs and LISA}",
      journal = {CQGra},
         year = 2011,
        month = may,
       volume = {28},
       number = {9},
          eid = {094019},
        pages = {094019},
          doi = {10.1088/0264-9381/28/9/094019},
archivePrefix = {arXiv},
       eprint = {1101.4970},
 primaryClass = {astro-ph.SR},
       adsurl = {https://ui.adsabs.harvard.edu/abs/2011CQGra..28i4019M}
}

@ARTICLE{Napiwotzki+2020,
       author = {{Napiwotzki}, R. and {Karl}, C.~A. and {Lisker}, T. and {Catal{\'a}n}, S. and {Drechsel}, H. and {Heber}, U. and {Homeier}, D. and {Koester}, D. and {Leibundgut}, B. and {Marsh}, T.~R. and {Moehler}, S. and {Nelemans}, G. and {Reimers}, D. and {Renzini}, A. and {Str{\"o}er}, A. and {Yungelson}, L.},
        title = "{The ESO supernovae type Ia progenitor survey (SPY). The radial velocities of 643 DA white dwarfs}",
      journal = {\aap},
         year = 2020,
        month = jun,
       volume = {638},
          eid = {A131},
        pages = {A131},
          doi = {10.1051/0004-6361/201629648},
archivePrefix = {arXiv},
       eprint = {1906.10977},
 primaryClass = {astro-ph.SR},
       adsurl = {https://ui.adsabs.harvard.edu/abs/2020A&A...638A.131N}
}

@ARTICLE{Munday+2025,
       author = {{Munday}, James and {Pelisoli}, Ingrid and {Tremblay}, Pier-Emmanuel and {Jones}, David and {Nelemans}, Gijs and {Kilic}, Mukremin and {Cunningham}, Tim and {Toonen}, Silvia and {Santos-Garc{\'\i}a}, Alejandro and {Dawson}, Harry and {Pinter}, Viktoria and {Godson}, Benjamin and {Martinez}, Llanos and {Chand}, Jaya and {Dobson}, Ross and {Jhass}, Kiran and {Shenoy}, Shravya},
        title = "{The DBL Survey II: towards a mass-period distribution of double white dwarf binaries}",
      journal = {\mnras},
         year = 2025,
        month = aug,
       volume = {541},
       number = {4},
        pages = {3494-3512},
          doi = {10.1093/mnras/staf1198},
archivePrefix = {arXiv},
       eprint = {2507.14123},
 primaryClass = {astro-ph.SR},
       adsurl = {https://ui.adsabs.harvard.edu/abs/2025MNRAS.541.3494M}
}

@ARTICLE{Bours+2015,
       author = {{Bours}, M.~C.~P. and {Marsh}, T.~R. and {G{\"a}nsicke}, B.~T. and {Tauris}, T.~M. and {Istrate}, A.~G. and {Badenes}, C. and {Dhillon}, V.~S. and {Gal-Yam}, A. and {Hermes}, J.~J. and {Kengkriangkrai}, S. and {Kilic}, M. and {Koester}, D. and {Mullally}, F. and {Prasert}, N. and {Steeghs}, D. and {Thompson}, S.~E. and {Thorstensen}, J.~R.},
        title = "{A double white dwarf with a paradoxical origin?}",
      journal = {\mnras},
         year = 2015,
        month = jul,
       volume = {450},
       number = {4},
        pages = {3966-3974},
          doi = {10.1093/mnras/stv889},
archivePrefix = {arXiv},
       eprint = {1505.05144},
 primaryClass = {astro-ph.SR},
       adsurl = {https://ui.adsabs.harvard.edu/abs/2015MNRAS.450.3966B}
}

@ARTICLE{Marsh+1995,
       author = {{Marsh}, T.~R.},
        title = "{The discovery of a short-period double-degenerate binary star}",
      journal = {\mnras},
         year = 1995,
        month = jul,
       volume = {275},
       number = {1},
        pages = {L1-L5},
          doi = {10.1093/mnras/275.1.L1},
archivePrefix = {arXiv},
       eprint = {astro-ph/9504088},
 primaryClass = {astro-ph},
       adsurl = {https://ui.adsabs.harvard.edu/abs/1995MNRAS.275L...1M}
}

@ARTICLE{Parsons+2016,
       author = {{Parsons}, S.~G. and {Rebassa-Mansergas}, A. and {Schreiber}, M.~R. and {G{\"a}nsicke}, B.~T. and {Zorotovic}, M. and {Ren}, J.~J.},
        title = "{The white dwarf binary pathways survey - I. A sample of FGK stars with white dwarf companions}",
      journal = {\mnras},
         year = 2016,
        month = dec,
       volume = {463},
       number = {2},
        pages = {2125-2136},
          doi = {10.1093/mnras/stw2143},
archivePrefix = {arXiv},
       eprint = {1604.01613},
 primaryClass = {astro-ph.SR},
       adsurl = {https://ui.adsabs.harvard.edu/abs/2016MNRAS.463.2125P}
}

@ARTICLE{Angel+1981,
       author = {{Angel}, J.~R.~P. and {Borra}, E.~F. and {Landstreet}, J.~D.},
        title = "{The magnetic fields of white dwarfs.}",
      journal = {\apjs},
         year = 1981,
        month = mar,
       volume = {45},
        pages = {457-474},
          doi = {10.1086/190720},
       adsurl = {https://ui.adsabs.harvard.edu/abs/1981ApJS...45..457A}
}

@ARTICLE{Tout+2008,
       author = {{Tout}, C.~A. and {Wickramasinghe}, D.~T. and {Liebert}, J. and {Ferrario}, L. and {Pringle}, J.~E.},
        title = "{Binary star origin of high field magnetic white dwarfs}",
      journal = {\mnras},
         year = 2008,
        month = jun,
       volume = {387},
       number = {2},
        pages = {897-901},
          doi = {10.1111/j.1365-2966.2008.13291.x},
archivePrefix = {arXiv},
       eprint = {0805.0115},
 primaryClass = {astro-ph},
       adsurl = {https://ui.adsabs.harvard.edu/abs/2008MNRAS.387..897T}
}

@ARTICLE{Schreiber+2021,
       author = {{Schreiber}, Matthias R. and {Belloni}, Diogo and {G{\"a}nsicke}, Boris T. and {Parsons}, Steven G. and {Zorotovic}, Monica},
        title = "{The origin and evolution of magnetic white dwarfs in close binary stars}",
      journal = {Nature Astronomy},
         year = 2021,
        month = apr,
       volume = {5},
        pages = {648-654},
          doi = {10.1038/s41550-021-01346-8},
archivePrefix = {arXiv},
       eprint = {2104.14607},
 primaryClass = {astro-ph.SR},
       adsurl = {https://ui.adsabs.harvard.edu/abs/2021NatAs...5..648S}
}

@ARTICLE{Camisassa+2024,
       author = {{Camisassa}, M. and {Fuentes}, J.~R. and {Schreiber}, M.~R. and {Rebassa-Mansergas}, A. and {Torres}, S. and {Raddi}, R. and {Dominguez}, I.},
        title = "{Main sequence dynamo magnetic fields emerging in the white dwarf phase}",
      journal = {\aap},
         year = 2024,
        month = nov,
       volume = {691},
          eid = {L21},
        pages = {L21},
          doi = {10.1051/0004-6361/202452539},
archivePrefix = {arXiv},
       eprint = {2411.02296},
 primaryClass = {astro-ph.SR},
       adsurl = {https://ui.adsabs.harvard.edu/abs/2024A&A...691L..21C}
}

@ARTICLE{Burleigh+1999,
       author = {{Burleigh}, M.~R. and {Jordan}, S. and {Schweizer}, W.},
        title = "{Phase-resolved Far-Ultraviolet Hubble Space Telescope Spectroscopy of the Peculiar Magnetic White Dwarf RE J0317-853}",
      journal = {\apjl},
         year = 1999,
        month = jan,
       volume = {510},
       number = {1},
        pages = {L37-L40},
          doi = {10.1086/311794},
archivePrefix = {arXiv},
       eprint = {astro-ph/9810109},
 primaryClass = {astro-ph},
       adsurl = {https://ui.adsabs.harvard.edu/abs/1999ApJ...510L..37B}
}

@ARTICLE{Hollands+2020,
       author = {{Hollands}, M.~A. and {Tremblay}, P.-E. and {G{\"a}nsicke}, B.~T. and {Camisassa}, M.~E. and {Koester}, D. and {Aungwerojwit}, A. and {Chote}, P. and {C{\'o}rsico}, A.~H. and {Dhillon}, V.~S. and {Gentile-Fusillo}, N.~P. and {Hoskin}, M.~J. and {Izquierdo}, P. and {Marsh}, T.~R. and {Steeghs}, D.},
        title = "{An ultra-massive white dwarf with a mixed hydrogen-carbon atmosphere as a likely merger remnant}",
      journal = {Nature Astronomy},
         year = 2020,
        month = mar,
       volume = {4},
        pages = {663-669},
          doi = {10.1038/s41550-020-1028-0},
archivePrefix = {arXiv},
       eprint = {2003.00028},
 primaryClass = {astro-ph.SR},
       adsurl = {https://ui.adsabs.harvard.edu/abs/2020NatAs...4..663H}
}

@ARTICLE{Gaensicke+2001,
       author = {{G{\"a}nsicke}, Boris T. and {Schmidt}, Gary D. and {Jordan}, Stefan and {Szkody}, Paula},
        title = "{Phase-resolved Hubble Space Telescope/STIS Spectroscopy of the Exposed White Dwarf in the High-Field Polar AR Ursae Majoris}",
      journal = {\apj},
         year = 2001,
        month = jul,
       volume = {555},
       number = {1},
        pages = {380-392},
          doi = {10.1086/321464},
archivePrefix = {arXiv},
       eprint = {astro-ph/0103131},
 primaryClass = {astro-ph},
       adsurl = {https://ui.adsabs.harvard.edu/abs/2001ApJ...555..380G}
}

@ARTICLE{Wilson+2021,
       author = {{Wilson}, David J. and {Toloza}, Odette and {Landstreet}, John D. and {G{\"a}nsicke}, Boris T. and {Drake}, Jeremy J. and {Hermes}, J.~J. and {Koester}, Detlev},
        title = "{Discovery of a young pre-intermediate polar}",
      journal = {\mnras},
         year = 2021,
        month = nov,
       volume = {508},
       number = {1},
        pages = {561-574},
          doi = {10.1093/mnras/stab2458},
archivePrefix = {arXiv},
       eprint = {2108.11414},
 primaryClass = {astro-ph.SR},
       adsurl = {https://ui.adsabs.harvard.edu/abs/2021MNRAS.508..561W}
}

@ARTICLE{Caiazzo+2021,
       author = {{Caiazzo}, Ilaria and {Burdge}, Kevin B. and {Fuller}, James and {Heyl}, Jeremy and {Kulkarni}, S.~R. and {Prince}, Thomas A. and {Richer}, Harvey B. and {Schwab}, Josiah and {Andreoni}, Igor and {Bellm}, Eric C. and {Drake}, Andrew and {Duev}, Dmitry A. and {Graham}, Matthew J. and {Helou}, George and {Mahabal}, Ashish A. and {Masci}, Frank J. and {Smith}, Roger and {Soumagnac}, Maayane T.},
        title = "{A highly magnetized and rapidly rotating white dwarf as small as the Moon}",
      journal = {\nat},
         year = 2021,
        month = jun,
       volume = {595},
       number = {7865},
        pages = {39-42},
          doi = {10.1038/s41586-021-03615-y},
archivePrefix = {arXiv},
       eprint = {2107.08458},
 primaryClass = {astro-ph.SR},
       adsurl = {https://ui.adsabs.harvard.edu/abs/2021Natur.595...39C}
}

@ARTICLE{DiStefano+2023,
       author = {{Di Stefano}, Rosanne and {Kruckow}, Matthias U. and {Gao}, Yan and {Neunteufel}, Patrick G. and {Kobayashi}, Chiaki},
        title = "{SCATTER: A New Common Envelope Formalism}",
      journal = {\apj},
         year = 2023,
        month = feb,
       volume = {944},
       number = {1},
          eid = {87},
        pages = {87},
          doi = {10.3847/1538-4357/acae9b},
archivePrefix = {arXiv},
       eprint = {2212.06770},
 primaryClass = {astro-ph.HE},
       adsurl = {https://ui.adsabs.harvard.edu/abs/2023ApJ...944...87D}
}

@ARTICLE{Barraza-Jorquera+2025,
       author = {{Barraza-Jorquera}, Joaqu{\'\i}n A. and {Schreiber}, Matthias R. and {Belloni}, Diogo},
        title = "{Further evidence of saturated, boosted, and disrupted magnetic braking from evolutionary tracks of cataclysmic variables}",
      journal = {\aap},
         year = 2025,
        month = apr,
       volume = {696},
          eid = {A92},
        pages = {A92},
          doi = {10.1051/0004-6361/202553757},
archivePrefix = {arXiv},
       eprint = {2503.18884},
 primaryClass = {astro-ph.SR},
       adsurl = {https://ui.adsabs.harvard.edu/abs/2025A&A...696A..92B}
}

@ARTICLE{Werner+2024,
       author = {{Werner}, Klaus and {El-Badry}, Kareem and {G{\"a}nsicke}, Boris T. and {Shen}, Ken J.},
        title = "{Ultraviolet spectroscopy of the supernova Ia hypervelocity runaway white dwarf J0927{\ensuremath{-}}6335}",
      journal = {\aap},
         year = 2024,
        month = sep,
       volume = {689},
          eid = {L6},
        pages = {L6},
          doi = {10.1051/0004-6361/202451635},
archivePrefix = {arXiv},
       eprint = {2408.08397},
 primaryClass = {astro-ph.SR},
       adsurl = {https://ui.adsabs.harvard.edu/abs/2024A&A...689L...6W}
}

@ARTICLE{Rebassa-Mansergas+2023,
       author = {{Rebassa-Mansergas}, A. and {Maldonado}, J. and {Raddi}, R. and {Torres}, S. and {Hoskin}, M. and {Cunningham}, T. and {Hollands}, M.~A. and {Ren}, J. and {G{\"a}nsicke}, B.~T. and {Tremblay}, P.-E. and {Camisassa}, M.},
        title = "{Main-sequence companions to white dwarfs - II. The age-activity-rotation relation from a sample of Gaia common proper motion pairs}",
      journal = {\mnras},
         year = 2023,
        month = dec,
       volume = {526},
       number = {3},
        pages = {4787-4800},
          doi = {10.1093/mnras/stad3050},
archivePrefix = {arXiv},
       eprint = {2310.02125},
 primaryClass = {astro-ph.SR},
       adsurl = {https://ui.adsabs.harvard.edu/abs/2023MNRAS.526.4787R}
}

@ARTICLE{Raddi+2022,
       author = {{Raddi}, Roberto and {Torres}, Santiago and {Rebassa-Mansergas}, Alberto and {Maldonado}, Jes{\'u}s and {Camisassa}, Mar{\'\i}a E. and {Koester}, Detlev and {Gentile Fusillo}, Nicola Pietro and {Tremblay}, Pier-Emmanuel and {Dimpel}, Markus and {Heber}, Ulrich and {Cunningham}, Tim and {Ren}, Juan-Juan},
        title = "{Kinematic properties of white dwarfs. Galactic orbital parameters and age-velocity dispersion relation}",
      journal = {\aap},
         year = 2022,
        month = feb,
       volume = {658},
          eid = {A22},
        pages = {A22},
          doi = {10.1051/0004-6361/202141837},
archivePrefix = {arXiv},
       eprint = {2111.01145},
 primaryClass = {astro-ph.SR},
       adsurl = {https://ui.adsabs.harvard.edu/abs/2022A&A...658A..22R}
}

@ARTICLE{Chiti+2024,
       author = {{Chiti}, Federica and {van Saders}, Jennifer L. and {Heintz}, Tyler M. and {Hermes}, J.~J. and {Ong}, J.~M. Joel and {Hey}, Daniel R. and {Ramirez-Weinhouse}, Michele M. and {Dugas}, Alison},
        title = "{Rotation at the Fully Convective Boundary: Insights from Wide WD + MS Binary Systems}",
      journal = {\apj},
         year = 2024,
        month = dec,
       volume = {977},
       number = {1},
          eid = {15},
        pages = {15},
          doi = {10.3847/1538-4357/ad856c},
archivePrefix = {arXiv},
       eprint = {2403.12129},
 primaryClass = {astro-ph.SR},
       adsurl = {https://ui.adsabs.harvard.edu/abs/2024ApJ...977...15C}
}

@ARTICLE{Amaro-Seoane+2023,
       author = {{Amaro-Seoane}, Pau and {Andrews}, Jeff and {Arca Sedda}, Manuel and {Askar}, Abbas and {Baghi}, Quentin and {Balasov}, Razvan and {Bartos}, Imre and {Bavera}, Simone S. and {Bellovary}, Jillian and {Berry}, Christopher P.~L. and {Berti}, Emanuele and {Bianchi}, Stefano and {Blecha}, Laura and {Blondin}, St{\'e}phane and {Bogdanovi{\'c}}, Tamara and {Boissier}, Samuel and {Bonetti}, Matteo and {Bonoli}, Silvia and {Bortolas}, Elisa and {Breivik}, Katelyn and {Capelo}, Pedro R. and {Caramete}, Laurentiu and {Cattorini}, Federico and {Charisi}, Maria and {Chaty}, Sylvain and {Chen}, Xian and {Chru{\'s}li{\'n}ska}, Martyna and {Chua}, Alvin J.~K. and {Church}, Ross and {Colpi}, Monica and {D'Orazio}, Daniel and {Danielski}, Camilla and {Davies}, Melvyn B. and {Dayal}, Pratika and {De Rosa}, Alessandra and {Derdzinski}, Andrea and {Destounis}, Kyriakos and {Dotti}, Massimo and {Du{\c{t}}an}, Ioana and {Dvorkin}, Irina and {Fabj}, Gaia and {Foglizzo}, Thierry and {Ford}, Saavik and {Fouvry}, Jean-Baptiste and {Franchini}, Alessia and {Fragos}, Tassos and {Fryer}, Chris and {Gaspari}, Massimo and {Gerosa}, Davide and {Graziani}, Luca and {Groot}, Paul and {Habouzit}, Melanie and {Haggard}, Daryl and {Haiman}, Zoltan and {Han}, Wen-Biao and {Istrate}, Alina and {Johansson}, Peter H. and {Khan}, Fazeel Mahmood and {Kimpson}, Tomas and {Kokkotas}, Kostas and {Kong}, Albert and {Korol}, Valeriya and {Kremer}, Kyle and {Kupfer}, Thomas and {Lamberts}, Astrid and {Larson}, Shane and {Lau}, Mike and {Liu}, Dongliang and {Lloyd-Ronning}, Nicole and {Lodato}, Giuseppe and {Lupi}, Alessandro and {Ma}, Chung-Pei and {Maccarone}, Tomas and {Mandel}, Ilya and {Mangiagli}, Alberto and {Mapelli}, Michela and {Mathis}, St{\'e}phane and {Mayer}, Lucio and {McGee}, Sean and {McKernan}, Berry and {Miller}, M. Coleman and {Mota}, David F. and {Mumpower}, Matthew and {Nasim}, Syeda S. and {Nelemans}, Gijs and {Noble}, Scott and {Pacucci}, Fabio and {Panessa}, Francesca and {Paschalidis}, Vasileios and {Pfister}, Hugo and {Porquet}, Delphine and {Quenby}, John and {Ricarte}, Angelo and {R{\"o}pke}, Friedrich K. and {Regan}, John and {Rosswog}, Stephan and {Ruiter}, Ashley and {Ruiz}, Milton and {Runnoe}, Jessie and {Schneider}, Raffaella and {Schnittman}, Jeremy and {Secunda}, Amy and {Sesana}, Alberto and {Seto}, Naoki and {Shao}, Lijing and {Shapiro}, Stuart and {Sopuerta}, Carlos and {Stone}, Nicholas C. and {Suvorov}, Arthur and {Tamanini}, Nicola and {Tamfal}, Tomas and {Tauris}, Thomas and {Temmink}, Karel and {Tomsick}, John and {Toonen}, Silvia and {Torres-Orjuela}, Alejandro and {Toscani}, Martina and {Tsokaros}, Antonios and {Unal}, Caner and {V{\'a}zquez-Aceves}, Ver{\'o}nica and {Valiante}, Rosa and {van Putten}, Maurice and {van Roestel}, Jan and {Vignali}, Christian and {Volonteri}, Marta and {Wu}, Kinwah and {Younsi}, Ziri and {Yu}, Shenghua and {Zane}, Silvia and {Zwick}, Lorenz and {Antonini}, Fabio and {Baibhav}, Vishal and {Barausse}, Enrico and {Bonilla Rivera}, Alexander and {Branchesi}, Marica and {Branduardi-Raymont}, Graziella and {Burdge}, Kevin and {Chakraborty}, Srija and {Cuadra}, Jorge and {Dage}, Kristen and {Davis}, Benjamin and {de Mink}, Selma E. and {Decarli}, Roberto and {Doneva}, Daniela and {Escoffier}, Stephanie and {Gandhi}, Poshak and {Haardt}, Francesco and {Lousto}, Carlos O. and {Nissanke}, Samaya and {Nordhaus}, Jason and {O'Shaughnessy}, Richard and {Portegies Zwart}, Simon and {Pound}, Adam and {Schussler}, Fabian and {Sergijenko}, Olga and {Spallicci}, Alessandro and {Vernieri}, Daniele and {Vigna-G{\'o}mez}, Alejandro},
        title = "{Astrophysics with the Laser Interferometer Space Antenna}",
      journal = {LRR},
         year = 2023,
        month = dec,
       volume = {26},
       number = {1},
          eid = {2},
        pages = {2},
          doi = {10.1007/s41114-022-00041-y},
archivePrefix = {arXiv},
       eprint = {2203.06016},
 primaryClass = {gr-qc},
       adsurl = {https://ui.adsabs.harvard.edu/abs/2023LRR....26....2A}
}

@ARTICLE{Luo+2016,
       author = {{Luo}, Jun and {Chen}, Li-Sheng and {Duan}, Hui-Zong and {Gong}, Yun-Gui and {Hu}, Shoucun and {Ji}, Jianghui and {Liu}, Qi and {Mei}, Jianwei and {Milyukov}, Vadim and {Sazhin}, Mikhail and {Shao}, Cheng-Gang and {Toth}, Viktor T. and {Tu}, Hai-Bo and {Wang}, Yamin and {Wang}, Yan and {Yeh}, Hsien-Chi and {Zhan}, Ming-Sheng and {Zhang}, Yonghe and {Zharov}, Vladimir and {Zhou}, Ze-Bing},
        title = "{TianQin: a space-borne gravitational wave detector}",
      journal = {Classical and Quantum Gravity},
         year = 2016,
        month = feb,
       volume = {33},
       number = {3},
          eid = {035010},
        pages = {035010},
          doi = {10.1088/0264-9381/33/3/035010},
archivePrefix = {arXiv},
       eprint = {1512.02076},
 primaryClass = {astro-ph.IM},
       adsurl = {https://ui.adsabs.harvard.edu/abs/2016CQGra..33c5010L}
}

@ARTICLE{Ajith+2025,
       author = {{Ajith}, Parameswaran and {Seoane}, Pau Amaro and {Arca Sedda}, Manuel and {Arcodia}, Riccardo and {Badaracco}, Francesca and {Banerjee}, Biswajit and {Belgacem}, Enis and {Benetti}, Giovanni and {Benetti}, Stefano and {Bobrick}, Alexey and {Bonforte}, Alessandro and {Bortolas}, Elisa and {Braito}, Valentina and {Branchesi}, Marica and {Burrows}, Adam and {Cappellaro}, Enrico and {Della Ceca}, Roberto and {Chakraborty}, Chandrachur and {Subrahmanya}, Shreevathsa Chalathadka and {Coughlin}, Michael W. and {Covino}, Stefano and {Derdzinski}, Andrea and {Doshi}, Aayushi and {Falanga}, Maurizio and {Foffa}, Stefano and {Franchini}, Alessia and {Frigeri}, Alessandro and {Futaana}, Yoshifumi and {Gerberding}, Oliver and {Gill}, Kiranjyot and {Di Giovanni}, Matteo and {Giudice}, Ines Francesca and {Giustini}, Margherita and {Gl{\"a}ser}, Philipp and {Harms}, Jan and {van Heijningen}, Joris and {Iacovelli}, Francesco and {Kavanagh}, Bradley J. and {Kawamura}, Taichi and {Kenath}, Arun and {Keppler}, Elisabeth-Adelheid and {Kobayashi}, Chiaki and {Komatsu}, Goro and {Korol}, Valeriya and {Krishnendu}, N.~V. and {Kumar}, Prayush and {Longo}, Francesco and {Maggiore}, Michele and {Mancarella}, Michele and {Maselli}, Andrea and {Mastrobuono-Battisti}, Alessandra and {Mazzarini}, Francesco and {Melandri}, Andrea and {Melini}, Daniele and {Menina}, Sabrina and {Miniutti}, Giovanni and {Mitra}, Deeshani and {Mor{\'a}n-Fraile}, Javier and {Mukherjee}, Suvodip and {Muttoni}, Niccol{\`o} and {Olivieri}, Marco and {Onori}, Francesca and {Papa}, Maria Alessandra and {Patat}, Ferdinando and {Perali}, Andrea and {Piran}, Tsvi and {Piranomonte}, Silvia and {Pol}, Alberto Roper and {Pookkillath}, Masroor C. and {Prasad}, R. and {Prasad}, Vaishak and {De Rosa}, Alessandra and {Chowdhury}, Sourav Roy and {Serafinelli}, Roberto and {Sesana}, Alberto and {Severgnini}, Paola and {Stallone}, Angela and {Tissino}, Jacopo and {Tkal{\v{c}}i{\'c}}, Hrvoje and {Tomasella}, Lina and {Toscani}, Martina and {Vartanyan}, David and {Vignali}, Cristian and {Zaccarelli}, Lucia and {Zeoli}, Morgane and {Zuccarello}, Luciano},
        title = "{The Lunar Gravitational-wave Antenna: mission studies and science case}",
      journal = {JCAP},
         year = 2025,
        month = jan,
       volume = {2025},
       number = {1},
          eid = {108},
        pages = {108},
          doi = {10.1088/1475-7516/2025/01/108},
archivePrefix = {arXiv},
       eprint = {2404.09181},
 primaryClass = {gr-qc},
       adsurl = {https://ui.adsabs.harvard.edu/abs/2025JCAP...01..108A}
}

@ARTICLE{Cooper+2023,
       author = {{Cooper}, Andrew P. and {Koposov}, Sergey E. and {Allende Prieto}, Carlos and {Manser}, Christopher J. and {Kizhuprakkat}, Namitha and {Myers}, Adam D. and {Dey}, Arjun and {G{\"a}nsicke}, Boris T. and {Li}, Ting S. and {Rockosi}, Constance and {Valluri}, Monica and {Najita}, Joan and {Deason}, Alis and {Raichoor}, Anand and {Wang}, M.-Y. and {Ting}, Y.-S. and {Kim}, Bokyoung and {Carrillo}, Andreia and {Wang}, Wenting and {Beraldo e Silva}, Leandro and {Han}, Jiwon Jesse and {Ding}, Jiani and {S{\'a}nchez-Conde}, Miguel and {Aguilar}, Jessica N. and {Ahlen}, Steven and {Bailey}, Stephen and {Belokurov}, Vasily and {Brooks}, David and {Cunha}, Katia and {Dawson}, Kyle and {de la Macorra}, Axel and {Doel}, Peter and {Eisenstein}, Daniel J. and {Fagrelius}, Parker and {Fanning}, Kevin and {Font-Ribera}, Andreu and {Forero-Romero}, Jaime E. and {Gazta{\~n}aga}, Enrique and {Gontcho a Gontcho}, Satya and {Guy}, Julien and {Honscheid}, Klaus and {Kehoe}, Robert and {Kisner}, Theodore and {Kremin}, Anthony and {Landriau}, Martin and {Levi}, Michael E. and {Martini}, Paul and {Meisner}, Aaron M. and {Miquel}, Ramon and {Moustakas}, John and {Nie}, Jundan J.~D. and {Palanque-Delabrouille}, Nathalie and {Percival}, Will J. and {Poppett}, Claire and {Prada}, Francisco and {Rehemtulla}, Nabeel and {Schlafly}, Edward and {Schlegel}, David and {Schubnell}, Michael and {Sharples}, Ray M. and {Tarl{\'e}}, Gregory and {Wechsler}, Risa H. and {Weinberg}, David H. and {Zhou}, Zhimin and {Zou}, Hu},
        title = "{Overview of the DESI Milky Way Survey}",
      journal = {\apj},
         year = 2023,
        month = apr,
       volume = {947},
       number = {1},
          eid = {37},
        pages = {37},
          doi = {10.3847/1538-4357/acb3c0},
archivePrefix = {arXiv},
       eprint = {2208.08514},
 primaryClass = {astro-ph.GA},
       adsurl = {https://ui.adsabs.harvard.edu/abs/2023ApJ...947...37C}
}

@ARTICLE{Pala+2025,
       author = {{Pala}, Anna F. and {Raddi}, Roberto and {Rebassa-Mansergas}, Alberto and {G{\"a}nsicke}, Boris T. and {Anderson}, Richard I. and {Belloni}, Diogo and {Binnenfeld}, Avraham and {Breedt}, Elm{\'e} and {Buckley}, David and {Cunningham}, Tim and {Ederoclite}, Alessandro and {Escorza}, Ana and {Korol}, Valeriya and {Kupfer}, Thomas and {de Martino}, Domitilla and {Merc}, Jaroslav and {Meza}, Joaquin and {Parsons}, Steven and {Pelisoli}, Ingrid and {Reindl}, Nicole and {Rodr{\'\i}guez-Gil}, Pablo and {Santos-Garc{\'\i}a}, Alejandro and {Scaringi}, Simone and {Szkody}, Paula and {Toloza}, Odette and {Torres}, Santiago and {Uzundag}, Murat and {Zorotovic}, Monica},
        title = "{White Dwarf Binaries: Probes of Future Astrophysics}",
      journal = {arXiv e-prints},
         year = 2025,
        month = dec,
          eid = {arXiv:2512.14800},
        pages = {arXiv:2512.14800},
          doi = {10.48550/arXiv.2512.14800},
archivePrefix = {arXiv},
       eprint = {2512.14800},
 primaryClass = {astro-ph.IM},
       adsurl = {https://ui.adsabs.harvard.edu/abs/2025arXiv251214800P}
}

@ARTICLE{Townsley+2009,
       author = {{Townsley}, Dean M. and {G{\"a}nsicke}, Boris T.},
        title = "{Cataclysmic Variable Primary Effective Temperatures: Constraints on Binary Angular Momentum Loss}",
      journal = {\apj},
         year = 2009,
        month = mar,
       volume = {693},
       number = {1},
        pages = {1007-1021},
          doi = {10.1088/0004-637X/693/1/1007},
archivePrefix = {arXiv},
       eprint = {0811.2447},
 primaryClass = {astro-ph},
       adsurl = {https://ui.adsabs.harvard.edu/abs/2009ApJ...693.1007T}
}

@ARTICLE{Patterson1984,
       author = {{Patterson}, J.},
        title = "{The evolution of cataclysmic and low-mass X-ray binaries.}",
      journal = {\apjs},
         year = 1984,
        month = apr,
       volume = {54},
        pages = {443-493},
          doi = {10.1086/190940},
       adsurl = {https://ui.adsabs.harvard.edu/abs/1984ApJS...54..443P}
}

\section*{Affiliations}
\noindent{
$^{1}$European Southern Observatory, Karl-Schwarzschild-Str.\ 2, 85748 Garching bei M\"unchen, Germany\\
$^{2}$Faculty of Physics, Ludwig Maximilian University, Scheinerstrasse 1, D-81679, Munich, Bavaria, Germany\\
$^{3}$Munich Center for Geoastronomy, Ludwig Maximilian University, Theresienstrasse 41, D-80333, Munich, Bavaria, Germany\\
$^{4}$University College London, Department of Physics \& Astronomy, Gower St, London, WC1E 6BT, United Kingdom\\
$^{5}$Department of Physics, University of Warwick, Coventry CV4 7AL, UK\\
$^{6}$Max-Planck-Institut f\"ur Extraterrestrische Physik, Giessenbachstrasse, D-85748 Garching-bei-M\"unchen, Germany\\
$^{7}$Space Telescope Science Institute, 3700 San Martin Drive, Baltimore, MD 21218, USA\\
$^{8}$Department of Physics, Presbyterian College, Clinton, SC 29325, USA\\
$^{9}$Department of Astronomy Astrophysics, The University of Chicago, Chicago, IL 60637, USA\\
$^{10}$INAF -- Osservatorio Astronomico di Trieste, Via Tiepolo 11, I-34143 Trieste, Italy\\
$^{11}$IFPU -- Institute for Fundamental Physics of the Universe, via Beirut 2, I-34151 Trieste, Italy\\
$^{12}$Laboratory for Atmospheric and Space Physics, University of Colorado Boulder, Boulder, CO 80309, USA\\
$^{13}$Department of Astrophysical and Planetary Sciences, University of Colorado Boulder, Boulder, CO 80309, USA\\
$^{14}$Center for Astrophysics and Space Astronomy, University of Colorado Boulder, Boulder, CO 80309, USA\\
$^{15}$AURA for ESA, Space Telescope Science Institute, 3700 San Martin Drive, Baltimore, MD 21218, USA\\
$^{16}$Division of Astrophysics, Department of Physics, Lund University, Box 118, SE-22 100 Lund, Sweden\\
$^{17}$Division of Science, National Astronomical Observatory of Japan, 2-21-1 Osawa, Mitaka, Tokyo 181-8588, Japan\\
$^{18}$Astrobiology Center, National Astronomical Observatory of Japan, 2-21-1 Osawa, Mitaka, Tokyo 181-8588, Japan\\
$^{19}$Department of Earth and Planetary Science, The University of Tokyo, Hongo, Bunkyo-ku, Tokyo 113-0033, Japan\\
$^{20}$Graduate Institute for Advanced Studies, SOKENDAI, 2-21-1 Osawa, Mitaka, Tokyo 181-8588, Japan\\
$^{21}$California Institute of Technology, 1200 East California Boulevard, Pasadena, USA\\
$^{22}$Department of Astronomy, The University of Michigan, 1109 Geddes Avenue, Ann Arbor, MI 48109, USA\\
$^{23}$Rikkyo University, College of Science, Department of Physics, Tokyo, Japan\\
$^{24}$JAXA, ISAS, Kanagawa, Japan\\
$^{25}$University of South Carolina, Department of Physics \& Astronomy, Columbia, SC, 29208\\
$^{26}$Anton Pannekoek Institute for Astronomy, University of Amsterdam, Science Park 904, NL-1098 XH Amsterdam, The Netherlands\\
$^{27}$Institute for Astronomy, University of Edinburgh, Royal Observatory, Edinburgh EH9 3HJ, UK\\
$^{28}$Institute for Physics and Astronomy, University of Potsdam, Karl-Liebknecht-Str 24/25, 14476 Potsdam, Germany\\
$^{29}$Department of Physics and Astronomy, North Carolina State University, 2401 Stinson Dr, Box 8202, Raleigh, NC 27695, USA\\
$^{30}$ Hamburger Sternwarte, Universität Hamburg, Gojenbergsweg 112, 21029 Hamburg, Germany\\
$^{31}$Centre for Astrophysics Research, Department of Physics, Astronomy and Mathematics, College Lane Campus, University of Hertfordshire, Hatfield AL10 9AB, UK\\
$^{32}$European Southern Observatory, Alonso de Cordova 3107, Santiago RM, Chile\\
$^{33}$Space Science Institute, 4750 Walnut Street, Suite 205, Boulder, CO 80301, USA\\
$^{34}$Institute of Astronomy and Astrophysics, Academia Sinica, 11F of Astronomy-Mathematics Building, No.1, Sec. 4, Roosevelt Rd., Taipei 106319, Taiwan\\
$^{35}$SRON Space Research Organisation Netherlands, Niels Bohrweg 4, 2333 CA Leiden, The Netherlands\\
$^{36}$Centro de Astrobiolog\'ia, Ctra.\ de Torrej\'on a Ajalvir, km 4, 28850 Torrej\'on de Ardoz, Madrid, Spain\\
$^{37}$Universit\'e Claude Bernard Lyon 1, Centre de Recherche Astrophysique de Lyon UMR5574, 9 Av.\ Charles Andr\'e, 69230 Saint-Genis-Laval, France\\
$^{38}$ Kavli Institute for Cosmology, University of Cambridge, Madingley Road, Cambridge CB3 0HA, UK\\
$^{39}$ Cavendish Laboratory, University of Cambridge, 19 JJ Thomson Avenue, Cambridge CB3 0HE, UK\\
$^{40}$ Department of Physics and Astronomy, University College London, Gower Street, London WC1E 6BT, UK 
$^{41}$Institute of Astronomy, KU Leuven, Celestijnenlaan 200D, 3001 Leuven, Belgium\\
$^{42}$Wits Centre for Astrophysics, School of Physics, University of the Witwatersrand, 1 Jan Smuts Avenue, Johannesburg 2000, South Africa\\
$^{43}$Zentrum für Astronomie der Universität Heidelberg, Astronomisches Rechen-Institut, Mönchhofstr. 12-14, 69120 Heidelberg, Germany\\
$^{44}$Institut für Theoretische Physik und Astrophysik, Christian-Albrechts-Universität zu Kiel, Leibnizstr. 15, 24118 Kiel, Germany\\
$^{45}$The School of Physics and Astronomy, Tel Aviv University, Tel Aviv 6997801, Israel\\
$^{46}$Instituto de Astrof\'isica de Canarias, C.\ V\'ia L\'actea, s/n, 38205 La Laguna, Santa Cruz de Tenerife, Spain\\
$^{47}$Universidad de La Laguna, Dpto.\ Astrof\'isica, Av.\ Astrof\'isico Francisco S\'anchez, 38206 La Laguna, Santa Cruz de Tenerife, Spain\\
$^{48}$Department of Physics and Astronomy, University of Utah, 270 S 1400 E, Salt Lake City, UT 84112, USA\\
$^{49}$Institute of Astronomy, Kharkiv National University, 4 Svobody Sq., Kharkiv 61022, Ukraine\\
}

\end{document}